\documentclass[11pt, a4paper]{article}

\usepackage{jheppub}

\usepackage[english]{babel}
\usepackage{amsmath, amssymb, physics, slashed, multirow,booktabs,bbold,bm,longtable,pdflscape,array}
\allowdisplaybreaks[4]
\usepackage{comment,multirow,placeins,subcaption,makecell,cancel,mathtools}

\usepackage{tikz}
\usetikzlibrary{arrows.meta, positioning}

\usepackage{tikz-feynman}
\usetikzlibrary{positioning,decorations.pathmorphing, decorations.markings,calc}

\tikzset{
coil it/.style={
decorate,
decoration={coil, aspect=1, amplitude=1 mm, segment length=2mm}
}
}
\usepackage[table]{xcolor}
\definecolor{grey}{rgb}{0.5,0.5,0.5}

\tikzfeynmanset{compat=1.1.0}

\usepackage{graphicx}
\usepackage[verbose]{wrapfig}
\usepackage{caption}

\def\Am{{\mathcal{A}}}
\def\Bm{{\mathcal{B}}}
\def\Cm{{\mathcal{C}}}
\def\Dm{{\mathcal{D}}}
\def\Em{{\mathcal{E}}}
\def\Fm{{\mathcal{F}}}
\def\Gm{{\mathcal{G}}}
\def\Hm{{\mathcal{H}}}
\def\Im{{\mathcal{I}}}
\def\Jm{{\mathcal{J}}}
\def\Km{{\mathcal{K}}}
\def\Lm{{\mathcal{L}}}
\def\Mm{{\mathcal{M}}}
\def\Nm{{\mathcal{N}}}
\def\Om{{\mathcal{O}}}
\def\Pm{{\mathcal{P}}}
\def\Qm{{\mathcal{Q}}}

\def\Sm{{\mathcal{S}}}

\def\Wm{{\mathcal{W}}}
\def\Xm{{\mathcal{X}}}
\def\Ym{{\mathcal{Y}}}
\def\Zm{{\mathcal{Z}}}

\newcommand{\Ch}{\hat{\mathcal{C}}}

\newcommand{\ab}{\mathbf{a}}
\newcommand{\bb}{\mathbf{b}}
\newcommand{\cb}{\mathbf{c}}
\newcommand{\db}{\mathbf{d}}
\newcommand{\al}{\alpha}
\newcommand{\be}{\beta}
\newcommand{\ad}{\dot{\alpha}}
\newcommand{\bd}{\dot{\beta}}
\newcommand{\Zb}{\mathbb{Z}}

\newcommand{\LS}{h}

\title{Conformal manifolds of $\bm{\mathcal{N}=1}$ orbifold theories\\
Part I: Marginal deformations and the one-loop spectrum}

\author[a,b,c]{Fabio Mantegazza,}
\author[a,b,c]{Elli Pomoni,}
\author[a]{Torben Skrzypek}

\affiliation[a]{Deutsches Elektronen-Synchrotron DESY, Notkestr. 85, 22607 Hamburg, Germany}
\affiliation[b]{II. Institut f\"ur Theoretische Physik, Universit\"at Hamburg, Notkestr. 9, 22607 Hamburg, Germany}
\affiliation[c]{Zentrum f\"ur Mathematische Physik, Universit\"at Hamburg, Bundesstr. 55, 20146 Hamburg, \\Germany}

\emailAdd{fabio.mantegazza@desy.de}
\emailAdd{elli.pomoni@desy.de}
\emailAdd{torben.skrzypek@desy.de}

\abstract{
We consider orbifolds of 4d $\mathcal{N}=4$ super-Yang--Mills theory that preserve minimal supersymmetry and study their conformal manifolds, focusing on two families of Abelian orbifolds in the planar limit. We compute their superconformal indices, construct their complete single-trace BPS spectra and identify the conformal manifolds. These are generically spanned by a subset of the gauge-couplings, whose number we derive from both gauge and string theory, and a universal Leigh--Strassler-type superpotential deformation. We then construct the complete, planar one-loop dilatation operator on these conformal manifolds, giving perturbative access to the full non-BPS spectrum away from the orbifold point. This lays the groundwork for studying special loci of the conformal manifold in a forthcoming \textbf{Part~II}.
}

\begin{document} 
\begin{flushright} \small{DESY-26-155}\end{flushright}
\maketitle

\section{Introduction}

In the quest for establishing computational control of four-dimensional interacting quantum field theories, superconformal field theories (SCFTs) provide particularly well-understood starting points for which a plethora of technical tools have been developed in recent years. Once an example SCFT is constructed, one may investigate deformations of the theory that preserve the superconformal symmetry. Such exactly marginal deformations sweep out a ``conformal manifold'' of SCFTs in the space of quantum field theories, giving us well-controlled access to large families of interacting field theories. The boundaries of these conformal manifolds often reveal exciting new physical phenomena \cite{Argyres:2007cn, Gaiotto:2009we,Gadde:2009dj, Aharony:2015zea, Baume:2020dqd, Perlmutter:2020buo, Baume:2023msm, Calderon-Infante:2024oed, Calderon-Infante:2026rkj, Mantegazza:2026spd, Baume:2026svx, Calderon-Infante:2026zmd, Grimm:2026tiz}.

One of the best understood superconformal toy models is four-dimensional $\mathcal{N}=4$ super-Yang--Mills (SYM) theory, which has a holographic dual description in terms of type IIB string theory on $AdS_5\times S^5$ background \cite{Maldacena:1997re,Witten:1998qj}. The large amount of supersymmetry preserved by $\mathcal{N}=4$ SYM restricts the number of free parameters to one discrete rank $N$ of the $\text{SU}(N)$ gauge group and one continuous coupling constant $g_{\text{YM}}$. The holographic duality is best understood in the limit $N\to \infty$ at fixed 't Hooft coupling $\lambda=g_{\text{YM}}^2N$, where planar Feynman diagrams dominate. In this limit the theory is integrable, which allows, e.g., for a precise determination of the spectrum of $\mathcal{N}=4$ SYM at any coupling $\lambda$ \cite{Minahan:2002ve,Beisert:2003jj,Gromov:2013pga}. Thus the marginal deformation parameter $\lambda$ sweeps out a one-dimensional conformal manifold of exceptionally well-understood SCFTs.

This success story 
raises the question whether similar control can be established in theories with less supersymmetry. This question can be answered in the affirmative for so-called orbifold theories \cite{Kachru:1998ys,Lawrence:1998ja} which descend from $\mathcal{N}=4$ SYM and retain a lot of its structure, e.g. its integrability \cite{Ideguchi:2004wm,Beisert:2005he,Solovyov:2007pw,Arutyunov:2010gu,Beccaria:2011qd,Skrzypek:2022cgg}. The orbifold theories under consideration are constructed by identifying a discrete subgroup $\Gamma\subset \text{SU}(4)$ of the R-symmetry group of $\mathcal{N}=4$ SYM, embedding a representation of the same symmetry group $\Gamma$ in the gauge structure and then quotienting the theory by the combined action of $\Gamma$.\footnote{There are other ways to obtain SCFTs with lower susy. Compactifying the six-dimensional $(2,0)$ theory on punctured Riemann surfaces results in four-dimensional $\Nm=2$ theories of class $\mathcal{S}$ \cite{Gaiotto:2009we}. Related $\Nm=1$ compactifications of M5-brane theories yield class $\mathcal{S}_k$ for $\mathbb{C}^2/\mathbb{Z}_k$ singularities \cite{Gaiotto:2015usa} and class $\mathcal{S}_{\Gamma}$ for general ADE singularities \cite{Heckman:2016xdl}.} Depending on how much of the R-symmetry is broken by this identification, supersymmetry is reduced accordingly. 

In this paper we will focus on the $\mathcal{N}=1$ cases, where only a quarter of supersymmetry survives. These are generated by a subgroup $\Gamma\subset \text{SU}(3)\subset \text{SU}(4)$ and reduce the R-symmetry to a single $\text{U}(1)$ factor. The possible choices for $\Gamma$ include the Abelian families $\mathbb{Z}_k$ and $\mathbb{Z}_k\times\mathbb{Z}_\ell$ as well as several non-Abelian families and exceptional cases; see e.g. \cite{Hanany:1998sd}. In this paper we will restrict our attention to two families of Abelian orbifolds which we treat as the prototypical examples of $\mathcal{N}=1$ orbifold theories. They feature $\abs{\Gamma}$ vector multiplets, each mediating an $\text{SU}(N)$ gauge symmetry and various bifundamental chiral matter multiplets, whose coupling is easiest formalised in so-called ``quiver diagrams" (see Figure \ref{fig:Zkquiver} and \ref{fig:ZkZlquiver} below). We review these quiver gauge theories in Section~\ref{sec: orbifolds}.

The orbifold theories under consideration are planar integrable at the ``orbifold point'' of their conformal manifold, which is actually a one-dimensional submanifold spanned by an overall gauge coupling $g_0$ familiar from $\mathcal{N}=4$ SYM. However, for most choices of $\Gamma$ there is a much larger conformal manifold also spanned by further relative gauge coupling deformations and superpotential deformations \cite{Leigh:1995ep,Green:2010da,Cordova:2016xhm}, which we aim to explore. In particular the additional appearance of superpotential deformations is a novel feature of $\mathcal{N}=1$ theories, which was not present in previous discussions of $\mathcal{N}=2$ scenarios \cite{Gadde:2009dj,Gadde:2010zi,Mantegazza:2026spd} but connects to the Leigh-Strassler or $\beta$-deformation of $\mathcal{N}=4$ SYM studied e.g. in \cite{Leigh:1995ep,Lunin:2005jy,Frolov:2005dj}.
In further contrast to the $\mathcal{N}=2$ case, the $\mathcal{N}=1$ quivers are generically chiral so not all gauge couplings are exactly marginal: A relative gauge-coupling deformation of an $\mathcal{N}=1$ orbifold theory may be subject to a quantum obstruction in the form of a Konishi anomaly, and only anomaly-free combinations survive.

Since standard integrability is broken away from the orbifold point,\footnote{See \cite{Gadde:2010zi,Pomoni:2011jj,Pomoni:2013poa} for early studies away from the orbifold point, \cite{Pomoni:2019oib} for a review, and \cite{Pomoni:2021pbj,Bozkurt:2024tpz,Bertle:2024djm,Bozkurt:2025exl,AnaPaul} for recently discovered novel types of hidden symmetries.} we currently have to rely on superconformal symmetry and perturbative gauge-theory computations in order to analyse the deformed theories. In this paper we will develop the necessary tools for these investigations, including the construction of the BPS spectra, a survey of the conformal manifolds and the complete one-loop dilatation operator. We will use these insights in the forthcoming \textbf{Part II} \cite{PartII} to study boundaries of the conformal manifolds of $\mathcal{N}=1$ orbifold theories.

\paragraph{The BPS spectrum}
The operator content of a four-dimensional $\mathcal{N}=1$ SCFT is organised into superconformal multiplets. A multiplet is short/semishort when some of the descendants generated by acting with Poincar\'e supercharges on the superconformal primary are null states \cite{Dolan:2002zh,Cordova:2016emh}. The shortening condition fixes the exact scaling dimension of the primary in terms of its Lorentz spins and superconformal $R$-charge, saturating a BPS inequality \cite{Witten:1978mh}. Shortening at a particular point on the conformal manifold does not, however, guarantee protection as exactly marginal couplings are varied. Compatible short multiplets can recombine into long multiplets, whose dimensions are no longer fixed by the BPS bound \cite{Dolan:2002zh,Cordova:2016emh}(see also Appendix~\ref{App:SCA}). Multiplets that cannot participate in any recombination rule are called absolutely protected. Other multiplets may remain short along a given deformation even when recombination is allowed by representation theory. Short multiplets independent of the overall coupling $g_0$ then constitute a natural first observable to match across the holographic duality since their spectrum is accessible from both perturbative gauge theory and the dual string theory at strong coupling. A first characterisation of this sector is given by the superconformal index \cite{Kinney:2005ej,Romelsberger:2005eg,Rastelli:2016tbz}, which is a protected quantity along the conformal manifold and is therefore particularly suited for holographic comparisons (see e.g. \cite{Nakayama:2005mf} for $\Zb_k$ orbifold theories). The index, however, only computes a net degeneracy of short multiplets. The contributions of multiplets that can recombine into long representations cancel against each other, and as a consequence the index does not determine the complete BPS spectrum. 
 
In Section \ref{sec: orbifolds} we resolve this ambiguity by determining the full single-trace BPS spectrum at the orbifold point explicitly and show that string and gauge theory agree on the BPS operator content. As consistency check we match the superconformal index and perform a perturbative check of the vanishing anomalous dimension at weak coupling. Away from the boundaries of the conformal manifold, the BPS spectrum remains largely constant, with only the explicit field-theoretic expression of the BPS operators varying. As we will demonstrate in Section \ref{sec:applications}, these expressions may be determined perturbatively and in the holomorphic sector even non-perturbatively \cite{Mitev:2014yba}. A change in the spectrum only occurs when approaching or departing from special loci on the conformal manifold where BPS multiplet recombinations occur \cite{Mantegazza:2026spd}, which will be the subject of \textbf{Part II} \cite{PartII}. 

\paragraph{The conformal manifold}
In order to investigate the conformal manifold of the orbifold theories at hand we need to determine the exactly marginal deformations which span the conformal manifold. The conformal manifold can be described locally as the quotient of the space of marginal couplings by the complexified flavour symmetry group \cite{Green:2010da},
\begin{equation}
    \Mm_\Cm = \{\lambda_i\}/G_F^{\mathbb C}\, .
    \label{Eq:M_C}
\end{equation}

The presence of a marginal operator does not automatically imply the existence of a corresponding exactly marginal direction. To determine which operators really parameterise neighbouring CFTs, one requires a detailed understanding of the short multiplets containing the marginal deformations and the conserved global-symmetry currents \cite{Dolan:2002zh,Cordova:2016emh, Razamat:2020pra}. In particular, every individual gauge coupling in our quiver gauge theories is classically marginal, but its deformation may be obstructed by a chiral (Konishi) anomaly. As a result we can only deform combinations of gauge couplings associated to a \emph{balanced} partition of the quiver diagram as discussed in Section \ref{sec: U(1)s}. Unbalanced combinations face a quantum obstruction.

At weak overall coupling $g_0\ll 1$, we can characterise the conformal manifold by requiring the one-loop beta-function to vanish \cite{Novikov:1983uc,Novikov:1985rd,Arkani-Hamed:1997qui,Aharony:2002tp}. This provides us with a weak-coupling parametrisation of the conformal manifold in terms of relative couplings $\kappa_i$ and a Leigh-Strassler deformation $\LS$ of the superpotential. For the two Abelian orbifold families considered in this paper, we find the generic dimensions
\begin{equation}
\begin{split}
    \Zb_k[a_X,a_Y,a_Z]: \qquad
    &\dim_{\mathbb C}\Mm_\Cm =
    \gcd(k,a_X)+\gcd(k,a_Y)+\gcd(k,a_Z)-1\,, \\
    \Zb_k \times \Zb_\ell: \qquad
    &\dim_{\mathbb C}\Mm_\Cm =
    k+\ell+\gcd(k,\ell)-1\,.
\end{split}
\end{equation}
The only theories for which the conformal manifold is further enlarged are the exceptional cases $\Zb_3[1,1,1]$ and $\Zb_3\times\Zb_3$, whose conformal manifolds have dimensions $3$ and $15$, respectively. This characterisation of conformal manifolds will be the subject of Section~\ref{sec: marginal}.

\paragraph{The one-loop dilatation operator} 
With the explicit parametrisation of the conformal manifold in place, we can solve the one-loop spectral problem anywhere on the conformal manifold. To this end we construct the planar one-loop dilatation operator in Section \ref{sec: orbidila}, starting from the nearest-neighbour Hamiltonian of $\Nm=4$ SYM \cite{Beisert:2003jj} and its orbifold descendants \cite{Beisert:2005he,Gadde:2010zi,Skrzypek:2022cgg}, and modifying it to include arbitrary relative gauge couplings $\kappa_i$ together with the Leigh--Strassler deformation $\LS$.\footnote{In this discussion, we exclude the special $\Zb_3[1,1,1]$ and $\Zb_3\times\Zb_3$ orbifold deformations.} This gives us perturbative control over the full single-trace operator spectrum across the conformal manifold and in particular allows us to directly observe multiplet recombination phenomena as we approach special loci. A similar construction was performed in \cite{Liendo:2011xb} for $\mathcal{N}=2$ orbifold theories, but here we had to fix a significantly larger number of coefficients due to the reduced amount of supersymmetry. Similar constructions for $\Nm=1$ SCQCD and  $\beta$-deformed $\Nm=4$ theory were previously given in \cite{Liendo:2011wc} and \cite{Fokken:2013mza}, respectively. 

\paragraph{}In \textbf{Part II} \cite{PartII} we will use this technology to study special loci of the conformal manifold. Boundaries at which gauge couplings vanish as well as loci with enhanced global symmetries can lead to substantial changes in the spectrum, to which we now have direct perturbative access. These limits are also natural places to investigate the relation between infinite-distance limits on conformal manifolds and the appearance of towers of light operators, as suggested by the CFT Distance Conjecture \cite{Perlmutter:2020buo,Baume:2020dqd}, see also \cite{Baume:2023msm,Calderon-Infante:2024oed,Calderon-Infante:2026rkj,Baume:2026svx,Calderon-Infante:2026zmd,Fenati:2026hkn, Grimm:2026tiz} for recent developments. With the results of the current paper at hand we have gained the ability to investigate emergent BPS towers in these limits, both in the decoupling and in the remaining sector.

\paragraph{}In Section \ref{sec:Branes} we make contact to useful geometric pictures of the gauge-coupling deformations in terms of dual brane-box and brane-tiling descriptions, where the deformations are encoded in geometric separations of the corresponding brane configuration \cite{Hanany:1997tb,Hanany:1998it,Imamura:2007dc,Hanany:2005ve,Kennaway:2007tq}. These descriptions illustrate why certain gauge couplings can only be deformed simultaneously and that decoupling limits may be obstructed geometrically, in line with gauge theory observations. Furthermore, we will make use of these constructions in \textbf{Part II}  \cite{PartII} to gain insights into possible string-dual descriptions of decoupling limits. We close with a brief summary and a discussion of further applications of our gauge-theory technology in Section~\ref{Sec:Conclusion}. Technical details on superconformal representation theory, explicit orbifold examples and coefficients of the dilatation operator are collected in three appendices. We provide two Wolfram Mathematica notebooks as ancillary files to the arXiv submission of this paper. The first one may be used to generate a list of BPS multiplets in the orbifold theories at hand, while the second one implements the one-loop dilatation operator and provides application examples.

\section{\texorpdfstring{$\bm{\mathcal{N}=1}$ }~orbifold theories}\label{sec: orbifolds}

The canonical example of the holographic correspondence between type IIB string theory on $AdS_5\times S^5$ background and $\mathcal{N}=4$ SYM arises when considering a stack of $N$ D3-branes \cite{Maldacena:1997re,Witten:1998qj}. On one hand, the branes deform the surrounding spacetime and in the near horizon regime generate an $AdS_5\times S^5$ geometry in which closed strings propagate. On the other hand, one may consider the 4d worldvolume low-energy theory which consists of the superconformal $\mathcal{N}=4$ SYM theory with gauge group $\text{U}(N)$. The central $\text{U}(1)$ factor decouples completely and may therefore be neglected, resulting in an $\text{SU}(N)$ theory. The duality between these two theories can be checked rigorously in the planar large-$N$ limit \cite{tHooft:1973alw}. 

In order to break supersymmetry we may now consider D3-branes on an orbifold background $\mathbb{R}^{1,3}\times \mathbb{R}^{6} /\Gamma$, where $\Gamma$ is a discrete subgroup of $\text{SO}(6)$ \cite{Kachru:1998ys,Lawrence:1998ja}. The near horizon limit of a stack of D3-branes placed on the fixed locus under $\Gamma$ then results in an  $AdS_5\times S^5/\Gamma$ orbifold background. In the dual worldvolume theory, the discrete group $\Gamma$ acts on both the R-symmetry and the $\text{SU}(\abs{\Gamma}N)$ gauge group, where the order $\abs{\Gamma}$ is the number of elements in $\Gamma$. Projecting onto $\Gamma$-invariant fields results in a decomposition of the $\mathcal{N}=4$ $\text{SU}(\abs{\Gamma}N)$ gauge multiplet into smaller gauge and charged matter multiplets.

The amount of conserved supersymmetry depends on the precise form of the orbifold action. If $\Gamma\subset \text{SU}(2)\subset \text{SO}(6)$ acts on a $\mathbb{C}^2 $ subspace, the resulting $\mathbb{R}^{1,5}\times\mathbb{C}^2/\Gamma$  orbifold background preserves half-maximal supersymmetry. On the dual CFT side, we end up with $\mathcal{N}=2$  superconformal quiver gauge theories. These orbifolds are organised by the famous ADE classification \cite{McKay} and the simplest examples are necklace quiver gauge theories with $\Gamma=\mathbb{Z}_k$ ($A_{k-1}$ in the ADE classification), which have been studied extensively in the literature \cite{Ideguchi:2004wm,Gadde:2009dj,Gadde:2010zi,Zarembo:2020tpf,Pomoni:2021pbj,Billo:2022fnb,Bozkurt:2025exl,BarredoMartinez:2026wzn}.\footnote{More complicated ADE quiver theories have recently been considered in \cite{Bath:2026yeu}. A full survey of infinite distance limits for Lagrangian $\Nm=2$ theories at large $N$, including all ADE quivers, was presented in \cite{Baume:2026svx}.} If we want to further break supersymmetry down to $\mathcal{N}=1$ we have to investigate orbifold actions $\Gamma\subset \text{SU}(3)\subset \text{SO}(6)$ acting on the full $\mathbb{C}^3 $ subspace without breaking the complex structure. The resulting $\mathbb{R}^{1,3}\times \mathbb{C}^3/\Gamma$ orbifold backgrounds preserve a quarter of maximal supersymmetry. The discrete subgroups of $\text{SU}(3)$ have also been classified but feature a greater wealth of options, including Abelian families of $\mathbb{Z}_k$ and $\mathbb{Z}_k\times \mathbb{Z}_\ell $ orbifolds as well as three non-Abelian families and numerous exceptional cases (see e.g. \cite{Hanany:1998sd} for a discussion in the context of quiver gauge theories). Finally a generic subgroup of $\Gamma\subset \text{SO}(6)$ not preserving any complex structure will break supersymmetry completely. \footnote{$\mathcal{N}=0$ orbifold theories are generally non-conformal: while the single-trace beta functions vanish in the planar limit, double-trace couplings may be generated and run already at leading order in large $N$. As a consequence, conformal invariance is lost, with the corresponding holographic description exhibiting tachyonic instabilities \cite{Klebanov:1998yya,Pomoni:2009joh, Skrzypek:2021eue}.}

\subsection{Two families of Abelian \texorpdfstring{$\mathcal{N}=1$}{N=1} orbifold theories}

\begin{table}[ht]
\centering
\renewcommand{\arraystretch}{1.2}
\begin{tabular}{c|cccc|c|c|c}
\hline
\rowcolor{gray!20} & $0$ & $1$ & $2$ & $3$ &   $\,\,4\quad5\,\,$ &  $\,\,6\quad7\,\,$  & $\,\,8\quad9\,\,$  \\
\hline
$N 
\times $D3-branes & $\times$ & $\times$ & $\times$ & $\times$ & -- & -- & -- \\
\hline
$\mathbb{Z}_k$ &--
 &--
 &--
 &--
 &
$e^{\frac{2\pi i}{k} a_X}$ &
$e^{\frac{2\pi i}{k} a_Y}$ &
$e^{\frac{2\pi i}{k} a_Z}$\\ 
\hline
\end{tabular}

\phantom{word}

\begin{tabular}{c|cccc|c|c|c}
\hline
\rowcolor{gray!20} & $0$ & $1$ & $2$ & $3$ &  $\,\,4\quad5\,\,$ &  $\,\,6\quad7\,\,$  & $\,\,8\quad9\,\,$ \\
\hline
$N\times$D3-branes & $\times$ & $\times$ & $\times$ & $\times$ & -- & -- & -- \\
\hline
$\mathbb{Z}_k$ &--
 &--
 &--
 &--
 &
$e^{\frac{2\pi i}{k} }$ &
-- &
$e^{-\frac{2\pi i}{k} }$\\
$\mathbb{Z}_\ell$ &--
 &--
 &--
 &--
 &--
&
$e^{\frac{2\pi i}{\ell} }$ &
$e^{-\frac{2\pi i}{\ell} }$  \\
\hline
\end{tabular}
\caption{Worldvolume directions of the D3-branes and action of $\mathbb{Z}_k[a_X,a_Y,a_Z]$ and
$\mathbb{Z}_k\times\mathbb{Z}_\ell$ orbifolds on the transverse complex coordinates. In the near horizon limit after backreaction, the radius $r^2=\abs{X}^2+\abs{Y}^2+\abs{Z}^2$ joins the flat coordinates 0123 to form an $AdS_5$ geometry. The remaining angular coordinates become the compact $S^5/\Gamma$ spacetime component. The three ``configuration numbers'' $[a_X,a_Y,a_Z]$ satisfy the relation \eqref{Eq: restriction}.
\label{tab:orbifold_action}}
\end{table}

In this paper we will restrict our attention to two families of Abelian $\mathcal{N}=1$ orbifold theories without discrete torsion, which arise from a projection by either $\Gamma=\mathbb{Z}_k\subset \text{SU}(3)$ or  $\Gamma=\mathbb{Z}_k\times \mathbb{Z}_\ell\subset \text{SU}(3)$. The D3-brane setup for their string-theory construction has been sketched in Table \ref{tab:orbifold_action}. We will review both families in turn.

\subsubsection{The \texorpdfstring{$\mathbb{Z}_k$}{Zk} family}\label{sec: Z_k}

We may parameterise $\mathbb{C}^3$ by three complex coordinates $(X,Y,Z)$, which in the dual gauge theory correspond to scalar fields. The orbifold group $\Gamma=\mathbb{Z}_k$ consists of powers $\gamma^l$ of the fundamental generator $\gamma$, which acts as
\begin{equation}\label{Eq: orbaction}
    \gamma: \,\, (X,Y,Z) \to \left(\omega_k^{a_X}X,\omega_k^{a_Y}Y,\omega_k^{a_Z}Z\right)\,,\qquad \omega_k=\exp(\frac{2\pi i}{k})\,.
\end{equation}
The triplet of integer ``configuration numbers'' $[a_X,a_Y,a_Z]$ defines the precise action of the orbifold group. For $\gamma$ to be an element of $\text{SU}(3)$ we have to restrict 
\begin{equation}\label{Eq: restriction}
    a_Z=(-a_X-a_Y)\text{ mod }k\,.
\end{equation}
Without loss of generality, we may assume $a_X,a_Y\in \{0,\dots,k-1\}$ and that there is no common divisor $m$ of $k$ and all the configuration numbers $a_I$, which would otherwise result in a simpler $\mathbb{Z}_{k/m}$ orbifold theory.
In case one of the configuration numbers vanishes, we restore $\mathcal{N}=2$ supersymmetry and make contact with the $\mathcal{N}=2$ necklace quiver gauge theories mentioned above. We will not discuss this case much further in this paper but use it as a consistency check for our one-loop dilatation operator in Section \ref{sec: orbidila}. 

In addition to restricting $a_X,a_Y\in \{1,\dots,k-1\}$ and $a_Z=(-a_X-a_Y)\text{ mod }k$, there are further equivalence relations between a priori different sets of configuration numbers:
\begin{itemize}
    \item Permutations of $[a_X,a_Y,a_Z]$ will result in the same theory with renamed coordinates.
    \item Global multiplications $[m \cdot a_X, m \cdot a_Y,m\cdot  a_Z]\text{ mod }k$ by a factor $m\in \{1,\dots,k-1\}$ with $\gcd(m,k)=1$ may be absorbed in the definition of the fundamental generator $\gamma$. This establishes an equivalence between configurations related by such global multiplications.
\end{itemize}
The set of physically distinct orbifold theories should only feature one representative per equivalence class. To each representative we associate an orbifold background $AdS_5\times S^5/\mathbb{Z}_k[a_X,a_Y,a_Z]$ of type IIB string theory, where the action \eqref{Eq: orbaction} acts on the embedding space coordinates of $S^5\subset \mathbb{C}^3$.

\begin{figure}[t]
\centering

\begin{tikzpicture}[
    vertex/.style={circle, draw, minimum size=6mm, thick},
    arrow/.style={->, thick},
    scale=0.8
]
\def\off{3pt}
\def\offone{2.6pt}
\def\offtwo{1.5pt}

\node[vertex] (a) at (1.5,2.59) {};
\node[vertex] (b) at (3,0) {};
\node[vertex] (c) at (0,0) {};

\draw[arrow, transform canvas={yshift=\offtwo, xshift=\offone}] (a) -- (b);
\draw[arrow, transform canvas={{yshift=-\offtwo, xshift=-\offone}}] (a) -- (b);
\draw[arrow] (a) -- node[right] {\,\,$\Xm,\Ym,\Zm$}(b);

\draw[arrow, transform canvas={yshift=\off}] (b) -- (c);
\draw[arrow, transform canvas={yshift=-\off}] (b) -- (c);
\draw[arrow] (b) -- (c);

\draw[arrow, transform canvas={yshift=-\offtwo, xshift=\offone}] (c) -- (a);
\draw[arrow, transform canvas={{yshift=\offtwo, xshift=-\offone}}] (c) -- (a);
\draw[arrow] (c) -- (a);

\end{tikzpicture}
\hspace{1cm}
\begin{tikzpicture}[
    vertex/.style={circle, draw, minimum size=6mm,thick},
    arrow/.style={->, thick},
    scale=0.8
]
\def\off{2.5pt}
\def\offd{1.7pt}

\node[vertex] (p1) at (0,0) {};
\node[vertex] (p2) at (3,0) {};
\node[vertex] (p3) at (3,3) {};
\node[vertex] (p4) at (0,3) {};

\draw[arrow, transform canvas={yshift=\off}] (p2) -- (p1);
\draw[arrow, transform canvas={yshift=-\off}] (p2) -- (p1);

\draw[arrow, transform canvas={yshift=\off}] (p4) --  (p3);
\draw[arrow, transform canvas={yshift=-\off}] (p4) -- (p3);

\draw[arrow, transform canvas={xshift=-\off}] (p3) --  (p2);
\draw[arrow, transform canvas={xshift=\off}] (p3) -- node[right] {$\Xm,\Ym$}(p2);

\draw[arrow, transform canvas={xshift=-\off}] (p1) -- (p4);
\draw[arrow, transform canvas={xshift=\off}] (p1) -- (p4);

\draw[arrow, transform canvas={xshift=-\offd,yshift=\offd}] (p3) --  (p1);
\draw[arrow, transform canvas={xshift=\offd,yshift=-\offd}] (p1) --   (p3);
\draw[draw=none,transform canvas={xshift=-\offd/2,yshift=\offd/2}] (p3) --  node[right,, yshift=0.5pt] {\,\,\, $\Zm$}(p1);

\draw[arrow, transform canvas={xshift=-\offd,yshift=-\offd}] (p4) -- (p2);
\draw[arrow, transform canvas={xshift=\offd,yshift=\offd}] (p2) -- (p4);
\end{tikzpicture}
\hspace{1cm}
\begin{tikzpicture}[
    vertex/.style={circle, draw, minimum size=6mm, thick},
    arrow/.style={->, thick},
    scale=0.8
]
\def\off{2.5pt}
\def\offone{2.38pt}
\def\offtwo{0.77pt}
\def\offthree{1.47pt}
\def\offfour{2.02pt}

\node[vertex] (v1) at (0,0) {};
\node[vertex] (v2) at (-0.93,2.85) {};
\node[vertex] (v3) at (1.5,4.62) {};
\node[vertex] (v4) at (3.93,2.85) {};
\node[vertex] (v5) at (3,0) {};

\draw[arrow, transform canvas={yshift=\off}] (v5) -- (v1);
\draw[arrow, transform canvas={yshift=-\off}] (v5) -- (v1);

\draw[arrow, transform canvas={xshift=\offone,yshift=\offtwo}] (v1) -- (v2);
\draw[arrow, transform canvas={xshift=-\offone,yshift=-\offtwo}] (v1) -- (v2);

\draw[arrow, transform canvas={xshift=\offone,yshift=-\offtwo}] (v4) -- (v5);
\draw[arrow, transform canvas={xshift=-\offone,yshift=\offtwo}] (v4) -- node[right, yshift=-3pt] {\,\,$\Xm,\Ym$}(v5);

\draw[arrow, transform canvas={xshift=-\offthree,yshift=\offfour}] (v2) -- (v3);
\draw[arrow, transform canvas={xshift=\offthree,yshift=-\offfour}] (v2) -- (v3);

\draw[arrow, transform canvas={xshift=-\offthree,yshift=-\offfour}] (v3) -- (v4);
\draw[arrow, transform canvas={xshift=\offthree,yshift=\offfour}] (v3) -- (v4);

\draw[arrow] (v3) -- (v1);
\draw[arrow] (v5) -- node[left, yshift=-3pt] {$\Zm$}(v3);
\draw[arrow] (v2) -- (v5);
\draw[arrow] (v4) -- (v2);
\draw[arrow] (v1) -- (v4);

\end{tikzpicture}

\caption{Quiver diagrams for the simplest examples of $\mathcal{N}=1$ $\mathbb{Z}_k$ orbifold theories, from left to right: $\mathbb{Z}_3[1,1,1]$, $\mathbb{Z}_4[1,1,2]$ and $\mathbb{Z}_5\,[1,1,3]$. Circles denote vector multiplets and arrows bifundamental chiral multiplets, which we label by their scalar components. Conjugate anti-chiral multiplets are implicitly included in the arrows. At these low orders of $\mathbb{Z}_k$ the orbifold configurations are unique, but at $\mathbb{Z}_6$ we would first observe two separate configurations $\mathbb{Z}_6[1,1,4]$ and $\mathbb{Z}_6[1,2,3]$. }\label{fig:Zkquiver}
\end{figure}
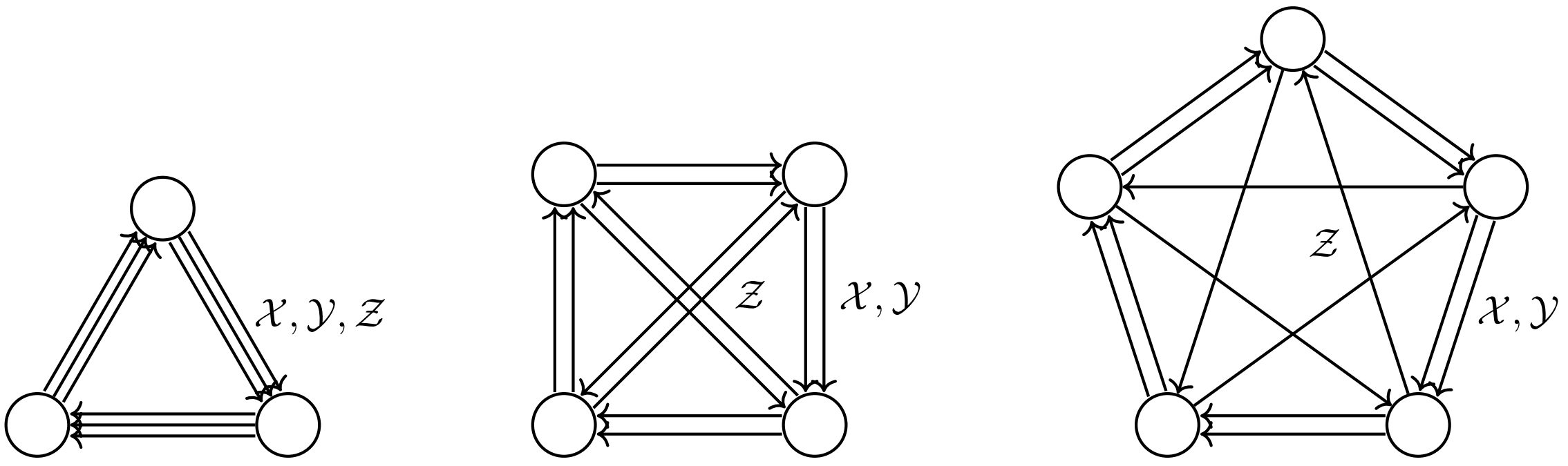
On the gauge side we project $\mathcal{N}=4$ SYM with gauge group $\text{SU}(kN)$ to invariant fields under the simultaneous action of \eqref{Eq: orbaction} and a gauge group automorphism, which acts on the scalar fields as
\begin{equation}\label{Eq: gauge-twist}
    \gamma: \,\, (X,Y,Z) \to \Omega^\dagger \left(\omega_k^{a_X}X,\omega_k^{a_Y}Y,\omega_k^{a_Z}Z\right)\Omega\,,\qquad \Omega=\text{diag}(1, \omega_k ,\dots,\omega^{k-1}_k)\otimes \mathbb{1}_N\,.
\end{equation}
Other fields are treated analogously according to their $R$-symmetry and gauge representations. In particular this breaks the overall gauge group $\text{SU}(kN)$ down to $\text{SU}(N)^k\times \text{U}(1)^{k-1}$. The $\text{U}(1)$ factors have a non-vanishing $\beta$-function and thus have to be decoupled. We will postpone a discussion of this step to Section \ref{sec: U(1)s}. After this decoupling we end up in an $\mathcal{N}=1$ superconformal quiver gauge theory with $k$ $\text{SU}(N)$ gauge nodes as well as $3k$ chiral and $3k$ antichiral multiplets in bifundamental representations. We introduce chiral superfields $(\Xm,\Ym,\Zm)$ for the multiplets descending from the scalar fields $(X,Y,Z)$. Labelling the gauge nodes by integers $i\in\{1,\dots,k\}$, the $k$ chiral multiplets $\Xm$ live in the fundamental representation of gauge node $i$ and in the antifundamental representation of gauge node $i+a_X$. We may represent these relations in a circular quiver diagram with $k$ gauge nodes forming a circle and the $\Xm$ chiral multiplets represented by arrows connecting each $i$-th node to the $(i+a_X)$-th one. Similarly, we have to construct $\Ym$ and $\Zm$ superfields in appropriate representations, resulting in additional arrows as depicted in Figure \ref{fig:Zkquiver}. For a full characterisation of the theory, the $\mathcal{N}=1$ quiver diagram depicting all gauge interactions has to be complemented by an explicit superpotential $W$ and in the case of orbifold theories, the superpotential is inherited from $\mathcal{N}=4$ SYM:
\begin{equation}\label{Eq: superpotential}
  W=g_0 \sum_{i=1}^k \Big(\Tr_i(\Xm\Ym\Zm)-\Tr_i(\Xm\Zm\Ym)\Big)\,,
\end{equation}
where we introduced the overall gauge coupling $g_0$, which simultaneously fixes the Yukawa couplings at the orbifold point. 

At this stage we should comment on our trace conventions. The $\mathcal{N}=1$ quiver gauge theories we discuss here feature fields in adjoint and bifundamental representations, such as e.g. $F^{A_i}{}_{B_i}$ and $ X^{A_i}{}_{B_{i+a_X}}$, where a raised index $A_i$ is a fundamental $\text{SU}(N)$ index of the $i$\textsuperscript{th} gauge node and a lowered one the associated anti-fundamental index. Here we took into account that $ X^{A_i}{}_{B_{i+a_X}}$ connects separate gauge nodes according to the shift $a_X$. When writing down traces of fields, we assume that neighbouring fields are contracted along their adjacent indices, e.g. $(FX)^{A_i}{}_{C_{i+a_X}}=F^{A_i}{}_{B_i}X^{B_i}{}_{C_{i+a_X}}$. This allows us to use an almost index-free notation for single-trace operators where we simply indicate the initial gauge group and infer the correct gauge contractions along the chain of fields from the quiver diagram, e.g. $\Tr_{i}(FXXXY\dots)$. In the large-$N$ limit, single-trace operators form a closed sector, while correlation functions of multi-trace operators factorise at leading order into products of single-trace correlators. The single-trace spectrum therefore captures the basic spectral data which also governs the multi-trace spectrum. One may worry that the initial and final gauge index of a single trace operator could not coincide but closure of the trace is in fact enforced by the orbifold invariance condition 
\begin{equation}
    a_X(\#(\Xm)-\#(\bar{\Xm})-\#(\Zm)+\#(\bar{\Zm}))+a_Y(\#(\Ym)-\#(\bar{\Ym})-\#(\Zm)+\#(\bar{\Zm}))\in k\mathbb{Z}\,.
    \label{Eq:Zk_invariance}
\end{equation}
In other words, each consistently closing single-trace operator is a physical state of the orbifold theory. The orbifold projection is thus completely encoded in the gauge structure. With these conventions in place, we can check directly that the traces in the superpotential \eqref{Eq: superpotential} close appropriately.

A useful alternative basis for the single-trace Hilbert space may be constructed by introducing the quiver trace 
\begin{equation}
    \Tr(FXXXY\dots)=\sum_{i=1}^k\Tr_i(FXXXY\dots)\,,
\end{equation}
which is the symmetric sum over all gauge nodes. Operators of this form are referred to as untwisted sector operators while we can also insert powers of the twist-matrix $\Omega$ from \eqref{Eq: gauge-twist} to generate twisted sector operators $\Tr(\Omega^a FXXXY\dots)$. We note that the full set of untwisted and twisted sector operators spans the same Hilbert space as the individual traces $\Tr_i( FXXXY\dots)$. However, the twist-basis makes manifest the charge $a$ under the $\hat{\mathbb{Z}}_k$ symmetry that rotates the quiver diagram by an angle $\frac{2\pi}{k}$. This discrete symmetry is the  Pontryagin dual of the original $\Gamma=\mathbb{Z}_k$ group that we have ``gauged'' in order to construct the orbifold theory.\footnote{This dual symmetry is sometimes referred to as the ``quantum symmetry'' of the orbifold \cite{Vafa:1989ih}.} We will observe below that the dilatation operator of the theory preserves the twist $a$, which is therefore a good quantum number according to which we can decompose the Hilbert space.

\subsubsection{The \texorpdfstring{$\mathbb{Z}_k\times\mathbb{Z}_\ell$}{Zk x Zl} family}\label{sec: Z_kZ_l}
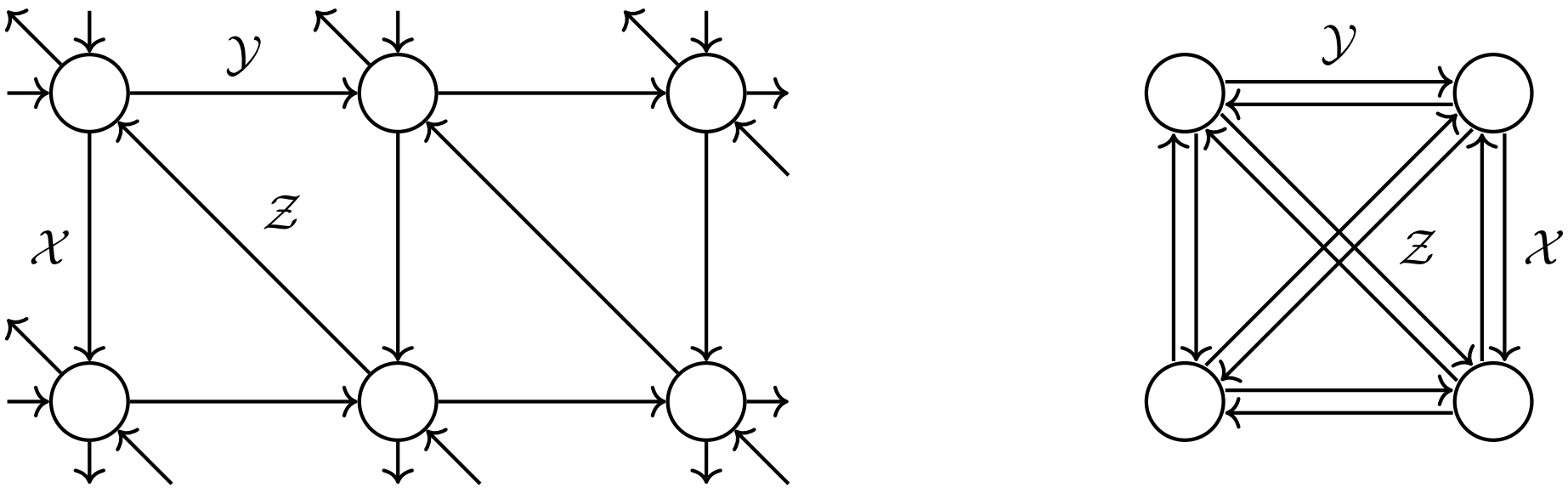
\begin{figure}[t]
\centering

\begin{tikzpicture}[
    vertex/.style={circle, draw, minimum size=6mm,thick},
    arrow/.style={->, thick},
    scale=0.8
]

\node[vertex] (a1) at (0,3) {};
\node[vertex] (a2) at (3,3) {};
\node[vertex] (a3) at (6,3) {};

\node[vertex] (b1) at (0,0) {};
\node[vertex] (b2) at (3,0) {};
\node[vertex] (b3) at (6,0) {};

\draw[arrow] (a1) -- node[above] {$\Ym$} (a2);
\draw[arrow] (a2) -- (a3);

\draw[arrow] (b1) -- (b2);
\draw[arrow] (b2) -- (b3);

\draw[arrow] (a1) -- node[left] {$\Xm$} (b1);
\draw[arrow] (a2) -- (b2);
\draw[arrow] (a3) -- (b3);

\draw[arrow] (b2) -- node[above right] {$\Zm$} (a1);
\draw[arrow] (b3) -- (a2);

\draw[arrow] (a1) ++(-0.8,0) -- (a1); \draw[arrow] (b1) ++(-0.8,0) -- (b1); \draw[arrow] (a3) (a3) -- ++(0.8,0); \draw[arrow] (b3) (b3) -- ++(0.8,0); \draw[arrow] (a3) ++(0.8,-0.8)-- (a3); \draw[arrow] (b1) (b1) -- ++(-0.8,0.8);

\foreach \n in {a1,a2,a3}{ \draw[arrow] (\n) ++(0,0.8) -- (\n); \draw[arrow] (\n) (\n) -- ++(-0.8,0.8); } 
\foreach \n in {b1,b2,b3}{ \draw[arrow] (\n) (\n)-- ++(0,-0.8); \draw[arrow] (\n) ++(0.8,-0.8)-- (\n); }
\end{tikzpicture}
\hspace{2.5cm}
\begin{tikzpicture}[
    vertex/.style={circle, draw, minimum size=6mm,thick},
    arrow/.style={->, thick},
    scale=0.8
]
\def\off{2.5pt}
\def\offd{1.7pt}

\node[vertex] (p1) at (0,0) {};
\node[vertex] (p2) at (3,0) {};
\node[vertex] (p3) at (3,3) {};
\node[vertex] (p4) at (0,3) {};

\draw[arrow, transform canvas={yshift=\off}] (p1) -- (p2);
\draw[arrow, transform canvas={yshift=-\off}] (p2) -- (p1);

\draw[arrow, transform canvas={yshift=\off}] (p4) -- node[above] {$\Ym$} (p3);
\draw[arrow, transform canvas={yshift=-\off}] (p3) -- (p4);

\draw[arrow, transform canvas={xshift=-\off}] (p2) --  (p3);
\draw[arrow, transform canvas={xshift=\off}] (p3) -- node[right] {$\Xm$} (p2);

\draw[arrow, transform canvas={xshift=-\off}] (p1) -- (p4);
\draw[arrow, transform canvas={xshift=\off}] (p4) -- (p1);

\draw[arrow, transform canvas={xshift=-\offd,yshift=\offd}] (p1) -- (p3);
\draw[arrow, transform canvas={xshift=\offd,yshift=-\offd}] (p3) -- (p1);
\draw[draw=none] (p3) --  node[right] {\,\,\, $\Zm$}(p1);

\draw[arrow, transform canvas={xshift=-\offd,yshift=-\offd}] (p2) -- (p4);
\draw[arrow, transform canvas={xshift=\offd,yshift=\offd}] (p4) -- (p2);

\phantom{\draw[arrow] (p1) (p1) -- ++(0,-0.8);}
\end{tikzpicture}

\caption{Left: Generic structure of the $\mathcal{N}=1$ $\mathbb{Z}_k\times \mathbb{Z}_\ell$ quiver diagram, which forms a repeating lattice with toroidal identification after $k$ rows and $\ell$ columns. Right: The simplest example of the $\mathbb{Z}_2\times \mathbb{Z}_2$ orbifold theory. The toroidal identifications result in doubled chiral multiplets on each edge. Note the subtle difference in arrow directions to the $\mathbb{Z}_4$ case depicted in Figure \ref{fig:Zkquiver}.}
\label{fig:ZkZlquiver}
\end{figure}

The orbifold group $\Gamma=\mathbb{Z}_k\times \mathbb{Z}_\ell$ is generated by two fundamental generators $\gamma_X$ and $\gamma_Y$ which we choose to act on the coordinates $(X,Y,Z)$ as
\begin{equation}\label{Eq: orbaction2}\begin{split}
    \gamma_X: &\,\, (X,Y,Z) \to \left(\omega_kX,Y,\omega_k^{-1}Z\right)\,,\qquad \omega_k=\exp(\frac{2\pi i}{k})\,,\\
    \gamma_Y: &\,\, (X,Y,Z) \to \left(X,\omega_\ell Y,\omega_\ell^{-1}Z\right)\,,\,\,\qquad \omega_\ell=\exp(\frac{2\pi i}{\ell})\,.\end{split}
\end{equation}
Without loss of generality we may impose $k\leq \ell$. We further note that if $\gcd(k,\ell)=1$, we can identify the $\mathbb{Z}_k\times \mathbb{Z}_\ell$ orbifold with a $\mathbb{Z}_{k \ell}[\ell,k,k \ell-k-\ell]$ orbifold theory from the previous family.\footnote{We should note that in order to  truly capture all Abelian orbifolds as counted in \cite{Hanany:2010ne} one should allow for more general embeddings of the form $\mathbb{Z}_k[a_X,a_Y,a_Z]\times\mathbb{Z}_\ell[b_X,b_Y,b_Z]$. The lowest-order orbifold theory not captured by our two families has twelve gauge nodes and arises in the embedding $\mathbb{Z}_6[1,1,4]\times\mathbb{Z}_2[0,1,1]$.}

Again the gauge-side orbifold projection enforces invariance under the simultaneous action with \eqref{Eq: orbaction2} and a gauge group automorphism on the parent $\text{SU}(k\ell N)$ gauge group encoded in the matrices
\begin{equation}
    \Omega_X=\text{diag}(1, \omega_k ,\dots,\omega^{k-1}_k)\otimes \mathbb{1}_\ell\otimes \mathbb{1}_N\,,\qquad \Omega_Y=\mathbb{1}_k\otimes\text{diag}(1, \omega_\ell ,\dots,\omega^{\ell-1}_\ell)\otimes \mathbb{1}_N\,.
\end{equation}
The gauge group $\text{SU}(k \ell N)$ breaks down to $\text{SU}(N)^{k\ell }\times \text{U}(1)^{k\ell-1}$, where we again have to decouple the non-conformal $\text{U}(1)$ factors. As depicted in Figure \ref{fig:ZkZlquiver}, the resulting quiver diagram may be represented as a 2d square lattice of $k\times \ell$ gauge nodes, which are connected by $\Xm$ and $\Ym$ chiral multiplets along the vertical and horizontal directions, respectively. The last gauge node in each row and column is connected to the respective first gauge node, resulting in an overall
toroidal topology of the quiver diagram.\footnote{Viewing
these theories as $\mathbb{Z}_{\kappa}$ orbifolds of
$\mathcal{N}=2$ necklace quivers with $k$ gauge nodes and $\kappa=\ell$ gives
a class-$\mathcal{S}_{\kappa}$ description associated with a genus-one surface with $k$ punctures \cite{Gaiotto:2015usa,Coman:2015bqq,Mitev:2017jqj,Bourton:2017pee,Bourton:2020rfo}.}
The $\Zm$ chiral multiplets connect gauge nodes diagonally,
such that $\Xm\Ym\Zm$ and $\Xm\Zm\Ym$ cycles close onto the
same gauge node, allowing for the superpotential
\begin{equation}
    W=g_0 \sum_{i=1}^k \sum_{j=1}^\ell \left(\Tr_{(i,j)}(\Xm\Ym\Zm)-\Tr_{(i,j)}(\Xm\Zm\Ym)\right)\,,
\end{equation}
to descend from $\mathcal{N}=4$ SYM. Here we labelled gauge nodes in the lattice by $(i,j)$ and followed essentially the same trace conventions as discussed above. Closure of traces again encodes the orbifold invariance conditions
\begin{equation}
    \#(\Xm)-\#(\bar{\Xm})-\#(\Zm)+\#(\bar{\Zm})\in k\mathbb{Z}\,,\qquad \#(\Ym)-\#(\bar{\Ym})-\#(\Zm)+\#(\bar{\Zm})\in \ell \mathbb{Z}\,,
    \label{Eq:ZkZl_invariance}
\end{equation}
and twisted sector single-trace operators are now generated by the insertion of both $\Omega_X$ and $\Omega_Y$ twist matrices in the quiver trace.
 
\subsection{The BPS spectrum}\label{sec: BPS}

Having constructed Abelian $\mathcal{N}=1$ orbifolds of $\mathcal{N}=4$ SYM of both $\mathbb{Z}_k$ and $\mathbb{Z}_k\times \mathbb{Z}_\ell$ type, we first want to establish control of their single-trace BPS spectra.\footnote{Here we assume a large-$N$ limit has been taken and multi-trace operators may be decomposed in their single-trace components, with no additional mixing between single- and multi-trace operators.} As reviewed in Appendices \ref{App:N1Reps} and \ref{App:N2Reps}, BPS operators furnish short representations of the superconformal algebra and saturate a BPS-bound which fixes their conformal dimensions in terms of their discrete quantum numbers. BPS operators therefore cannot acquire an anomalous dimension unless some recombination into long multiplets occurs. In particular, when a recombination is forbidden or does not happen dynamically, the quantum numbers of BPS operators have no coupling dependence, allowing us to match the BPS spectrum across the holographic duality, which is a strong-weak duality by nature. 

A first characterisation of the BPS spectrum is provided by the superconformal index, which counts BPS multiplets up to possible recombinations and therefore captures only part of the full BPS spectrum. We then discuss how BPS states arise from the KK reduction of 10d supergravity on $S^5/\Gamma$ and from additional twisted sectors. Superconformal representation theory allows us to present an algorithm for determining the full BPS spectrum of each Abelian $\mathcal{N}=1$ orbifold theory constructed above. A Wolfram Mathematica implementation of this algorithm is supplied as ancillary file to the arXiv submission.

As consistency check, we confirm that the BPS spectrum determined this way matches the superconformal index, but also note that the index severely undercounts the full BPS spectrum, counting only a protected net degeneracy.

\subsubsection{Superconformal index}

\begin{table}[ht]
    \centering
    \renewcommand{\arraystretch}{1.3}
    \begin{tabular}{c||c|c|c|c}
         Letter & $\Delta$ & $r$ & $j_1$ & $j_2$\\\hline\hline
         $\lambda$ & $\frac{3}{2}$ & $1$ & $\frac{1}{2}$  & $0$\\
         $\bar \lambda$ & $\frac{3}{2}$ & $-1$ & $0$  & $\frac{1}{2}$\\
         $ F$ & $2$ & $0$ & $1$  & $0$ \\
         $\bar F$ & $2$ & $0$ & $0$  & $1$ \\\hline
         $\phi$, $\tilde{\phi}$ & $1$ & $\frac{2}{3}$ & $0$ & $0$ \\
         $\bar \phi$, $\bar{\tilde{ \phi}}$ & $1$ & $-\frac{2}{3}$ & $0$ & $0$ \\
         $\psi_{\alpha}$, $\tilde{\psi}_{\alpha}$ & $\frac{3}{2}$ & $-\frac{1}{3}$ & $\frac{1}{2}$ & $0$\\
         $\bar{\psi}_{\dot \alpha}$, $\bar{\tilde{\psi}}_{\dot \alpha}$ & $\frac{3}{2}$ & $\frac{1}{3}$ & $0$& $\frac{1}{2}$ 
         \\\hline
         $\partial$ & 1 & 0 &  $\tfrac{1}{2}$ & $\tfrac{1}{2}$\\
    \end{tabular}
    \caption{Free field letters for building operators in $\mathcal{N}=1$ theories. $\partial$  denotes a derivative while all other letters are fields. We list their contributions to the various quantum numbers. Note that the $j_1, j_2$ quantum numbers are not strictly additive but obey the familiar $\text{SU}(2)$ tensor-product decomposition rules. The chiral multiplet fields with equal quantum numbers are distinguished by their gauge representation. In addition to these quantum numbers, fields can be charged under additional global symmetries.
    \label{Tab:N1_Letters}}
\end{table}

A generic state in $\Nm=1$ SCFT can be labelled by the quantum numbers $(\Delta, j_1,j_2,r)$ associated to the maximal bosonic sub-algebra of the superconformal algebra
\begin{equation}
    \mathfrak{u}(1)_{\Delta} \oplus \mathfrak{su}(2)_{j_1} \oplus \mathfrak{su}(2)_{j_2}  \oplus \mathfrak{u}(1)_{r} \subset \mathfrak{su}(2,2|1).
\end{equation}
Here $\Delta$ is the scaling dimension of the operator, $j_1, j_2$ are the spins and $r$ is the R-symmetry charge.
The right\footnote{The index is defined with respect to a supercharge $Q$ and the Hermitian conjugate $S= Q^{\dagger}$. For the right index we pick $Q= \tilde{\mathcal{Q}}_{\dot -}$, while for the left one we pick $Q= \mathcal{Q}_{-}$. The two definitions are equivalent and ultimately give the same index as function of the fugacities, provided one exchanges $j_1 \longleftrightarrow j_2$ and $r \longleftrightarrow -r$.} superconformal index is defined as \cite{Kinney:2005ej, Romelsberger:2005eg, Rastelli:2016tbz}
\begin{align}
    \mathcal{I}(t,x,\{a_i\})&=\text{Tr}_{\text{ker} \cal H} (-1)^{F} t^{\cal R} x^{2 j_1} \prod_{i}a_i^{\alpha_i} \,,
    \label{Eq:Index}
\end{align}
where the fugacities $t,x, \{a_i\}$ and the exponent  $\mathcal{R} = r + 2 j_2  + \frac{2}{3} \mathcal H $  were introduced; the fugacities $a_i$ take extra global symmetries commuting with the $\Nm=1$ superalgebra into account. \eqref{Eq:Index} counts states with $\mathcal H=0$, where
\begin{equation}
        \mathcal H = \Delta -2 j_2 - \frac{3}{2}r\,. \\
\end{equation}
The index is invariant under marginal deformations, so it can be computed at weak coupling starting from the free field letters, see Table \ref{Tab:N1_Letters}. For a generic chiral multiplet $\Phi=(\phi,  \psi_{\alpha})$ and its anti-chiral partner the letters contributing to the index are
\begin{equation}
    \phi\,, \quad \bar{\psi}_{\dot +}\,, \quad \partial_{+ \dot +}\,, \quad \partial_{- \dot +}\,.
\end{equation}
For the vector multiplet $\mathcal{W}= (\lambda_{\alpha}, F)$ and its conjugate the letters are 
\begin{equation}
    \lambda_+\,, \quad \lambda_-\,, \quad \bar{F}_{\dot + \dot +}\,, \quad \partial_{+ \dot +}\,, \quad \partial_{- \dot +}\,, \quad - \epsilon^{\alpha \beta} \partial_{\alpha \dot +} \lambda_{\beta}\,,
\end{equation}
where the last entry is an equation of motion. The unrefined single-letter indices of chiral and 
vector fields read accordingly
\begin{equation}
    i_{\Phi}(t,x)  =   \frac{t^{\frac{2}{3}}-t^{\frac{4}{3}}}{(1-t x )(1 - t x^{-1})}\,, \qquad 
    i_{\mathcal W}(t,x)  =   \frac{-t (x+x^{-1}) + 2 t^2}{(1-t x )(1 - t x^{-1})}\,. 
\end{equation}

\begin{table}
\begin{centering}
\begin{tabular}{c||c|c|c|c|c}
 & $\text{U}(1)_{\alpha}$ & $\text{U}(1)_{\beta}$ & $\text{U}(1)_{r}$\tabularnewline
\hline
\hline 
$F_{ii}$ & $0$ & $0$ &  $0$ \\
$X_{ij}$ & $1$& $\frac{1}{2}$   & $\frac{2}{3}$ \\
$Y_{ij}$ & $-1$& $\frac{1}{2}$  & $\frac{2}{3}$ \\
$Z_{ij}$  & $0$& $-1$  & $\frac{2}{3}$ \\
\end{tabular}
\par\end{centering}
\caption{\label{Tab:Orbifold} Untwisted global symmetries of a generic orbifold theory. We list the quantum numbers under the flavour and $R-$symmetries, the representation under the gauge groups can be inferred from the subscripts. These charges are related to the  bulk angular momenta of Section \ref{sec: strings} below as $q_\alpha=J_X-J_Y=2L_p$, $q_\beta=\tfrac{J_X+J_Y}{2}-J_Z=3 R_p-\tfrac{3}{2} r $ and $r = \tfrac{2}{3}(J_X+J_Y+J_Z)$.}
\end{table}

We can now focus on the orbifold theories introduced in the previous section and refine the index by also taking into account the global symmetries present in these theories. In addition to the $\mathcal{N}=1$ superconformal algebra we generically observe two further global $\text{U}(1)$ symmetries, which we name $\text{U}(1)_{\alpha}$ and $\text{U}(1)_\beta$. $\text{U}(1)_{\beta}$ \eqref{Eq:N=1_SCA} is a by-product of the reduction of the $\Nm=2$ superconformal algebra, see the discussion around $\mathbf{2.3}$ in \cite{Bourton:2020rfo} for more details,\footnote{\label{foot: Bourton}In \cite{Bourton:2020rfo} this symmetry was called $\mathfrak{u(1)_{q_t}}$ or $\mathfrak{u(1)_{t}}$. Here, to avoid confusion with the index fugacity $t$ we call it $\mathfrak{u(1)}_\beta$ and $\beta$ the associated fugacity.  It is interesting to note that this additional $\mathfrak{u}(1)_\beta$ symmetry allows for the definition of Higgs and Coulomb branches for these $\Nm=1$ theories \cite{Bourton:2020rfo}. For class-$\mathcal{S}_k$ theories, the Coulomb-branch geometry can also be investigated through spectral curves \cite{Coman:2015bqq}.} and is fixed by the Cartan projections of the $\Nm=2$ $\text{SU}(2)_R\times \text{U}(1)_r$ R-symmetry. The chiral superfields $\Xm,\Ym,\Zm$ are charged under these symmetries and we list the quantum numbers of their scalar components in Table \ref{Tab:Orbifold}, where we also list the uncharged vector multiplets. We will refine the index just by these generic $\text{U}(1)$ symmetries, even though additional global symmetries may appear in certain special cases, to be discussed below. The refined index can be written as 
\begin{equation}
    \Im(t,x,\alpha, \beta) = \Tr_{\ker \Hm} (-1)^F t^{\frac{2}{3}(\Delta + j_2)} x^{2 j_1} \alpha^{q_{\alpha}} \beta^{q_\beta} \, ,
    \label{Eq:Index_Alpha_T}
\end{equation}
where $\alpha, \beta$ are the fugacities associated, respectively, to $\text{U}(1)_{\alpha}$ and $\text{U}(1)_{\beta}$.
We can then write the refined single-letter index for the vector and chiral fields as 
\begin{equation}
    \begin{split}
        f_{\Wm}(t,x,U_i, U^{\dagger}_i) &= i_{\Wm}(t,x) (U_i U^{\dagger}_i-1) \, , \\
        f_{\Xm_{ij}}(t,x,\alpha, \beta,U_i, U_j^{\dagger}) &= \frac{( \alpha \beta^{\frac{1}{2}} t ^\frac{2}{3}) U_i U_j^{\dagger} - ( \alpha^{-1}\beta^{-\frac{1}{2}}  t ^\frac{4}{3}) U_i^{\dagger} U_j}{(1-t x )(1 - t x^{-1})} \eqcolon  \frac{f_{\Xm} U_i U^{\dagger}_j+ \bar{f}_{\Xm} U^{\dagger}_i U_j}{(1-t x )(1 - t x^{-1})} \, ,\\
        f_{\Ym_{ij}}(t,x,\alpha,\beta,U_i, U_j^{\dagger}) &= \frac{(\alpha^{-1} \beta^{\frac{1}{2}} t ^\frac{2}{3}) U_i U_j^{\dagger} - (\alpha\beta^{-\frac{1}{2}}  t ^\frac{4}{3}) U_i^{\dagger} U_j}{(1-t x )(1 - t x^{-1})} \eqcolon  \frac{f_{\Ym} U_i U^{\dagger}_j+ \bar{f}_{\Ym} U^{\dagger}_i U_j}{(1-t x )(1 - t x^{-1})} \, ,\\
        f_{\Zm_{ij}}(t,x,\alpha, \beta,U_i, U_j^{\dagger}) &= \frac{(\beta^{-1}  t ^\frac{2}{3}) U_i U_j^{\dagger} - (\beta t ^\frac{4}{3}) U_i^{\dagger} U_j}{(1-t x )(1 - t x^{-1})} \eqcolon  \frac{f_{\Zm} U_i U^{\dagger}_j+ \bar{f}_{\Zm} U^{\dagger}_i U_j}{(1-t x )(1 - t x^{-1})} \, ,
    \end{split}
    \label{Eq:Single_particle_f}
\end{equation}
where $U_i$ are the characters associated to the gauge representation of the field. We can then project onto gauge invariant states by performing the standard integration over the gauge groups $G_j$ with Haar measure $\dd\mu_j$
\begin{equation}
   \Im(t,x,\alpha, \beta)= \prod_{j=1}^{\abs{\Gamma}} \left(\int_{G_j} d\mu_j\right)\, \text{exp}\bigg[\sum_{n=1}^\infty\frac{1}{n}\Im(t^n,x^n,\alpha^n, \beta^n, \{U^n\}, \{(U^{\dagger})^n\})\bigg]\,,
\end{equation} 
where $\Im(t,x,\alpha, \beta, \{U\}, \{U^{\dagger}\})$ is the single-letter index given by summing appropriate contributions \eqref{Eq:Single_particle_f} with corresponding gauge fugacities $U_j$, as we will discuss momentarily. We will perform the integrals over the gauge groups in the large-$N$ approximation, which allows us to evaluate them explicitly.

\paragraph{$\mathbb{Z}_k$ orbifold theories}

As explained in Section \ref{sec: Z_k} the $\mathbb{Z}_k$ orbifold theories are characterised by three integer configuration numbers $[a_X, a_Y, a_Z]$ which determine the gauge representations of the bifundamental chiral multiplets as 
\begin{equation}
    \Xm_{i, i+a_X}, \quad \Ym_{i, i+a_Y}, \quad \Zm_{i, i+a_Z},
\end{equation}
where all the shifts have to be taken modulo $k$. To evaluate the gauge-group integral, we make use of standard large-$N$
techniques for unitary matrix models
\cite{Sundborg:1999ue,Aharony:2003sx,Kinney:2005ej}:
One first introduces an eigenvalue density for each gauge node and expands it in Fourier modes. Around the large-$N$
uniform saddle, the matrix integral becomes Gaussian in these modes.
For $\Nm=2$ and $\Nm=1$ $\mathbb{Z}_k$ orbifold quiver theories, this procedure was applied explicitly in
\cite{Nakayama:2005mf}. The resulting large-$N$ index can be expressed as the plethystic exponential (PE) of the corresponding single-trace index,
\begin{equation}
    \Im(t,x,\alpha, \beta) = \frac{\text{PE}[- k\,  i_{\Wm}(t,x)]}{\prod_{n=1}^{\infty}\text{det}[(1-i_{\Wm}(t^n,x^n)) \mathbb{1}-\Am(t^n,x^n,\alpha^n, \beta^n)]}\,.
\end{equation}
where $\Am(t,x,\alpha, \beta)$ is the adjacency matrix where each non-vanishing entry $\Am_{ij}$ is replaced by the single-letter index associated to the bifundamental multiplets connecting the corresponding gauge nodes $i$ and $j$.\footnote{See also a more detailed discussion in \cite{Baume:2026svx} in the context of free thermal partition functions.} For example, the matrices $\Am$ for the $\mathbb{Z}_3$ and $\mathbb{Z}_4$ quivers in Figure \ref{fig:Zkquiver} read 
\begin{equation}
\begin{split}
    & \mathbb{Z}_3 : \quad  \begin{pmatrix}
        0 & \frac{f_{\Xm} + f_{\Ym} +f_{\Zm}}{{(1-t x )(1 - t x^{-1})}} & \frac{\bar{f}_{\Xm} + \bar{f}_{\Ym} +\bar{f}_{\Zm}}{(1-t x )(1 - t x^{-1})} \\
        \frac{\bar{f}_{\Xm} + \bar{f}_{\Ym} +\bar{f}_{\Zm}}{(1-t x )(1 - t x^{-1})} & 0 & \frac{f_{\Xm} + f_{\Ym} +f_{\Zm}}{(1-t x )(1 - t x^{-1})} \\
        \frac{f_{\Xm} + f_{\Ym} +f_{\Zm}}{(1-t x )(1 - t x^{-1})} & \frac{\bar{f}_{\Xm} + \bar{f}_{\Ym} +\bar{f}_{\Zm}}{(1-t x )(1 - t x^{-1})}  & 0
    \end{pmatrix}\,, \\
    &\mathbb{Z}_4 : \quad  \begin{pmatrix}
        0 & \frac{f_{\Xm} + f_{\Ym} }{{(1-t x )(1 - t x^{-1})}} & \frac{f_{\Zm} + \bar{f}_{\Zm} }{{(1-t x )(1 - t x^{-1})}} & \frac{\bar{f}_{\Xm} + \bar{f}_{\Ym} }{(1-t x )(1 - t x^{-1})} \\
        \frac{\bar{f}_{\Xm} + \bar{f}_{\Ym} }{(1-t x )(1 - t x^{-1})} & 0 & \frac{f_{\Xm} + f_{\Ym} }{(1-t x )(1 - t x^{-1})} & \frac{f_{\Zm} + \bar{f}_{\Zm} }{(1-t x )(1 - t x^{-1})} \\
        \frac{f_{\Zm} + \bar{f}_{\Zm} }{(1-t x )(1 - t x^{-1})} & \frac{\bar{f}_{\Xm} + \bar{f}_{\Ym} }{(1-t x )(1 - t x^{-1})} & 0 & \frac{f_{\Xm} + f_{\Ym} }{(1-t x )(1 - t x^{-1})} \\
        \frac{\bar{f}_{\Xm} + \bar{f}_{\Ym} }{(1-t x )(1 - t x^{-1})} & \frac{f_{\Zm} + \bar{f}_{\Zm} }{(1-t x )(1 - t x^{-1})} & \frac{f_{\Xm} + f_{\Ym} }{(1-t x )(1 - t x^{-1})}  & 0\\
    \end{pmatrix}\,,
\end{split}
\end{equation}
where we suppressed the variable dependence for readability.
By performing an explicit computation of the determinant following \cite{Nakayama:2005mf}, and using the identities
\begin{equation}
		\text{PE}[f(x)] = \exp \left (\sum_{n=1}^\infty\frac{f(x^n)}{n} \right )\,,\quad 
		\text{PE}[a\, x] = \frac{1}{(1-x)^a}\,,\quad 
		\text{PE}[\frac{a\, x}{1-x}] = \prod_{n=1}^\infty\frac{1}{(1-x^n)^a}\,,
        \label{Eq:PE_Identities}
\end{equation}
we can derive the general form of the index as
\begin{equation}
    \Im(t,x,\alpha, \beta) = \text{PE}\left [\ell_X \frac{f_{\Xm}^{\frac{k}{\ell_X}}}{1- f_{\Xm}^{\frac{k}{\ell_X}}} + \ell_Y \frac{f_{\Ym}^{\frac{k}{\ell_Y}}}{1- f_{\Ym}^{\frac{k}{\ell_Y}}}  + \ell_Z \frac{f_{\Zm}^{\frac{k}{\ell_Z}}}{1- f_{\Zm}^{\frac{k}{\ell_Z}}} \right ]\,,
    \label{Eq:IZk}
\end{equation}
where we introduced $\ell_I = \text{gcd}(k,a_I)$. This result coincides with the result derived in \cite{Nakayama:2005mf}, once we set $\alpha=\beta=1$ and decouple all the $\text{U}(1)$ gauge factors. The single-trace index is then easily obtained by taking the plethystic logarithm of \eqref{Eq:IZk} which reads
\begin{equation}
    \Im^{s.t.}(t,x,\alpha,\beta)= \ell_X \frac{f_{\Xm}^{\frac{k}{\ell_X}}}{1- f_{\Xm}^{\frac{k}{\ell_X}}} + \ell_Y \frac{f_{\Ym}^{\frac{k}{\ell_Y}}}{1- f_{\Ym}^{\frac{k}{\ell_Y}}}  + \ell_Z \frac{f_{\Zm}^{\frac{k}{\ell_Z}}}{1- f_{\Zm}^{\frac{k}{\ell_Z}}}\,.
    \label{Eq:ST_IZk}
\end{equation}

\paragraph{$\mathbb{Z}_k \times \mathbb{Z}_\ell$ orbifold theories}

In this case, the gauge nodes $\Wm_{(i,j)}$ of the theory are distributed on a $k \times \ell$ grid (see Figure \ref{fig:ZkZlquiver}), and we can identify the gauge nodes through the indices $(i,j)$ representing a node at the $i$-th row and $j$-th column. We can then represent the bifundamental chiral multiplets as $\Phi_{(i,j),(m,n)}$, indicating that they connect the gauge node at position $(i,j)$ to the one at position $(m,n)$ on the grid. The $\Xm,\Ym,\Zm$ superfields connect
\begin{equation}
    \Xm_{(i,j), (i+1,j)}\,, \quad \Ym_{(i,j), (i,j+1)}\,, \quad  \Zm_{(i,j), (i-1,j-1)}\,,
    \label{Eq:ZkZl_fields}
\end{equation}
where the shifts on the row index are taken modulo $k$ and shifts on the column index modulo $\ell$. For index calculations, it is useful to label all gauge nodes by a single index through the identification
\begin{equation}
    \Phi_{(a,b),(c,d)}\equiv \Phi_{b+ \ell (a-1), d + \ell  (c-1)}\,.
\end{equation}
We can then perform the gauge integration using the standard tools of the large-$N$ limit and find an expression
\begin{equation}
    \Im(t,x,\alpha, \beta) = \frac{\text{PE}[- k \ell \,  i_{\Wm}(t,x)]}{\prod_{n=1}^{\infty}\text{det}[(1-i_{\Wm}(t^n,x^n)) \mathbb{1}-\Am(t^n,x^n,\alpha^n, \beta^n)]}
\end{equation}
for the index, where $\Am(t,x,\alpha, \beta)$ is again the adjacency matrix where each non-vanishing entry $\Am_{ij}$ is replaced by the single-letter index associated to the bifundamental multiplets connecting the gauge nodes $i$ and $j$. For example, the matrix $\Am$ for the $\mathbb{Z}_2 \times \mathbb{Z}_2$ quiver in Figure \ref{fig:ZkZlquiver} reads 
\begin{equation}
    \begin{pmatrix}
        0 & \frac{f_{\Ym} + \bar{f}_{\Ym}}{{(1-t x )(1 - t x^{-1})}} & \frac{f_{\Xm} + \bar{f}_{\Xm}}{{(1-t x )(1 - t x^{-1})}} & \frac{f_{\Zm} + \bar{f}_{\Zm}}{{(1-t x )(1 - t x^{-1})}} \\
        \frac{f_{\Ym} + \bar{f}_{\Ym}}{{(1-t x )(1 - t x^{-1})}} & 0 & \frac{f_{\Zm} + \bar{f}_{\Zm}}{{(1-t x )(1 - t x^{-1})}} & \frac{f_{\Xm} + \bar{f}_{\Xm}}{{(1-t x )(1 - t x^{-1})}} \\
        \frac{f_{\Xm} + \bar{f}_{\Xm}}{{(1-t x )(1 - t x^{-1})}} &\frac{f_{\Zm} + \bar{f}_{\Zm}}{{(1-t x )(1 - t x^{-1})}} & 0 & \frac{f_{\Ym} + \bar{f}_{\Ym}}{{(1-t x )(1 - t x^{-1})}} \\
        \frac{f_{\Zm} + \bar{f}_{\Zm}}{{(1-t x )(1 - t x^{-1})}} & \frac{f_{\Xm} + \bar{f}_{\Xm}}{{(1-t x )(1 - t x^{-1})}} & \frac{f_{\Ym} + \bar{f}_{\Ym}}{{(1-t x )(1 - t x^{-1})}} & 0
    \end{pmatrix}\,,
\end{equation}
where we suppressed the variable dependence for readability.
After an explicit computation of the determinant and using the identities in \eqref{Eq:PE_Identities}, the general form of the index turns out as
\begin{equation}
    \Im(t,x,\alpha, \beta) = \text{PE}\left [\ell \frac{f_{\Xm}^{k}}{1- f_{\Xm}^{k}} + k \frac{f_{\Ym}^{\ell}}{1- f_{\Ym}^{\ell}}  + \text{gcd}(k,\ell) \frac{f_{\Zm}^{\frac{k \, \ell}{\text{gcd}(k,\ell)}}}{1- f_{\Zm}^{\frac{k \, \ell}{\text{gcd}(k,\ell)}}} \right ]\,.
    \label{Eq:IZkZl}
\end{equation}
The single-trace index is then easily obtained taking the plethystic logarithm
\begin{equation}
    \Im^{s.t.}(t,x,\alpha,\beta)= \ell \frac{f_{\Xm}^{k}}{1- f_{\Xm}^{k}} + k \frac{f_{\Ym}^{\ell}}{1- f_{\Ym}^{\ell}}  + \text{gcd}(k,\ell) \frac{f_{\Zm}^{\frac{k \, \ell}{\text{gcd}(k,\ell)}}}{1- f_{\Zm}^{\frac{k \, \ell}{\text{gcd}(k,\ell)}}} \,.
    \label{Eq:ST_IZkZl}
\end{equation}
As a consistency check we may note that for $\mathbb{Z}_k\times\mathbb{Z}_\ell$ orbifolds with $\text{gcd}(k,\ell)=1$, \eqref{Eq:ST_IZkZl} coincides with \eqref{Eq:ST_IZk} for a $\mathbb{Z}_{k \ell} [\ell, k, k \ell - \ell -k]$ orbifold, since these theories are equivalent.

We conclude that the superconformal indices in
\eqref{Eq:ST_IZk} and \eqref{Eq:ST_IZkZl} account for towers of BPS
operators with increasing charges under the three Abelian symmetries
$\text{U}(1)_\alpha$, $\text{U}(1)_\beta$, and $\text{U}(1)_r$. In the following subsections, we
will interpret the appearance of these towers in the two indices from both
the string-theory and gauge-theory perspectives.

Before doing so, we note that the twisted and untwisted sector contributions to the
indices \eqref{Eq:ST_IZk} and \eqref{Eq:ST_IZkZl} can be isolated separately. Projecting the $\Nm=4$ parent theory index onto its orbifold-invariant sector yields only a single copy of each tower, matching the untwisted indices
\eqref{Eq:ST_Untw_IZk} and \eqref{Eq:ST_Untw_IZkZl}, which we derive by applying the orbifold
projection directly to the $\Nm=4$ BPS
multiplets. The remaining towers in \eqref{Eq:ST_IZk} and \eqref{Eq:ST_IZkZl} arise from the twisted sectors, with every additional tower associated to a distinct twisted sector. 

\subsubsection{String theory perspective}\label{sec: strings}

\paragraph{Untwisted sector} On the string side of the holographic duality the orbifold projection identifies points in $AdS_5\times S^5$ which are related by the chosen discrete subgroup $\Gamma\subset \text{SU}(3)\subset P\text{SU}(2,2|4)$. By projecting type IIB string states on $AdS_5\times S^5$ to $\Gamma$-invariant states we create the so-called untwisted sector of the spectrum. In the large-$N$ limit, the BPS spectrum of $\mathcal{N}=4$ SYM is captured precisely by the KK-reduction of 10d type IIB supergravity on $S^5$ \cite{Kim:1985ez}. From a string-theory perspective, these are ``massless'' states without massive string excitations at the string scale $M_\text{s}=(\alpha')^{-1/2}$, although the $AdS_5\times S^5$ background geometry may still generate a mass at the $AdS_5$-scale $M_{\text{AdS}}=R_{\text{AdS}}^{-1}$. Semiclassically, we may envision pointlike strings moving on the $AdS_5\times S^5$ background.

The KK-modes on the $S^5$ are characterised by angular momenta $(J_X,J_Y, J_Z)$ corresponding to the three Cartan elements of the isometry group $\text{SU}(4)$. These quantum numbers are mapped to the R-symmetry representation of the dual single-trace operators. If we now apply the projection to $\Gamma$-invariant states, we have to select KK-modes whose angular-momentum quantum numbers satisfy 
\begin{equation}\label{Eq: Jkprojection}
    a_X(J_X-J_Z)+a_Y(J_Y-J_Z)\in k\mathbb{Z}\,.
\end{equation}
for $\Gamma=\mathbb{Z}_k$ in analogy to \eqref{Eq:Zk_invariance} or 
\begin{equation}\label{Eq: JkJlprojection}
    J_X-J_Z\in k\mathbb{Z}\,,\qquad J_Y-J_Z\in \ell\mathbb{Z}\,,
\end{equation}
for $\Gamma=\mathbb{Z}_k\times\mathbb{Z}_\ell $ in analogy to \eqref{Eq:ZkZl_invariance}. 

These orbifold actions break the $\text{SU}(4)$ isometry down to its Cartan subgroup $\text{U}(1)^3$ and we observe that the superconformal indices \eqref{Eq:ST_IZk} and $\eqref{Eq:ST_IZkZl}$ receive contributions from three types of KK-towers, each corresponding to one of the Cartan elements.\footnote{Another basis for these $\text{U}(1)^3$ Cartan elements was given in Table \ref{Tab:Orbifold}.} The orbifold projection \eqref{Eq: Jkprojection}, \eqref{Eq: JkJlprojection} to multiples of $k$, $\ell$, $k\ell$ or a suitable fraction depending on the divisor structure is precisely captured by the exponents in $\eqref{Eq:ST_IZk}$ and $\eqref{Eq:ST_IZkZl}$. However, the KK-mode towers only contribute one copy of each tower, so we have not accounted for the full index yet. To complete the matching we need to take twisted sectors into account. 

We note in passing that the projections \eqref{Eq: Jkprojection}, \eqref{Eq: JkJlprojection}  allow for a plethora of additional states in the full KK-decomposition of 10d supergravity. The superconformal index is blind to these modes due to cancellations among their contributions. We will work out the full untwisted BPS spectrum in the next section using superconformal representation theory and here just ascertain that all states found there can be matched one-to-one to KK-modes of 10d supergravity fields reduced on $S^5/\Gamma$. This is supported by the perfect matching in the $\mathcal{N}=4$ parent theory \cite{Witten:1998qj,Ferrara:1998ej} and the identification of the orbifold projections across the holographic duality. 

\paragraph{Twisted sectors} The untwisted sector of the orbifold spectrum alone is not modular invariant and we have to furthermore introduce $\abs{\text{Conj}(\Gamma)}-1$ twisted sectors, which consist of strings that close up to a $\Gamma$-action and are thus proper closed strings in the orbifold theory. Since there are no global fixed points of $\Gamma$ on $S^5/\Gamma$, these twisted strings are generically stretched out, which generates an energy at the string scale $M_\text{s}$ according to the intrinsic tension of the string. One may thus expect naively that there are no pointlike twisted sector strings that could account for additional single-trace BPS states in the superconformal index. 

However, this reasoning is too fast due to the presence of \emph{special twisted sectors} for which the closing $\Gamma$-action reduces to an $\text{SU}(2)$ action, such as e.g.
\begin{equation}\label{eq: special twist}
    \gamma_{\text{special}}: (X,Y,Z)\to (\omega^{a} X, \omega^{-a} Y, Z)\,.
\end{equation}
These special twisted sectors allow for a fixed $AdS_5\times S^1$ submanifold at $X=Y=0$, on which pointlike twisted strings states can be constructed. These may in turn be captured by an effective 6d $\mathcal{N}=1$ quantum field theory description.\footnote{\label{foot:6dN=2} In fact, since the local geometry is that of an $\mathcal{N}=2$ orbifold theory, we may consider the local 6d $\mathcal{N}=(2,0)$ description discussed in \cite{Gukov:1998kk, Pomoni:2009joh,Skrzypek:2023fkr,Martinez:2025jjq,Mantegazza:2026spd} as intermediate step. Here the additional fields furnish a tensor multiplet on which we have to act with a further $\mathcal{N}=1$ orbifold projection. This projection allows us to match the full BPS spectrum at the orbifold point.}
The KK-reduction on the $S^1$ generates a tower of BPS states similar to the untwisted sector, but restricted to a single $\text{U}(1)_Z$ quantum number. Similar special twisted sectors can also arise for $\text{U}(1)_X$ and $\text{U}(1)_Y$ depending on the orbifold under consideration. By counting the possible special twisted sectors, we will observe precisely the right number of KK-towers to complete the superconformal indices $\eqref{Eq:ST_IZk}$ and $\eqref{Eq:ST_IZkZl}$. The twisted BPS states generated this way match the representation theory decomposition discussed below. 

Let us first count special twisted sectors in the $\mathbb{Z}_k$ family of orbifolds generated by \eqref{Eq: orbaction}. We realise that the equation $\omega_k^{a_I n}=1$ has non-trivial solutions whenever $\gcd(a_I,k)>1$. In precisely these cases we observe special twisted sectors as in \eqref{eq: special twist} fixing the $S^1_i$ circle and thus generating a $\text{U}(1)_i$ KK-tower. We expect a total number of
\begin{equation}\label{eq:NZk}
     \text{gcd}(a_X,k)+\text{gcd}(a_Y,k)+\text{gcd}(a_Z,k)-3.
\end{equation}
special twisted sectors. Combined with the three untwisted sector $\text{U}(1)$-towers we therefore recover precisely the superconformal index \eqref{Eq:ST_IZk}.

In case of the  $\mathbb{Z}_k\times \mathbb{Z}_\ell$ orbifold theories, we can immediately determine $k+\ell-2$ special twisted sectors whenever one of the two subgroups is left untwisted, i.e. in the partially twisted sectors $(a,0)$ and $(0,b)$, which fix the $Y$ and $X$ circles, respectively. Additional special twisted sectors arise when $\gcd(k,\ell)>1$ which allows for twisted sectors fixing the $Z$ circle. The total number of special twisted sectors is thus 
\begin{equation}\label{eq:NZkZl}
    k + \ell + \text{gcd}(k,\ell) -3
\end{equation}
 Combined with the three untwisted sector $\text{U}(1)$-towers we again recover the superconformal index \eqref{Eq:ST_IZkZl}.

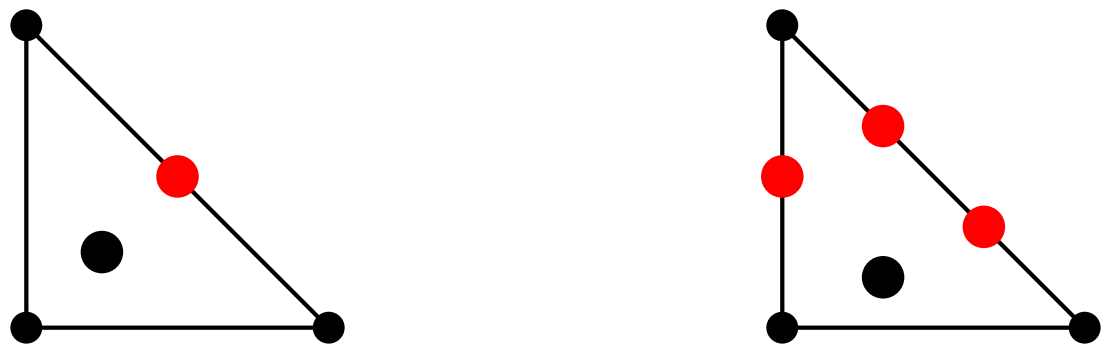
\begin{figure}[ht]
\centering
\begin{tikzpicture}[
    scale=2,
    every node/.style={font=\small}
]

\begin{scope}[xshift=0cm]

\draw[thick] (0,0) -- (1,0) -- (0,1) -- cycle;

\fill (0,0) circle (1.5pt);
\fill (1,0) circle (1.5pt);
\fill (0,1) circle (1.5pt);

\fill[black] (1/4,1/4) circle (2pt);
\fill[red]  (1/2,1/2) circle (2pt);

\node at (0.5,-0.25) {$\mathbb{Z}_4[1,1,2]$};

\end{scope}

\begin{scope}[xshift=2.5cm]

\draw[thick] (0,0) -- (1,0) -- (0,1) -- cycle;

\fill (0,0) circle (1.5pt);
\fill (1,0) circle (1.5pt);
\fill (0,1) circle (1.5pt);

\fill[black] (1/3,1/6) circle (2pt);
\fill[red] (1/3,2/3) circle (2pt);
\fill[red] (0,1/2) circle (2pt);
\fill[red] (2/3,1/3) circle (2pt);

\node at (0.5,-0.25) {$\mathbb{Z}_6[1,2,3]$};

\end{scope}

\end{tikzpicture}

\caption{Toric diagrams of $\mathbb{Z}_4[1,1,2]$ and $\mathbb{Z}_6[1,2,3]$ orbifolds of $\mathbb{C}^3$. Lattice points on the boundaries (in red) correspond to non-compact exceptional divisors, which intersect the $S^5/\Gamma$ subspace and carve out exceptional two-cycles.}
\label{fig:Zktoric}
\end{figure}

\paragraph{Geometric perspective} The special twisted sectors can be associated to the presence of exceptional two-cycles which resolve the local orbifold singularity of \eqref{eq: special twist} at the fixed point. Wrapping 10d supergravity fields on these two-cycles results in a local effective field theory (EFT) on $AdS_5\times S^1$ capturing the additional twisted BPS states. In the $\mathbb{C}^3/\Gamma$ embedding space the relevant two-cycles arise at the intersection of non-compact exceptional divisors with the $S^5/\Gamma$ submanifold. In the associated toric diagram, such divisors correspond to boundary points. Let us give a quick prescription of the toric diagram construction, referring the reader to \cite{Fulton:1436535,Closset:2009sv} for further details on toric geometry. $\mathbb{C}^3$ is represented by the fundamental triangle $(1,0,1)$, $(0,1,1)$, $(0,0,1)$ in the $\mathbb{Z}^3$ lattice. These points are restricted to a $(i,j,1)$ hypersurface, identifying $\mathbb{C}^3$ as a (non-compact) Calabi-Yau three-fold. Taking a  $\mathbb{Z}_k$ orbifold corresponds to introducing a fractional refined lattice 
\begin{equation}
    \mathbb{Z}^3+\frac{\mathbb{Z}}{k}(a_X,a_Y,a_X+a_Y+a_Z)= \mathbb{Z}^3+\frac{\mathbb{Z}}{k}(a_X,a_Y,0)\,,
\end{equation}
which we restrict to the $(i,j,1)$ hypersurface  
\begin{equation}
    (\mathbb{Z},\mathbb{Z},1)+\frac{\mathbb{Z}}{k}(a_X,a_Y,0)\,.
\end{equation}
If this procedure introduces additional points on the boundary of the fundamental triangle, we associate to these non-compact divisors of $\mathbb{C}^3/\Gamma$ which in turn generate exceptional two-cycles on $S^5/\Gamma$. As instructive example, we draw the toric diagrams associated to the $\mathbb{Z}_4[1,1,2]$ and $\mathbb{Z}_6[1,2,3]$ orbifold theory in Figure \ref{fig:Zktoric}. The total number of twisted BPS sectors determined this way matches \eqref{eq:NZk} precisely and the edge at which the boundary point intersects the fundamental triangle determines which $S^1$ is fixed in this special twisted sector. The interior lattice points correspond to compact exceptional divisors in the crepant resolution of $\mathbb{C}^3/\Gamma$, where ``crepant'' means that the resolution preserves the Calabi--Yau condition. Since these compact divisors are localised at the origin of $\mathbb{C}^3/\Gamma$ they do not intersect the $S^5/\Gamma$ submanifold.

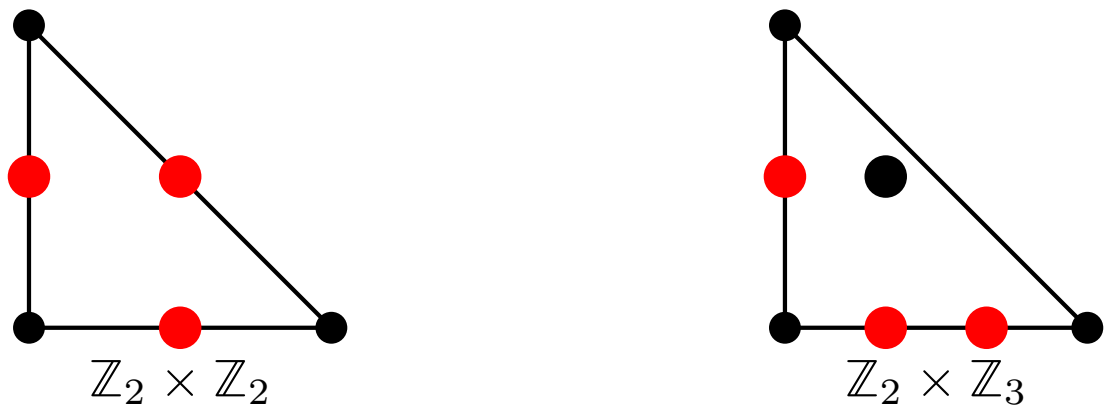
\begin{figure}[ht]
\centering
\begin{tikzpicture}[
    scale=2,
    every node/.style={font=\small}
]
\begin{scope}[xshift=0cm]

\draw[thick] (0,0) -- (1,0) -- (0,1) -- cycle;

\fill (0,0) circle (1.5pt);
\fill (1,0) circle (1.5pt);
\fill (0,1) circle (1.5pt);

\fill[red] (1/2,0) circle (2pt);
\fill[red] (0,1/2) circle (2pt);
\fill[red] (1/2,1/2) circle (2pt);

\node at (0.5,-0.18) {$\mathbb{Z}_2\times\mathbb{Z}_2$};

\end{scope}

\begin{scope}[xshift=2.5cm]

\draw[thick] (0,0) -- (1,0) -- (0,1) -- cycle;

\fill (0,0) circle (1.5pt);
\fill (1,0) circle (1.5pt);
\fill (0,1) circle (1.5pt);

\fill[red]   (0,1/2)   circle (2pt);
\fill[red]   (1/3,0)   circle (2pt);
\fill[red]   (2/3,0)   circle (2pt);
\fill[black] (1/3,1/2) circle (2pt);

\node at (0.5,-0.18) {$\mathbb{Z}_2\times\mathbb{Z}_3$};

\end{scope}

\end{tikzpicture}

\caption{Toric diagrams of $\mathbb{Z}_2\times \mathbb{Z}_2$ and $\mathbb{Z}_2\times \mathbb{Z}_3$ orbifolds of $\mathbb{C}^3$. Lattice points on the boundaries (in red) correspond to non-compact exceptional divisors, which intersect the $S^5/\Gamma$ subspace and carve out exceptional two-cycles.}
\label{fig:ZkZltoric}
\end{figure}

For $\mathbb{Z}_k\times\mathbb{Z}_\ell$ orbifolds the appropriate lattice refinement is
\begin{equation}
    (\mathbb{Z},\mathbb{Z},1)+\frac{\mathbb{Z}}{k}(1,0,0)+\frac{\mathbb{Z}}{\ell}(0,1,0)\,,
\end{equation}
which again recovers the correct count \eqref{eq:NZkZl} of boundary points and identifies the fixed circles of the special twisted sectors. We provide some example toric diagrams for the $\mathbb{Z}_2\times \mathbb{Z}_2$ and $\mathbb{Z}_2\times \mathbb{Z}_3$ orbifold theories in Figure \ref{fig:ZkZltoric}.

\subsubsection{Gauge theory perspective}\label{sec: gauge theory}

\paragraph{Untwisted sector} In this paragraph we adopt a representation theory approach to compute the BPS spectrum of orbifold theories. The necessary representation theory details for $4$d $\mathcal{N}=1$ and $\mathcal{N}=2$ superconformal algebras are reviewed in Appendices \ref{App:N1Reps} and \ref{App:N2Reps}, respectively. In particular we can use character decomposition of the superconformal algebra to decompose the single-trace BPS spectrum of $\mathcal{N}=4$ SYM into $\mathcal{N}=1$ multiplets. From these we select the orbifold-invariant multiplets which form the untwisted sector of the theory. Despite the fact that the orbifold projection conditions are theory-dependent, we will write down an algorithm that works for a generic orbifold theory. The construction of the twisted sectors is less straight-forward and will be discussed further below. The full BPS spectrum can be constructed using the Wolfram Mathematica notebook we supply as ancillary file to the arXiv submission of this paper.

The single-trace BPS spectrum of $\mathcal{N}=4$ SYM is generated by the chiral primaries $\Tr \left [ \Phi^{\{i_1} \dots \Phi^{i_p \}} \right ]$, where $\Phi^i$ are scalars in the anti-symmetric representation of the $\text{SU}(4)_R$ R-symmetry or, equivalently, in the fundamental representation of $\text{SO}(6)$ from which we derive the index $i$. These states generate half-BPS multiplets $\mathcal{B}^{\frac{1}{2}, \frac{1}{2}}_{[0,p,0]}$ of the $\mathcal{N}=4$ superconformal algebra. Here we adopt the Dolan-Osborn notation \cite{Dolan:2002zh},\footnote{The translation to C\'ordova-Dumitrescu-Intriligator notation \cite{Cordova:2016emh} is given in Appendices \ref{App:N1Reps} and \ref{App:N2Reps}.} with $\Bm_{[0,p,0]}^{\frac{1}{2},\frac{1}{2}}$ indicating that the highest-weight state of the multiplet is annihilated by half of the $\Qm$ and $\bar \Qm$ supercharges and sits in the $p$-index symmetric traceless representation of $\text{SO}(6)$. The decomposition of these $\mathcal{N}=4$ multiplets into $\mathcal{N}=2$ multiplets was already stated in \cite{Dolan:2002zh} and we report it in equation \eqref{Eq:N4toN2} of Appendix \ref{App:Decompositions}. At this point we can further halve supersymmetry, breaking the superalgebra $\mathfrak{su}(2,2|2)$ of $\Nm=2$ down to $\mathfrak{su}(2,2|1) \oplus \mathfrak{u}(1)_{\beta}$, where this additional $\mathfrak{u}(1)_\beta$ factor is a global symmetry commuting with the supercharges.\footnote{As stated in footnote \ref{foot: Bourton}, this flavour symmetry is an essential ingredient for the definition of Higgs and Coulomb branches in $\Nm=1$ theories \cite{Bourton:2020rfo}.} We can label the $\Nm=2$ states via the quantum numbers $(\Delta, j_1,j_2,R, r^{\Nm=2})$ labelling the maximal bosonic subalgebra
\begin{equation}
    \mathfrak{u}(1)_{\Delta} \oplus \mathfrak{su}(2)_{j_1} \oplus \mathfrak{su}(2)_{j_2} \oplus \mathfrak{su}(2)_{R} \oplus \mathfrak{u}(1)_{r^{\Nm=2}} \subset \mathfrak{su}(2,2|2)\,.
\end{equation}
After halving the supersymmetry, the states are labelled by the quantum numbers $(\Delta, j_1,j_2, \allowbreak  r, q_\beta)$ of the maximal $\Nm=1$ bosonic algebra
\begin{equation}
     \mathfrak{u}(1)_{\Delta} \oplus \mathfrak{su}(2)_{j_1} \oplus \mathfrak{su}(2)_{j_2}  \oplus \mathfrak{u}(1)_{r^{\Nm=1}} \oplus \mathfrak{u}(1)_{\beta} \subset \mathfrak{su}(2,2|1)\oplus \mathfrak{u}(1)_{\beta}\,.
     \label{Eq:N=1_SCA}
\end{equation}
The Cartan elements of the $\mathfrak{u}(1)_{r^{\Nm=1}} \oplus \mathfrak{u}(1)_{\beta}$ symmetries are given by a particular linear combination of the Cartan elements of the $\mathcal{N}=2$ R-symmetry algebra, resulting in the following relations between the quantum numbers
\begin{equation}
    r^{\Nm=1}= \frac{2}{3}(2 R_p-r^{\Nm=2})\,, \quad q_\beta= R_p + r^{\Nm=2}\,.
    \label{Eq:N2toN1_rules}
\end{equation}
Here, the subscript $p$ in $R_p \in \{-R, \dots, R\}$ indicates the projection of the $\mathfrak{su}(2)_R$ spin onto the Cartan direction.
At this point the decomposition of an $\Nm=2$ multiplet into $\Nm=1$ multiplets is straightforward: by writing the character of the $\Nm=2$ multiplet and applying the identifications in \eqref{Eq:N2toN1_rules} one obtains characters of the $\Nm=1$ algebra. By systematic comparison with $\Nm=1$ characters, the decompositions stated in \eqref{Eq:N2toN1} are obtained. In this way we have generated all the possible BPS operators descending from the $\Nm=4$ parent theory.

What remains to do is to determine which multiplets survive the orbifold projection and compose the untwisted sector. This requires keeping track of their original $\text{SU}(4)$ R-symmetry quantum numbers. To this end, let us choose a Cartan basis of charges ($J_X,J_Y, J_Z$) according to the fundamental action 
\begin{equation}
    U(\xi_X,\xi_Y,\xi_Z)=\begin{pmatrix}
        e^{i \xi_X} & 0 & 0\\
        0& e^{i\xi_Y} &0\\
        0 & 0 & e^{i\xi_Z}
    \end{pmatrix}
\end{equation}
on the fields $(X,Y,Z)$, which have charges $(1,0,0)$, $(0,1,0)$ and $(0,0,1)$, respectively. This is a valid choice of quantum numbers for the $\text{SU}(4)_R$ R-symmetry, related to the Dynkin labels by $[J_Y-J_Z,J_X-J_Y,J_Y+J_Z]$. 

Once we break supersymmetry to $\mathcal{N}=2$, we retain an $\text{SU}(2)_{R}\times \text{U}(1)_{r^{\Nm=2}}$ R-symmetry, decomposing representations of the broken $\text{SU}(2)_L$ symmetry. In terms of these groups the quantum numbers are $(J_X,J_Y, J_Z)=(R_p+L_p,R_p-L_p, -r^{\Nm=2}) $, where the subscript $p$ again indicates projection in the Cartan direction. Further breaking to $\mathcal{N}=1$ the $R_p$ and $r^{\Nm=2}$ quantum numbers are combined as $r^{\Nm=1}=\tfrac{2}{3}(2R_p-r^{\Nm=2})=\tfrac{2}{3}(J_X+J_Y+ J_Z)$, which suggests that the original $\text{SU}(4)$ charges are ultimately given by
\begin{equation}
    (J_X,J_Y, J_Z)=(\tfrac{3 r^{\Nm=1}+2r^{\Nm=2}}{4}+L_p,\tfrac{3 r^{\Nm=1}+2r^{\Nm=2}}{4}-L_p, -r^{\Nm=2})\,. 
    \label{Eq:Jxyz}
\end{equation}
Only $r^{\Nm=1}$ is now a true R-symmetry quantum number, while the labels $r^{\Nm=2}$ and $L_p$ simply recall which decomposition the state in question came from. We may keep track of these labels during the decomposition procedure outlined above.

The orbifold invariance conditions now descend from \eqref{Eq: Jkprojection} and \eqref{Eq: JkJlprojection}. An $\mathcal{N}=1$ $\mathbb{Z}_k[a_X,a_Y,a_Z]$ orbifold imposes the invariance condition 
\begin{equation}
    a_X(J_X-J_Z)+a_Y(J_Y-J_Z)=\frac{3}{4}(r^{\Nm=1}+2r^{\Nm=2})(a_X+a_Y)+L_p(a_X-a_Y)\in k\mathbb{Z}\,, 
    \label{Eq:Orb_proj_Zk}
\end{equation}
while an $\mathcal{N}=1$ $\mathbb{Z}_k\times \mathbb{Z}_\ell$ orbifold imposes the conditions
\begin{equation}
    J_X- J_Z=\frac{3}{4}(r^{\Nm=1}+2r^{\Nm=2})+L_p\in k\mathbb{Z}\,,\qquad
    J_Y-J_Z=\frac{3}{4}(r^{\Nm=1}+2r^{\Nm=2})-L_p\in \ell \mathbb{Z}\,.
    \label{Eq:Orb_proj_ZkZl}
\end{equation}
Enforcing these constraints after the $\mathcal{N}=1$ decomposition, we can determine the full untwisted BPS spectrum. The result of the projection is orbifold-dependent, but the algorithm explained here is completely general and implemented in the ancillary Mathematica notebook. 

As consistency check, we can read off the untwisted contribution to the superconformal index which yields the expected expressions
\begin{equation}
    \Im^{s.t.; \text{untwisted}}_{\mathbb{Z}_k}(t,x,\alpha,\beta)=  \frac{f_{\Xm}^{\frac{k}{\ell_X}}}{1- f_{\Xm}^{\frac{k}{\ell_X}}} +  \frac{f_{\Ym}^{\frac{k}{\ell_Y}}}{1- f_{\Ym}^{\frac{k}{\ell_Y}}}  +  \frac{f_{\Zm}^{\frac{k}{\ell_Z}}}{1- f_{\Zm}^{\frac{k}{\ell_Z}}}\,,\qquad \ell_I=\text{gcd}(a_I,k)\,,
    \label{Eq:ST_Untw_IZk}
\end{equation}
for the $\mathbb{Z}_k$ family, and 
\begin{equation}
    \Im^{s.t.; \text{untwisted}}_{\mathbb{Z}_k \times \mathbb{Z}_\ell}(t,x,\alpha,\beta)=  \frac{f_{\Xm}^{k}}{1- f_{\Xm}^{k}} + \frac{f_{\Ym}^{\ell}}{1- f_{\Ym}^{\ell}}  + \frac{f_{\Zm}^{\frac{k \, \ell}{\text{gcd}(k,\ell)}}}{1- f_{\Zm}^{\frac{k \, \ell}{\text{gcd}(k,\ell)}}}\,,
    \label{Eq:ST_Untw_IZkZl}
\end{equation}
for the $\mathbb{Z}_k\times \mathbb{Z}_{\ell}$ family. Note that we could have arrived at these results independently by directly projecting the parent $\Nm=4$ index to the orbifold invariants.

We observe that, generically, the untwisted BPS spectrum alone does not saturate the full superconformal index of the orbifold theory at hand, signalling the presence of further twisted sector BPS states. The only cases in which the index is matched straight away are $\mathbb{Z}_k$ orbifold theories where all $a_I$ are coprime with $k$, i.e. $\text{gcd}(k,a_I)=1$ and thus no special twisted sectors arise. 

We further notice that due to various cancellations the index for both families of orbifold theories is only counting operators with a scalar superconformal primary of the form $\Tr[ X^{p_X}]$, $\Tr[ Y^{p_Y}]$, $\Tr[ Z^{p_Z}]$, where the powers $p_I$ are taken from the sets
\begin{equation}
    \begin{split}
        & p_I = n \cdot \frac{k}{\gcd(a_I,k)} \, , \qquad \qquad\qquad\qquad\qquad \quad n \in \mathbb{Z}_{> 0} \, ,\quad \text{for } \mathbb{Z}_k\,, \\
        & p_X = n \cdot k \, , \; p_Y = n \cdot \ell \, , \; p_Z= n \cdot \frac{k \ell}{\gcd(k,\ell)} \, , \quad n \in \mathbb{Z}_{> 0} \, ,\quad \text{for } \mathbb{Z}_k \times \mathbb{Z}_\ell \, .
    \end{split}
    \label{Eq:Invariance_Orbifold_Single_Operators}
\end{equation}
This is only a small subset of the full set of BPS states we observed in the multiplet decomposition, which for example also features all possible symmetrised combinations of the states
\begin{equation}
    \Tr[ X^{p_X} Y^{p_Y} Z^{p_Z}] + \text{perms.} \, ,
\end{equation}
subject to the orbifold invariance conditions \eqref{Eq:Zk_invariance} and \eqref{Eq:ZkZl_invariance}
\begin{equation}
\begin{split}
    & a_X(p_X - p_Z) + a_Y (p_Y - p_Z)  \in k \mathbb{Z}\,, \qquad \text{for } \mathbb{Z}_k \, , \\ 
    & p_X - p_Z \in k \mathbb{Z} \, , \quad p_Y - p_Z \in \ell \mathbb{Z}\,, \qquad \quad\,\,\text{for } \mathbb{Z}_k \times \mathbb{Z}_{\ell} \, .
\end{split}
\label{Eq:Orbi_Invariance_Operators}
\end{equation}
It is possible to further dress these states with global symmetry current multiplets as well as with the stress-energy tensor multiplet:
\begin{equation}
    \Tr[  \Mm_F X^{p_X} Y^{p_Y} Z^{p_Z}] + \text{perms.}   \, ,\qquad  \Tr[ \Jm_r X^{p_X} Y^{p_Y} Z^{p_Z}]+ \dots \, .
\end{equation}
Here $\Mm_F$ denotes the mesonic state sitting at the top of the flavour symmetry current multiplet and $\Jm_r$ the R-symmetry current sitting at the top of the stress-energy tensor multiplet. In the latter case the dots indicate that the completion generally involves more than just permutations of fields, due to the presence of a derivative in $\Jm_r$. We summarise these states and the corresponding multiplets in Table \ref{Tab:Orbifold_Untwisted_States}. Despite the index not counting these states, their protection is ensured from the multiplet decomposition. We explicitly checked the protections for the first members of the towers with the one-loop dilatation operator of Section \ref{sec: orbidila}. In Section \ref{sec:applications} we will furthermore see that the explicit operator expressions are deformed as we move through the conformal manifold or cease to be protected altogether (see e.g. \eqref{eq:deformedop}).

\begin{table}[t]
\begin{centering}
\renewcommand{\arraystretch}{1.3}
\begin{tabular}{c|c}
 Multiplet & Orbifold operator \\
\hline
$\bar{\Bm}_{\frac{2p_X}{3}(0,0)}$ & $\Tr [ X^{p_X}] + \text{perms.}$\\
$\bar{\Bm}_{\frac{2p_Y}{3}(0,0)}$ & $\Tr [ Y^{p_Y}] + \text{perms.}$\\
$\bar{\Bm}_{\frac{2p_Z}{3}(0,0)}$ & $\Tr [ Z^{p_Z}] + \text{perms.}$\\
$\bar{\Bm}_{\frac{2(p_X+p_Y+p_Z)}{3}(0,0)}$ & $\Tr [ X^{p_X}Y^{p_Y}Z^{p_Z}] + \text{perms.}$\\
$\bar{\Cm}_{\frac{2(p_X+p_Y+p_Z)}{3}(0,0)}$ & $\Tr [ \Mm_F X^{p_X}Y^{p_Y}Z^{p_Z}] + \text{perms.}$\\
$\bar{\Cm}_{\frac{2(p_X+p_Y+p_Z)}{3}(\frac{1}{2},\frac{1}{2})}$ & $\Tr [ \mathcal{J}_r X^{p_X}Y^{p_Y}Z^{p_Z}] + \dots $
\end{tabular}
\par\end{centering}
\caption{\label{Tab:Orbifold_Untwisted_States} Superconformal BPS primary operators built from scalar components in the untwisted sector of the $\mathbb{Z}_k [a_X,a_Y,a_Z]$ and $\mathbb{Z}_k \times \mathbb{Z}_{\ell}$ orbifold theories. The different exponents have to obey the conditions written in equations \eqref{Eq:Invariance_Orbifold_Single_Operators} and \eqref{Eq:Orbi_Invariance_Operators}, and we sum over all symmetric permutations of the component fields within the single-trace operators. Conjugate operators can be constructed analogously.}
\end{table}

\paragraph{Twisted sectors} 

The twisted sector BPS multiplets cannot be inferred from decomposing $\Nm=4$ multiplets. They arise in twisted sectors generated by a special orbifold action \eqref{eq: special twist} which leaves one of the complex scalar fields invariant, e.g. $Z$ in \eqref{eq: special twist}. Twisted single-trace operators built from this scalar then turn out to be protected. To perform a detailed investigation of the twisted BPS spectrum, we make use of the comment in footnote \ref{foot:6dN=2} and consider the effective $\mathcal{N}=2$ orbifold theory around the special orbifold loci fixed by \eqref{eq: special twist}. This theory features $\mathcal{N}=2$ tensor multiplets in the twisted sector \cite{Gukov:1998kk, Gadde:2009dj}, which generate the $\Nm=2$ multiplets 
\begin{equation}\label{eq: twisted N=2}
    \mathcal{N}=2:\qquad  \hat{\Bm}_1\,,\qquad \Em_{p+2(0,0)}\,,\quad \bar{\Em}_{-p-2(0,0)}\,,\qquad p\geq 0\,,
\end{equation} in the dual theory.\footnote{The $\hat{\Bm}_1$ multiplet in \eqref{eq: twisted N=2} is present due to the decoupling of non-conformal $\text{U}(1)$ factors to be discussed in Section \ref{sec: U(1)s}. This was explained in detail in \textbf{Appendix B} of \cite{Gadde:2009dj}.}  We may decompose this operator content in terms of $\Nm=1$ multiplets according to \eqref{Eq:N2toN1} and impose the orbifold invariance projection \eqref{Eq:Orb_proj_Zk} or \eqref{Eq:Orb_proj_ZkZl}.
In this projection, we have to respect that the embedding of the $\Nm=2$ subalgebra inside $\Nm=4$ changes depending on the fixed point under consideration. The charge assignment given in \eqref{Eq:Jxyz} should be rotated accordingly: If the special twisted sector features trivial twist action on $X\,(Y)$ instead of $Z$, we should use the projections \eqref{Eq:Orb_proj_Zk} and \eqref{Eq:Orb_proj_ZkZl} only after exchanging $ J_X \leftrightarrow J_Z\,(J_Y \leftrightarrow J_Z)$.

The number of special twisted sectors is given by \eqref{eq:NZk} and \eqref{eq:NZkZl} and each twisted sector contributes with the respective projection of the $\Nm=2$ multiplets as outlined above. Summing the contribution of all these multiplets to the index and adding the untwisted contribution \eqref{Eq:ST_Untw_IZk} or \eqref{Eq:ST_Untw_IZkZl}, we indeed reproduce the full indices in \eqref{Eq:ST_IZk} and \eqref{Eq:ST_IZkZl}. We take this as strong evidence that the construction of the BPS spectrum is complete. The twisted BPS multiplets and their highest-weight operators are presented in  Table \ref{Tab:Orbifold_Twisted_States}.

\begin{table}[t]
\begin{centering}
\renewcommand{\arraystretch}{1.3}
\begin{tabular}{c|c}
 Multiplet & Orbifold operator \\
\hline
$\bar{\Bm}_{\frac{2p_I}{3}(0,0)}$ & $\Tr [\Omega_{\text{special}} \phi^{p_I}]$\\
$\hat{\Cm}_{(0,0)}$ & $\Tr [\Omega_{\text{special}} \Mm_F^{\text{twist}} ]$
\end{tabular}
\par\end{centering}
\caption{\label{Tab:Orbifold_Twisted_States} Superconformal BPS primary operators built from the scalar components in the special twisted sector of the $\mathbb{Z}_k [a_X,a_Y,a_Z]$ and $\mathbb{Z}_k \times \mathbb{Z}_{\ell}$ orbifold theories. The  exponent $p_I$ has to obey the appropriate condition \eqref{Eq:Invariance_Orbifold_Single_Operators}. The twist matrix $\Omega_{\text{special}}$ encodes the $\text{SU}(2)$ action \eqref{eq: special twist} in the gauge structure of the quiver gauge theory. Conjugate operators can be constructed analogously.}
\end{table}

The index is again counting states of the form  $\Tr[ \Omega_{\text{special}} \phi^{p_i}]$, where the scalar field $\phi$ is the one invariant under the special twist action \eqref{eq: special twist} and the $p_I$'s satisfy the relations \eqref{Eq:Invariance_Orbifold_Single_Operators}. Contrary to the untwisted sector we do not observe additional states dressed by corresponding twisted flavour current primaries $\Mm_F^{\text{twist}}$. This is compatible with the result for the $\Nm=2$ case in \cite{Gadde:2009dj,Gadde:2010zi} and due to the more involved protection mechanism of the twisted flavour current. As we will see in the next Section \ref{sec: U(1)s}, these twisted global symmetry operators are protected due to double-trace corrections acting only on the length-two operator space. Longer operators are not affected by this mechanism and generically receive perturbative corrections. We  further observe an absence of twisted towers mixing the $X,Y,Z$ fields, which has a clear interpretation at the level of the local geometry near the fixed points of the special twist action. As discussed in Section \ref{sec: strings}, the operators of the twisted sectors are captured by a local EFT on $AdS_5 \times S^1$. To each field $\phi = X,Y,Z$ we would usually associate a $\text{U}(1)$ charge and towers of increasing charge can be interpreted as KK modes of the corresponding $S^1$. The presence of only one $S^1$ in each twisted sector explicitly isolates a single tower of uniform field content.

\subsection{Decoupling of U(1)-factors and twisted current multiplets}\label{sec: U(1)s}

We now return to the discussion of $\text{U}(1)$ factors arising during the orbifold projection, which we postponed in the previous sections. Remember that the orbifold theories we consider in this paper descend from 4d $\mathcal{N}=4$ SYM theory with gauge group $\text{U}(\abs{\Gamma}N)$, which we project down to fields and operators invariant under the orbifold action, e.g. $\gamma$ in \eqref{Eq: gauge-twist}. In particular, the gauge group is broken down to $\text{U}(N)^\abs{\Gamma}=\text{U}(1)^\abs{\Gamma}\times \text{SU}(N)^{\abs{\Gamma}}$. An overall $\text{U}(1)$ factor decouples as in $\mathcal{N}=4$ SYM and can therefore be dropped but the $\abs{\Gamma}-1$ remaining $\text{U}(1)$ factors couple to the charged matter that arises in off-diagonal terms of the original $\text{U}(\abs{\Gamma}N)$ representation. Explicitly, consider an arbitrary gauge group $\text{U}(N)_i$ in the orbifold theory, represented by a node in the associated quiver diagram. It factorises into $\text{SU}(N)_i\times \text{U}(1)_i$, such that the vector multiplets $\mathcal{W}_{\text{SU}(N)_i}$ and $\mathcal{V}_{U(1)_i}$ decouple. However the charged chiral multiplets couple via the current
\begin{equation}\label{Eq: currents}
    \mathcal{J}_i=\sum_{j}\left(\tr(\bar{\Phi}_{ji}\mathcal{D}\Phi_{ij})-\tr(\Dm(\bar{\Phi}_{ji})\Phi_{ij})\right)-\sum_{k}\left(\tr(\bar{\Phi}_{ik}\mathcal{D}\Phi_{ki})-\tr(\Dm(\bar{\Phi}_{ik})\Phi_{ki})\right)
\end{equation}
to the $\text{U}(1)$ factor. Here, $j$ and $k$ run over appropriately connected adjacent gauge nodes in the quiver diagram. These currents sit in superconformal multiplets descending from the highest-weight operators
\begin{equation}
    \mathcal{M}_i=\sum_{j}\tr(\bar{\Phi}_{ji}\Phi_{ij})-\sum_{k}\tr(\Phi_{ki}\bar{\Phi}_{ik})\,.
\end{equation}
The coupling term in the Lagrangian may be extracted from the gauge-invariant kinetic term
\begin{equation}
    \mathcal{L}\supset \int \dd^4\theta \,\left(\sum_{j}\tr(\bar{\Phi}_{ji}e^{g_i\mathcal{V}_i}\Phi_{ij})+\sum_{k}\tr(\Phi_{ki}e^{-g_i\mathcal{V}_i}\bar\Phi_{ik})\right) \supset  \int \dd^4 \theta \,g_i\mathcal{V}_i\mathcal{M}_i +\dots
\end{equation}

Alternatively, we may reorganise the current multiplets into twisted sectors by using quiver traces. For the $\mathbb{Z}_k$ family of orbifold theories, we may define
 \begin{equation}\label{Eq: current multiplets}
    \mathcal{M}_a=\Tr \left(\Omega^a({\mathcal{X}}\bar{\mathcal{X}}+{\mathcal{Y}}\bar{\mathcal{Y}}+{\mathcal{Z}}\bar{\mathcal{Z}}-\bar{\mathcal{X}}{\mathcal{X}}-\bar{\mathcal{Y}}{\mathcal{Y}}-\bar{\mathcal{Z}}{\mathcal{Z}})\right)\,,
    \end{equation}
while the analogous multiplets $\mathcal{M}_{(a,b)}$ in the $\mathbb{Z}_k\times \mathbb{Z}_\ell$ family carry two indices $(a,b)$ associated to insertions of $\Omega_X^a\Omega_Y^b$\,.  The currents in these twisted multiplets couple to twisted combinations of $\text{U}(1)$ gauge factors. $\mathcal{M}_0$ vanishes, yielding the expected decoupling of the overall (untwisted) $\text{U}(1)$.

We may now compute the one-loop $\beta$-function of the Abelian $\text{U}(1)$ couplings which is given by \cite{Novikov:1983uc,Novikov:1985rd,Arkani-Hamed:1997qui} (see also \eqref{Eq: beta1} below)
\begin{equation}
    \beta_{U(1)_i}=\frac{g^3}{16\pi^2}\times 3N>0\,,
\end{equation} 
while the non-Abelian $\text{SU}(N)$ components receive another contribution from the gauge-field itself and thus satisfy the conformality condition 
\begin{equation}
    \beta_{\text{SU}(N)_i}=\frac{g^3}{16\pi^2}\times(3N-3N)=0\,.
\end{equation}
This suggests that the $\text{U}(1)$ factors decouple in the IR and we may analyse the IR conformal fixed point in which only an $\text{SU}(N)^\abs{\Gamma}$ gauge group remains. In fact, this fixed point is what is usually considered in the literature as the orbifold CFT without reference to any $\text{U}(1)$ factors that may have been present. For most purposes it is sufficient to just consider this theory as a starting point.

The immediate result of this decoupling is that we have to use different rules for contracting gauge representations. The $\text{SU}(N)$ trace rules are  
\begin{equation}
    (T_A)^{i}{}_j(T^A)^{k}{}_{l}=\delta^i_l\delta^k_j-\frac{1}{N}\delta^i_j\delta^k_l\,,
\end{equation}
where the first term is the standard $\text{U}(N)$ expression and the second term precisely annihilates the $\text{U}(1)$ trace part. 

When computing the one-loop dilatation operator $\Delta$ in Section \ref{sec: orbidila}, we will rely on a spin-chain description in which the one-loop correction to the bare dimension $\Delta_0$ is generated by a nearest-neighbour Hamiltonian $\mathcal{H}_{i,i+1}$ \cite{Beisert:2003jj}, which acts on neighbouring pairs of fields in the single-trace operators. However, this prescription makes implicit use of the $\text{U}(N)$ trace rules, assuming a suppression of the $\text{U}(1)$ part by $\tfrac{1}{N}$ in the large-$N$ limit. Acting for example on the first and second site of an arbitrary state, we would assume
\begin{equation}\begin{split}
    \mathcal{H}_{1,2}\Tr(\Phi_1\Phi_2\Phi_3\dots)=&\Tr(\mathcal{H}_{U(N)}(\Phi_1\Phi_2)\Phi_3\dots)-\frac{1}{N}\Tr(\mathcal{H}_{U(1)}(\Phi_1\Phi_2))\Tr(\Phi_3\dots)\\
    \sim &\Tr(\mathcal{H}_{U(N)}(\Phi_1\Phi_2)\Phi_3\dots)\,,
\end{split}\end{equation}
where $\mathcal{H}_{U(1)}$ denotes the Hamiltonian contribution that arises from an exchanged $\text{U}(1)$ vector multiplet component in $\mathcal{V}_{U(1)_i}$. For most operators, the large-$N$ limit thus makes no distinction between $\text{U}(N)$ and $\text{SU}(N)$ gauge groups. This logic only breaks down for operators at length two, for which the double-trace term comes with an additional $\Tr \mathbb{1} =N$ which precisely cancels the $\tfrac{1}{N}$ factor, e.g.
\begin{equation}\label{Eq: U(1)-contrib}
    \begin{split}
    \mathcal{H}_{1,2}\Tr(\Phi_1\Phi_2)&=\Tr(\mathcal{H}_{U(N)}(\Phi_1\Phi_2))-\frac{\Tr(\mathbb{1})}{N}\Tr(\mathcal{H}_{U(1)}(\Phi_1\Phi_2))\\
   &=\Tr(\mathcal{H}_{U(N)}(\Phi_1\Phi_2))-\Tr(\mathcal{H}_{U(1)}(\Phi_1\Phi_2))\,.\\
    \end{split}
\end{equation}
At length two we thus have to cancel explicitly the
$\text{U}(1)$-exchange processes that would otherwise contribute
to the planar Hamiltonian.\footnote{Related finite-size
effects arising from the distinction between $\text{SU}(N)$ and
$\text{U}(N)$ were analysed in real $\beta$-deformed $\mathcal{N}=4$ SYM, where they were called prewrapping effects \cite{Fokken:2013mza}.}

After decoupling the $\text{U}(1)$ factors, the Hilbert space of single-trace operators is constructed from $\text{SU}(N)$ adjoint vector multiplets and $\text{SU}(N)\times \text{SU}(N)$ bifundamental chiral fields only. Since the $\text{U}(1)$ factors decouple from $\text{SU}(N)$ vector multiplet self-interactions, any nearest-neighbour interaction that involves an $\text{SU}(N)$ vector multiplet field cannot generate intermediate $\text{U}(1)$-fields. Thus the only processes that feature intermediate $\text{U}(1)$-channels arise from chiral-antichiral bilinears such as the schematic process
 \begin{equation}\label{eq: U(1)channel}
\begin{tikzpicture}[baseline={(current bounding box.center)}]

\begin{feynman}

\vertex (g1) at (0,2) {$\Phi_{ij}$};
\vertex (f1) at (0,0) {$\bar{\Phi}_{ji}$};

\vertex (v1) at (0.7,1);
\vertex (v2) at (1.3,1);

\vertex (g2) at (2,2) {$\Phi'_{ik}$};
\vertex (f2) at (2,0) {$\bar{\Phi'}_{ki}$};

\diagram*{
(g1) -- [fermion] (v1),
(v1) -- [fermion] (f1),

(v2) -- [gluon] (v1),

(v2) -- [fermion] (g2),
(f2) -- [fermion] (v2),
};

\end{feynman}
\end{tikzpicture}:\qquad \Tr(\Phi_{ij}\bar{\Phi}_{ji})\to \Tr(\Phi'_{ik}\bar{\Phi}'_{ki})\subset\Tr(\mathcal{H}_{U(1)}(\Phi_{ij}\bar{\Phi}_{ji}))\,,
\end{equation}
where $\Phi_{ij}$, $\Phi'_{ik}$ are two chiral multiplets transforming under the same gauge group $\text{SU}(N)_i$. Accounting for the $\text{SU}(N)$-trace rule \eqref{Eq: U(1)-contrib} requires us to cancel these processes in the one-loop Hamiltonian if and only if it is acting on operators of length two. As we will explain in Section \ref{sec: orbidila}, we have accounted for this eventuality in our Wolfram Mathematica implementation supplemented to the arXiv submission of this paper.

A direct result of this modification is the one-loop protection of the current multiplets $\mathcal{M}_a$ in \eqref{Eq: current multiplets}
\begin{equation}
    \Dm \mathcal{M}_a=(\Delta_0+g_0^2\Hm+\order{g_0^4}) \mathcal{M}_a = 2 \mathcal{M}_a+\order{g_0^4}\,,
\end{equation}
where we point out the absence of a quadratic term in $g_0$, i.e. $\Hm \Mm_a=0$. We thus find $k-1$ ($k\ell-1$) additional $\hat{\Cm}_{(0,0)}$ BPS multiplets, one for each twisted sector in the $\mathbb{Z}_k\,(\mathbb{Z}_k\times\mathbb{Z}_\ell)$ orbifold theory, as discussed above. The appearance of such protected global symmetry current multiplets signals precisely that the $\text{U}(1)$ gauge symmetries of the naive orbifold projection have become classical global symmetries of the theory after decoupling their gauge fields.

\subsubsection{The Konishi anomaly}
\label{Subsub:konishi}

We hasten to add that the double-trace contributions protect the current multiplets $\mathcal{M}_a$ only at one loop. At higher-loops they can still receive anomalous dimensions due to quantum anomalies. We will now show that this in fact is the generic behaviour and only a subset of the current multiplets remain exactly BPS at higher loop orders.
\begin{figure}[b]
\centering

\begin{tikzpicture}[thick]

\coordinate (A) at (180:1);
\coordinate (B) at (60:1);
\coordinate (C) at (-60:1);

\draw[
postaction={decorate},
decoration={
markings,
mark=at position 0.6 with {\arrow{Triangle[length=3mm,width=2mm]}}
}
]
(A)--(B);

\draw[
postaction={decorate},
decoration={
markings,
mark=at position 0.6 with {\arrow{Triangle[length=3mm,width=2mm]}}
}
]
(B)--(C);

\draw[
postaction={decorate},
decoration={
markings,
mark=at position 0.6 with {\arrow{Triangle[length=3mm,width=2mm]}}
}
]
(C)--(A);


\fill (A) circle (2pt);

\draw[coil it] (B) -- +(1,0);
\draw[coil it] (C) -- +(1,0);


\node at (-1.5,0)
{$\Jm^\mu_{i}$};

\node at (1.8,0.8)
{$\Wm_j$};

\node at (1.8,-0.9)
{$\Wm_j$};


\node at (0,0)
{$\Phi_{ij}$};

\node at (5.5,0)
{$\Rightarrow\quad D_\mu \Jm^\mu_i
\propto \sum_j  C_{ij} g_j^2
\mathrm{Tr}(\Wm_j^2)\, .
$};

\end{tikzpicture}

\caption{
Triangle diagram generating the Konishi anomaly for the global
$\text{U}(1)$ currents arising from decoupling.}
\label{fig: anomaly}

\end{figure}
 To that end, let us investigate the triangle diagram in Figure \ref{fig: anomaly}, which generates a Konishi anomaly \cite{Konishi:1983hf,Konishi:1985tu,
Cachazo:2002ry} for the global $\text{U}(1)_i$ symmetry current $\Jm_i^\mu$ in \eqref{Eq: currents} associated to a specific gauge node in the quiver diagram. The Konishi anomaly is the supersymmetric version of the chiral (ABJ) anomaly, arising because $\text{U}(1)_i$ acts chirally on the fermions in the chiral multiplets. Summing over these chiral fermions, we receive an anomalous contribution from every adjacent gauge node, with a sign depending on whether the connecting chiral multiplet is in the fundamental or anti-fundamental representation of $\text{U}(1)_{i}$. This results in an anomaly matrix which is proportional to the adjacency matrix $C_{ij}$ of the quiver, a matrix specifying which gauge nodes are connected and by how many chiral multiplets. For example we can consider the theories in Figure \ref{fig:Zkquiver} and read off the adjacency matrices
\begin{equation}\label{Eq: adjacency}
    C(\mathbb{Z}_3)=\begin{pmatrix}
        0 & 3 & -3\\
        -3 & 0 & 3\\
        3 &-3 & 0
    \end{pmatrix},\,\, C(\mathbb{Z}_4)=\begin{pmatrix}
        0 & 2 & 0 &-2\\
         -2 & 0 & 2 &0\\
         0 & -2 & 0 &2\\
         2 & 0 & -2 &0
    \end{pmatrix},\,\, C(\mathbb{Z}_5)=\begin{pmatrix}
        0 & 2 & -1 & 1 &-2\\
        -2& 0 & 2 & -1 & 1 \\
         1 &-2 & 0 & 2 & -1 \\
         -1 & 1 &-2& 0 & 2 \\
           2 & -1 & 1 &-2 &0
    \end{pmatrix}\,.
\end{equation}
Thus only linear combinations $a^i \Jm_i$ that satisfy $a^iC_{ij}=0$, i.e. sit in the kernel of the adjacency matrix can be exact global symmetry currents satisfying $D_\mu (a^i \Jm_i^\mu)=0$. For the other combinations, the non-conservation equation
\begin{equation}\label{eq: non-cons}
D_\mu \Jm^\mu_i
\propto \sum_j  C_{ij} g_j^2
\mathrm{Tr}(F_j\tilde F_j)\,
\end{equation}
is signalling a recombination of the $\hat{\Cm}_{(0,0)}$ multiplets descending from $\mathcal{M}_i$ with appropriate $\bar{\mathcal{B}}_{2(0,0)}$ \ multiplets to which the operators $\mathrm{Tr}(F_j\tilde F_j)$ belong. This recombination is described at the level of the superconformal algebra by the threshold decomposition \eqref{Eq:RecN13}
\begin{equation}
    \Am^{2}_{0(0,0)} \simeq \hat{\Cm}_{(0,0)} \oplus \Bm_{-2(0,0)} \oplus \bar{\Bm}_{2(0,0)}\, ,
\end{equation}
involving the conjugate marginal deformation as well.
The resulting long multiplets pick up an anomalous dimension at order $\order{g_0^4}$ and are thus not exactly BPS.

The number $n_{\text{ex}}$ of exactly protected global $\text{U}(1)$ symmetry current multiplets $\hat{\mathcal{C}}_{(0,0)}$ depends on the orbifold theory at hand and is determined by the kernel dimension of the associated adjacency matrix. For the two families of $\mathbb{Z}_k[a_X,a_Y,a_Z]$ and $\mathbb{Z}_k\times \mathbb{Z}_\ell$ orbifolds we find, respectively
\begin{align}
    \label{Eq: nrZk}
    n_{\text{ex}}(\mathbb{Z}_k[a_X,a_Y,a_Z]) &= \text{gcd}(a_X,k)+\text{gcd}(a_Y,k)+\text{gcd}(a_Z,k)-3\,,\\
    \label{Eq: nrZkZl}
     n_{\text{ex}}(\mathbb{Z}_k\times\mathbb{Z}_\ell)&= k + \ell + \text{gcd}(k,\ell) -3\,.
\end{align}
These numbers exactly match the numbers \eqref{eq:NZk} and \eqref{eq:NZkZl} of special twisted sectors discussed in Section \ref{sec: strings}. This is no surprise since these global $\text{U}(1)$ symmetry currents are twisted BPS operators, which can only arise in special twisted sectors. In order to match these twisted global symmetry currents to fields in the local  6d EFT on the $AdS_5\times S^1$ fixed point set, we invoke the usual holographic dictionary, which posits that currents in the boundary theory correspond to vector fields in $AdS_5$. Such vector fields can arise from the 10d $C_4$ potential, which wraps both the exceptional two-cycle within the orbifold resolution of $AdS_5\times S^1\times \mathbb{C}^2/\Gamma_{\text{special}}$ and the $S^1$ circle. In the local 6d EFT this is a tensor mode. The exactly marginal deformation operators sitting in the (uncombined) BPS $\mathcal{B}_{-2(0,0)}\oplus \bar{\Bm}_{2(0,0)}$ multiplets on the other hand correspond to the 10d two-form potential $ C_2+i B_2$ wrapping the exceptional two-cycle and are appropriately described by scalar fields in the 6d EFT.

\begin{figure}[h]
\centering
\begin{tikzpicture}[
    vertex/.style={circle, draw, minimum size=6mm,thick},
    arrow/.style={->, thick},
    scale=0.8
]
\def\off{2.5pt}
\def\offd{1.7pt}

\node[vertex] (p1) at (0,0) {4};
\node[vertex] (p2) at (3,0) {3};
\node[vertex] (p3) at (3,3) {2};
\node[vertex] (p4) at (0,3) {1};

\draw[red, dashed, very thick, rotate around={45:(1.5,1.5)}]
    (1.5,1.5) ellipse [x radius=2.8, y radius=1];

\draw[arrow, transform canvas={yshift=\off}] (p2) -- (p1);
\draw[arrow, transform canvas={yshift=-\off}] (p2) -- (p1);

\draw[arrow, transform canvas={yshift=\off}] (p4) -- (p3);
\draw[arrow, transform canvas={yshift=-\off}] (p4) -- (p3);

\draw[arrow, transform canvas={xshift=-\off}] (p3) -- (p2);
\draw[arrow, transform canvas={xshift=\off}] (p3) -- (p2);

\draw[arrow, transform canvas={xshift=-\off}] (p1) -- (p4);
\draw[arrow, transform canvas={xshift=\off}] (p1) -- (p4);

\draw[arrow, transform canvas={xshift=-\offd,yshift=\offd}] (p3) -- (p1);
\draw[arrow, transform canvas={xshift=\offd,yshift=-\offd}] (p1) -- (p3);

\draw[arrow, transform canvas={xshift=-\offd,yshift=-\offd}] (p4) -- (p2);
\draw[arrow, transform canvas={xshift=\offd,yshift=\offd}] (p2) -- (p4);
\end{tikzpicture}
\caption{Balanced partition of the $\mathbb{Z}_4[1,1,2]$ orbifold quiver. One may check node-by-node that the arrows crossing the boundary balance. Thus the combined global $\text{U}(1)$ symmetry of the gauge nodes $2$ and $4$ is preserved and similar for nodes $1$ and $3$. If, on the other hand, we were to single out an individual gauge node, the resulting partition would be unbalanced.
\label{Fig:partition}}
\end{figure}
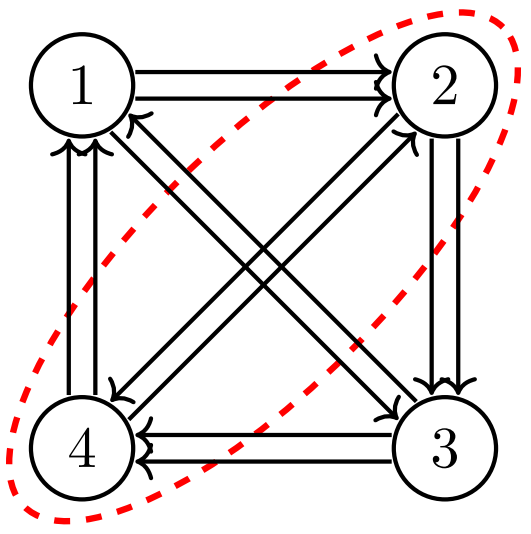

Let us close this section with a few comments regarding these exactly BPS global symmetry currents and the Konishi anomaly:
\begin{itemize}
    \item Note that the vector $a^i=1\,,\,\, \forall i\,, $ which is always in the kernel of $C_{ij}$ does not generate a current multiplet since $a^i\Jm_i=0$. This vector corresponds to the  untwisted overall $\text{U}(1)$ that already decouples completely in $\mathcal{N}=4$ SYM. 
   \item In the following sections, we restrict to approximate one-loop results, e.g.\ for
the anomalous dimensions. In contrast, the Konishi anomaly is one-loop
exact in the holomorphic scheme and receives no higher-order corrections. See also \cite{Green:2010da} for the link to exact
marginality and \cite{Aharony:2002tp} for higher-order analyses.
    \item The Konishi anomaly suggests that the non-conformal orbifold theory before decoupling of $\text{U}(1)$ factors requires an additional anomaly cancellation mechanism in order to remain a consistent quantum field theory. Although it would be interesting to investigate this mechanism further, such considerations go beyond the scope of the current paper and we refer the reader to a related discussion in \cite{Ibanez:1998qp}.
    \item \textbf{Intuitive picture:} If we partition the gauge nodes into two subsets, we say that the arrows \emph{balance} at a node if the node receives as many arrows from the \emph{other} subset as it sends to it. Balance at every node is precisely the absence of a Konishi anomaly for the combined global $\text{U}(1)$ current of either subset: this current is then exactly conserved and its multiplet does not recombine with the associated marginal deformation multiplet. As we will see in Section \ref{sec: gauge manifold}, this will allow us to vary the gauge couplings of one subset relative to those of the other while staying on the conformal manifold. Conversely, if we single out a set of gauge nodes for which the arrows do not balance, the corresponding $\text{U}(1)$ current is anomalous and recombines with a would-be marginal deformation multiplet. This poses a quantum obstruction to varying their gauge couplings relative to the rest of the quiver. An example of a balanced partition in the $\mathbb{Z}_4[1,1,2]$ orbifold theory is given in Figure \ref{Fig:partition}.

\end{itemize}

\section{The conformal manifold}\label{sec: marginal}

Having established the basic characteristics of Abelian $\mathcal{N}=1$ orbifolds of $\mathcal{N}=4$ SYM, we now investigate deformations of these theories that keep their superconformal structure intact. The parameter space of such exactly marginal deformations is referred to as the conformal manifold of the respective theory.\footnote{For theories obtained by compactifying six-dimensional $(1,0)$ SCFTs, conformal-manifold data can receive contributions from both the complex structure of the compactification surface and background flavour-bundle data \cite{Razamat:2016dpl}.}
At a given superconformal fixed point, supersymmetric marginal
deformations are generated by operators $\mathcal{O}_{\text{marg.}}$
in short $\bar\Bm_{2(0,0)}$ multiplets and their associated
conjugate multiplets $\Bm_{-2(0,0)}$
\cite{Cordova:2016xhm,Rastelli:2016tbz,Cordova:2016emh},
\begin{equation}
    \mathcal{L}\quad \to \quad
    \mathcal{L}
    +\lambda \int \dd^2\theta\, \mathcal{O}_{\text{marg.}}
    +\bar\lambda\int \dd^2\bar\theta \,
    \bar{\mathcal{O}}_{\text{marg.}}.
\end{equation}
The existence of such a marginal operator does not, however, guarantee that the associated deformation is exactly marginal.

In particular, these deformations may not be independent, as they can be related by global symmetries, which in turn are generated by conserved currents in short $\hat{\Cm}_{(0,0)}$ multiplets. Turning on a marginal deformation may explicitly break the global symmetries under which the corresponding operator is charged. As a result, the associated currents cease to be conserved and pick up an anomalous dimension. This is only possible if they become part of long multiplets, which they do by absorbing the corresponding marginal deformation operators through the recombination
\begin{align}\label{Eq: deformationrecombination}
    \Am^{2}_{0,(0,0)}
    \simeq
    \hat{\Cm}_{(0,0)}
    \oplus
    \bar\Bm_{2(0,0)}
    \oplus
    \Bm_{-2(0,0)}\,.
\end{align}
This recombination mechanism relates the number of exactly marginal directions to the pattern of preserved and broken global symmetries. In particular, at symmetry-enhancement loci on the conformal manifold additional current and marginal multiplets may become short, while at a generic point some of them recombine into long multiplets.

The index \eqref{Eq:Index} provides a first rough estimate of the possible recombinations of this type. The coefficient at order $t^2$ in its expansion counts with opposite sign the net number \cite{Rastelli:2016tbz}
\begin{equation}\label{eq:Net degeneracy}
    \text{ND}[0,0]
    =
    \#\bigl(\hat{\Cm}_{(0,0)}\bigr)
    -
    \#\bigl(\bar\Bm_{2(0,0)}\bigr)\,,
\end{equation}
namely, the number of conserved-current multiplets minus the number of marginal chiral multiplets.
However, this net count does not determine the dimension of the conformal manifold, since not all kinematically allowed recombinations necessarily occur.\footnote{As we will see in Section \ref{sec: special} below, the net degeneracy $-\text{ND}[0,0]$ counts the number of $\mathbb{Z}_3$ orbits in the quiver gauge theory and can be associated to special marginal BPS multiplets.} In particular, conserved currents associated with flavour symmetries that remain unbroken along the conformal manifold must be accounted for separately. Moreover, some marginal chiral operators may fail to generate exactly marginal directions altogether. A complete characterisation of the conformal manifold therefore requires a more careful analysis. As explained in \cite{Green:2010da} the conformal manifold may be described as a K\"ahler quotient of the space of marginal couplings by the complexified flavour symmetry group 

\begin{equation}
    \Mm_\Cm
    \simeq
    \frac{\{\lambda^i\}}{G_F^{\mathbb{C}}}\,,
    \label{Eq:complexified_flavour_quotient}
\end{equation}
where the local marginal coupling parametrisation has to be chosen such that the quotient is non-trivial.

\subsection{Generic marginal deformations}\label{sec: generic}

We will first consider marginal deformations and global symmetries generic to all Abelian $\mathcal{N}=1$ orbifold theories discussed in Section \ref{sec: orbifolds}. We then perform a $\beta$-function analysis in order to determine which one-loop protected $\bar\Bm_{2(0,0)}$ operators generate exact marginal deformations. In certain orbifold theories additional superpotential deformations and enhanced global symmetries can appear, as we will discuss in the next section.

\subsubsection{Candidate multiplets around the orbifold point}\label{sec: one-loop}

As a starting point, let us list the generic $\bar\Bm_{2(0,0)}$ and $\hat{\Cm}_{(0,0)}$ multiplets present in the orbifold theories discussed in the previous section. We list operators that appear to be protected at one-loop, as can be checked by a Hamiltonian analysis according to Section \ref{sec: orbidila}. These multiplets provide us with a set of candidates for exactly marginal deformations and preserved global symmetries, but could still participate in recombination if they pick up an anomalous dimension at higher-loop order. In $\mathbb{Z}_k[a_X,a_Y,a_Z]$ orbifold theories we generically find the following one-loop protected operators
\begin{itemize}
    \item $k\times\bar\Bm_{2(0,0)}$ multiplets with highest-weight operators
    \begin{equation}\label{Eq: gauge-deformations}
        \mathcal{G}_a=\Tr (\Omega^a \lambda^\alpha\lambda_\alpha)\,,\qquad a\in\{0,\dots,k-1\}\,.
    \end{equation}
        The untwisted operator $\mathcal{G}_0$ generates a change of the overall gauge coupling $g_0$, while the twisted multiplets generate relative deformations of the individual gauge-coupling constants $g_i$ within the quiver gauge theory. In the previous section we saw that some of these multiplets participate in recombination with the $\mathcal{M}_a$ global symmetry current multiplets in \eqref{Eq: current multiplets} and are thus not exactly marginal.
        
    \item$1\times\bar\Bm_{2(0,0)}$ multiplet with highest-weight operator
    \begin{equation}\label{Eq: LS}
        \mathcal{F}_{XYZ}=\Tr (XYZ)+\Tr (XZY)\,.
    \end{equation}
        This untwisted operator generates the Leigh-Strassler deformation \cite{Leigh:1995ep} of the superpotential \eqref{Eq: superpotential}.
    \item$(k-1)\times \hat{\Cm}_{(0,0)}$ multiplets with highest-weight operators 
    \begin{equation}
    \mathcal{M}_a=\Tr \left(\Omega^a(X\bar X+Y\bar Y+Z\bar Z-\bar X X-\bar YY-\bar ZZ)\right)\,,\qquad a\in\{1,\dots,k-1\}\,,
    \label{Eq:CH0_dec}
    \end{equation}
    which arise after decoupling the non-conformal $\text{U}(1)$ vector multiplets as discussed in Section \ref{sec: U(1)s}. The associated $\text{U}(1)^{k-1}$ gauge group is promoted to a global symmetry for which the $\Mm_a$ multiplets supply the conserved charges. Some (or all) of these symmetries are broken by the Konishi anomaly in Figure \ref{fig: anomaly} and recombine with their associated $\mathcal{G}_a$ multiplets \eqref{Eq: gauge-deformations}, picking up an anomalous dimension at two-loop level. The number $n_{\text{ex}}$ of exactly BPS current multiplets unaffected by the anomaly was given in equations \eqref{Eq: nrZk} and \eqref{Eq: nrZkZl}.
    
    \item $2\times \hat{\Cm}_{(0,0)}$ multiplets with highest-weight operators
    \begin{equation}
        \mathcal{J}_{(\alpha)}= \Tr(X\bar{X})- \Tr(Y\bar{Y})\,,\qquad \mathcal{J}_{(\beta)}= \frac{1}{2}\Tr(X\bar{X})+\frac{1}{2}\Tr(Y\bar{Y})- \Tr(Z\bar{Z})\,.
         \label{Eq:CH0_ab}
    \end{equation}
    Within these untwisted multiplets sit the conserved currents of two global $\text{U}(1)$ symmetries under which the chiral multiplets $\Xm,\Ym,\Zm$ have charges $(1,\tfrac{1}{2})$, $(-1,\tfrac{1}{2})$ and $(0,-1)$, respectively.
 \end{itemize}
In the untwisted sector of $\mathbb{Z}_k\times \mathbb{Z}_\ell$ orbifold theories we also find the Leigh-Strassler $\bar\Bm_{2(0,0)}$ multiplet generated by $\Fm_{XYZ}$ as well as the global $\text{U}(1)_\alpha\times \text{U}(1)_\beta$ current multiplets $\hat{\Cm}_{(0,0)}$ generated by $\Jm_{(\alpha)}$ and $\Jm_{(\beta)}$. For the other candidate multiplets we have to take into account the labelling of twisted sectors by two separate twist matrices $\Omega_X^a,\Omega_Y^b$. We find $k\ell\times\bar\Bm_{2(0,0)}$ multiplets generated by $\mathcal{G}_{a,b}$ and also $(k\ell-1)\times\hat{\Cm}_{(0,0)}$ multiplets generated by $\Mm_{a,b}$. 

We see that the generically present $\bar\Bm_{2(0,0)}$ and $\hat{\Cm}_{(0,0)}$ multiplets appear in equal numbers $(\abs{\Gamma}+1)$ and therefore generate no net-degeneracy $\text{ND}[0,0]$ \eqref{eq:Net degeneracy}. Any non-trivial value of $\text{ND}[0,0]$ is thus due entirely to special BPS operators to be discussed in Section \ref{sec: special} below.

\subsubsection{Beta-function analysis and parametrisation} \label{sec: gauge manifold}

We now want to determine which of the candidate marginal deformations extend to exactly marginal deformations spanning the conformal manifold and how we can parametrise this manifold, at least in some coordinate patch around the orbifold point. We will start this discussion agnostic of the recombination effects due to the Konishi anomaly discussed in Section \ref{sec: U(1)s} and only require vanishing $\beta$-functions for gauge-coupling deformations. We will find the dimension of the thus generated conformal submanifold to match the expected $n_{\text{ex.}}+1$ from \eqref{Eq: nrZk} and \eqref{Eq: nrZkZl}, which serves as an internal consistency check. 

Deforming the Lagrangian by the $\mathcal{G}_a$-multiplets \eqref{Eq: gauge-deformations} changes the individual coupling constants $g_i$ of the different gauge nodes in the quiver gauge theory away from a uniform value $g_0$. We first define an overall gauge coupling strength 
\begin{equation}
    g_0^2:=\frac{1}{\abs{\Gamma}}\sum_{i\in\mathcal{A}} g_i^2\,,
    \end{equation}
where $\mathcal{A}$ is the set of indices labelling the gauge groups in the theory at hand. The relative coupling constants may be parameterised by the homogeneous ratios 
\begin{equation}\label{eq:kappa}
    \kappa_i=\frac{g_i}{g_0}\in[0,\sqrt{\abs{\Gamma}}]\,, \qquad \sum_i \kappa^2_i=\abs{\Gamma}\,.
\end{equation}
We also introduce a generic parametrisation for the superpotential  
\begin{equation}\label{Eq: superpot}
    W=g_0\sum_{i\in\mathcal{A}} \left(f^+_i\Tr_{i}(\Xm\Ym\Zm)-f^-_i\Tr_{i}(\Xm\Zm\Ym)\right)\,.
\end{equation}
Here we only considered terms that were present in the original orbifold theory. As we will see, the coefficients $f_i^\pm$ depart from their orbifold-point value $f_i^\pm=1$ once relative gauge coupling deformations $\mathcal{G}_a$ or the Leigh-Strassler deformation $\Fm_{XYZ}$ are turned on. In the special cases of $\mathbb{Z}_3[1,1,1]$ and $\mathbb{Z}_3\times\mathbb{Z}_3$ orbifolds discussed in Section \ref{sec: special} below, additional terms may be added to this superpotential, but here we will restrict our attention to the generic subspace of the conformal manifold.

At the orbifold point $\kappa_i=f^\pm_i=1$
the gauge theory is conformal after the decoupling of the $\text{U}(1)$ factors we discussed in Section \ref{sec: U(1)s}. This may be confirmed explicitly by computing the $\beta$-functions of the gauge and superpotential couplings $g_i$ and $f_i^\pm$, which all vanish. After deforming the theory, we need to ensure that the $\beta$-functions remain unperturbed. Let us investigate the resulting conformality constraints at leading order in $g_0^2$, referring to \cite{Aharony:2002tp} for a discussion at higher-loop levels. 

The gauge-coupling beta functions are given by the NSVZ equation \cite{Novikov:1983uc,Novikov:1985rd}
\begin{equation}\label{Eq: beta1}
    \beta_g=-\frac{g^3}{16\pi^2}\frac{3 T(\text{adj.})-\sum_i T(R_i)(1-\gamma_i)}{1-\frac{g^2}{8\pi^2}T(\text{adj.})}\,,
\end{equation} 
which is exact in a particular
renormalisation scheme \cite{Arkani-Hamed:1997qui} that we shall apply in the following. Here, we sum over all multiplets in a non-trivial representation $R_i$ of the respective gauge group. $T(R_i)$ is the Dynkin index of the gauge representation which enters the trace form $\Tr(T^A_iT^B_i)=T(R_i)\delta^{AB}$. For $\text{SU}(N)$, the Dynkin indices of the adjoint and {(anti-)fundamental} representations are $T(\text{adj.})=N$ and $T(\text{fund.})=\tfrac{1}{2}$, respectively. In addition, we impose conformality of the superpotential couplings $f^\pm$. Due to the non-renormalisation of the superpotential, the associated beta function is simply given by the anomalous dimensions of the coupled chiral multiplets 
\begin{equation}\label{Eq: beta2}
    \beta_{f}=\frac{f}{2}\sum_i \gamma_i\,.
\end{equation}
The $\gamma_i$ denote the anomalous dimension of the chiral multiplets interacting with the respective gauge node or superpotential term. At one loop (i.e. first order in $g_0^2$), these anomalous dimensions are given by 
\begin{equation}\label{Eq:anomdims2} 
    \gamma_i=\frac{g_0^2}{4\pi^2}\left(-\sum_{A\perp i} \kappa ^2_AC_2(R_i^A)+\frac{N}{2}\sum_{B \perp i}\abs{f_B}^2\right)\underset{N\gg 1}{\longrightarrow}\frac{g_0^2N}{8\pi^2}\left(-\sum_{A \perp i} \kappa ^2_A+\sum_{B \perp i}\abs{f_B}^2\right)\,,
\end{equation}
where $\perp$ indicates adjacency. The first term in this expression represents contributions from the gauge representation of the chiral multiplet in question and is proportional to the quadratic Casimir $C_2$, which takes the value $\tfrac{N^2-1}{2N}$ in the fundamental representation of $\text{SU}(N)$. The second term represents the superpotential couplings involving the multiplet, which can be read off \eqref{Eq: superpot}.\footnote{Note here that an individual chiral multiplet enters a gauge trace over its own and one further gauge index, thus resulting in $N$ equal contributions from each term in \eqref{Eq: superpot}.} 
Vanishing of the $\beta$-functions \eqref{Eq: beta1} and \eqref{Eq: beta2} simply implies cancellation of anomalous dimensions around every node and closed $(X,Y,Z)$ triangle in the associated quiver diagram (e.g. Figures \ref{fig:Zkquiver} and \ref{fig:ZkZlquiver}). Multiplying \eqref{Eq:anomdims2} by $\gamma_i$ and summing over all chiral multiplets, we can use this cancellation to argue
\begin{equation}
    \sum_i\gamma_i^2\sim -\sum_A \kappa ^2_A\underbrace{\sum_{i\perp A} \gamma_i}_{=0}+\sum_{B}\abs{f_B}^2 \underbrace{\sum_{i\perp B} \gamma_i}_{=0}=0\,.
\end{equation}
Unitarity now enforces that all anomalous dimensions vanish at one-loop order
\begin{equation}
    \gamma_i=0\,.
\end{equation}
The vanishing of anomalous dimensions at one-loop was also observed in \cite{Aharony:2002tp}. It imposes constraints on the possible combinations of $\kappa_i$ and $f^\pm_i$, which we can solve algorithmically for the two families of orbifold theories:

\paragraph{The $\mathbb{Z}_k[a_X,a_Y,a_Z]$ orbifold family} The gauge nodes can be labelled by integers $\mathcal{A}=\{1,\dots k\}$. Vanishing of anomalous dimensions $\gamma_X$, $\gamma_Y$ and $\gamma_Z$ results in the equations
\begin{align}
\kappa_{i}^2+\kappa_{i+a_X}^2&=|f^+_{i}|^2+|f^-_{i}|^2\,,\label{Eq: Xfield}\\
\kappa_{i}^2+\kappa_{i+a_Y}^2&=|f^+_{i-a_X}|^2+|f^-_{i+a_Y}|^2\,,\label{Eq: Yfield}\\
\kappa_{i}^2+\kappa_{i+a_Z}^2&=|f^+_{i+a_Z}|^2+|f^-_{i-a_X}|^2\,.\label{Eq: Zfield}
\end{align}
Solving the constraints \eqref{Eq: Xfield} and $\eqref{Eq: Yfield}$ for $f^-$ and $f^+$ we end up with
\begin{equation}
    f^{+}_i=\sqrt{\kappa_{i+a_X}^2+c_{i\text{ mod }\text{gcd}(a_Z,k)}}\,,\qquad f^{-}_i=\sqrt{\kappa_{i}^2-c_{i\text{ mod }\text{gcd}(a_Z,k)}}\,,
    \label{Eq:f+f-_beta}
\end{equation}
which feature $\text{gcd}(a_Z,k)$ unfixed parameters $c_{i}$. The square roots arise since the equations \eqref{Eq: Xfield}-\eqref{Eq: Zfield} define circles and relationships between their radii. For each pair of superpotential couplings $f^\pm$ we could likewise introduce a polar parametrisation of the schematic form 
\begin{equation}\label{eq: angle form}
    f^+=\kappa \cos \phi\,,\qquad f^-=\kappa \sin \phi\,,
\end{equation}
but this would in turn complicate the other couplings, so we stick with the square root parametrisation for the time being. 

For coprime $a_Z$ and $k$ we only have one deformation parameter $c_0$ and the remaining equations \eqref{Eq: Zfield} result in
\begin{equation}
    \kappa_{i}^2+\kappa_{i+a_X+a_Y}^2=\kappa_{i+a_Y}^2+\kappa_{i+a_X}^2\,.
\end{equation}
If furthermore $\text{gcd}(a_X,k)=\text{gcd}(a_Y,k)=1$ this fixes all couplings to equal values $\kappa_i=1$ and leaves behind only two deformation parameters $\{c_0,g_0\}$. In case we have $\text{gcd}(a_Y,k)>1$ these equations decompose into $\text{gcd}(a_Y,k)$ sectors, in which the gauge couplings can be chosen independently. The same effect occurs with $\text{gcd}(a_X,k)>1$. If $\text{gcd}(a_Z,k)>1$, the additional $c_i$ factors entering \eqref{Eq: Zfield} can likewise be used to reduce the number of constraints imposed on the gauge couplings. The overall count of free gauge coupling parameters then boils down to
\begin{equation}
    \dim \mathcal{M}_{k,\text{ gauge}}^{a_X,a_Y} = \text{gcd}(a_X,k)+\text{gcd}(a_Y,k)+\text{gcd}(a_Z,k)-2=n_{\text{ex}}+1\,.
\end{equation}
This reproduces the number \eqref{Eq: nrZk} of exactly marginal $\bar\Bm_{2(0,0)}$ multiplets in addition to the overall gauge coupling deformation $\Gm_0$. On the way, we have also determined an explicit parametrisation of the superpotential couplings $f^\pm$ as functions of the gauge couplings and an additional deformation parameter $c_0$ which we identify with a global Leigh--Strassler-type deformation
\begin{equation}
    c_0:=h\,,
\end{equation}
 that only affects the superpotential and raises the dimension of the conformal manifold to 
\begin{equation}\label{Eq: dimZk}
    \dim \mathcal{M}_{k}^{a_X,a_Y} =\text{gcd}(a_X,k)+\text{gcd}(a_Y,k)+\text{gcd}(a_Z,k)-1= n_{\text{ex}}+2\,.
\end{equation}
This parametrisation covers a real slice of the conformal manifold which is limited to $\abs{\LS}< \min(\kappa_i^2)$, where a separate coordinate patch is required to smoothly connect to a different square-root branch of the superpotential couplings $f^\pm$ in \eqref{Eq:f+f-_beta}. The resulting sign flips of the superpotential couplings $f^\pm$ may be absorbed in field-redefinitions and do not affect the spectrum of the theory. We will discuss an explicit example of this patching in \textbf{Part II} \cite{PartII}. For $\mathbb{Z}_k$ orbifold theories with up to nine gauge nodes, we worked out the coupling constant dependencies of $g_i$ and $f_i^\pm$ in Appendix \ref{sec: Zkexamples}.

\paragraph{The $\mathbb{Z}_k\times \mathbb{Z}_\ell$ orbifold family}   \quad Here the gauge nodes are labelled by two integers $\mathcal{A}=\{(1,1),\dots,(k,\ell)\}$. The vanishing of anomalous dimensions $\gamma_X$, $\gamma_Y$ and $\gamma_Z$ results in the constraints
\begin{align}
\kappa_{i,j}^2+\kappa_{i+1,j}^2&=|f_{i,j}^+|^2+|f_{i,j}^-|^2\,,\label{Eq: X-field2}\\
\kappa_{i,j}^2+\kappa_{i,j+1}^2&=|f_{i-1,j}^+|^2+|f_{i,j+1}^-|^2\,,\label{Eq: Y-field2}\\
\kappa_{i,j}^2+\kappa_{i+1,j+1}^2 &=|f_{i,j}^+|^2+|f_{i,j+1}^-|^2\,.\label{Eq: Z-field2}
\end{align}
We note an immediate restriction on the (real) gauge couplings which can be read off the linear combination \eqref{Eq: X-field2}+\eqref{Eq: Y-field2}--\eqref{Eq: Z-field2}
\begin{equation}\begin{split}\label{eq:obstruction}
    \kappa_{i,j}^2+\kappa_{i+1,j}^2+\kappa_{i,j+1}^2 -\kappa_{i+1,j+1}^2&=|f_{i-1,j}^+|^2+|f_{i,j}^-|^2\geq0 \\
   \Rightarrow  \quad \kappa_{i+1,j+1}^2&\leq\kappa_{i,j}^2+\kappa_{i+1,j}^2+\kappa_{i,j+1}^2 \,.
\end{split}\end{equation}
From this follows, for example, that no single gauge coupling can be parametrically larger than all other couplings in the quiver. As we will discuss in \textbf{Part II} \cite{PartII} this is an obstruction to obtaining  $\mathcal{N}=1$ SQCD as decoupling limit of $\mathbb{Z}_k\times \mathbb{Z}_\ell$ orbifold theories.

From the first two sets of constraints \eqref{Eq: X-field2}, \eqref{Eq: Y-field2} we may deduce a parametrisation for the superpotential couplings of the form
\begin{equation}\label{eq: square root ZkZl}
    f_{i,j}^{+}=\sqrt{\kappa_{i+1,j}^2+c_{(i-j)\text{ mod }\text{gcd}(k,\ell)}}\,,\quad f_{i,j}^{-}=\sqrt{\kappa_{i,j}^2-c_{(i-j)\text{ mod }\text{gcd}(k,\ell)}}\,,
\end{equation}
which feature $\text{gcd}(k,\ell)$ unfixed parameters $c_{i}$. We first consider the case of coprime $k,\ell$ where only one constant $c_0$ appears and the remaining set \eqref{Eq: Z-field2} of constraints take the form
\begin{equation}
 \kappa_{i,j}^2+\kappa_{i+1,j+1}^2=\kappa_{i+1,j}^2+\kappa_{i,j+1}^2\,.
 \end{equation}
These restrict the gauge couplings such that they have to balance across each rectangular cell in Figure \ref{fig:ZkZlquiver}. We may solve these equations to fix $(k-1)(\ell-1)$ couplings in terms of $k + \ell -1$ arbitrary gauge couplings. A possible choice for the conformal manifold coordinates is thus the set $\{c_0, g_0, \kappa_{1,2}\,\dots, \kappa_{1,\ell}, \kappa_{2,1},\dots \kappa_{k,1}\}$ with $k+\ell$ elements. 

Let us now consider the case where $\text{gcd}(k,\ell)>1$. We gain additional $c_i$ parameters which enter \eqref{Eq: Z-field2}. We may thus use the constraints \eqref{Eq: Z-field2} to determine these $c_i$ parameters at the benefit of introducing additional unfixed gauge coupling parameters. This adds another $\text{gcd}(k,\ell)-1$ dimensions to the conformal manifold, resulting in a total of 
\begin{equation}
    \dim \mathcal{M}_{k,\ell, \text{ gauge}}= k + \ell + \text{gcd}(k,\ell) -2=n_{\text{ex}}+1
\end{equation}
gauge coupling constants, again in agreement with \eqref{Eq: nrZkZl}. We again find an additional Leigh-Strassler deformation parameter $c_0:=\LS$, which only affects the superpotential and raises the conformal manifold dimension to
\begin{equation}\label{Eq: dimZkZl}
    \dim\Mm_{k,\ell} =k + \ell + \text{gcd}(k,\ell) -1= n_{\text{ex}}+2\,.
\end{equation}
For $\mathbb{Z}_k\times \mathbb{Z}_\ell$ orbifold theories with up to nine gauge nodes, we worked out the coupling constant dependencies of $g_i$ and $f_i^\pm$ in Appendix \ref{sec: ZkZlexamples}. Again we would like to qualify that this parametrisation holds on the real patch up to the branch points of the various square-roots in \eqref{eq: square root ZkZl}, where a coordinate change becomes necessary.

We have now determined the dimension and explicit parametrisation of the conformal manifolds of generic $\mathcal{N}=1$ orbifold theories in the $\mathbb{Z}_k[a_X,a_Y,a_Z]$ and $\mathbb{Z}_k\times \mathbb{Z}_\ell$ families introduced in Section \ref{sec: orbifolds}. The number of gauge coupling deformations could be matched explicitly to the Konishi anomaly discussion in Section \ref{sec: U(1)s}. The additional generic Leigh-Strassler deformation $\LS$ generated by $\mathcal{F}_{XYZ}$ in \eqref{Eq: LS} was taken into account. We will make use of this parametrisation when studying the one-loop spectrum of the marginally deformed theories through the construction of the one-loop dilatation operator in Section~\ref{sec: orbidila}. 

The preceding analysis determines the dimension of the generic conformal manifold, but does not exhaust the candidate marginal operators present for special orbifold configurations. In particular, certain orbifold actions admit additional gauge-invariant cubic operators. As we will show in Section \ref{sec: special}, most of these additional operators are obstructed from generating exactly marginal directions by the flavour-symmetry quotient \eqref{Eq:complexified_flavour_quotient}. An enlargement of the conformal manifold occurs only for the $\mathbb{Z}_3[1,1,1]$ and $\mathbb{Z}_3\times\mathbb{Z}_3$ theories. Before moving on to these, however, let us close our discussion of beta functions by considering the immediate environment of the conformal manifold in the wider moduli space of non-conformal theories.

\paragraph{Conformal manifold as attractor} One may wonder what would happen if we disregard the conformality constraints and just choose to change an individual gauge coupling by hand. This deformation generally breaks the conformal symmetry and induces a renormalisation group flow, which we should follow to the IR. For a small departure from conformality we observe that this renormalisation group (RG) flow brings us back to the conformal manifold. 

Let us for simplicity consider the $\mathbb{Z}_4[1,1,2]$ orbifold theory with four gauge couplings $g_1,\dots, g_4$. On the conformal manifold we require $g_3=g_1$ and $g_4=g_2$ but we may now for example choose to perturb $g_3$ as 
\begin{equation}
    g_3=g_1+\delta g_3\,.
\label{Eq:Beta_Attractor}
\end{equation}
This perturbation generates anomalous dimensions for the chiral multiplets
\begin{equation}
\gamma(X_{23})=\gamma(X_{34})=\gamma(Y_{23})=\gamma(Y_{34})=\gamma(Z_{13})=\gamma(Z_{31})\sim-\frac{N}{4\pi^2} g_1 \delta g_3\,,
\end{equation}
which in turn result in the beta functions
\begin{equation}\label{eq:beta-function}
    \beta_{g_1}\sim \frac{g_1^4N^2}{64\pi^4}\delta g_3\,,\qquad \beta_{g_2}=\beta_{g_4}\sim \frac{g_2^3g_1N^2}{64\pi^4}\delta g_3\,,\qquad \beta_{g_3}\sim 3 \times  \frac{g_1^4N^2}{64\pi^4}\delta g_3\,.
\end{equation}

\begin{figure}[t]
    \centering
    \begin{tikzpicture}[
        scale=0.8,
        >=Latex,
        flow/.style={->, black, thick},
        fixedline/.style={red!75!black, very thick},
        every node/.style={font=\small}
    ]

        \draw[->, thick]
            (0,0) -- (5.4,0)
            node[right] {$g_1$};

        \draw[->, thick]
            (0,0) -- (0,5.4)
            node[above] {$g_3$};

        \draw[fixedline]
            (0,0) -- (5,5);

        \foreach \x in {1.4,2.7,4.0}{
            \draw[flow]
                ({\x+0.35},{\x+1.05}) -- (\x,\x);
        }

        \foreach \x in {1.4,2.7,4.0}{
            \draw[flow]
                ({\x-0.35},{\x-1.05}) -- (\x,\x);
        }

        \node[red!75!black, rotate=45] at (0.7,1)
            {CFT };

    \end{tikzpicture}
    \caption{
        Schematic RG flow in the $(g_1,g_3)$ plane of the $\mathbb{Z}_4[1,1,2]$ orbifold theory. The red diagonal
        denotes the CFT fixed line at $g_1=g_3$. The RG flow around this line is determined by \eqref{eq:beta-function} and drives the theory towards the CFT fixed line in the IR.
    }
    \label{fig:rg-flow}
\end{figure}
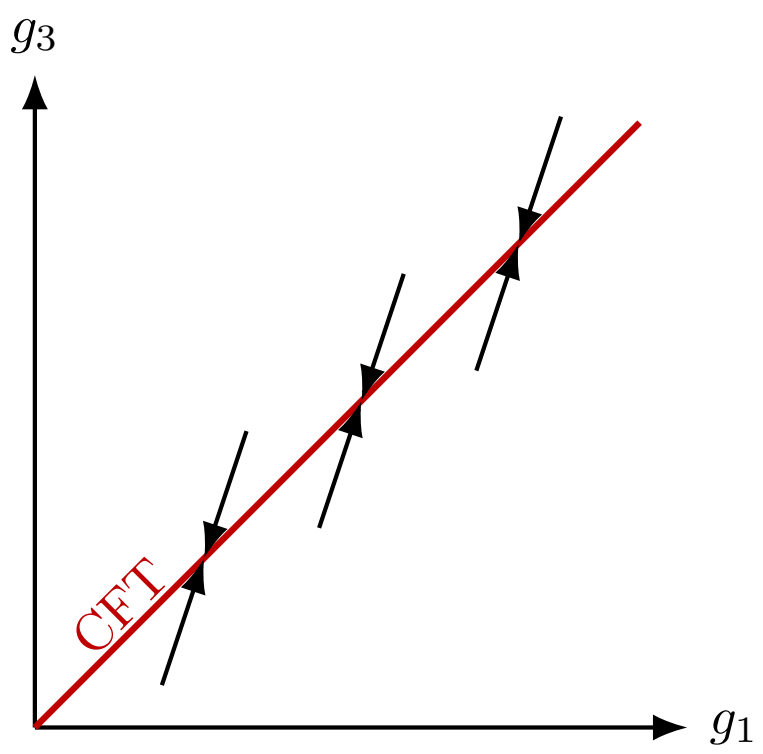
If $\delta g_3$ was positive, the RG flow reduces the coupling constants but crucially $g_3$ is shrinking faster than $g_1$ bringing us back to the conformal manifold $g_1=g_3$. Similarly, if $\delta g_3$ was negative, the RG flow increases the coupling constants, but again $g_3$ grows quicker than $g_1$, bringing us back to the conformal manifold again. We depicted these RG-flows close to the CFT fixed line in Figure \ref{fig:rg-flow}. We conclude that the conformal manifold is a local attractor of the RG flow and stable under small deformations of gauge couplings. Thus imposing $g_1=g_3$ is not a fine-tuned condition but a local minimum of the ``moduli space'' of couplings. This matches the discussion in \cite{Green:2010da}, which showed that in general 4d $\mathcal{N}=1$ SCFTs, non-exactly marginal deformations recombine with flavour currents to give irrelevant operators. They then cannot drive the theory to another fixed point under RG flow. Of course, this perturbative argument may fail for finite deformations, e.g. $\delta g_3 \sim g_1$, where we would have to employ seriously non-perturbative technology.

\subsection{Special cases}\label{sec: special}

We now turn to orbifold theories in which the gauge structure allows for the construction of additional cubic scalar operators which serve as candidate primaries for $\bar{\Bm}_{2(0,0)}$ multiplets. At one-loop these operators are protected, but the decision whether they remain exactly marginal at higher-loop orders requires a more careful analysis. We will see that in many cases, a simultaneous symmetry enhancement generates further $\hat{\Cm}_{(0,0)}$ multiplets that can recombine with the special $\bar{\Bm}_{2(0,0)}$ multiplets and render them irrelevant. Furthermore, even if the $\bar{\Bm}_{2(0,0)}$ multiplets are protected at all loop-orders, they may not generate a marginal deformation of the theory, since they are obstructed by the triviality of the associated flavour-symmetry quotient \eqref{Eq:complexified_flavour_quotient}. This means their explicit realisation does not satisfy the beta-function constraints and thus does not define a proper marginal direction.

We classify the special cases within our two orbifold families and determine which of their additional marginal operators give rise to independent exactly marginal directions. It will turn out that only the $\mathbb{Z}_3[1,1,1]$ and the $\mathbb{Z}_3\times\mathbb{Z}_3$ orbifold theories feature such an extension of the conformal manifold.

\paragraph{1. $\bm{\Zb_k}$ orbifolds with coincident configuration numbers $\bm{a_I=a_J}$} \phantom{a}\\
The first special case occurs when two configuration numbers coincide \cite{Aharony:2002tp,Green:2010da}. Without loss of generality, we may consider 
    \begin{equation}
        a_X=a_Y=a \, , \qquad a_Z= -2a\text{ mod } k\, ,
    \end{equation}
    with $3 a_I \neq 0\text{ mod }k \, ,  \forall I$. In this setup, the requirement $\gcd(k,a_X,a_Y,a_Z)=1$ described in Section \eqref{sec: Z_k} reduces to the requirement
    \begin{equation}
        \gcd (k,a)=1 \, .
    \end{equation}
    In these theories the orbifold point (and with it the entire submanifold spanned by gauge-coupling deformations) constitutes a symmetry enhancement locus with global $\text{SU}(2)\times \text{U}(1)$ symmetry, resulting in additional $\bar{\Bm}_{2(0,0)}$ deformation multiplets generated by
    \begin{equation}
        \mathcal{F}_{XXZ}=\Tr (XXZ)\,,\qquad \mathcal{F}_{YYZ}=\Tr (YYZ)\,,
    \end{equation}
    and additional $\hat{\Cm}_{(0,0)}$ current multiplets generated by
    \begin{equation}
        \mathcal{J}_{(\alpha)}^+= \Tr(Y\bar{X})\,,\qquad \mathcal{J}_{(\alpha)}^-= \Tr(X\bar{Y})\,.
         \label{Eq:CH0_enh}
    \end{equation}
    Turning on a generic non-zero combination of the $\Fm$ deformations breaks the enhanced symmetry down to its $\text{U}(1)\times \text{U}(1)$ Cartan subgroup, triggering a recombination of all but one marginal deformation and two global symmetry currents. In this case we thus again expect only a single independent exactly marginal superpotential deformation, which is a (possibly $\text{SU}(2)$-rotated) Leigh-Strassler deformation $\Fm_{XYZ}$. The two untwisted symmetries, which survive after turning on an this appropriately $\text{SU}(2)$-rotated Leigh–Strassler deformation, may be identified with $\text{U}(1)_{\alpha} \times \text{U}(1)_\beta$ in Table \ref{Tab:Orbifold}. 
    The dimension of the conformal manifold for these theories is then given as before by \eqref{Eq: dimZk} which reduces to
    \begin{equation}
        \dim \Mm_\Cm = \begin{cases}
            3 \,, \quad k \text{ even} \\
            2 \,,\quad k \text{ odd} \, .
        \end{cases}
    \end{equation}

\paragraph{2. $\bm{\Zb_{3k}}$ orbifolds with $\bm{a_I=k}$ or $\bm{a_I=2k}$}
\phantom{a}\\
For concreteness we consider the case
$a_X=k$. One finds $k$ independent
operators of the form
\begin{equation}
    \mathcal{F}_{XXX,3m}
    =
    \Tr\!\left(\Omega^{3m} XXX\right)\, ,\quad
    m=0,\ldots,k-1 \, .
    \label{eq:XXX}
\end{equation}
Operators of the form \eqref{eq:XXX} contribute to the index \eqref{Eq:ST_IZk} and generate a net degeneracy in the equivalence class $\text{ND}[0,0]$ \eqref{eq:Net degeneracy}. To see this, recall that the index counts operators of the schematic form $\Tr\Omega^m\Phi_I^{3nk/\ell_I}$, where $\ell_I=\gcd(k,a_I)$. Here, $\ell_X=k$, so indeed the operators \eqref{eq:XXX} contribute at order $t^2$ and are responsible for the additional marginal multiplets contributing to the net degeneracy $\text{ND}[0,0]$ reported in Table~\ref{tab:Dim Marginal} of Appendix~\ref{App:survey}. 

One may now be tempted to conclude that these additional multiplets generate additional exactly marginal deformations of the superpotential. However, in \cite{Aharony:2002tp} this expectation was shown to be erroneous. An explicit perturbative analysis shows that no additional direction is added to the conformal manifold.
This can be understood directly from the Abelian flavour quotient \eqref{Eq:complexified_flavour_quotient}. Here it is convenient to introduce the basis
\begin{equation}\label{eq:symmetrycombo}
    Q_{\mathrm{br}} = Q_{\alpha} \, ,
    \qquad
    Q_{\mathrm{pres}} = Q_\beta - \frac{1}{2}Q_{\alpha} \,,
\end{equation}
for the $\text{U}(1)_\alpha\times \text{U}(1)_\beta$ flavour-symmetry generators. In this basis the ordinary cubic operators are neutral,
\begin{equation}
    q\!\left(\mathcal{F}_{XYZ}\right)
    =
    q\!\left(\mathcal{F}_{XZY}\right)
    =
    (0,0) \, ,
\end{equation}
whereas all the additional cubic operators carry the same charge
\begin{equation}
    q\!\left(\mathcal{F}_{XXX,\ell}\right)
    =
    (3,0) \, ,
    \qquad
    \ell=3m,\quad
    m=0,\ldots,k-1 \, .
\end{equation}
Thus the $XXX$ deformations preserve the linear combination
$\text{U}(1)_{\mathrm{pres}}$ while breaking $\text{U}(1)_{\mathrm{br}}$.
Denoting the corresponding superpotential couplings by $h_\ell$, the latter transform with the opposite charge under the complexified broken symmetry,
\begin{equation}
    h_\ell \longrightarrow \lambda^{-3} h_\ell \, ,
    \qquad
    \lambda\in U(1)_{\mathrm{br}}^{\mathbb{C}} \simeq \mathbb{C}^{\ast} \, .
\end{equation}
Since all the $h_\ell$ carry the same non-zero charge, while the $XYZ$, $XZY$, and gauge-coupling deformations are neutral under $\text{U}(1)_{\mathrm{br}}$, no non-constant regular holomorphic invariant involving the $h_\ell$ can be constructed in a neighbourhood of the orbifold point. Equivalently, the associated moment-map, which determines the beta-function, is positive definite in the $XXX$ directions and obstructs these marginal operators from sourcing exactly marginal deformations \cite{Green:2010da}.
This is analogous to the obstruction discussed in \textbf{Appendix B.1.2} of \cite{Razamat:2020pra} and, more directly, to the situation of \textbf{Appendix B.2.2}, where a conformal manifold exists but additional marginal operators are marginally irrelevant because they are charged under an Abelian symmetry for which no non-trivial Kähler quotient exists.
Consequently the appearance of the additional $\mathcal{F}_{XXX,\ell}$ multiplets does not enlarge the dimension of the conformal manifold beyond \eqref{Eq: dimZk}.

As discussed above, the superconformal index detects these operators in the net degeneracy $\text{ND}[0,0]$ \eqref{eq:Net degeneracy}. This is not in contradiction with the fact that the corresponding deformations are marginally irrelevant. At the orbifold fixed point the $\Fm_{XXX,\ell}$ operators are genuine protected $\bar{\Bm}_{2(0,0)}$ multiplets and therefore contribute to the index. Turning on their couplings, however, does not lead to a neighbouring superconformal fixed point: the Abelian moment-map condition obstructs the deformation and the corresponding RG flow returns to the orbifold theory. Away from the fixed point the theory is not superconformal, and hence the classification in terms of superconformal BPS multiplets does not apply.
\\

The two special-case families discussed above overlap in two particular points, the $\Zb_3[1,1,1]$ orbifold discussed below and the $\Zb_6[1,1,4]$ orbifold. While the latter case shows no novel behaviour and both special effects just occur simultaneously, the $\Zb_3[1,1,1]$ case is genuinely different:

\paragraph{3. The $\bm{\Zb_{3}[1,1,1]}$ orbifold} \phantom{a}\\
This theory features a global symmetry enhancement to $\text{SU}(3)$ and ten superpotential deformations generated by all untwisted cubic combinations of $X$, $Y$ and $Z$ fields. Turning on first the (possibly $\text{SU}(3)$-rotated) Leigh–Strassler deformation $\Fm_{XYZ}$ breaks $\text{SU}(3)$ down to its $\text{U}(1) \times \text{U}(1) $ Cartan, such that six $\Fm$ and $\Jm$ multiplets recombine, leaving behind $4\times \Fm$ and $2\times \Jm$ multiplets. This allows us to further deform by another marginal deformation, in which case the global symmetry is broken completely, recombining the remaining two $\Jm$ current multiplets with two $\Fm$ multiplets. Here the flavour quotient does not obstruct any direction. We thus find a two-dimensional conformal manifold of superpotential deformations. Combined with the overall gauge coupling deformation $\mathcal{G}_0$ this results in a three-dimensional conformal manifold \cite{Aharony:2002tp}
\begin{equation}
    \dim \mathcal{M}_C=3\,.
\end{equation}
A convenient representative of the additional exactly marginal superpotential direction is the cubic Leigh--Strassler deformation
\begin{equation}
    \Fm_{\Zb_3}
    =
    \Tr(XXX+YYY+ZZZ)\,.
\end{equation}

\paragraph{4. $\bm{\mathbb{Z}_3 \times \mathbb{Z}_{\ell}}$ orbifolds}\phantom{a}\\
    These theories exhibit the same behaviour that we already observed in the second special case above. There are $\ell$ additional marginal operators of the form 
    \begin{equation}
    \Fm_{XXX,b} = \Tr [\Omega_Y^b XXX] \, \quad b=0,1,\dots , \ell -1 \, ,
    \end{equation}
    but again they do not form new exactly marginal directions of the conformal manifold. This is due to the same Abelian obstruction presented above. Since these operators all have the same charge under a combination $Q_{\text{br}}$ from \eqref{eq:symmetrycombo} with respect to which the other superpotential terms are neutral, it is not possible to construct a non-trivial quotient for them. They are thus not sourcing exactly marginal deformations. Consequently the dimension of the conformal manifold is just given by \eqref{Eq: dimZkZl}\,. 
    The same comment made above about the relation between the superconformal index and the marginal directions applies here as well.

\paragraph{5. The $\bm{\mathbb{Z}_3 \times \mathbb{Z}_{3}}$ orbifold}\phantom{a}\\
For this theory we have $9$ additional marginal deformations
    \begin{equation}
    \begin{split}
         & \Fm_{XXX,b} = \Tr [\Omega_Y^b XXX], \quad \Fm_{YYY,b} = \Tr [\Omega_X^b YYY], \\
         & \Fm_{ZZZ,b} = \Tr [\Omega_X^b \Omega_Y^{-b} ZZZ], \quad b=0,1,2 \, .
    \end{split}
    \label{Eq:Additional_W_Z3Z3}
    \end{equation}
    A generic deformation involving these three families breaks the two global $\text{U}(1)$ symmetries. In contrast to the previous cases, however, the corresponding Abelian flavour quotient is non-empty. Indeed, the charges of the three families under $\text{U}(1)_{\alpha}\times \text{U}(1)_\beta$ are
    \begin{equation}
    q(\Fm_{XXX})=\left(3,\frac32\right),\qquad
    q(\Fm_{YYY})=\left(-3,\frac32\right),\qquad
    q(\Fm_{ZZZ})=(0,-3).
    \end{equation}
    Denoting the associated couplings by $h^X_a,h^Y_b,h^Z_c$, non-trivial holomorphic invariants can therefore be constructed, for example \begin{equation}
    I_{abc}=h^X_a h^Y_b h^Z_c \, .
    \end{equation} 
    Hence the same-sign Abelian obstruction encountered for a single order-three family is absent. The complexified flavour quotient therefore removes precisely the two broken untwisted Abelian transformations from the nine additional cubic couplings. Thus the full conformal manifold has complex dimension
    \begin{equation}
    \dim_{\mathbb C}\Mm_\Cm
    =
    \underbrace{7}_{\text{gauge-coupling submanifold}}
    +
    \underbrace{1}_{\text{Leigh--Strassler}}
    +
    \underbrace{7}_{XXX,\,YYY,\,ZZZ\text{ quotient}}
    =15 \, .
    \end{equation}
    The last seven dimensions should be understood as a quotient of the nine special cubic couplings by the two complexified untwisted $\text{U}(1)$ transformations. In particular, they are correlated combinations of the $XXX$, $YYY$, and $ZZZ$ couplings rather than seven arbitrary independent linear combinations at the orbifold point.\\

To summarise, special orbifold actions can give rise to additional marginal chiral operators through coincident configuration numbers or $\mathbb{Z}_3$ orbits in the quiver gauge theory. However, the appearance of these operators does not in general enlarge the conformal manifold. 
In the cyclic $\mathbb{Z}_k$ family, the additional operators are either accounted for by the quotient under an enhanced non-Abelian flavour symmetry or obstructed by an Abelian moment-map condition, signalling the non-vanishing of the beta function. Similarly, in the product  $\mathbb{Z}_k\times \mathbb{Z}_\ell$ family, an isolated
order-three orbifold factor produces additional marginal operators but no new exactly marginal directions. The two exceptional theories $\mathbb{Z}_3[1,1,1]$ and $\mathbb{Z}_3\times\mathbb{Z}_3$ instead admit additional exactly marginal deformations, yielding conformal manifolds of complex dimensions three and fifteen, respectively.

\section{The complete one-loop dilatation operator}\label{sec: orbidila}

In Section \ref{sec: BPS} we presented a construction of the protected single-trace BPS spectrum of the Abelian $\mathcal{N}=1$ orbifold theories under consideration. Of course the BPS spectrum, albeit interesting in its own right, is only a small subset of the full single-trace spectrum of our $\mathcal{N}=1$ orbifold theories. All other states pick up anomalous dimensions, which can in principle be computed by integrability (see \cite{Beisert:2010jr} for a general review and \cite{Skrzypek:2022cgg} for one in the orbifold context). In this paper we will also consider marginal deformations away from the orbifold point where integrability techniques have not been developed so far. Therefore we restrict our attention to a perturbative calculation of anomalous dimensions to one-loop order $\mathcal{O}(g^2_0)$. 

We will follow the strategy developed in \cite{Beisert:2003jj} for $\mathcal{N}=4$ SYM theory and further generalised to $\mathcal{N}=2$ theories and $\mathcal{N}=1$ SQCD in \cite{Liendo:2011wc,Liendo:2011xb}. We may expand the dilatation operator 
\begin{equation}\label{Eq: dila}
    \Dm=\Dm_0+g_0^2\mathcal{H}+\mathcal{O}(g_0^4)\,,
\end{equation}
where $\Dm_0$ extracts the bare dimension of the operator $\mathcal{O}_L$ it acts upon while the one-loop correction $\mathcal{H}$ decomposes into a nearest-neighbour interaction $\mathcal{H}_{i,i+1}$ between each successive pair of fields within the single-trace operator $\mathcal{O}_L$
\begin{equation}
    \mathcal{H}=\sum_{i=1}^L \mathcal{H}_{i,i+1}\,,\qquad \mathcal{H}_{L,L+1}:=\mathcal{H}_{L,1}\,.
\end{equation}
Here the length $L$ of $\mathcal{O}_L$ counts the number of fundamental fields inserted in the trace. This is reminiscent of the action of a nearest-neighbour Hamiltonian on a spin chain and this analogy has led to great progress in the integrability program \cite{Minahan:2002ve,Beisert:2003jj}. At each ``site'' of the spin-chain (position in the single-trace operator) we can insert a representation of $\mathfrak{psu}(2,2|4)$, which may be decomposed into an oscillator basis of two sets of bosonic oscillators $(\ab^\al,\ab^\dagger_\al)$, $(\bb^{\ad},\bb^\dagger_{\ad})$ and one set of fermionic oscillators
$(\cb^{\Im},\cb^\dagger_{\Im})$, where $(\al,\ad)$ are Lorentz indices and $\Im$ is a fundamental $\text{SU}(4)$ index. The non-zero (anti-)commutation relations are
\begin{align} 
[\ab^\al,\ab^\dagger_\be] = \delta^\al_\be\,,\qquad 
[\bb^{\ad},\bb^\dagger_{\bd}] = \delta^{\ad}_{\bd}\,, \qquad \{ \cb^{\Im},\cb^\dagger_{\Jm} \} = \delta^{\Im}_{\Jm}\,.
\end{align}
The one-loop Hamiltonian acts on neighbouring sites by shuffling the oscillators as observed in \cite{Beisert:2003jj}. Schematically,
\begin{equation}\label{Eq: shuffle}
\mathcal{H} \ket{A}\otimes \ket{B}=\sum_{\sigma}c_\sigma\,\delta_{\Cm\ket{C(\sigma)}}\,\delta_{\Cm\ket{D(\sigma)}}\ket{C(\sigma)}\otimes \ket{D(\sigma)}\,,
\end{equation}
where $\sigma$ denotes all reshufflings of the oscillators in $\ket{A}\otimes \ket{B}$ and $\ket{C(\sigma)}\otimes \ket{D(\sigma)}$ are the resulting states. The Kronecker-deltas ensure that the resulting states are physical, which requires the central charge
\begin{equation}
    \Cm = 1 - \frac{1}{2}\ab_{\al}^{\dagger}\ab^\al + \frac{1}{2}\bb_{\ad}^{\dagger}\bb^{\ad}-\frac{1}{2}\cb^\dagger_{\Jm}\cb^{\Jm}
\end{equation}
to vanish at each site. The coefficients $c_\sigma$ can be fixed by $\text{PSU}(2,2|4)$ invariance and crossing symmetry and depend on the total number $n$ of oscillators $(\ab^\dag,\bb^\dag,\cb^\dagger)$ present in a pair of fields, as well as the numbers $n_{12}$ and $n_{21}$ of such oscillators moved from position one to two and vice versa
\begin{equation}\begin{split}\label{Eq: coeff}
 c_\sigma(n,n_{12},n_{21})&=(-1)^{1+n_{12}n_{21}}\frac{\Gamma(\tfrac{1}{2}(n_{12}+n_{21}))\Gamma(1+\tfrac{1}{2}(n-n_{12}-n_{21}))}{\Gamma(1+\tfrac{1}{2}n)}\,,\\\quad c_\sigma(n,0,0)&=h(\tfrac{n}{2})\,,
\end{split}\end{equation}
where $h(m)=\sum_{k=1}^m\tfrac{1}{k}$ is the $m$\textsuperscript{th} harmonic number.

Let us now adapt this strategy for $\mathcal{N}=1$ orbifold theories. As we observed above, the orbifolding procedure predominantly changes the gauge structure of fields, which does not play a role in the one-loop Hamiltonian \eqref{Eq: shuffle}. We may thus simply apply \emph{the same one-loop dilatation operator} to a basis of orbifold invariant states, which is given by all closed single-quiver-trace operators. What does change, however is the interpretation of the oscillator basis, since the preserved superconformal group is just $\text{SU}(2,2|1)$, leaving three of the $(\cb^{\Im},\cb^\dagger_{\Im})$ oscillator pairs unassigned. We interpret these oscillators as generators of the broken $\text{SU}(3)$ R-symmetry, which now distinguish the real and chiral multiplets in the quiver gauge theory \cite{Liendo:2011wc,Liendo:2011xb}. To be explicit, let us rename these ``auxiliary'' fermionic generators by $(\db^{\Im},\db^\dagger_{\Im})_{\Im\in\{1,2,3\}}$ and list all the physical fields that can appear at the sites of our spin-chain (compare also to Table \ref{Tab:N1_Letters}):
\begin{align}
    \partial^k\Fm &= (\ab^\dagger)^{k+2}(\bb^\dagger)^{k}\ket{0}\,,\nonumber\\
    \partial^k\lambda &= (\ab^\dagger)^{k+1}(\bb^\dagger)^{k}\cb^\dagger\ket{0}\,,\nonumber\\
     \partial^k \psi_X &= (\ab^\dagger)^{k+1}(\bb^\dagger)^{k}\db_1^\dagger\ket{0}\,,\nonumber\\
    \partial^k \psi_Y &= (\ab^\dagger)^{k+1}(\bb^\dagger)^{k}\db_2^\dagger\ket{0}\,,\nonumber\\
    \partial^k \psi_Z &= (\ab^\dagger)^{k+1}(\bb^\dagger)^{k}\db_3^\dagger\ket{0}\,,\nonumber\\
    \partial^k X &= (\ab^\dagger)^{k}(\bb^\dagger)^{k}\cb^\dagger\db_1^\dagger\ket{0}\,,\nonumber\\
    \partial^k Y &= (\ab^\dagger)^{k}(\bb^\dagger)^{k}\cb^\dagger\db_2^\dagger\ket{0}\,,\nonumber\\
    \partial^k Z &= (\ab^\dagger)^{k}(\bb^\dagger)^{k}\cb^\dagger\db_3^\dagger\ket{0}\,,\nonumber\\
    \partial^k \bar X &= (\ab^\dagger)^{k}(\bb^\dagger)^{k}\db_2^\dagger\db_3^\dagger\ket{0}\,,\nonumber\\
    \partial^k \bar Y &= (\ab^\dagger)^{k}(\bb^\dagger)^{k}\db_3^\dagger\db_1^\dagger\ket{0}\,,\nonumber\\
    \partial^k \bar Z &= (\ab^\dagger)^{k}(\bb^\dagger)^{k}\db_1^\dagger\db_2^\dagger\ket{0}\,,\nonumber\\
    \partial^k \bar \psi_X &= (\ab^\dagger)^{k}(\bb^\dagger)^{k+1}\cb^\dagger\db_2^\dagger\db_3^\dagger\ket{0}\,,\nonumber\\
    \partial^k \bar \psi_Y &= (\ab^\dagger)^{k}(\bb^\dagger)^{k+1}\cb^\dagger\db_3^\dagger\db_1^\dagger\ket{0}\,,\nonumber\\
    \partial^k \bar \psi_Z &= (\ab^\dagger)^{k}(\bb^\dagger)^{k+1}\cb^\dagger\db_1^\dagger\db_2^\dagger\ket{0}\,,\nonumber\\
    \partial^k \bar \lambda &= (\ab^\dagger)^{k}(\bb^\dagger)^{k+1}\db_1^\dagger\db_2^\dagger\db_3^\dagger\ket{0}\,,\nonumber\\
    \partial^k \bar \Fm &= (\ab^\dagger)^{k}(\bb^\dagger)^{k+2}\cb^\dagger\db_1^\dagger\db_2^\dagger\db_3^\dagger\ket{0}\,.\label{Eq: fields}
\end{align}
The gauge representation of a field depends implicitly on the $\db^\dagger$ insertions and can be read off the quiver diagram. Since the Hamiltonian conserves the overall gauge representation of the pair of states it acts on, we neglect these details for now. The $\text{SU}(2,2|1)$ symmetry generators can be written in this oscillator representation as
\begin{equation}\label{eq:sca}\begin{split}
\Qm_{\al} &=  \ab^{\dagger}_{\al}\cb\, ,
\qquad
 \Sm^{\al}  = \cb^{\dagger}\ab^{\al}\, ,
 \qquad
\bar{\Qm}_{\ad}  = \bb^{\dagger}_{\ad}\cb^{\dagger}\, ,
\qquad 
\bar{\Sm}^{\ad}  =  \bb^{\ad}\cb\, ,
\\
\Pm_{\al \bd} &= \ab^{\dagger}_{\al} \bb^{\dagger}_{\bd}\, ,
\qquad
 \Km^{\al \bd} = \ab^\al \bb^{\bd}\, ,\qquad
\mathcal{D} = 1 + \frac{1}{2}\ab_{\al}^{\dagger}\ab^\al + \frac{1}{2}\bb_{\ad}^{\dagger}\bb^{\ad}\, ,
\\
\Lm^{\phantom{\be}\al}_{\be} &= \ab^\dagger_\be\ab^\al-\frac{1}{2}\delta^\al_\be \ab^\dagger_\gamma\ab^\gamma\, ,
\qquad
\dot{\Lm}^{\phantom{\bd}\ad}_{\bd} = \bb^\dagger_{\bd}\bb^{\ad}-\frac{1}{2}\delta^{\ad}_{\bd} \bb^\dagger_{\dot{\gamma}}\bb^{\dot{\gamma}}\, ,
\\
r &= \cb^\dagger\cb-\frac{1}{3}\db_1^\dagger\db_1-\frac{1}{3}\db_2^\dagger\db_2-\frac{1}{3}\db_3^\dagger\db_3\, ,
\\
\Cm &= 1 - \frac{1}{2}\ab_{\al}^{\dagger}\ab^\al + \frac{1}{2}\bb_{\ad}^{\dagger}\bb^{\ad}-\frac{1}{2}\cb^\dagger\cb
-\frac{1}{2}\db_1^\dagger\db_1-\frac{1}{2}\db_2^\dagger\db_2-\frac{1}{2}\db_3^\dagger\db_3\,,
\end{split}\end{equation}
where the central charge $\Cm$ has to vanish on physical states, such as the ones in \eqref{Eq: fields}. By construction, the one-loop Hamiltonian \eqref{Eq: shuffle} commutes with the superconformal symmetry generators.

With this setup in place we can analyse the perturbative spectrum of $\mathcal{N}=1$ orbifold theories by acting with the dilatation operator \eqref{Eq: dila} on the Hilbert-space of closed single-trace operators and diagonalising the resulting mixing matrix. Only operators with the same discrete spin and R-symmetry quantum numbers can mix, so we may subdivide the full spectral problem into sectors of equal global charges. At one-loop order, a further conserved quantum number is the length $L$ which denotes the number of fields from \eqref{Eq: fields} that are inserted in the trace. This allows for a further decomposition of the spectral problem. We implemented the one-loop Hamiltonian in a  Wolfram Mathematica notebook, which we supply as ancillary file to the arXiv submission of this paper. As consistency check for our Hamiltonian and its implementation, we confirmed that the BPS spectrum we constructed in Section \ref{sec: BPS} could be reproduced by extracting the kernel of the one-loop Hamiltonian in some sample sectors of the orbifold theories listed in Appendix \ref{App:survey}. An example for such consistency checks in the $\mathbb{Z}_4[1,1,2]$ and $\mathbb{Z}_2\times \mathbb{Z}_2$ orbifold theories is outlined in the ancillary Mathematica notebook. 

Extra care had to be taken for operators of length $L=2$ where the $\text{U}(1)$-subtractions \eqref{Eq: U(1)-contrib} discussed in Section \ref{sec: U(1)s} had to be taken into account. This was necessary since the planar Hamiltonian in \eqref{Eq: dila} implicitly assumes $\text{U}(N)$ trace relations instead of the correct $\text{SU}(N)$ ones. Since this difference between the trace relations only affects length-two operators, the Mathematica notebook automatically distinguishes the cases $L=2$ and $L>2$. We then accounted for the $\text{U}(1)$ subtraction by manually cancelling any chiral-antichiral interactions \eqref{eq: U(1)channel} that would result from an intermediate $\text{U}(1)$-channel. In order to  isolate these channels, we had to implement a case disambiguation between different reshufflings of $\mathbf{d}^\dagger_{\mathcal{I}}$ oscillators, which will also be useful for the following discussion of marginal deformations.

\subsection{Accounting for marginal deformations}\label{sec: diladef}

In Section \ref{sec: generic}, we discussed which marginal deformations are exact and we also gave a parametrisation at one-loop level, where we could express all couplings $g_i$ and $f_i^\pm$ in terms of an overall coupling scale $g_0$, a number of ratios $\kappa_i$ and the Leigh-Strassler deformation parameter $\LS$ (see appendix \ref{App:survey} for some examples). We exclude the special deformations occurring in the $\mathbb{Z}_3[1,1,1]$ and $\mathbb{Z}_3\times\mathbb{Z}_3$ orbifold theories which we discussed in Section \ref{sec: special} since they introduce new terms to the superpotential which would have to be included by hand.

The benefit of restricting to the subset $(\kappa_i,\LS)$ of marginal deformations is that we can adapt the one-loop dilatation operator \eqref{Eq: dila} from the previous section to these deformed models. At the orbifold point $\kappa_i=f^\pm_i=1$ the one-loop Hamiltonian $\mathcal{H}$ acting as in \eqref{Eq: shuffle} captures all planar tree-level interactions between neighbouring fields in a single-trace operator. In $\mathcal{N}=1$ superspace language, the supergraphs contributing to this nearest neighbour interactions are essentially only $s$- and $t$-channel diagrams involving two cubic vertices of the following form\footnote{There are also quartic (and higher) vertices for the vector multiplets, but since we are only interested in the coupling constants, we may treat them as successive insertion of two (or more) cubic vertices.}
\begin{figure}[ht]
\centering
\begin{minipage}{0.23\textwidth}
\centering
\begin{tikzpicture}[scale=0.7]
\begin{feynman}
\vertex (v) at (0,0);
\vertex[label=above:$\mathcal{W}_i$]   (g1) at (90:1);
\vertex[label=below left:$\mathcal{W}_i$]  (g2) at (210:1);
\vertex[label=below right:$\mathcal{W}_i$] (g3) at (330:1);
\diagram*{
(g1) -- [gluon] (v),
(g2) -- [gluon] (v),
(g3) -- [gluon] (v),
};
\end{feynman}
\end{tikzpicture}
\vspace{-0.5cm}
\[
g_0 \kappa_i
\]
\vspace{-0.2cm}
\end{minipage}
\begin{minipage}{0.23\textwidth}
\centering
\begin{tikzpicture}[scale=0.7]
\begin{feynman}
\vertex (v) at (0,0);
\vertex[label=above:$\mathcal{W}_i$] (g) at (90:1);
\vertex[label=below left:$\Phi_{ij}$]  (f1) at (210:1);
\vertex[label=below right:$\bar{\Phi}_{ji}$] (f2) at (330:1);
\diagram*{
(g)  -- [gluon] (v),
(f1) -- [fermion] (v),
(v)  -- [fermion] (f2),
};
\end{feynman}
\end{tikzpicture}
\vspace{-0.5cm}
\[
g_0 \kappa_i
\]
\vspace{-0.2cm}
\end{minipage}\\
\begin{minipage}{0.23\textwidth}
\centering
\begin{tikzpicture}[scale=0.7]
\begin{feynman}
\vertex (v) at (0,0);
\vertex[label=above:$\mathcal{X}_{ij}$]      (f1) at (90:1);
\vertex[label=below left:$\mathcal{Y}_{jk}$] (f2) at (210:1);
\vertex[label=below right:$\mathcal{Z}_{ki}$] (f3) at (330:1);
\diagram*{
(f1) -- [fermion] (v),
(f2) -- [fermion] (v),
(f3) -- [fermion] (v),
};
\end{feynman}
\end{tikzpicture}
\vspace{-0.5cm}
\[
g_0f^+_{i}
\]
\end{minipage}
\begin{minipage}{0.23\textwidth}
\centering
\begin{tikzpicture}[scale=0.7]
\begin{feynman}
\vertex (v) at (0,0);
\vertex[label=above:$\mathcal{X}_{ij}$]      (f1) at (90:1);
\vertex[label=below left:$\mathcal{Z}_{jk}$] (f2) at (210:1);
\vertex[label=below right:$\mathcal{Y}_{ki}$] (f3) at (330:1);
\diagram*{
(f1) -- [fermion] (v),
(f2) -- [fermion] (v),
(f3) -- [fermion] (v),
};
\end{feynman}
\end{tikzpicture}
\vspace{-0.5cm}
\[
-g_0f^-_{i}
\]
\end{minipage}
\begin{minipage}{0.23\textwidth}
\centering
\begin{tikzpicture}[scale=0.7]
\begin{feynman}
\vertex (v) at (0,0);
\vertex[label=above:$\bar{\mathcal{X}}_{ij}$]      (f1) at (90:1);
\vertex[label=below left:$\bar{\mathcal{Y}}_{jk}$] (f2) at (210:1);
\vertex[label=below right:$\bar{\mathcal{Z}}_{ki}$] (f3) at (330:1);
\diagram*{
(v) -- [fermion] (f1),
(v) -- [fermion] (f2),
(v) -- [fermion] (f3),
};
\end{feynman}
\end{tikzpicture}
\vspace{-0.5cm}
\[
-g_0f^-_{j}
\]
\end{minipage}
\begin{minipage}{0.23\textwidth}
\centering
\begin{tikzpicture}[scale=0.7]
\begin{feynman}
\vertex (v) at (0,0);
\vertex[label=above:$\bar{\mathcal{X}}_{ij}$]      (f1) at (90:1);
\vertex[label=below left:$\bar{\mathcal{Z}}_{jk}$] (f2) at (210:1);
\vertex[label=below right:$\bar{\mathcal{Y}}_{ki}$] (f3) at (330:1);
\diagram*{
(v) -- [fermion] (f1),
(v) -- [fermion] (f2),
(v) -- [fermion] (f3),
};
\end{feynman}
\end{tikzpicture}\,.
\vspace{-0.5cm}
\[
g_0f^+_{j}
\]
\end{minipage}
\end{figure}

In this schematic representation, we denote the entire tower of superconformal descendants by a single superfield, anticipating that the superconformal structure is captured by the oscillator representation resulting in the coefficients $c_\sigma$ \eqref{Eq: coeff}. $\mathcal{W}_i$ denotes the superfield associated to the $i$-th vector multiplet. $\Phi_{ij}$ denotes an arbitrary chiral multiplet connecting the $i$-th and $j$-th gauge node in the orbifold quiver, namely one of the $\mathcal{X}$, $\mathcal{Y}$ or $\mathcal{Z}$ chiral multiplets, which also enter cubic interactions governed by the superpotential \eqref{Eq: superpot}. For the vertices displayed above we interpreted all fields as ``ingoing'', but depending on context in the full diagram, one may have to conjugate ``outgoing'' legs. In the following we will read supergraphs from left to right, so outgoing fields are to the right and appropriately conjugated.

The coupling constants are all equal at the orbifold point, but once we turn on marginal deformations we have to insert the parametrisation which we determined in the Section \ref{sec: gauge manifold}. We therefore have to introduce a new factor $d_\sigma$ in the one-loop Hamiltonian \eqref{Eq: shuffle} \begin{equation}\label{eq: hamdef}
\mathcal{H}' \ket{A}\otimes \ket{B}=\sum_{\sigma}c_\sigma d_\sigma\,\delta_{C\ket{C(\sigma)},0}\,\delta_{C\ket{D(\sigma)},0}\ket{C(\sigma)}\otimes \ket{D(\sigma)}\,,
\end{equation}
which accounts for the change in coupling constants. The correct choice for the factor $d_{\sigma}$ is determined by the configurations of auxiliary $\mathbf{d}_i^\dagger$ oscillators before and after the reshuffling. In particular, this dependence on auxiliary oscillators alone guarantees that the Hamiltonian \eqref{eq: hamdef} commutes with the $\mathcal{N}=1$ superconformal algebra in \eqref{eq:sca}.\footnote{A similar construction for $\mathcal{N}=1$ SQCD was performed in \cite{Liendo:2011wc}. In our scenario we have to take into account non-vanishing superpotential interactions and a larger set of fields, but on a technical level the approaches agree.} There are a total of 216 inequivalent reshuffling processes that can occur and for each of them we need to determine an appropriate coefficient $d_{\sigma}$ depending on which cubic vertices participate.\footnote{To confirm this number of coefficients, assign occupation numbers $(s_1,s_2,s_3)$ and $(t_1,t_2,t_3)$ to the two sites and count the number $n=\sum_{i}\abs{s_i-t_i}$ of occupation mismatches where a reshuffling can occur. The number of possible reshufflings for a fixed initial configuration is $2^n$. Summing over initial configurations gives the total
\begin{equation*}
    \sum_{s_i,t_i=0}^1 2^{n(s_i,t_i)}=2^3\sum_{n=0}^3  \binom{3}{n}2^n = 6^3=216\,,
\end{equation*}
where we first summed over $2^3$ arbitrary choices for $s_i$ and then counted the number $\binom{3}{n}$ of $t^i$ configurations with fixed $n(s_i,t_i)$. In $\mathcal{N}=2$ orbifolds we only need $6^2=36$ coefficients while $\mathcal{N}=0$ would require $6^4=1296$. For the latter, conformal representation theory may provide a more direct approach \cite{Liendo:2017wsn}. }

Take for example the process that moves one $\mathbf{d}_1^\dagger$ oscillator across in the exchange $\mathcal{W}_i\otimes \mathcal{X}_{ij}\to  \mathcal{X}_{ij}\otimes \mathcal{W}_j $. At the level of auxiliary oscillators this process is simply the reshuffling $\mathbb{1}\otimes \mathbf{d}_1^\dagger \to  \mathbf{d}_1^\dagger \otimes\mathbb{1}$. The only planar tree-level supergraphs that mediate this exchange are the diagrams
\begin{equation}
\begin{tikzpicture}[baseline={(current bounding box.center)}]

\begin{feynman}

\vertex (g1) at (0,2) {$\mathcal{W}_i$};
\vertex (f1) at (0,0) {$\mathcal{X}_{ij}$};

\vertex (v1) at (0.7,1);
\vertex (v2) at (1.3,1);

\vertex (g2) at (2,0) {$\mathcal{W}_j$};
\vertex (f2) at (2,2) {$\mathcal{X}_{ij}$};

\diagram*{
(g1) -- [gluon] (v1),
(f1) -- [fermion] (v1),

(v1) -- [fermion] (v2),

(v2) -- [gluon] (g2),
(v2) -- [fermion] (f2),
};

\end{feynman}
\end{tikzpicture}
\;\sim\;
\begin{tikzpicture}[baseline={(current bounding box.center)}]

\begin{feynman}

\vertex (g1) at (0,2) {$\mathcal{W}_i$};
\vertex (f1) at (0,0) {$\mathcal{X}_{ij}$};

\vertex (v1) at (1,1.3);
\vertex (v2) at (1,0.7);

\vertex (g2) at (2,0) {$\mathcal{W}_j$};
\vertex (f2) at (2,2) {$\mathcal{X}_{ij}$};

\diagram*{
(g1) -- [gluon] (v1),
(f1) -- [fermion] (v2),

(v2) -- [fermion] (v1),

(v2) -- [gluon] (g2),
(v1) -- [fermion] (f2),
};

\end{feynman}

\end{tikzpicture}
\propto g_0^2 \kappa_i\kappa_j\,.
\end{equation}
We thus assign the factor $d_\sigma=\kappa_i\kappa_j$ to this process. It turns out that for 186 of the reshuffling processes a similarly unique combination of coupling constants can be identified since only certain cubic vertices can connect all incoming and outgoing multiplets. For the remaining processes, we compare our Hamiltonian with the deformed $\mathcal{N}=2$ Hamiltonian for the $\mathbb{Z}_2$ orbifold theory developed in \cite{Liendo:2011xb,Mantegazza:2026spd}. In our $\mathcal{N}=1$ nomenclature the theory in question corresponds to a $\mathbb{Z}_2[1,1,0]$ orbifold theory. We confirmed for several operators that our deformed $\mathcal{N}=1$ Hamiltonian correctly reproduces the spectrum of the $\mathcal{N}=2$ theory at any point of the conformal manifold, provided we fix the following relative factors:
\begin{itemize}
    \item $\mathcal{X}\mathcal{Y}\to \mathcal{X}\mathcal{Y}$ and eleven similar combinations have two contributing diagrams
    \begin{equation}
\begin{tikzpicture}[baseline={(current bounding box.center)}]

\begin{feynman}

\vertex (g1) at (0,2) {$\mathcal{X}_{ij}$};
\vertex (f1) at (0,0) {$\mathcal{Y}_{jk}$};

\vertex (v1) at (0.7,1);
\vertex (v2) at (1.3,1);

\vertex (g2) at (2,2) {$\mathcal{X}_{ij}$};
\vertex (f2) at (2,0) {$\mathcal{Y}_{jk}$};

\diagram*{
(g1) -- [fermion] (v1),
(f1) -- [fermion] (v1),

(v2) -- [fermion] (v1),

(v2) -- [fermion] (g2),
(v2) -- [fermion] (f2),
};

\end{feynman}

\end{tikzpicture}
\;\propto g_0^2(f_i^+)^2\,\,,\qquad \;
\begin{tikzpicture}[baseline={(current bounding box.center)}]

\begin{feynman}

\vertex (g1) at (0,0) {$\mathcal{Y}_{jk}$};
\vertex (f1) at (0,2) {$\mathcal{X}_{ij}$};

\vertex (v1) at (1,0.7);
\vertex (v2) at (1,1.3);

\vertex (g2) at (2,2) {$\mathcal{X}_{ij}$};
\vertex (f2) at (2,0) {$\mathcal{Y}_{jk}$};

\diagram*{
(g1) -- [fermion] (v1),
(f1) -- [fermion] (v2),

(v2) -- [gluon] (v1),

(v2) -- [fermion] (g2),
(v1) -- [fermion] (f2),
};

\end{feynman}
\end{tikzpicture}
\propto g_0^2 \kappa_j^2\,,
\end{equation}
which have to be weighted by the following factors
\begin{equation}
    d_\sigma=-\frac{c_\sigma(n,n_{12},n_{21}+2)}{c_\sigma(n,n_{12},n_{21})}
    (f_i^+)^2 +
    \frac{c_\sigma(n-2,n_{12},n_{21})}{c_\sigma(n,n_{12},n_{21})}\kappa_j^2\,,
\end{equation}
given in terms of the coefficients $c_\sigma$ defined in \eqref{Eq: coeff}. In the limit of equal couplings, the two coefficients sum to $d_\sigma=1$ and we recover the undeformed orbifold Hamiltonian. Note, in particular, that $n\geq 2$ in these processes, so evaluating \eqref{Eq: coeff} at $n-2$ does not pose a problem. The weight factors are the same for similar combinations ($\mathcal{X}\mathcal{Z}$ etc.) with appropriately adjusted coupling constants. 
 \item $\mathcal{X}\bar{\mathcal{X}}\to \mathcal{X}\bar{\mathcal{X}}$ and five similar combinations have two contributing diagrams
    \begin{equation}
\begin{tikzpicture}[baseline={(current bounding box.center)}]

\begin{feynman}

\vertex (g1) at (0,2) {$\mathcal{X}_{ij}$};
\vertex (f1) at (0,0) {$\bar{\mathcal{X}}_{ji}$};

\vertex (v1) at (0.7,1);
\vertex (v2) at (1.3,1);

\vertex (g2) at (2,2) {$\mathcal{X}_{ij}$};
\vertex (f2) at (2,0) {$\bar{\mathcal{X}}_{ji}$};

\diagram*{
(g1) -- [fermion] (v1),
(v1) -- [fermion] (f1),

(v2) -- [gluon] (v1),

(v2) -- [fermion] (g2),
(f2) -- [fermion] (v2),
};

\end{feynman}

\end{tikzpicture}
\;\propto g_0^2\kappa_i^2\,\,,\qquad \;
\begin{tikzpicture}[baseline={(current bounding box.center)}]

\begin{feynman}

\vertex (g1) at (0,0) {$\bar{\mathcal{X}}_{ji}$};
\vertex (f1) at (0,2) {$\mathcal{X}_{ij}$};

\vertex (v1) at (1,0.7);
\vertex (v2) at (1,1.3);

\vertex (g2) at (2,2) {$\mathcal{X}_{ij}$};
\vertex (f2) at (2,0) {$\bar{\mathcal{X}}_{ji}$};

\diagram*{
(v1) -- [fermion] (g1),
(f1) -- [fermion] (v2),

(v2) -- [gluon] (v1),

(v2) -- [fermion] (g2),
(f2) -- [fermion] (v1),
};

\end{feynman}
\end{tikzpicture}
\propto g_0^2 \kappa_j^2\,.
\end{equation}
In this case the diagrams have to be weighted by the following factors
\begin{equation}
    d_\sigma=-\frac{c_\sigma(n,n_{12},n_{21}+2)}{c_\sigma(n,n_{12},n_{21})}\kappa_i^2 +\frac{c_\sigma(n-2,n_{12},n_{21})}{c_\sigma(n,n_{12},n_{21})}\kappa_j^2\,.
\end{equation}
\item $\mathcal{X}\bar{\mathcal{X}}\to \bar{\mathcal{Y}}\mathcal{Y}$ and eleven similar combinations have two contributing diagrams
    \begin{equation}
\begin{tikzpicture}[baseline={(current bounding box.center)}]

\begin{feynman}

\vertex (g1) at (0,2) {$\mathcal{X}_{ij}$};
\vertex (f1) at (0,0) {$\bar{\mathcal{X}}_{ji}$};

\vertex (v1) at (0.7,1);
\vertex (v2) at (1.3,1);

\vertex (g2) at (2,2) {$\bar{\mathcal{Y}}_{ik}$};
\vertex (f2) at (2,0) {$\mathcal{Y}_{ki}$};

\diagram*{
(g1) -- [fermion] (v1),
(v1) -- [fermion] (f1),

(v2) -- [gluon] (v1),

(g2) -- [fermion] (v2),
(v2) -- [fermion] (f2),
};

\end{feynman}

\end{tikzpicture}
\;\propto g_0^2\kappa_i^2\,\,,\qquad \;
\begin{tikzpicture}[baseline={(current bounding box.center)}]

\begin{feynman}

\vertex (g1) at (0,0) {$\bar{\mathcal{X}}_{ji}$};
\vertex (f1) at (0,2) {$\mathcal{X}_{ij}$};

\vertex (v1) at (1,0.7);
\vertex (v2) at (1,1.3);

\vertex (g2) at (2,2) {$\bar{\mathcal{Y}}_{ik}$};
\vertex (f2) at (2,0) {$\mathcal{Y}_{ki}$};

\diagram*{
(v1) -- [fermion] (g1),
(f1) -- [fermion] (v2),

(v1) -- [fermion] (v2),

(g2) -- [fermion] (v2),
(v1) -- [fermion] (f2),
};

\end{feynman}
\end{tikzpicture}
\propto g_0^2 (f^-_i)^2\,.
\end{equation}
As in the previous cases these have to be weighted by the factors
\begin{equation}
    d_\sigma=-\frac{c_\sigma(n,n_{12},n_{21}+2)}{c_\sigma(n,n_{12},n_{21})}\kappa_i^2 +\frac{c_\sigma(n-2,n_{12},n_{21})}{c_\sigma(n,n_{12},n_{21})}(f_i^-)^2\,.
\end{equation}
\end{itemize}
The full list of 216 weight factors can be found in Appendix \ref{app: coeffs}. In the Mathematica notebook supplemented to the arXiv submission of this article we provide an implementation of the deformed one-loop dilatation operator. This allows us to determine the one-loop anomalous dimensions of operators at any value of the gauge coupling and Leigh-Strassler deformations. Our implementation of the Hamiltonian is generic for all orbifold theories in the $\mathbb{Z}_k$ and $\mathbb{Z}_k\times\mathbb{Z}_\ell$ families. In order to adapt it to the desired theory, one has to fix a parametrisation of the gauge and superpotential couplings, $\kappa_i$ and $f_i^{\pm}$ and can then act with the dilatation operator on arbitrary single-trace operators of the theory.\footnote{For length-two operators the implementation appropriately cancels the $\text{U}(1)$ double-trace contributions discussed in Section \ref{sec: U(1)s}.} This usually results in a mixing matrix, which should be diagonalised to find the one-loop anomalous dimensions $\gamma\sim\Om(g_0^2)$. By varying the coupling constants we can observe first-hand how operators mix as we move on the conformal manifold and in particular we see the number of BPS states increase drastically at decoupling and symmetry enhancement loci $\kappa\to 0$ which will be discussed in \textbf{Part II} \cite{PartII}. 

Let us conclude this section by stressing that the harmonic action described above is insensitive to higher-loop effects, which are, by construction, not captured by the one-loop dilatation operator. Thus, we observe vanishing anomalous dimension for all marginal gauge deformation multiplets $\mathcal{G}_{a}$ \eqref{Eq: gauge-deformations} and global symmetry current multiplets $\mathcal{M}_a$ \eqref{Eq:CH0_dec} discussed in Section \ref{sec: one-loop}, even though we know that recombinations occur for these multiplets due to the Konishi anomaly \eqref{eq: non-cons}, generating anomalous dimensions at higher-loop order.

\subsection{Example applications}
\label{sec:applications}

To present example applications of this construction, we will now construct explicit BPS operators and consider closed subsectors of unprotected states.

\paragraph{Explicit BPS states} We listed explicit BPS primary operators at the orbifold point in Table \ref{Tab:Orbifold_Untwisted_States}. These expressions are modified when we turn on marginal deformations and diagonalisation of the one-loop dilatation operator gives us direct access to these modified expressions. Take for example the primary operator $\mathcal{O}_{p_X,p_Z}=\Tr[X^{p_X}Z^{p_Z}]$~+~perms.\,\, of the short $\bar{\Bm}_{\frac{2(p_X+p_Z)}{3}(0,0)}$ multiplet in the $\mathbb{Z}_4[1,1,2]$ orbifold theory. As before, ``+~perms.'' denotes all possible orderings of the fields in the quiver trace. If we turn on the unique relative gauge coupling deformation $\kappa:=\kappa_1=\kappa_3=\sqrt{2-\kappa_2^2}=\sqrt{2-\kappa_4^2}$ in this theory, the protected BPS operators get modified to
\begin{equation}
    \tilde{\mathcal{O}}_{p_X,p_Z}=\Tr[X^{p_X}\tilde Z^{p_Z}]\,+\,\text{perms.}\,,
\end{equation}
where the $Z$-fields are rescaled as 
\begin{equation}
    \tilde{Z}_{24}=\kappa_1 Z_{24}\,,\quad \tilde{Z}_{42}=\kappa_1 Z_{42}  \,,\quad \tilde{Z}_{13}=\kappa_2 Z_{13}\,,\quad \tilde{Z}_{31}=\kappa_2 Z_{31}\,.
\end{equation}

The YZ sector is perturbed analogously and the XY sector receives no relative weights, since XY have an enhanced global $\text{SU}(2)$ flavour symmetry in the $\mathbb{Z}_4[1,1,2]$ orbifold theory. We can therefore determine the full $\bar{\Bm}_{\frac{2(p_X+p_Y+p_Z)}{3}(0,0)}$ tower of BPS primaries in the gauge-coupling deformation subspace of the conformal manifold\footnote{Note that the $\text{SU}(3)$ sector spanned by $X$, $Y$, $Z$ is not closed beyond one loop, since length-changing interactions allow mixing with gaugino bilinears.}
\begin{equation}
\tilde{\mathcal{O}}_{p_X,p_Y,p_Z}=\Tr[X^{p_X}Y^{p_Y}\tilde Z^{p_Z}]\,+\,\text{perms.}\,.
\end{equation}
A similar discussion for the $\mathbb{Z}_2\times \mathbb{Z}_2$ orbifold theory at $\LS=0$ yields the BPS operators
\begin{equation}\label{eq:deformedop}
    \tilde{\mathcal{O}}_{p_X,p_Y,p_Z}=\Tr[\tilde X^{p_X}\tilde Y^{p_Y}\tilde Z^{p_Z}]\,+\,\text{perms.}\,,\qquad \tilde{\Phi}_{ij}=\sqrt{2-\kappa_i^2}\Phi_{ij}\,.
\end{equation}
A generalisation to arbitrary members of our orbifold families requires an analysis of the appropriate superpotential couplings determining the holomorphic XYZ-sector interactions. These are included as wave-function renormalisation constants in the fields $\tilde\Phi$. In fact, we may conjecture that this polynomial form of the BPS primaries persists at finite coupling upon making the replacements
\begin{equation}
    \kappa_a \longrightarrow
    \kappa_a^{\mathrm{eff}}\,,
    \qquad
    \Phi_{ij}\longrightarrow\Phi_{ij}^{\mathrm{ren}}\,,
\end{equation}
where $\kappa_a^{\rm eff}$ denotes the effective couplings discussed in \cite{Mitev:2014yba} and $\Phi_{ij}^{\mathrm{ren}}$  the renormalised fields.
Note, however, that these operators cease to be protected when the Leigh-Strassler deformation $\LS$ is turned on, as will be discussed in \textbf{Part II} \cite{PartII}.

\paragraph{\text{SU}(2) sectors} To explore the state space beyond the BPS spectrum, it is useful to first consider simple subsectors of states, such as the $\text{SU}(2)$-sectors of single-trace operators which only consist of two scalar fields such as $X$ and $Y$. In the $\mathbb{Z}_k[a_X,a_Y,a_Z]$ orbifold family, the one-loop Hamiltonian acts on the nearest neighbour basis $(XX,XY,YX,YY)$ as 
\begin{equation}
    \mathcal{H}_{XY}=\begin{pmatrix}
        0&0&0&0\\0& (f^+_{i})^2 & -f^+_if^-_{i+a_Y}&0\\0 & -f^+_if^-_{i+a_Y} & (f^-_{i+a_Y})^2 &0\\0&0&0&0
    \end{pmatrix}\,,
\end{equation}
where $i$ denotes the leftmost gauge index and we observe an explicit dependence of the Hamiltonian on the position within the quiver diagram. As we move along a single-trace operator, this position changes, resulting in a ``dynamical'' change of the Hamiltonian \cite{Pomoni:2021pbj}. The resulting dynamical spin-chain model does not obey the rules of standard integrability, but we can nevertheless investigate its spectrum in close analogy. For example, we may consider low numbers of ``magnon'' excitations $Y$ on a vacuum of $X$ fields and determine their energy spectrum. The usual starting point is just a single $Y$ magnon moving on an infinite line of $X$ fields with quasi-momentum $p$
\begin{equation}
    \ket{p}=\sum_{n}u_n e^{ipn}|\dots XX\underset{\substack{\uparrow \\ n}}{Y}XX\dots\rangle\,.
\end{equation}
Since the $X$ fields periodically return to the same gauge node after $P=\tfrac{k}{\gcd(a_X,k)}$ steps, we may impose a Bloch-wave Ansatz with periodic coefficients $u_n=u_{n+P}$. The energy $E$ (mapped to the anomalous dimension as $\gamma\propto g^2_0E$) is then fixed via the magnon dispersion relation
\begin{equation}
\Tr\left[T_P(E)\dots T_2(E)T_1(E)\right]=2 \cos (Pp)\,,
\end{equation}
where the left-hand side is  given in terms of the transfer elements 
\begin{equation}
T_n(E)=
\begin{pmatrix}
\frac{\big(f^+_{1+(n-1)a_X}\big)^2+\big(f^-_{1+n a_X+a_Y}\big)^2-E}{f^+_{1+n a_X}\,f^-_{1+n a_X+a_Y}}
& -\frac{f^+_{1+(n-1)a_X}\,f^-_{1+(n-1)a_X+a_Y}}{f^+_{1+n a_X}\,f^-_{1+n a_X+a_Y}}\\[14pt]
1 & 0
\end{pmatrix}\,,
\end{equation}
which ultimately generate a polynomial of order $P$ in $E$.

For example, we may consider the case of the $\mathbb{Z}_4[1,1,2]$ orbifold theory in the parametrisation of Appendix \ref{App:survey} and find the solutions 
\begin{equation}
E=\kappa_1^2+\kappa_2^2\pm\sqrt{\kappa_1^4+\kappa_2^4-2\LS^2\pm2\sqrt{(\kappa_1^4-\LS^2)(\kappa_2^4-\LS^2)}\,\cos 2p}\,.
\end{equation}
We may then form single-trace operators of lengths $L\in \{4,8,12,\dots\}$ with a single magnon of momentum $p\in\{0,\tfrac{1}{2}\pi,\pi,\tfrac{3}{2}\pi\}$ and observe energies 
\begin{equation}
    E=\kappa_1^2+\kappa_2^2\pm\sqrt{\kappa_1^4-\LS^2}\pm\sqrt{\kappa_2^4-\LS^2}\,.
\end{equation}
These are directly confirmed by our Hamiltonian implementation. A similar analysis can be performed in the $XZ$ and $YZ$ sectors, where the nearest-neighbour Hamiltonians are given by
\begin{equation}
    \mathcal{H}_{XZ}=\begin{pmatrix}
        0&0&0&0\\0& (f^-_{i})^2 & -f^-_if^+_{i+a_Z}&0\\0 & -f^-_if^+_{i+a_Z} & (f^+_{i+a_Z})^2 &0\\0&0&0&0
    \end{pmatrix}\,,\quad \mathcal{H}_{YZ}=\begin{pmatrix}
        0&0&0&0\\0& (f^+_{i-a_X})^2 & -f^+_{i-a_X}f^-_{i-a_X}&0\\0 & -f^+_{i-a_X}f^-_{i-a_X} & (f^-_{i-a_X})^2 &0\\0&0&0&0
    \end{pmatrix}\,.
\end{equation}

The natural next step is a continuation to higher magnon numbers where explicit eigenstates can likewise be constructed, thus laying the groundwork for an analysis of hidden symmetry structures \cite{Pomoni:2019oib,Pomoni:2021pbj,Bozkurt:2024tpz,Bertle:2024djm,Bozkurt:2025exl}. We provide further application examples for the $\mathbb{Z}_4[1,1,2]$ and $\mathbb{Z}_2\times \mathbb{Z}_2$ orbifold theories in the ancillary Mathematica notebook.

\section{Dual brane systems}
\label{sec:Branes}

We have now investigated $\mathcal{N}=1$ orbifold theories mostly from a gauge-theory perspective, where we were able to turn on marginal deformations, explore the conformal manifold and construct the one-loop dilatation operator to generate perturbative spectral data. One may now wonder how these deformations translate to the string-theory side.

In the type IIB orbifold construction starting from the stack of D3 branes on the orbifold singularity of $\mathbb{R}^{1,3}\times \mathbb{R}^{6}/\Gamma$ presented in Table \ref{tab:orbifold_action}, the overall gauge coupling $g_0$ is directly related to the radius of the resulting $AdS_5\times S^{5}/\Gamma$ background. The Leigh-Strassler deformation $\LS$ is captured by a deformation of the $S^5$ geometry, which squashes the sphere while retaining a $\text{U}(1)\times \text{U}(1)$ isometry and turns on various RR and NSNS background fields \cite{Leigh:1995ep,Lunin:2005jy}. More obscurely, the relative gauge-coupling deformations are governed by $B_2$-fields wrapping collapsed exceptional two-cycles $E_i$ hidden in the fixed points of special twist actions \eqref{eq: special twist}
\begin{equation}
    \frac{1}{\kappa_i^2}\sim \int_{E_i} B_2\,.
\end{equation}
We briefly discussed these cycles in Section \ref{sec: strings}, but in order to investigate them in depth one either has to resolve the orbifold singularities or analyse the associated twisted sectors directly on the string worldsheet   \cite{Aspinwall:1994ev,Gukov:1998kk,Skrzypek:2023fkr,Martinez:2025jjq}.\footnote{Currently the resolution procedure is best understood for $\mathbb{C}^2/\mathbb{Z}_k$ singularities, which are resolved by Gibbons-Hawking backgrounds \cite{Gibbons:1978tef}. A complete resolution in curved backgrounds and in the $\mathcal{N}=1$ case is an interesting open problem, presumably best tackled after appropriate T-dualisation.} Alternatively, string dualities allow us to interpret the gauge-coupling deformations from an equivalent brane construction.

\begin{figure}[t]
    \centering

    \begin{tikzpicture}[
        line cap=round,
        line join=round,
        font=\small,
        ns5/.style={
            draw=grey!65!black,
            fill=grey!18,
            very thick
        },
        d4/.style={
            draw= black,
            line width=2.5pt
        }
    ]

    \begin{scope}[xshift=0cm, scale=0.8]

        %
        \draw[d4]
            (0,0)
            ellipse[
                x radius=2.8cm,
                y radius=0.82cm
            ];

        %
        \begin{scope}[xshift=-1.45cm]

            \filldraw[ns5]
                (-0.48,-2.10)
                -- ( 0.48,-1.70)
                -- ( 0.48, 2.10)
                -- (-0.48, 1.70)
                -- cycle;

        \end{scope}

        \begin{scope}[xshift=1.45cm]

            \filldraw[ns5]
                (-0.48,-2.10)
                -- ( 0.48,-1.70)
                -- ( 0.48, 2.10)
                -- (-0.48, 1.70)
                -- cycle;

        \end{scope}

        \node[
            grey!65!black
        ] at (-1.45,2.48)
        {$\mathrm{NS5}$};

        \node[
            grey!65!black
        ] at (1.45,2.48)
        {$\mathrm{NS5}$};

        \node[
            black
        ] at (0,-1.23)
        {$N\times\mathrm{D4}$};

        \node[
            black
        ] at (0,0.45)
        {$N\times\mathrm{D4}$};

    \draw[d4]
    ({2.8*cos(240)},{0.82*sin(240)})
    arc[
        start angle=240,
        end angle=250,
        x radius=2.8cm,
        y radius=0.82cm
    ];
    \draw[d4]
    ({2.8*cos(303)},{0.82*sin(303)})
    arc[
        start angle=303,
        end angle=360,
        x radius=2.8cm,
        y radius=0.82cm
    ];

    \end{scope}
    \end{tikzpicture}

    \caption{
        Type IIA Hanany--Witten brane configuration \cite{Witten:1997sc,Lykken:1997gy} T-dual to an $\mathcal{N}=2$ $\mathbb{Z}_2$ orbifold theory. D4-branes wrapping the T-duality circle are suspended between NS5 branes which arise as dual description of the orbifold geometry. The distance between the NS5 branes determines the extension of the D4 branes and thus the relative gauge coupling of the worldvolume gauge theory \cite{Gadde:2009dj}.
    }
    \label{fig:IIA-Hanany-Witten}

\end{figure}
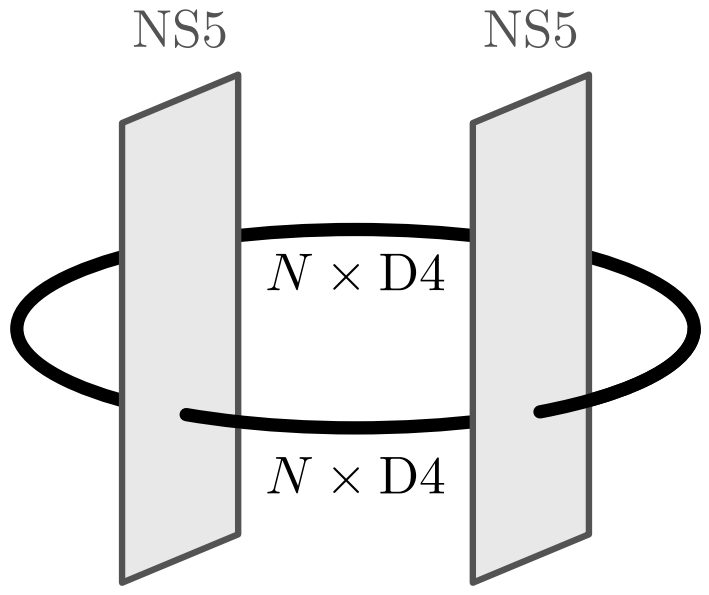

For $\mathcal{N}=2$ orbifold theories  this can be done by T-dualising to a type IIA Hanany-Witten setup \cite{Witten:1997sc, Lykken:1997gy, Gadde:2009dj} which features NS5 branes distributed along the T-duality circle and stacks of D4 branes suspended between them. We depict this setup in Figure \ref{fig:IIA-Hanany-Witten} for the simplest $\mathcal{N}=2$ $\mathbb{Z}_2$ orbifold theory. The marginal gauge-coupling deformations (which are the only marginal deformations in $\mathcal{N}=2$) are realised geometrically as the distances between the NS5 branes. 

\begin{table}[ht]
\centering
\renewcommand{\arraystretch}{1.3}
1) IIA:\quad \begin{tabular}{c|cccc|cccccc}
\hline
\rowcolor{gray!20}  & $0$ & $1$ & $2$ & $3$ & $4$ &  $5$ &$6$ &$7$ &$8$ &$9$  \\
\hline
$N\times $D4 & $\times$ & $\times$ & $\times$ & $\times$ & $\times$ & -- & -- & -- & -- & --\\
\hline
$k\times $NS5 & $\times$ & $\times$ & $\times$ & $\times$ &
--&
-- & $\times$ & $\times$ & -- & --\\
$\mathbb{Z}_\ell$ &--
 &--
 &--
 &--&
-- & --
& $\times$ & $\times$ &
 $\times$ & $\times$ \\ 
\hline
\end{tabular}

\phantom{word}

2) IIB:\quad \begin{tabular}{c|cccc|cccccc}
\hline
\rowcolor{gray!20}  & $0$ & $1$ & $2$ & $3$ & $4$ &  $5$ &$6$ &$7$ &$8$ &$9$  \\
\hline
$N\times $D5 & $\times$ & $\times$ & $\times$ & $\times$ & $\times$ & -- & $\times$ & -- & -- & --\\
\hline
$k\times $NS5 & $\times$ & $\times$ & $\times$ & $\times$ &
--&
-- & $\times$ & $\times$ & -- & --\\
$\ell\times $NS5&$\times$ & $\times$ & $\times$ & $\times$
& $\times$ & $\times$ &
--&
-- &-- & -- \\ 
\hline
\end{tabular}
\caption{Starting from the $\mathbb{Z}_k\times \mathbb{Z}_\ell$ orbifold configuration in Table \ref{tab:orbifold_action}, we may perform two successive T-dualities. The first one generates an orbifolded type IIA Hanany-Witten scenario while the second one results in a type IIB brane-box realisation.}
\label{tab:Tdualities}
\end{table}

\begin{figure}[t]
    \centering

    \begin{minipage}[c]{0.47\textwidth}
        \centering
        \begin{tikzpicture}[
            x=1.25cm,
            y=1.00cm,
            line width=0.8pt,
            every node/.style={font=\small}
        ]

        \foreach \i in {1,2,3}{
            \foreach \j in {1,2,3}{
                \draw (\j-1,-\i+1) rectangle (\j,-\i);
                \node at (\j-0.5,-\i+0.5) {$(\i,\j)$};
            }
        }

        \foreach \y in {0,-1,-2,-3}{
            \draw (-0.25,\y) -- (0,\y);
            \draw (3,\y) -- (3.25,\y);
        }

        \foreach \x in {0,1,2,3}{
            \draw (\x,0.25) -- (\x,0);
            \draw (\x,-3) -- (\x,-3.25);
        }

        \foreach \y in {-0.5,-1.5,-2.5}{
            \node at (3.50,\y) {$\cdots$};
        }

        \foreach \x in {0.5,1.5,2.5}{
            \node at (\x,-3.50) {$\vdots$};
        }

        \node at (3.50,-3.50) {$\ddots$};

        \end{tikzpicture}

        \vspace{0.4em}
       (a)
    \end{minipage}
    \hfill
    \begin{minipage}[c]{0.47\textwidth}
        \centering
        \begin{tikzpicture}[
    x=1.25cm,
    y=1.00cm,
    line width=0.8pt,
    every node/.style={font=\small}
]


\foreach \i/\yt/\yb in {
    1/0/-0.82,
    2/-0.82/-1.90,
    3/-1.90/-3
}{
    \foreach \j/\xl/\xr in {
        1/0/0.75,
        2/0.75/1.82,
        3/1.82/3
    }{
        \draw (\xl,\yt) rectangle (\xr,\yb);

        \pgfmathsetmacro{\xc}{(\xl+\xr)/2}
        \pgfmathsetmacro{\yc}{(\yt+\yb)/2}
        \node at (\xc,\yc) {$(\i,\j)$};
    }
}

\foreach \y in {0,-0.82,-1.90,-3}{
    \draw (-0.25,\y) -- (0,\y);
    \draw (3,\y) -- (3.25,\y);
}

\foreach \x in {0,0.75,1.82,3}{
    \draw (\x,0.25) -- (\x,0);
    \draw (\x,-3) -- (\x,-3.25);
}

\node at (3.50,-0.41) {$\cdots$};
\node at (3.50,-1.36) {$\cdots$};
\node at (3.50,-2.45) {$\cdots$};

\node at (0.375,-3.50) {$\vdots$};
\node at (1.285,-3.50) {$\vdots$};
\node at (2.41,-3.50) {$\vdots$};

\node at (3.50,-3.50) {$\ddots$};

\end{tikzpicture}

        \vspace{0.4em}
        (b)
    \end{minipage}

    \caption{
        Depiction of the $(46)$ torus in the brane-box construction of $\mathbb{Z}_k\times \mathbb{Z}_\ell$ quiver gauge theories. (a) displays the orbifold point configuration. Each parallelogram denotes a D5 patch hosting a worldvolume $\text{SU}(N)$ gauge theory with coupling determined by the area of the patch. The lines correspond to NS5 branes intersecting the $(46)$ torus.  Toroidal identification of the lattice is implied. (b) Relative gauge-coupling deformations correspond to moving the NS5 branes to create an inhomogeneous lattice. Note the similarity to the $\mathcal{N}=2$ scenario in Figure \ref{fig:IIA-Hanany-Witten}, where the distance between NS5 branes likewise determined the relative gauge couplings. }
    \label{fig:brane-box}
\end{figure}
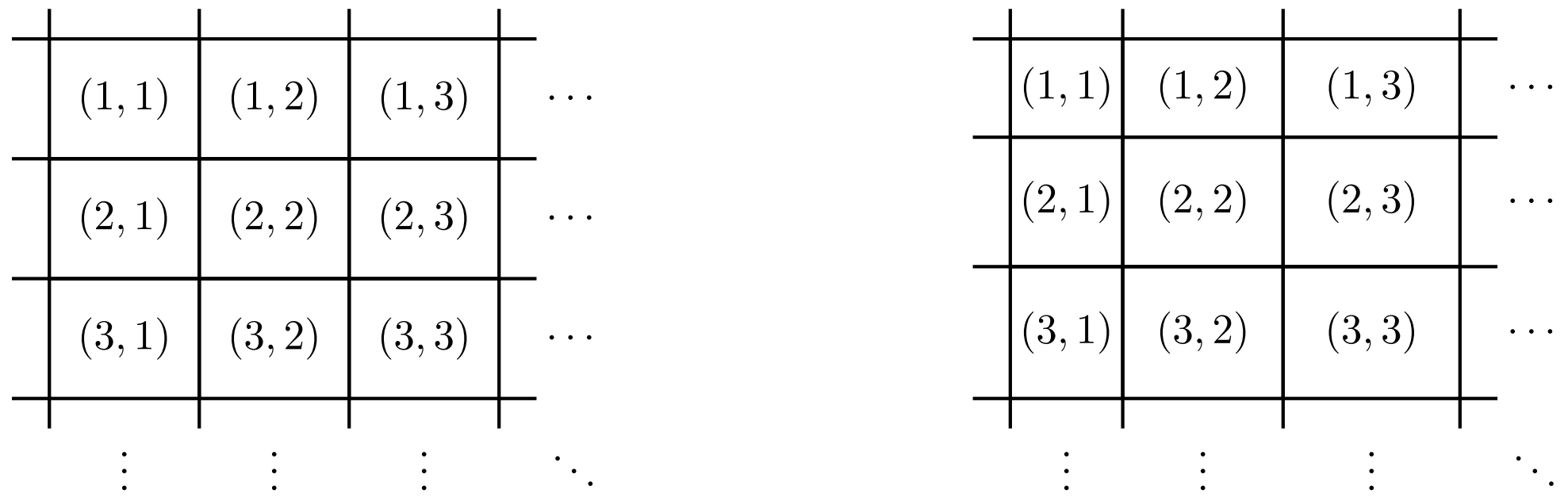

This brane construction naturally generalises to the $\mathcal{N}=1$ $\mathbb{Z}_k\times\mathbb{Z}_{\ell}$ orbifold theories, which can be regarded as $\mathbb{Z}_{\ell}$ orbifolds of an $\mathcal{N}=2$ $\mathbb{Z}_k$ necklace theory. Applying the same quotient to the type IIA Hanany--Witten setup gives an orbifolded D4/NS5-brane configuration of the kind used to construct class-$\mathcal{S}_{k}$ theories \cite{Hanany:2015pfa,Franco:2015jna}. If we want to completely resolve all orbifold actions, we have to perform yet another T-duality (see Table \ref{tab:Tdualities}) which brings us back to type IIB theory with now an intersecting grid of NS5 branes with D5 branes spanned between them.\footnote{Note in particular that T-duality along a longitudinal direction of an NS5 brane interchanges the type IIA/IIB NS5 branes without changing dimensionality, in contrast to the behaviour of D-branes \cite{Eyras:1998hn}.} These models are known as ``brane-box" models \cite{Hanany:1997tb,Hanany:1998it}. In Figure \ref{fig:brane-box} we depict the $(46)$ torus on which the intersecting NS5 branes are drawn as lines and the rectangular areas correspond to separate patches of D5 branes. Each D5 patch hosts its own worldvolume $\text{SU}(N)_{(i,j)}$ gauge theory with the gauge coupling $g_{(i,j)}$ determined by the area $\mathcal{A}_{(i,j)}$ of the patch
\cite{Imamura:2007dc}
\begin{equation}
    \frac{1}{g_{(i,j)}^2}\sim \frac{\mathcal{A}_{(i,j)}}{\alpha'g_s^{TT}}\sim\frac{\mathcal{A}_{(i,j)}}{\mathcal{A}_{T^2}}\frac{1}{g_s}\,,\label{eq:arealaw}
\end{equation}
where the new string coupling $g_s^{TT}\sim\frac{\mathcal{A}_{T^2}}{\alpha'}g_s $ is related to the orbifold string coupling $g_s$ by two successive T-dualisations.
This relation follows from the DBI action after compactification. At the orbifold point all D5 patches have the same area, but once we move the NS5 branes around as in Figure \ref{fig:brane-box}(b) the gauge couplings become unequal. We thus have a geometric realisation of the gauge coupling deformations in terms of brane-separations: For $\mathbb{Z}_k\times\mathbb{Z}_\ell$ orbifold theories we can move $k+\ell-2$ NS5-branes relative to each other.  Additional $\gcd(k,\ell)-1$ gauge coupling deformations correspond to more involved deformations of relative intersection angles of the NS5 branes, which can only be modified consistently, if the diagonals of the brane-box diagram factorise into $\gcd(k,\ell)$ separate strips.

In this geometric picture, we observe that gauge coupling modifications are not quite independent but usually affect several boxes at once. In particular it seems impossible to shrink one D5 patch to significantly smaller area than its neighbouring ones. An even stronger constraint was imposed by the $\beta$-functions in \eqref{eq:obstruction}, which translates to a bound
\begin{equation}
    \frac{1}{\Am_{i+1,j+1}}\leq\frac{1}{\Am_{i,j}}+\frac{1}{\Am_{i+1,j}}+\frac{1}{\Am_{i,j+1}}\,,
\end{equation}
obstructing each D5 box from becoming significantly smaller than its neighbours. 

At this stage we have only discussed brane constructions without any backreaction. Of course the holographic duality applies after a large number $N$ of D-branes have deformed the geometry to an $AdS_5\times \mathcal{M}$ spacetime. Determining the correct backreacted geometry of this intricate web of NS5 and D-branes is a challenge for future research. Another gap we need to fill is a parallel discussion of $\mathbb{Z}_k[a_X,a_Y,a_Z]$ orbifolds and although one may be able to construct appropriate brane-box models with twisted toroidal identifications for these theories, there is a much more direct approach available in terms of brane tilings on the T-duality torus.

\subsection{Brane tiling construction}

As we discussed above, for $\mathcal{N}=1$ orbifold theories a single T-duality is not enough to convert all orbifold actions into branes. For the $\mathbb{Z}_k\times\mathbb{Z}_
\ell$ family the previous brane-box discussion provided a first brane-realisation. An alternative construction which extends directly to the $\mathbb{Z}_k$ family is provided by type IIB brane-tiling models \cite{Hanany:2005ve,Kennaway:2007tq}, which arise after two T-dualities and again feature D5 branes on the resulting compact torus (see Table \ref{tab:brane-tiling-configuration}). An NS5 brane wraps a holomorphic curve $\Sigma$ and cuts the D5 branes into patches which at the level of topology are represented by the diagrams in Figure \ref{fig: Brane-tiling}. These diagrams encode the information of the quiver diagrams (e.g. Figures \ref{fig:Zkquiver} and \ref{fig:ZkZlquiver}) in the brane intersection structure on the torus. 
\begin{table}[ht]
\centering
\renewcommand{\arraystretch}{1.3}
\setlength{\tabcolsep}{6pt}
\begin{tabular}{c|cccc|cccccc}
\hline
 \rowcolor{gray!20} & $0$ & $1$ & $2$ & $3$
 & $4$ & $5$ & $6$ & $7$ & $8$ & $9$ \\
\hline

D5
 & $\times$ & $\times$ & $\times$ & $\times$
 & $\times$ & -- & $\times$ & -- & -- & -- \\

NS5
 & $\times$ & $\times$ & $\times$ & $\times$
 & \multicolumn{4}{c}{$\Sigma\subset(4567)$}
 & -- & -- \\
\hline
\end{tabular}
\caption{Brane configuration underlying the brane-tiling construction \cite{Hanany:2005ve,Kennaway:2007tq}.
The D5 branes wrap the compact $(46)$ torus, while the NS5-brane
wraps a holomorphic curve $\Sigma$ in the $(4567)$ directions.
The four-dimensional quiver gauge theory is supported on the common
$(0123)$ directions.}
\label{tab:brane-tiling-configuration}
\end{table}
\begin{figure}[htbp]
\centering

\begin{minipage}[t]{0.48\textwidth}
\centering

\begin{tikzpicture}[
    scale=1.00,
    line width=0.8pt,
    hexlabel/.style={
        font=\small,
        fill=white,
        inner sep=1.5pt
    },
    edgelabel/.style={
        font=\scriptsize,
        fill=white,
        inner sep=1pt
    },
    blackvertex/.style={
        circle,
        draw=black,
        fill=black,
        minimum size=5pt,
        inner sep=0pt
    },
    whitevertex/.style={
        circle,
        draw=black,
        fill=white,
        minimum size=5pt,
        inner sep=0pt
    },
    continuation/.style={
        line width=0.8pt
    }
]

\begin{scope}[rotate=90]


\coordinate (p1)  at ( 1.0,  0.000);
\coordinate (p2)  at ( 0.5,  0.866);
\coordinate (p3)  at (-0.5,  0.866);
\coordinate (p4)  at (-1.0,  0.000);
\coordinate (p5)  at (-0.5, -0.866);
\coordinate (p6)  at ( 0.5, -0.866);

\coordinate (p7)  at ( 2.5,  0.866);
\coordinate (p8)  at ( 2.0,  1.732);
\coordinate (p9)  at ( 1.0,  1.732);
\coordinate (p10) at ( 2.0,  0.000);

\coordinate (p11) at ( 2.5, -0.866);
\coordinate (p12) at ( 1.0, -1.732);
\coordinate (p13) at ( 2.0, -1.732);

\coordinate (p14) at (-1.0, -1.732);
\coordinate (p15) at (-0.5, -2.598);
\coordinate (p16) at ( 0.5, -2.598);


\draw
    (p1) -- (p2) -- (p3) --
    (p4) -- (p5) -- (p6) -- cycle;

\draw
    (p7) -- (p8) -- (p9) --
    (p2) -- (p1) -- (p10) -- cycle;

\draw
    (p11) -- (p10) -- (p1) --
    (p6) -- (p12) -- (p13) -- cycle;

\draw
    (p12) -- (p6) -- (p5) --
    (p14) -- (p15) -- (p16) -- cycle;


\draw[continuation] (p3)  -- ++(120:0.42);
\draw[continuation] (p4)  -- ++(180:0.42);

\draw[continuation] (p7)  -- ++(0:0.42);
\draw[continuation] (p8)  -- ++(60:0.42);
\draw[continuation] (p9)  -- ++(120:0.42);

\draw[continuation] (p11) -- ++(0:0.42);
\draw[continuation] (p13) -- ++(-60:0.42);

\draw[continuation] (p14) -- ++(180:0.42);
\draw[continuation] (p15) -- ++(-120:0.42);
\draw[continuation] (p16) -- ++(-60:0.42);


\node[hexlabel] at (0,0)        {$1$};
\node[hexlabel] at (1.5,0.866)  {$1+a_Y$};
\node[hexlabel] at (1.5,-0.866) {$1-a_Z$};
\node[hexlabel] at (0,-1.732)   {$1+a_X$};



\path (p1) -- (p2)
    node[midway, edgelabel] {$Y$};

\path (p2) -- (p3)
    node[midway, edgelabel] {$X$};

\path (p3) -- (p4)
    node[midway, edgelabel] {$Z$};

\path (p4) -- (p5)
    node[midway, edgelabel] {$Y$};

\path (p5) -- (p6)
    node[midway, edgelabel] {$X$};

\path (p6) -- (p1)
    node[midway, edgelabel] {$Z$};


\foreach \v in {p1,p3,p5,p7,p9,p11,p12,p15}{
    \node[blackvertex] at (\v) {};
}

\foreach \v in {p2,p4,p6,p8,p10,p13,p14,p16}{
    \node[whitevertex] at (\v) {};
}

\end{scope}
\end{tikzpicture}

\par\smallskip
{$\mathbb{Z}_k[a_X,a_Y,a_Z]$}

\end{minipage}
\hfill
\begin{minipage}[t]{0.48\textwidth}
\centering

\begin{tikzpicture}[
    scale=1.00,
    line width=0.8pt,
    hexlabel/.style={
        font=\small,
        fill=white,
        inner sep=1.5pt
    },
    edgelabel/.style={
        font=\scriptsize,
        fill=white,
        inner sep=1pt
    },
    blackvertex/.style={
        circle,
        draw=black,
        fill=black,
        minimum size=5pt,
        inner sep=0pt
    },
    whitevertex/.style={
        circle,
        draw=black,
        fill=white,
        minimum size=5pt,
        inner sep=0pt
    },
    continuation/.style={
        line width=0.8pt
    }
]

\begin{scope}[rotate=90]


\coordinate (p1)  at ( 1.0,  0.000);
\coordinate (p2)  at ( 0.5,  0.866);
\coordinate (p3)  at (-0.5,  0.866);
\coordinate (p4)  at (-1.0,  0.000);
\coordinate (p5)  at (-0.5, -0.866);
\coordinate (p6)  at ( 0.5, -0.866);

\coordinate (p7)  at ( 2.5,  0.866);
\coordinate (p8)  at ( 2.0,  1.732);
\coordinate (p9)  at ( 1.0,  1.732);
\coordinate (p10) at ( 2.0,  0.000);

\coordinate (p11) at ( 2.5, -0.866);
\coordinate (p12) at ( 1.0, -1.732);
\coordinate (p13) at ( 2.0, -1.732);

\coordinate (p14) at (-1.0, -1.732);
\coordinate (p15) at (-0.5, -2.598);
\coordinate (p16) at ( 0.5, -2.598);


\draw
    (p1) -- (p2) -- (p3) --
    (p4) -- (p5) -- (p6) -- cycle;

\draw
    (p7) -- (p8) -- (p9) --
    (p2) -- (p1) -- (p10) -- cycle;

\draw
    (p11) -- (p10) -- (p1) --
    (p6) -- (p12) -- (p13) -- cycle;

\draw
    (p12) -- (p6) -- (p5) --
    (p14) -- (p15) -- (p16) -- cycle;


\draw[continuation] (p3)  -- ++(120:0.42);
\draw[continuation] (p4)  -- ++(180:0.42);

\draw[continuation] (p7)  -- ++(0:0.42);
\draw[continuation] (p8)  -- ++(60:0.42);
\draw[continuation] (p9)  -- ++(120:0.42);

\draw[continuation] (p11) -- ++(0:0.42);
\draw[continuation] (p13) -- ++(-60:0.42);

\draw[continuation] (p14) -- ++(180:0.42);
\draw[continuation] (p15) -- ++(-120:0.42);
\draw[continuation] (p16) -- ++(-60:0.42);

\node[hexlabel] at (0,0)        {$(1,1)$};
\node[hexlabel] at (1.5,0.866)  {$(1,2)$};
\node[hexlabel] at (1.5,-0.866) {$(2,2)$};
\node[hexlabel] at (0,-1.732)   {$(2,1)$};


\path (p1) -- (p2)
    node[midway, edgelabel] {$Y$};

\path (p2) -- (p3)
    node[midway, edgelabel] {$X$};

\path (p3) -- (p4)
    node[midway, edgelabel] {$Z$};

\path (p4) -- (p5)
    node[midway, edgelabel] {$Y$};

\path (p5) -- (p6)
    node[midway, edgelabel] {$X$};

\path (p6) -- (p1)
    node[midway, edgelabel] {$Z$};

\foreach \v in {p1,p3,p5,p7,p9,p11,p12,p15}{
    \node[blackvertex] at (\v) {};
}

\foreach \v in {p2,p4,p6,p8,p10,p13,p14,p16}{
    \node[whitevertex] at (\v) {};
}

\end{scope}
\end{tikzpicture}

\par\smallskip
{$\mathbb{Z}_k\times \mathbb{Z}_\ell$}

\end{minipage}

\caption{Brane-tiling lattice of the two families of Abelian $\mathcal{N}=1$ orbifold theories. The hexagonal faces denote individual patches of $D5$ stacks which host the $\text{SU}(N)$ gauge groups. The lines denote intersections with the NS5 brane. Open strings connecting to the adjacent faces contribute the bifundamental chiral multiplets. At nodes of the diagram the chiral multiplets interact, representing the superpotential terms. We have indicated how appropriate gauge group labels are assigned in  both families of orbifold theories. The pattern repeats periodically and we are supposed to restrict our attention to a rectangular fundamental cell which forms the $(46)$ torus of the embedding background. }
\label{fig: Brane-tiling}
\end{figure}

\begin{figure}[htbp]
\centering


\begin{minipage}[t]{0.48\textwidth}
\vspace{0pt}
\centering

\begin{tikzpicture}[
    line width=0.8pt,
    hexlabel/.style={
        font=\small,
        inner sep=1.5pt
    },
    blackvertex/.style={
        circle,
        draw=black,
        fill=black,
        minimum size=5pt,
        inner sep=0pt
    },
    whitevertex/.style={
        circle,
        draw=black,
        fill=white,
        minimum size=5pt,
        inner sep=0pt
    },
    continuation/.style={
        line width=0.8pt
    },
    unitcell/.style={ 
        draw=red,
        densely dashed,
        line width=1.1pt
    }
]


\pgfmathsetmacro{\s}{0.72}
\pgfmathsetmacro{\dx}{sqrt(3)*\s}
\pgfmathsetmacro{\dy}{1.5*\s}


\path[use as bounding box]
    (-1.0,-3.16)
    rectangle
    ({4*\dx+1.0},1.0);


\coordinate (A) at (0,0);
\coordinate (B) at (\dx,0);
\coordinate (C) at ({2*\dx},0);
\coordinate (D) at ({3*\dx},0);

\coordinate (E) at ({0.5*\dx},{-\dy});
\coordinate (F) at ({1.5*\dx},{-\dy});
\coordinate (G) at ({2.5*\dx},{-\dy});
\coordinate (H) at ({3.5*\dx},{-\dy});

\coordinate (I) at ({1.0*\dx},{-2*\dy});
\coordinate (J) at ({2.0*\dx},{-2*\dy});
\coordinate (K) at ({3.0*\dx},{-2*\dy});
\coordinate (L) at ({4.0*\dx},{-2*\dy});


\foreach \c in {A,C,F,H,I,K}{
    \begin{scope}[shift={(\c)}]
        \fill[gray!15]
            (90:\s)   --
            (30:\s)   --
            (-30:\s)  --
            (-90:\s)  --
            (-150:\s) --
            (150:\s)  --
            cycle;
    \end{scope}
}


\foreach \c in {A,B,C,D,E,F,G,H,I,J,K,L}{
    \begin{scope}[shift={(\c)}]
        \draw
            (90:\s)   --
            (30:\s)   --
            (-30:\s)  --
            (-90:\s)  --
            (-150:\s) --
            (150:\s)  --
            cycle;
    \end{scope}
}


\foreach \c in {A,B,C,D}{
    \begin{scope}[shift={(\c)}]
        \draw[continuation]
            (90:\s)
            --
            (90:{1.30*\s});
    \end{scope}
}

\foreach \c in {I,J,K,L}{
    \begin{scope}[shift={(\c)}]
        \draw[continuation]
            (-90:\s)
            --
            (-90:{1.30*\s});
    \end{scope}
}

\begin{scope}[shift={(A)}]
    \draw[continuation]
        (150:\s)
        --
        (150:{1.30*\s});

    \draw[continuation]
        (-150:\s)
        --
        (-150:{1.30*\s});
\end{scope}

\begin{scope}[shift={(E)}]
    \draw[continuation]
        (-150:\s)
        --
        (-150:{1.30*\s});
\end{scope}

\begin{scope}[shift={(I)}]
    \draw[continuation]
        (-150:\s)
        --
        (-150:{1.30*\s});
\end{scope}

\begin{scope}[shift={(D)}]
    \draw[continuation]
        (30:\s)
        --
        (30:{1.30*\s});
\end{scope}

\begin{scope}[shift={(H)}]
    \draw[continuation]
        (30:\s)
        --
        (30:{1.30*\s});
\end{scope}

\begin{scope}[shift={(L)}]
    \draw[continuation]
        (30:\s)
        --
        (30:{1.30*\s});

    \draw[continuation]
        (-30:\s)
        --
        (-30:{1.30*\s});
\end{scope}


\coordinate (ucBL) at ($(E)+(0,-\s)$);
\coordinate (ucTL) at ($(B)+(30:\s)$);
\coordinate (ucTR) at ($(G)+(30:\s)$);
\coordinate (ucBR) at ($(J)+(0,-\s)$);

\draw[unitcell]
    (ucBL) -- (ucTL) -- (ucTR) -- (ucBR) -- cycle;


\node[hexlabel] at (A) {$1$};
\node[hexlabel] at (B) {$2$};
\node[hexlabel] at (C) {$3$};
\node[hexlabel] at (D) {$4$};

\node[hexlabel] at (E) {$4$};
\node[hexlabel] at (F) {$1$};
\node[hexlabel] at (G) {$2$};
\node[hexlabel] at (H) {$3$};

\node[hexlabel] at (I) {$3$};
\node[hexlabel] at (J) {$4$};
\node[hexlabel] at (K) {$1$};
\node[hexlabel] at (L) {$2$};


\foreach \c in {A,B,C,D,E,F,G,H,I,J,K,L}{
    \begin{scope}[shift={(\c)}]

        \node[blackvertex] at (90:\s) {};
        \node[blackvertex] at (-30:\s) {};
        \node[blackvertex] at (-150:\s) {};

        \node[whitevertex] at (30:\s) {};
        \node[whitevertex] at (-90:\s) {};
        \node[whitevertex] at (150:\s) {};

    \end{scope}
}

\end{tikzpicture}

\par\smallskip
\textnormal{(a)}

\end{minipage}
\hfill
\begin{minipage}[t]{0.48\textwidth}
\vspace{0pt}
\centering

\begin{tikzpicture}[
    line width=0.8pt,
    hexlabel/.style={
        font=\small,
        inner sep=1.5pt
    },
    blackvertex/.style={
        circle,
        draw=black,
        fill=black,
        minimum size=5pt,
        inner sep=0pt
    },
    whitevertex/.style={
        circle,
        draw=black,
        fill=white,
        minimum size=5pt,
        inner sep=0pt
    },
    continuation/.style={
        line width=0.8pt
    },
    unitcell/.style={
        draw=red,
        densely dashed,
        line width=1.1pt
    }
]


\pgfmathsetmacro{\s}{0.72}
\pgfmathsetmacro{\dx}{sqrt(3)*\s}
\pgfmathsetmacro{\dy}{1.5*\s}

%
%
%
%
\pgfmathsetmacro{\del}{0.3*\s}

\pgfmathsetmacro{\a}{0.5*\dx}
\pgfmathsetmacro{\ex}{sqrt(3)*\del/4}
\pgfmathsetmacro{\ey}{\del/4}

\pgfmathsetmacro{\stub}{0.30*\s}


\path[use as bounding box]
    (-1.0,-3.16)
    rectangle
    ({4*\dx+1.0},1.0);


\coordinate (A) at (0,0);
\coordinate (B) at (\dx,0);
\coordinate (C) at ({2*\dx},0);
\coordinate (D) at ({3*\dx},0);

\coordinate (E) at ({0.5*\dx},{-\dy});
\coordinate (F) at ({1.5*\dx},{-\dy});
\coordinate (G) at ({2.5*\dx},{-\dy});
\coordinate (H) at ({3.5*\dx},{-\dy});

\coordinate (I) at ({1.0*\dx},{-2*\dy});
\coordinate (J) at ({2.0*\dx},{-2*\dy});
\coordinate (K) at ({3.0*\dx},{-2*\dy});
\coordinate (L) at ({4.0*\dx},{-2*\dy});

%


\foreach \c in {A,C,F,H,I,K}{
    \begin{scope}[shift={(\c)}]

        \fill[gray!15]
            ({-\ex},{\s+\ey})
            --
            ({\a+\ex},{0.5*\s-\ey})
            --
            ({\a+\ex},{-0.5*\s-\ey})
            --
            ({\ex},{-\s-\ey})
            --
            ({-\a-\ex},{-0.5*\s+\ey})
            --
            ({-\a-\ex},{0.5*\s+\ey})
            --
            cycle;

    \end{scope}
}


\foreach \c in {A,C,F,H,I,K}{
    \begin{scope}[shift={(\c)}]

        \draw
            ({-\ex},{\s+\ey})
            --
            ({\a+\ex},{0.5*\s-\ey})
            --
            ({\a+\ex},{-0.5*\s-\ey})
            --
            ({\ex},{-\s-\ey})
            --
            ({-\a-\ex},{-0.5*\s+\ey})
            --
            ({-\a-\ex},{0.5*\s+\ey})
            --
            cycle;

    \end{scope}
}

%

\foreach \c in {B,D,E,G,J,L}{
    \begin{scope}[shift={(\c)}]

        \draw
            ({\ex},{\s-\ey})
            --
            ({\a-\ex},{0.5*\s+\ey})
            --
            ({\a-\ex},{-0.5*\s+\ey})
            --
            ({-\ex},{-\s+\ey})
            --
            ({-\a+\ex},{-0.5*\s-\ey})
            --
            ({-\a+\ex},{0.5*\s-\ey})
            --
            cycle;

    \end{scope}
}



\foreach \c in {A,C}{
    \begin{scope}[shift={(\c)}]
        \draw[continuation]
            ({-\ex},{\s+\ey})
            --
            ++(0,\stub);
    \end{scope}
}

\foreach \c in {B,D}{
    \begin{scope}[shift={(\c)}]
        \draw[continuation]
            ({\ex},{\s-\ey})
            --
            ++(0,\stub);
    \end{scope}
}


\foreach \c in {I,K}{
    \begin{scope}[shift={(\c)}]
        \draw[continuation]
            ({\ex},{-\s-\ey})
            --
            ++(0,-\stub);
    \end{scope}
}

\foreach \c in {J,L}{
    \begin{scope}[shift={(\c)}]
        \draw[continuation]
            ({-\ex},{-\s+\ey})
            --
            ++(0,-\stub);
    \end{scope}
}


\begin{scope}[shift={(A)}]

    \draw[continuation]
        ({-\a-\ex},{0.5*\s+\ey})
        --
        ++(150:\stub);

    \draw[continuation]
        ({-\a-\ex},{-0.5*\s+\ey})
        --
        ++(-150:\stub);

\end{scope}

\begin{scope}[shift={(E)}]

    \draw[continuation]
        ({-\a+\ex},{-0.5*\s-\ey})
        --
        ++(-150:\stub);

\end{scope}

\begin{scope}[shift={(I)}]

    \draw[continuation]
        ({-\a-\ex},{-0.5*\s+\ey})
        --
        ++(-150:\stub);

\end{scope}


\begin{scope}[shift={(D)}]

    \draw[continuation]
        ({\a-\ex},{0.5*\s+\ey})
        --
        ++(30:\stub);

\end{scope}

\begin{scope}[shift={(H)}]

    \draw[continuation]
        ({\a+\ex},{0.5*\s-\ey})
        --
        ++(30:\stub);

\end{scope}

\begin{scope}[shift={(L)}]

    \draw[continuation]
        ({\a-\ex},{0.5*\s+\ey})
        --
        ++(30:\stub);

    \draw[continuation]
        ({\a-\ex},{-0.5*\s+\ey})
        --
        ++(-30:\stub);

\end{scope}


\coordinate (ucBL) at ($(E)+({-\ex},{-\s+\ey})$);
\coordinate (ucTL) at ($(B)+({\a-\ex},{0.5*\s+\ey})$);
\coordinate (ucTR) at ($(G)+({\a-\ex},{0.5*\s+\ey})$);
\coordinate (ucBR) at ($(J)+({-\ex},{-\s+\ey})$);

\draw[unitcell]
    (ucBL) -- (ucTL) -- (ucTR) -- (ucBR) -- cycle;


\node[hexlabel] at (A) {$1$};
\node[hexlabel] at (B) {$2$};
\node[hexlabel] at (C) {$3$};
\node[hexlabel] at (D) {$4$};

\node[hexlabel] at (E) {$4$};
\node[hexlabel] at (F) {$1$};
\node[hexlabel] at (G) {$2$};
\node[hexlabel] at (H) {$3$};

\node[hexlabel] at (I) {$3$};
\node[hexlabel] at (J) {$4$};
\node[hexlabel] at (K) {$1$};
\node[hexlabel] at (L) {$2$};



\foreach \c in {A,C,F,H,I,K}{
    \begin{scope}[shift={(\c)}]

        \node[blackvertex]
            at ({-\ex},{\s+\ey}) {};

        \node[blackvertex]
            at ({\a+\ex},{-0.5*\s-\ey}) {};

        \node[blackvertex]
            at ({-\a-\ex},{-0.5*\s+\ey}) {};

        \node[whitevertex]
            at ({\a+\ex},{0.5*\s-\ey}) {};

        \node[whitevertex]
            at ({\ex},{-\s-\ey}) {};

        \node[whitevertex]
            at ({-\a-\ex},{0.5*\s+\ey}) {};

    \end{scope}
}


\foreach \c in {B,D,E,G,J,L}{
    \begin{scope}[shift={(\c)}]

        \node[blackvertex]
            at ({\ex},{\s-\ey}) {};

        \node[blackvertex]
            at ({\a-\ex},{-0.5*\s+\ey}) {};

        \node[blackvertex]
            at ({-\a+\ex},{-0.5*\s-\ey}) {};

        \node[whitevertex]
            at ({\a-\ex},{0.5*\s+\ey}) {};

        \node[whitevertex]
            at ({-\ex},{-\s+\ey}) {};

        \node[whitevertex]
            at ({-\a+\ex},{0.5*\s-\ey}) {};

    \end{scope}
}

\end{tikzpicture}

\par\smallskip
\textnormal{(b)}

\end{minipage}

\caption{
Brane tiling diagram for the \(\mathbb{Z}_4[1,1,2]\) example.
\textnormal{(a)} displays the orbifold point configuration. The dashed
rectangle marks a sample unit cell representing the physical
\((46)\) torus. The shaded and unshaded \(60^\circ\) diagonals
partition the gauge groups into the sets \(\{1,3\}\) and \(\{2,4\}\),
allowing a relative scaling of the D5 brane areas as in \textnormal{(b)} and hence the
gauge-coupling deformation $\kappa:=\kappa_1=\kappa_3=\sqrt{2-\kappa_2^2}=\sqrt{2-\kappa_4^2}$.}
\label{fig:brane-tiling-and-decoupling}
\end{figure}
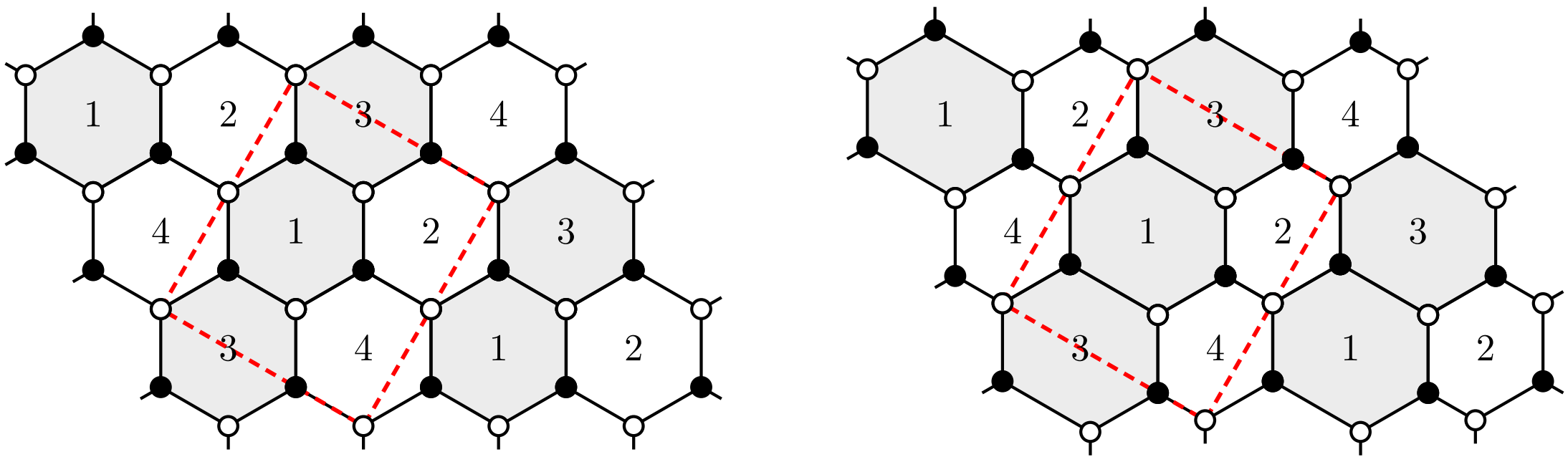

Furthermore, we can identify marginal gauge coupling deformations by noting that the area $\mathcal{A}_i$ of the D5 patches determines the gauge couplings \cite{Imamura:2007dc} as in \eqref{eq:arealaw}. After large-$N$ rescaling, one overall gauge coupling deformation is just given by the overall size of the compact geometry, as expected from the holographic dictionary. For any brane-tiling lattice that allows for a separation of gauge groups into distinct strips along one of the three principal lattice directions, we can furthermore scale the width of the strips against each other, resulting in a relative difference in D5 areas and thus gauge couplings. Take for example the $\mathbb{Z}_4[1,1,2]$ orbifold, for which the brane-tiling diagram is presented in Figure \ref{fig:brane-tiling-and-decoupling}(a). This theory has one relative gauge coupling $\kappa:=\kappa_1=\kappa_3=\sqrt{2-\kappa_2^2}=\sqrt{2-\kappa_4^2}$ and in fact, we observe that the gauge groups $\{1,3\}$ and $\{2,4\}$ arrange themselves in distinct strips along the $Z$-direction of the lattice. We can now introduce a relative scaling between these strips which corresponds precisely to the $\kappa$ deformation.

At this superficial level, we can make contact to one of the running themes of this paper: By following the construction of brane tilings outlined in Figure \ref{fig: Brane-tiling}, we can convince ourselves that the number of possible relative area scalings depends on the orbifold projection in question and is given by
\begin{equation}
     \text{gcd}(a_X,k)+\text{gcd}(a_Y,k)+\text{gcd}(a_Z,k)-3
\end{equation}
for $\mathbb{Z}_k[a_X,a_Y,a_Z]$ orbifolds and 
\begin{equation}
    k + \ell + \text{gcd}(k,\ell) -3
\end{equation}
for $\mathbb{Z}_k\times \mathbb{Z}_\ell$ orbifolds. These are of course the numbers of special twisted sectors \eqref{eq:NZk} and \eqref{eq:NZkZl} we have encountered time and time again.

\section{Conclusion}
\label{Sec:Conclusion}

In this work we performed a systematic analysis of the Abelian $\Nm=1$ orbifolds of $\Nm=4$ SYM, belonging to two $\Zb_k$ and $\Zb_k \times \Zb_{\ell}$ families. These theories provide a comparatively tractable class of holographic four-dimensional $\mathcal N=1$ SCFTs, while exhibiting a considerably richer structure of exactly marginal deformations than their $\mathcal N=2$ counterparts. We combined techniques ranging from superconformal representation theory, perturbative gauge theory and the dual string/brane description to provide a characterisation of the BPS spectrum, the weak coupling conformal manifold and the one-loop spectrum.

Working in the planar limit, we were able to determine the full single-trace BPS spectrum of these theories. From the string-theory perspective, the untwisted sector is
obtained by an orbifold projection of the Kaluza--Klein spectrum of type IIB supergravity on $S^5$ \cite{Kim:1985ez}, while additional BPS states arise in special twisted sectors with fixed $AdS_5\times S^1$ submanifolds. The same states can then be identified on the gauge theory side by decomposing the parent $\Nm=4$ and $\mathcal{N}=2$ multiplets, by means of the branching rules stated in \cite{Dolan:2002zh, Cordova:2016emh} and in \eqref{Eq:N2toN1}. The orbifold projection ensures the correct invariance condition of the multiplets. The untwisted sector of the orbifold theory is given by decomposition and projection of the BPS spectrum of $\Nm=4$ SYM, while the twisted spectrum is obtained by decomposing and projecting the $\Nm=2$ twisted-sector states of $\mathcal{N}=2$ orbifold theories. This follows from the fact that the localised supergravity modes on the twist fixed point organise in a 6d $\Nm=(2,0)$ tensor multiplet \cite{Gukov:1998kk,Skrzypek:2023fkr,Martinez:2025jjq,Mantegazza:2026spd}, which can in turn be acted upon with a further $\Nm=1$ orbifold. The superconformal index \cite{Kinney:2005ej,Romelsberger:2005eg,Rastelli:2016tbz} provides an important consistency check on this construction, but only captures protected net degeneracies. It is therefore insensitive to large parts of the complete BPS spectrum. We provided an ancillary Wolfram Mathematica notebook for computing the BPS spectrum of all members of the two orbifold families discussed in this paper.

We then turned to the conformal manifolds of these theories. Following the general relation between marginal operators, global symmetries and multiplet recombination \cite{Green:2010da,Cordova:2016xhm}, we identified the candidate marginal deformations near the orbifold point and determined which of them extend to exactly marginal directions. Generically, the conformal manifold is spanned by an overall gauge coupling $g_0$, $n_{\rm ex}$ independent relative gauge-coupling deformations $\kappa_i$, where $n_{\rm ex}$ is given in \eqref{Eq: nrZk} and \eqref{Eq: nrZkZl}, and the universal Leigh--Strassler superpotential deformation $\LS$ \cite{Leigh:1995ep}. The number $n_{\rm ex}$ has a simple diagrammatic origin. Although every gauge coupling is classically marginal, the couplings of a subset of gauge nodes in the quiver diagram can be varied relative to the rest only if the arrows connecting the subset to the complement balance at every node. For all other combinations the Konishi anomaly discussed in Section \ref{sec: U(1)s} recombines the associated global $\text{U}(1)$ current with the would-be marginal deformation multiplet. This quantum obstruction is invisible to the one-loop dilatation operator, but it is enforced by the vanishing of the beta functions and reduces the $\abs{\Gamma}-1$ relative gauge couplings to $n_{\rm ex}$ exactly marginal ones, matching the string-theory count of special twisted sectors.

We obtained an explicit parametrisation of the weak-coupling domain ($g_0\ll1$) of the conformal manifolds by imposing the vanishing of the gauge and superpotential beta functions. The counting of exactly marginal directions agrees with the flavour-symmetry quotient description of conformal manifolds and with the general analysis of supersymmetric orbifold descendants in the literature
\cite{Aharony:2002tp,Green:2010da}. Special orbifold actions may admit additional cubic marginal operators, but these do not generically provide new exactly marginal directions: they are typically obstructed by the complexified flavour-symmetry quotient or by the corresponding moment-map conditions, signalling a non-vanishing beta function. The only exceptions are the $\mathbb Z_3[1,1,1]$ and $\mathbb Z_3\times\mathbb Z_3$ theories, for which the conformal manifold is enlarged to complex dimensions three and fifteen, respectively.

Having parametrised the weak-coupling conformal manifold, we constructed the one-loop dilatation operator for all the marginally deformed orbifold theories discussed above, with the exception of the two special cases $\mathbb Z_3[1,1,1]$ and $\mathbb Z_3\times\mathbb Z_3$, for which the additional cubic marginal deformations were neglected. Starting from the harmonic action of $\Nm=4$ SYM \cite{Beisert:2003jj}, we assigned different relative gauge couplings and superpotential couplings to the various contributing processes, allowing for a universal treatment of the spectral problem which takes a parametrisation of the couplings as an input. The resulting Hamiltonian applies both to the $\mathbb Z_k$ and $\mathbb Z_k\times\mathbb Z_\ell$ family and allows us to construct the mixing matrix and one-loop anomalous dimensions of arbitrary single-trace operators as functions of the coordinates on the conformal manifold. Together with the ancillary implementation, this provides a practical tool for following the spectrum away from the orbifold point. One should keep in mind that this description is intrinsically restricted to one loop: in particular, multiplet recombinations whose anomalous dimensions first arise at higher orders, such as the cases discussed in Section \ref{sec: U(1)s} are not visible to the harmonic action.

Finally, we related the gauge-theory description of the marginal
deformations to dual brane constructions. For the
$\mathbb Z_k\times\mathbb Z_\ell$ theories, the brane-box picture provides a geometric interpretation of relative gauge couplings in terms of the areas of D5 brane patches, while brane tilings extend this description naturally to the $\mathbb Z_k$ family
\cite{Hanany:1997tb,Hanany:1998it,Hanany:2005ve,Kennaway:2007tq}.
In accordance with the general discussion of exactly marginal deformations in brane tilings \cite{Imamura:2007dc}, relative gauge-coupling deformations are realised by area deformations of the corresponding brane configuration. \\

\begin{figure}[t]
  \centering
  \includegraphics[width=0.70\textwidth]{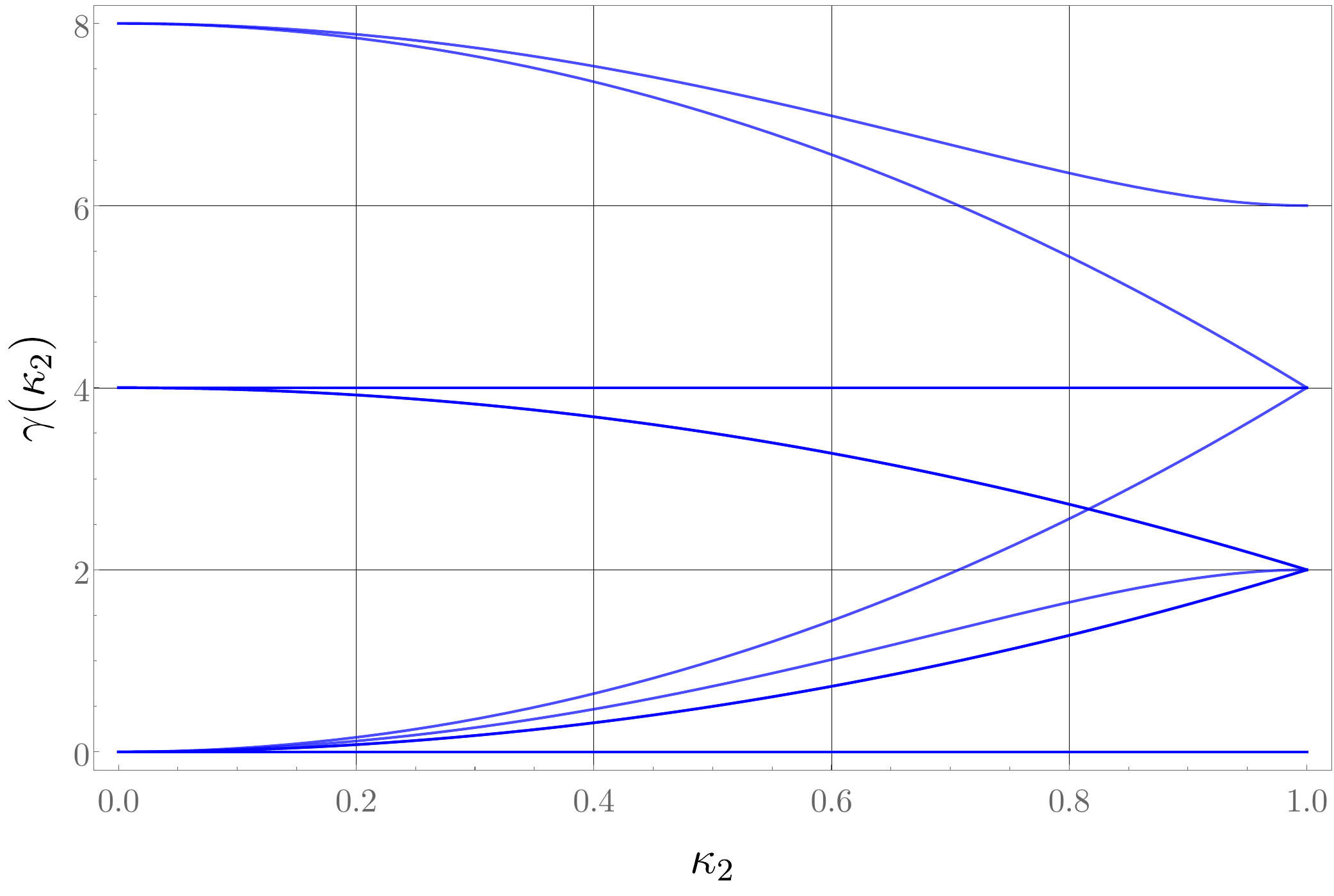}\\
  \phantom{a}\\
  \includegraphics[width=0.70\textwidth]{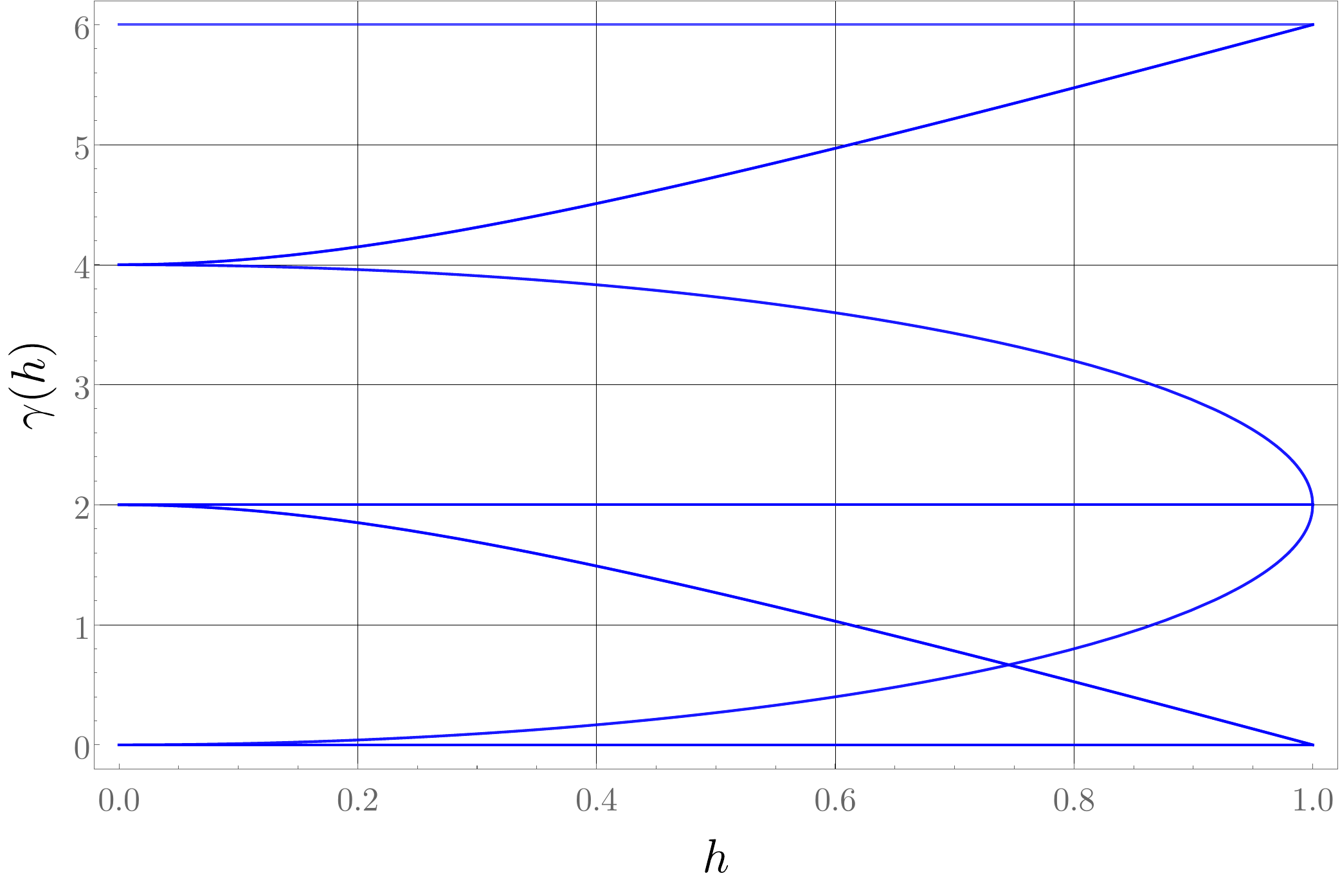}
  \caption{One-loop spectrum of scalar length-two single-trace operators with classical dimension $\Delta_0=2$. We present the anomalous dimensions (in units of $g_0^2$) along two slices of the conformal manifold, the first at $\LS=0$ and varying $\kappa_2$, the second at fixed $\kappa_2=1$ and varying $\LS$. We observe operators with vanishing anomalous dimensions emerge at $\kappa_2\to 0$ as well as for the extremal values $\LS\to0$ and $\LS\to 1$. }
  \label{fig:spectra}
\end{figure}

The framework developed in this paper provides the perturbative ingredients to study how the spectrum varies on the conformal manifold. In particular, the explicit coupling  parametrisation and the deformed one-loop dilatation operator allow us to approach loci at which gauge couplings vanish or symmetries are enhanced, where additional multiplets may shorten and the BPS spectrum can jump. 

As an example we may consider varying the unique relative gauge coupling $\kappa_2$ or the Leigh-Strassler parameter $\LS$ in the $\mathbb{Z}_4[1,1,2]$ orbifold theory and consider the sector of scalar length-two operators with classical dimension $\Delta_0=2$. Such operators can form the highest-weight operators for short current multiplets $\hat{\Cm}_{(0,0)}$ if their anomalous dimensions vanish. In figure \ref{fig:spectra} we see such possibilities arise at the decoupling limit $\kappa\to0$, and at the extremal points $\LS\to0$ and $\LS\to 1$. We will discuss these special loci in \textbf{Part~II} \cite{PartII}, where we will place them in the context of the CFT distance conjecture \cite{Perlmutter:2020buo, Baume:2020dqd} and corresponding applications \cite{Baume:2023msm, Calderon-Infante:2024oed, Calderon-Infante:2026rkj, Baume:2026svx, Calderon-Infante:2026zmd, Fenati:2026hkn, Grimm:2026tiz}.\\

In Section \ref{sec:applications} we demonstrated how the one-loop dilatation operator, which is one of the key results of this paper, could be used to generate spectral data in deformed orbifold theories. This has direct applications in the context of integrability, dynamical spin chains and the study of quantum chaos. For example, a precise investigation of twisted correlators in orbifold theories requires explicit information about the eigenstates of the dilatation operator, which our implementation can provide. In fact, a preliminary version of our code was used in \cite{lePlat:2026ndd,lePlat:2026bvt} to explicitly construct BPS operators in the untwisted sector. 

Beyond the orbifold point, integrability gives way to the more intricate structure of dynamical spin chains, which are under active investigation \cite{Pomoni:2019oib,Pomoni:2021pbj,Bozkurt:2024tpz,Bertle:2024djm,Bozkurt:2025exl,Bath:2026yeu}. In the context of marginally deformed $\mathcal{N}=2$ orbifold theories the structure underlying these dynamical spin chains was recently mapped to a dynamical generalisation of the familiar Temperley-Lieb algebra \cite{AnaPaul}. Similar structures should organise the spectrum of the marginally deformed $\mathcal{N}=1$ theories we discussed in this paper.

Finally, recent developments in the study of level-spacing statistics have made the transition from integrable to chaotic models visible \cite{McLoughlin:2020siu,McLoughlin:2020zew,McLoughlin:2022jyt,Baume:2026jyt,Casey:2026vbz}. The transition from the orbifold point to the larger conformal manifold breaks standard integrability and may result in similar chaotic statistics. To clearly characterise this transition and look for potential traces of hidden symmetries, sectors with large numbers of eigenvectors have to be made accessible, a task for which our Hamiltonian implementation provides an excellent starting point, but presumably can be streamlined further by restricting to the theory and sector of interest.

\paragraph{Acknowledgments}
We are particularly grateful to Jarryd Bath, Dennis le Plat, Shlomo Razamat, Timo Weigand and Konstantinos Zoubos for valuable discussions on this work. EP is supported by ERC-2021-CoG - BrokenSymmetries 101044226. 
This work is funded in part by Deutsche Forschungsgemeinschaft under Germany’s Excellence Strategy – EXC 2121 Quantum Universe – 390833306 and by Deutsche Forschungsgemeinschaft under SFB 1624 – “Higher structures, moduli spaces and integrability” –506632645.

\appendix 

\section{Background on superconformal representation theory}
\label{App:SCA}

In this appendix we collect the necessary details about the superconformal algebras in 4d $\mathcal{N}=1$ and $\mathcal{N}=2$ theories as well as the relevant decomposition rules.

\subsection[Representation theory of 4d \texorpdfstring{$\mathcal{N}=1$}{N=1} superconformal algebra]{Representation theory of 4d $\bm{\mathcal{N}=1}$ superconformal algebra}
\label{App:N1Reps}

In this appendix we review some basic facts regarding the superconformal algebra in $4$d $\mathcal{N}=1$ theories, following the ``Dolan-Osborn'' (DO) conventions of \cite{Rastelli:2016tbz, Bourton:2020rfo}. A generic long multiplet is generated by the action of the four Poincar\'e supercharges $\Qm$ and $\tilde \Qm$ on a superconformal primary, a state annihilated by all the superconformal charges $\Sm$ and $\tilde \Sm$, and is denoted as $\Am^{\Delta}_{r(j_1,j_2)}$ ($L\bar{L}[2j_1,2j_2]^{(-2r)}_{\Delta}$ in C\'ordova-Dumitrescu-Intriligator (CDI) notation \cite{Cordova:2016emh}). It may happen that some additional supercharge $\Qm$ (or a linear combination of supercharges) annihilates the primary as well, generating a shorter representation. In the cases where no recombination is possible, the conformal dimension of each state in the multiplet is protected against quantum corrections. Multiple such shortenings can occur at the same time. We summarise the possible shortening conditions in Table \ref{Tab:shortening}. In the first and last column we list the names of the multiplets following CDI and DO notation, respectively. In the second column the null states are given and in the other two columns we give the protected quantum numbers of the highest-weight state. A multiplet denoted as $\Xm_{r(j_1, j_2)}$ will have a highest-weight state obeying the shortening condition $\Xm$, with $\text{U}(1)_r$-charge $r$, and Lorentz spin $(j_1,j_2)$.

\begin{table}
\begin{centering}
\setlength{\tabcolsep}{4pt} 
\renewcommand{\arraystretch}{1.3} 
\resizebox{\textwidth}{!}{
\begin{tabular}{|c|l|l|l|l|}
\hline 
CDI & Null states & \multicolumn{2}{|c|}{Shortening Conditions} & DO \tabularnewline
\hline
\hline 
$B_1 \bar{L}[0,2j_2]^{(r)}_{\Delta}$ & $[1,2j_2]^{(r-1)}_{\Delta + \frac{1}{2}}$ & $j_1=0, r < -\frac{2}{3}(j_2 +1)$ & $\Delta=-\frac{3}{2}r$ & $\Bm_{r(0,j_2)}$\tabularnewline
\hline 
$L \bar{B}_1[2j_1,0]^{(r)}_{\Delta}$ & $[2j_1,1]^{(r+1)}_{\Delta + \frac{1}{2}}$ & $j_2=0, r > \frac{2}{3}(j_1+1)$ & $\Delta=\frac{3}{2}r$ & $\bar{\Bm}_{r(j_1,0)}$\tabularnewline
\hline 
$B_1 \bar{B}_1[0,0]^{(r)}_{\Delta}$ & $[1,0]^{(r-1)}_{\Delta + \frac{1}{2}}, [0,1]^{(r+1)}_{\Delta + \frac{1}{2}}$ & $r=0$, $j_1,j_2=0$ & $\Delta=0$ & $\hat{\Bm}$\tabularnewline
\hline
\hline 
$A_{1} \bar{L}[2j_1,2j_2]^{(r)}_{\Delta}$ & $[2j_1-1,2j_2]^{(r-1)}_{\Delta + \frac{1}{2}}$& $r<\frac{2}{3}(j_1-j_2)$ & $\Delta=2+2j_1-\frac{3}{2}r$ & $\Cm_{r(j_1,j_2)}$\tabularnewline
$A_{2} \bar{L}[0,2j_2]^{(r)}_{\Delta}$ & $[0,2j_2]^{(r-2)}_{\Delta + 1}$& $r<-\frac{2}{3}j_2, j_1=0$ & $\Delta=2-\frac{3}{2}r$ & $\Cm_{r(0,j_2)}$\tabularnewline
\hline 
$L \bar{A}_{1}[2j_1,2j_2]^{(r)}_{\Delta}$ &  $[2j_1, 2j_2-1]^{(r+1)}_{\Delta + \frac{1}{2}}$ & $r>\frac{2}{3}(j_1-j_2)$ & $\Delta=2+2j_2+\frac{3}{2}r$ & $\bar\Cm_{r(j_1,j_2)}$\tabularnewline
$L \bar{A}_{2}[2j_1,0]^{(r)}_{\Delta}$ &  $[2j_1,0]^{(r+2)}_{\Delta + 1}$ & $r>\frac{2}{3}j_1, j_2=0$ & $\Delta=2+\frac{3}{2}r$ & $\bar\Cm_{r(j_1,0)}$\tabularnewline
\hline 
$A_{1} \bar{A}_{1}[2j_1,2j_2]^{(r)}_{\Delta}$ & $[2j_1-1,2j_2]^{(r-1)}_{\Delta + \frac{1}{2}}, [2j_1, 2j_2-1]^{(r+1)}_{\Delta + \frac{1}{2}}$ & $r=\frac{2}{3}(j_1-j_2)$ & $\Delta=2+j_1+j_2$ & $\hat{\Cm}_{(j_1,j_2)}$\tabularnewline
$A_{1} \bar{A}_{2}[2j_1,0]^{(r)}_{\Delta}$ & $[2j_1-1,0]^{(r-1)}_{\Delta + \frac{1}{2}}, [2j_1, 0]^{(r+2)}_{\Delta + 1}$ & $r=\frac{2}{3}j_1 , j_2=0$ & $\Delta=2+j_1$ & $\hat{\Cm}_{(j_1,0)}$\tabularnewline
$A_{2} \bar{A}_{1}[0,2j_2]^{(r)}_{\Delta}$ & $[0,2j_2]^{(r-2)}_{\Delta + 1}, [0, 2j_2-1]^{(r+1)}_{\Delta + \frac{1}{2}}$ & $r=-\frac{2}{3}j_2, j_1=0$ & $\Delta=2+j_2$ & $\hat{\Cm}_{(0,j_2)}$\tabularnewline
$A_{2} \bar{A}_{2}[0,0]^{(r)}_{\Delta}$ & $[0,0]^{(r-2)}_{\Delta + 1}, [0, 0]^{(r+2)}_{\Delta + 1}$ & $r=j_1=j_2=0$ & $\Delta=2$ & $\hat{\Cm}_{(0,0)}$\tabularnewline
\hline 
\hline 
$B_1 \bar{A}_{1}[0,2j_2]^{(r)}_{\Delta}$ & $[1,2j_2]^{(r-1)}_{\Delta + \frac{1}{2}},[0, 2j_2-1]^{(r+1)}_{\Delta + \frac{1}{2}}$ & $r=-\frac{2}{3}(j_2+1)$ & $\Delta=1+j_2$ & $\mathcal{D}_{(0,j_2)}$\tabularnewline
$B_1 \bar{A}_{2}[0,0]^{(r)}_{\Delta}$ & $[1,0]^{(r-1)}_{\Delta + \frac{1}{2}},[0, 0]^{(r+2)}_{\Delta + 1}$ & $r=-\frac{2}{3}$ & $\Delta=1$ & $\mathcal{D}_{(0,0)}$\tabularnewline
\hline 
$A_{1}\bar{B}_1[2j_1,0]^{(r)}_{\Delta}$ & $[2j_1-1,0]^{(r-1)}_{\Delta + \frac{1}{2}},[2j_1,1]^{(r+1)}_{\Delta + \frac{1}{2}}$ & $r=\frac{2}{3}(j_1+1)$ & $\Delta=1+j_1$ & $\bar{\mathcal{D}}_{(j_1,0)}$\tabularnewline
$A_{2}\bar{B}_1[0,0]^{(r)}_{\Delta}$ & $[0,1]^{(r+1)}_{\Delta + \frac{1}{2}},[0,0]^{(r-2)}_{\Delta + 1}$ & $r=\frac{2}{3}$ & $\Delta=1$ & $\bar{\mathcal{D}}_{(0,0)}$\tabularnewline
\hline 
\end{tabular}
}
\par\end{centering}
\caption{\label{Tab:shortening}Shortening conditions
and short multiplets of 4d $\Nm=1$ superconformal algebra.}
\end{table}

We also collect here the recombination rules under which long multiplets hitting the unitarity bound (i.e. satisfying a shortening condition) split into shorter ones:
\begin{align}
    & \Am^{2 + 2j_1-\frac{3}{2}r}_{r <\frac{2}{3}(j_1-j_2), (j_1,j_2)} \simeq \Cm_{r (j_1,j_2)} \oplus \Cm_{r-1 (j_1-\frac{1}{2}, j_2)}\,,
    \label{Eq:RecN11}\\
    & \Am^{2 + 2j_2+\frac{3}{2}r}_{r >\frac{2}{3}(j_1-j_2), (j_1,j_2)} \simeq \bar{\Cm}_{r (j_1,j_2)} \oplus  \bar{\Cm}_{r+1 (j_1, j_2-\frac{1}{2})} \,,
    \label{Eq:RecN12}\\
    & \Am^{2 + j_1+j_2}_{\frac{2}{3}(j_1-j_2), (j_1,j_2)} \simeq \hat{\Cm}_{(j_1,j_2)} \oplus  \Cm_{\frac{2}{3}(j_1-j_2)-1, (j_1-\frac{1}{2},j_2)} \oplus \bar{\Cm}_{\frac{2}{3}(j_1-j_2)+1, (j_1, j_2-\frac{1}{2})}\, ,
    \label{Eq:RecN13}
\end{align}
where the following identifications have been used
\begin{equation}
    \Cm_{r(-\frac{1}{2},j_2)} \simeq \Bm_{r-1(0,j_2)}\,, \quad \bar{\Cm}_{r(j_1,-\frac{1}{2})} \simeq \bar{\Bm}_{r+1(j_1,0)}\,.
\end{equation}
These recombination rules imply that the following equivalence classes
\begin{equation}
\begin{split}
   & [\tilde{r},j_1]_+= \{\bar{\Cm}_{r,(j_1, \frac{\tilde{r}-r}{2})} | \tilde{r}-r \in 2 \mathbb{Z}_{\geq 0} \} \,  \cup \, \{\hat{\Cm}_{(j_1, \frac{3\tilde{r}-2 j_1}{4})} | 3\tilde{r}-2 j_1 \in 4 \mathbb{Z}_{\geq 0} \}\,, \\
   & [\tilde{r},j_1]_-= \{\bar{\Cm}_{r,(j_1, \frac{\tilde{r}-r}{2})} | \tilde{r}-r \in -1+2 \mathbb{Z}_{\geq 0} \} \,  \cup \, \{\hat{\Cm}_{(j_1, \frac{3\tilde{r}-2 j_1}{4})} | 3\tilde{r}-2 j_1 \in 2+4 \mathbb{Z}_{\geq 0} \}\,,
\end{split}
\label{Eq:N1_Equivalence_Classes}
\end{equation}
up to a sign factor, contribute equivalently to the superconformal index \eqref{Eq:Index}\,. In total, the index \eqref{Eq:Index} counts the net degeneracies $\text{ND}[\tilde{r},j_1]$ given by \cite{Rastelli:2016tbz}
\begin{equation}
    \text{ND}[\tilde{r},j_1]=\#[\tilde{r},j_1]_+ - \#[\tilde{r},j_1]_-\,.
\end{equation}

\subsection[Representation theory of 4d \texorpdfstring{$\mathcal{N}=2$}{N=2}~superconformal algebra]{Representation theory of 4d $\bm{\mathcal{N}=2}$ superconformal algebra}
\label{App:N2Reps}

In this appendix we review some basic facts regarding the superconformal algebra in $4$d $\mathcal{N}=2$ theories, following the conventions of \cite{Dolan:2002zh}. A generic long multiplet is generated by the action of the eight Poincar\'e supercharges $\Qm$ and $\tilde \Qm$ on a superconformal primary, a state annihilated by all the superconformal charges $\Sm$ and $\tilde \Sm$, and is denoted as $\Am^{\Delta}_{R,r(j_1,j_2)}$ ($L\bar{L}[2j_1,2j_2]^{(2R;-2r)}_{\Delta}$ in CDI \cite{Cordova:2016emh} notation). It may happen that some additional supercharge $\Qm$ (or a linear combination of supercharges) annihilates the primary as well, generating a shorter representation. As before, when no recombination rule is allowed, the conformal dimension of each state in the multiplet is protected against quantum corrections. Multiple such shortenings can occur at the same time. All possible shortening conditions have been investigated in \cite{Dolan:2002zh,Cordova:2016emh}, and we summarise their findings in Table \ref{Tab:shorteningN2}. In the first and last column we list the names of the multiplets following CDI and DO notation, respectively. In the second column the null states are given and in the other two columns we give the protected quantum numbers of the highest-weight state. A multiplet denoted as $\Xm_{R,r(j_1, j_2)}$ will have a highest-weight state obeying the shortening condition $\Xm$, with $\text{SU}(2)_R$-charge $R$, $\text{U}(1)_r$-charge $r$, and Lorentz spin $(j_1,j_2)$.

\begin{table}
\begin{centering}
\setlength{\tabcolsep}{4pt} 
\renewcommand{\arraystretch}{1.3} 
\resizebox{\textwidth}{!}{
\begin{tabular}{|c|l|l|l|l|}
\hline 
CDI & Null states & \multicolumn{2}{|c|}{Shortening Conditions} & DO \tabularnewline
\hline
\hline 
$B_1 \bar L[0, 2j_2]^{(2R>0;-2r)}$ & $[1,2j_2]^{2R+1; -2r-1}_{\Delta+ \frac{1}{2}}$ & $j_1=0, r>j_2+1$ & $\Delta=2R+r$ & $\Bm_{R,r(0,j_2)}$\tabularnewline
\hline 
$L \bar B_1 [2j_1, 0]^{(2R>0;-2r)}$ & $[2j_1,1]^{2R+1; -2r+1}_{\Delta+ \frac{1}{2}}$ & $j_2=0, r<-(j_1+1)$ & $\Delta=2R-r$ & $\bar{\Bm}_{R,r(j_1,0)}$\tabularnewline
\hline 
$B_1 \bar L[0, 2j_2]^{(0;-2r)}$ & $[1,2j_2]^{1; -2r-1}_{\Delta+ \frac{1}{2}}$ & $R=0,r>j_2+1$ & $\Delta=r$ & $\Em_{r(0,j_2)}$\tabularnewline
\hline 
$L \bar B_1 [2j_1, 0]^{(0;-2r)}$ & $[2j_1,1]^{1; -2r+1}_{\Delta+ \frac{1}{2}}$ & $R=0, r<-(j_1+1)$ & $\Delta=-r$ & $\bar \Em_{r(j_1,0)}$\tabularnewline
\hline 
$B_1 \bar B_1 [0, 0]^{(2R;0)}$ & $[1,0]^{2R+1; -1}_{\Delta+ \frac{1}{2}}, [0,1]^{2R+1; 1}_{\Delta+ \frac{1}{2}}$ & $r=0$, $j_1,j_2=0$ & $\Delta=2R$ & $\hat{\Bm}_{R}$\tabularnewline
\hline
\hline 
$A_1 \bar L [2j_1, 2j_2]^{(2R,-2r)}$ & $[2j_1-1,2j_2]^{2R+1; -2r-1}_{\Delta+ \frac{1}{2}}$ & $r> j_2-j_1$ & $\Delta=2+2j_1+2R+r$ & $\Cm_{R,r(j_1,j_2)}$\tabularnewline
 $A_2 \bar L [0, 2j_2]^{(2R,-2r)}$ & $[0,2j_2]^{2R+2; -2r-2}_{\Delta+ 1}$  & $r>j_2$& $\Delta=2+2R+r$ & $\Cm_{R,r(0,j_2)}$\tabularnewline
\hline 
$L \bar A_1 [2j_1, 2j_2]^{(2R,-2r)}$ & $[2j_1,2j_2-1]^{2R+1; -2r+1}_{\Delta+ \frac{1}{2}}$ & $r< j_2 -j_1$& $\Delta=2+2j_2+2R-r$ & $\bar\Cm_{R,r(j_1,j_2)}$\tabularnewline
$L \bar A_2 [2j_1, 0]^{(2R,-2r)}$ & $[2j_1,0]^{2R+2; -2r+2}_{\Delta+ 1}$ & $r<-j_1$ & $\Delta=2+2R-r$ & $\bar\Cm_{R,r(j_1,0)}$\tabularnewline
\hline 
$A_1 \bar A_1 [2j_1, 2j_2]^{(2R,-2r)}$ & $[2j_1-1,2j_2]^{2R+1; -2r-1}_{\Delta+ \frac{1}{2}}, [2j_1,2j_2-1]^{2R+1; -2r+1}_{\Delta+ \frac{1}{2}}$ & $r=j_2-j_1$ & $\Delta=2+2R+j_1+j_2$ & $\hat{\Cm}_{R(j_1,j_2)}$\tabularnewline
$A_1 \bar A_2 [2j_1, 0]^{(2R,-2r)}$ & $[2j_1-1,0]^{2R+1; -2r-1}_{\Delta+ \frac{1}{2}}, [2j_1,0]^{2R+2; -2r+2}_{\Delta+ 1}$ & $r=-j_1$ & $\Delta=2+2R+j_1$ & $\hat{\Cm}_{R(j_1,0)}$\tabularnewline
$A_2 \bar A_1 [0, 2j_2]^{(2R,-2r)}$ & $[0,2j_2]^{2R+2; -2r-2}_{\Delta+ 1}, [0,2j_2-1]^{2R+1; -2r+1}_{\Delta+ \frac{1}{2}}$ & $r=j_2$ & $\Delta=2+2R+j_2$ & $\hat{\Cm}_{R(0,j_2)}$\tabularnewline
$A_2 \bar A_2 [0,0]^{(2R,-2r)}$ & $[0,0]^{2R+2; -2r-2}_{\Delta+ 1}, [0,0]^{2R+2; -2r+2}_{\Delta+ 1}$ & $r=0$ & $\Delta=2+2R$ & $\hat{\Cm}_{R(0,0)}$\tabularnewline
\hline
\hline 
$B_1 \bar A_1[0,2j_2]^{2R,-2r}$ & $[1,2j_2]^{2R+1; -2r-1}_{\Delta+ \frac{1}{2}}, [0,2j_2-1]^{2R+1; -2r+1}_{\Delta+ \frac{1}{2}}$ & $r=j_2+1$ & $\Delta=1+2R+j_2$ & $\mathcal{D}_{R(0,j_2)}$\tabularnewline
$B_1 \bar A_2[0,0]^{2R,-2r}$ & $[1,0]^{2R+1; -2r-1}_{\Delta+ \frac{1}{2}}, [0,0]^{2R+2; -2r+2}_{\Delta+ 1}$ & $r=1$ & $\Delta=1+2R$ & $\mathcal{D}_{R(0,0)}$\tabularnewline
\hline 
$A_1 \bar B_1[2j_1,0]^{2R,-2r}$ & $[2j_1,1]^{2R+1; -2r+1}_{\Delta+ \frac{1}{2}}, [2j_1-1,0]^{2R+1; -2r-1}_{\Delta+ \frac{1}{2}}$ & $-r=j_1+1$ & $\Delta=1+2R+j_1$ & $\bar{\mathcal{D}}_{R(j_1,0)}$\tabularnewline
$A_2 \bar B_1[0,0]^{2R,-2r}$ & $[0,1]^{2R+1; -2r+1}_{\Delta+ \frac{1}{2}}, [0,0]^{2R+2; -2r-2}_{\Delta+ 1}$ & $r=-1$ & $\Delta=1+2R$ & $\bar{\mathcal{D}}_{R(0,0)}$\tabularnewline
\hline 
\end{tabular}
}
\par\end{centering}
\caption{\label{Tab:shorteningN2}Shortening conditions
and short multiplets of 4d $\Nm=2$ superconformal algebra.}
\end{table}

For completeness, we also collect here the recombination rules for long multiplets hitting the unitarity bound 
\begin{align}
 & \mathcal{A}^{2R + r +2j_1 +2}_{R,r(j_1,j_2)} \simeq \mathcal{C}_{R,r(j_1,j_2)} \oplus \mathcal{C}_{R+\frac{1}{2},r+\frac{1}{2}(j_1-\frac{1}{2},j_2)}\,,  \label{Eq:Recombination_Rules1} \\
 & \mathcal{A}^{2R - r +2j_2 +2}_{R,r(j_1,j_2)} \simeq \bar{\mathcal{C}}_{R,r(j_1,j_2)} \oplus \bar{\mathcal{C}}_{R+\frac{1}{2},r-\frac{1}{2}(j_1,j_2-\frac{1}{2})} \,,  \label{Eq:Recombination_Rules2}\\
 & \mathcal{A}^{2R + j_1 + j_2 +2}_{R,j_2-j_1(j_1,j_2)} \simeq \hat{\mathcal{C}}_{R(j_1,j_2)} \oplus \hat{\mathcal{C}}_{R+\frac{1}{2}(j_1-\frac{1}{2},j_2)} \oplus \hat{\mathcal{C}}_{R+\frac{1}{2}(j_1,j_2-\frac{1}{2})} \oplus \hat{\mathcal{C}}_{R+1(j_1-\frac{1}{2},j_2-\frac{1}{2})}\,,
 \label{Eq:Recombination_Rules3}
\end{align}
where we considered the following identifications, formally allowing for the $j_1,j_2$ quantum numbers in the $\Cm/\hat{\Cm}$ multiplet to also take the value $-\tfrac{1}{2}$:
\begin{equation}
\begin{split}
& \Cm_{R,r(-\frac{1}{2},j_2)}\simeq\Bm_{R+\frac{1}{2},r+\frac{1}{2}(0,j_2)}\,, \\
& \hat{\Cm}_{R(-\frac{1}{2},j_2)}\simeq \mathcal{D}_{R+\frac{1}{2}(0,j_2)}\,,\qquad\hat{\Cm}_{R(j_1,-\frac{1}{2})}\simeq\bar{\mathcal{D}}_{R+\frac{1}{2}(j_1,0)}\, , \\
&\hat{\Cm}_{R(-\frac{1}{2},-\frac{1}{2})}\simeq \mathcal{D}_{R+\frac{1}{2}(0,-\frac{1}{2})} \simeq
\bar{\mathcal{D}}_{R+\frac{1}{2}(-\frac{1}{2},0)} \simeq \hat{\Bm}_{R+1}\,.
\label{translation}
\end{split}
\end{equation}

\subsection{Decomposition of superconformal multiplets}
\label{App:Decompositions}

In Section \ref{sec: gauge theory} we used the decomposition of the single-trace BPS spectrum of $\Nm =4$ SYM, which consists of the multiplets $\Bm^{(\frac{1}{2},\frac{1}{2})}_{[0,p,0]}$ with $p \geq 2$, to $\Nm=2$ and $\Nm=1$ superconformal algebras. The decomposition from $\Nm=4$ to $\Nm=2$ was stated in \cite{Dolan:2002zh} 
\begin{eqnarray}
\Bm_{[0,p,0]}^{\frac{1}{2},\frac{1}{2}} & \simeq & (p+1)\hat{\Bm}_{\frac{1}{2}p}\oplus\Em_{p(0,0)}\oplus\bar{\Em}_{-p(0,0)}\nonumber \\
 & & \oplus(p-1)\hat{\Cm}_{\frac{1}{2}p-1(0,0)}\oplus p(\mathcal{D}_{\frac{1}{2}(p-1)(0,0)}\oplus\bar{\mathcal{D}}_{\frac{1}{2}(p-1)(0,0)})\nonumber \\
 & & \oplus\bigoplus_{k=1}^{p-2}(k+1)(\Bm_{\frac{1}{2}k,p-k(0,0)}\oplus\bar{\Bm}_{\frac{1}{2}k,k-p(0,0)})\nonumber \\
 & & \oplus\bigoplus_{k=0}^{p-3}(k+1)(\Cm_{\frac{1}{2}k,p-k-2(0,0)}\oplus\bar{\Cm}_{\frac{1}{2}k,k-p+2(0,0)})\nonumber \\
 & & \oplus\bigoplus_{k=0}^{p-4}\bigoplus_{l=0}^{p-k-4}(k+1)\Am_{\frac{1}{2}k,p-k-4-2l(0,0)}^{p} \, .
 \label{Eq:N4toN2}
\end{eqnarray}
These multiplets further decompose into $\Nm=1$ multiplets through the following decomposition rules\footnote{To generate these decomposition rules, we adapted the code for $\Nm=2$ SCFTs presented in \cite{Agarwal:2018zqi} to our $\Nm=1$ case.} 
\allowdisplaybreaks
\begin{align}
    & \hat{\Bm}_{1} \simeq{}_{[0]}\!\bar \Bm_{\frac{4}{3} (0,0)} \oplus{}_{[0]}\Bm_{-\frac{4}{3} (0,0)} \oplus{}_{[0]}\hat{\Cm}_{(0,0)} \,,\nonumber \\
    & \hat{\Bm}_{\frac{1}{2}p} \simeq{}_{[0]}\bar \Bm_{\frac{2p}{3} (0,0)} \oplus{}_{[0]} \Bm_{-\frac{2p}{3} (0,0)} \oplus{}_{[0]} \bar \Cm_{2\frac{p-2}{3} (0,0)}\oplus{}_{[0]} \Cm_{-2\frac{p-2}{3} (0,0)} \oplus  \bigoplus_{k=0}^{p-4} {}_{[0]} \Am^{p}_{\frac{8-2p+4k}{3}(0,0)}\;, (p>2)\,,\nonumber \\
    & \Em_{p(0,0)} \simeq {}_{[p]} \Bm_{-\frac{2p}{3} (0,0)} \oplus {}_{[p-\frac{1}{2}]} \Bm_{-\frac{2p+1}{3} (0,\frac{1}{2})}  \oplus{}_{[p-1]} \Bm_{-2\frac{p+1}{3} (0,0)} \quad (p \geq 2)\,, \nonumber \\
    & \hat{\Cm}_{0 (0,0)} \simeq{}_{[0]}  \hat \Cm_{(0, 0)} \oplus{}_{[-\frac{1}{2}]} \hat\Cm_{( 0, \frac{1}{2})} \oplus{}_{[\frac{1}{2}]} \hat \Cm_{(\frac{1}{2}, 0)} \oplus{}_{[0]} \hat \Cm_{( \frac{1}{2}, \frac{1}{2})} \,,\nonumber \\
   & \hat{\Cm}_{\frac{1}{2}p -1 (0,0)} \simeq \bigoplus_{l,k=0}^{1} \left (_{[\frac{l-k}{2}]}\bar \Cm_{\frac{-4+2p+l-k}{3}(\frac{l}{2}, \frac{k}{2})} \oplus{}_{[\frac{l-k}{2}]} \Cm_{\frac{4-2p-l +k}{3}(\frac{k}{2}, \frac{l}{2})} \right ) \oplus \nonumber \\
   & \phantom{ \hat{\Cm}_{\frac{1}{2}p -1 (0,0)} \simeq} \bigoplus_{s=0}^{p-4} \bigoplus_{l,k=0}^1 {}_{[\frac{l-k}{2}]} \Am^{p+ \frac{l+k}{2}}_{2\frac{4-p+2s -l/2 +k/2}{3}(\frac{l}{2},\frac{k}{2})} \quad (p>2) \,,\nonumber \\
   & \Dm_{\frac{1}{2}(0,0)} \simeq {}_{[1]}\Bm_{-\frac{4}{3}(0,0)} \oplus{}_{[\frac{1}{2}]}\Bm_{-\frac{5}{3}(0,\frac{1}{2})} \oplus{}_{[1]} \hat{\cal C}_{(0,0)} \oplus{}_{[\frac{1}{2}]}\hat{\cal C}_{(0,\frac{1}{2})} \,,\nonumber \\
   & \Dm_{\frac{1}{2}(p-1)(0,0)} \simeq \bigoplus_{l=0}^{1} {}_{[1-\frac{l}{2}]} \left ( \Bm_{-\frac{2p+l}{3}(0,\frac{l}{2})} \oplus   \bar \Cm_{-2\frac{2-p+l/2}{3}(0,\frac{l}{2})} \oplus \Cm_{2\frac{2-p-l/2}{3}(0,\frac{l}{2})}\right ) \oplus \nonumber \\
   & \phantom{\Dm_{\frac{1}{2}(p-1)(0,0)} \simeq} \bigoplus_{l=0}^{p-4} \bigoplus_{k=0}^{1} {}_{[1-\frac{k}{2}]} \Am^{p+\frac{k}{2}}_{-2\frac{4-p+2l+k/2}{3}(0,\frac{k}{2})} \quad (p>2) \,,\nonumber \\
   & \Bm_{\frac{m}{2},{n} (0,0)} \simeq \bigoplus_{l=0}^2 {}_{[n-\frac{l}{2}]} \left ( \Bm_{-2\frac{m+n}{3} - \frac{l}{3} (0,\frac{l-2\delta_{l,2}}{2})} \oplus \Cm_{-2\frac{m+n-2}{3} - \frac{l}{3} (0,\frac{l-2\delta_{l,2}}{2})} \right ) \oplus \nonumber\\
   & \phantom{\Bm_{\frac{m}{2},{n} (0,0)} \simeq} \bigoplus_{s=0}^{m-2}\bigoplus_{l=0}^2 {}_{[n-\frac{l}{2}]} \Am^{m+n+\frac{l}{2}}_{-2\frac{n+m-4-2s}{3}-\frac{l}{3}(0,\frac{l-2\delta_{l,2}}{2})} \,,\nonumber \\
   & \Cm_{\frac{m}{2},n (0,0)} \simeq \bigoplus_{l,k=0}^1 {}_{[n+\frac{l-k}{2}]} \Cm_{-2\frac{n+m-l/2+k/2}{3} (\frac{l}{2}, \frac{k}{2})} \oplus \bigoplus_{l=0}^1 {}_{[n-1+\frac{l}{2}]} \Cm_{-2\frac{n+m+1-l/2}{3}(\frac{l}{2},0) } \oplus \nonumber \\
   & \phantom{\Cm_{\frac{m}{2},n (0,0)} \simeq} \bigoplus_{s=0}^{m-1}\bigoplus_{l,k=0}^1 {}_{[n+\frac{l-k}{2}]} \Am^{n+m+2+\frac{l}{2}+ \frac{k}{2}}_{-2\frac{-2+n+m-2s}{3}+\frac{l}{3}-\frac{k}{3}(\frac{l}{2}, \frac{k}{2})} \oplus \nonumber \\
   & \phantom{\Cm_{\frac{m}{2},n (0,0)} \simeq} \bigoplus_{s=0}^{m-1} \bigoplus_{l=0}^1 {}_{[n-1+\frac{l}{2}]} \Am^{m+n+3+\frac{l}{2}}_{-2\frac{m+n-1-2s}{3}+\frac{l}{3}(\frac{l}{2},0)} \,,\nonumber\\
   & \Am^{p}_{\frac{m}{2},n (0,0)} \simeq \bigoplus_{s=0}^m \bigoplus_{l,k=0}^1 {}_{[n+\frac{l-k}{2}]} \Am^{p+\frac{l}{2}+ \frac{k}{2}}_{2\frac{n+m-2s}{3}+\frac{l}{3}-\frac{k}{3}(\frac{l}{2},\frac{k}{2})} \oplus \bigoplus_{s=0}^m \bigoplus_{l=0}^1 \biggl ({}_{[n-1+\frac{l}{2}]} \Am^{p+1+\frac{l}{2}}_{-2\frac{1-n-m+2s}{3}+\frac{l}{3}(\frac{l}{2},0)} \oplus \nonumber \\
   & \phantom{\Am^{p}_{\frac{m}{2},n (0,0)} \simeq} {}_{[n+1-\frac{l}{2}]} \Am^{p+1+\frac{l}{2}}_{-2\frac{-1-n-m+2s}{3}-\frac{l}{3}(0,\frac{l}{2})} \biggr ) \oplus  \bigoplus_{s=0}^m {}_{[n]} \Am^{p+2}_{-2\frac{-m-n+2s}{3}(0,0)}\,,
   \label{Eq:N2toN1}
\end{align}
where the conjugate decomposition rules are obtained by a simple conjugation of the multiplets and exchange of $r\leftrightarrow -r$ and $j_1 \leftrightarrow j_2$. 
Here the LHS specifies the multiplet in $\Nm=2$ notation as explained in Appendix \ref{App:N2Reps}, while we list the $\Nm=1$ multiplets into which it decomposes on the RHS, following the notation in Appendix \ref{App:N1Reps}. In square-bracket subscripts in front of every $\Nm=1$ multiplet we report also the $r^{\Nm=2}$ Abelian r-charge of the $\Nm=2$ parent to the highest-weight state. This is crucial information for the correct orbifold projection in Section \ref{sec: gauge theory}.

\section{Survey of \texorpdfstring{$\bm{\mathcal{N}=1}$}{N=1}~orbifolds with up to nine gauge nodes }
\label{App:survey}
In this appendix we gather details on marginal deformations of the simplest $\mathcal{N}=1$  orbifold theories with up to nine gauge nodes.\footnote{At these low numbers of gauge nodes, our two families of $\mathcal{N}=1$ orbifold theories cover all possible Abelian orbifolds counted in \cite{Hanany:2010ne} apart from the ones with enlarged $\mathcal{N}=2$ supersymmetry.} For each theory we specify the number of independent gauge couplings, list the dependent ones as well as the superpotential couplings $f^{\pm}_a$ in the superpotential
\begin{equation}
    \mathcal{W}=g_0\sum_{i\in\mathcal{A}} \left(f^+_i\Tr_{i}(\Xm\Ym\Zm)-f^-_i\Tr_{i}(\Xm\Zm\Ym)\right)\,.
    \label{Eq:W_generic}
\end{equation}
Here the index set $\mathcal{A}$ labels the gauge nodes with respect to which the trace is taken. For $\mathbb{Z}_k$ orbifolds $\mathcal{A}=\{1,\dots,k\}$ while for $\mathbb{Z}_k\times \mathbb{Z}_\ell$ orbifolds the gauge nodes are labelled by two integers $\mathcal{A}=\{(1,1),\dots,(k,\ell)\}$. In this survey we fix a particular parametrisation of the conformal manifold by a maximal set of independent gauge couplings, but of course a different parametrisation may be chosen. In line with the main text we define an overall gauge coupling $g_0$ and relative ratios $\kappa_i$
\begin{equation}\label{eq: definecouplings}
    g_0^2=\frac{1}{\abs{\Gamma}}\sum_{i\in \mathcal{A}}g_i^2\,,\qquad \kappa_i=\frac{g_i}{g_0}\in [0,\sqrt{\abs{\Gamma}}]\,.
\end{equation}

For readability, the following survey is restricted to the gauge-coupling deformations parameterised by the $\kappa_i$ ratios, but we can always include the universal Leigh-Strassler deformation by replacing
\begin{equation}\label{eq: roots}
    f^+_i\to\sqrt{(f^+_i)^2+\LS}\,,\qquad  f^-_i\to\sqrt{(f^-_i)^2-\LS}\,,
\end{equation}
where $\LS$ corresponds to the superpotential deformation \eqref{Eq: LS} that does not require the addition of new terms to the superpotential \eqref{Eq:W_generic}. Special marginal deformations for the $\mathbb{Z}_3[1,1,1]$ and $\mathbb{Z}_3\times \mathbb{Z}_3$ orbifold theories require a separate treatment as discussed in Section \ref{sec: special}. Note, that this parametrisation is restricted to the domain $\LS\in(-c_{\text{max}},c_{\text{max}})$ where $c_{\text{max}}=\min((f^+_i)^2,(f^-_i)^2)$\,, outside of which we have to switch to a different parametrisation. The square roots should be thought of as Cartesian parametrisation of circles traced out by the superpotential couplings $f^\pm$ under Leigh-Strassler deformation.

\subsection{The \texorpdfstring{$\mathbb{Z}_k$}{Zk} orbifold family}\label{sec: Zkexamples}

Here we consider all inequivalent (according to the identifications discussed below \eqref{Eq: restriction}) $\mathbb{Z}_k$ orbifold theories with up to nine gauge nodes.

\begin{itemize}
\item \textbf{Theories without relative gauge-coupling deformations:} If the configuration numbers $a_X$, $a_Y$ and $a_Z=(-a_X-a_Y)\text{ mod }k$ are all coprime with $k$, there is merely one overall gauge coupling $g_0$, such that $f^+_i=f^-_i=\kappa_i=1$, $\forall i\in\mathcal{A}$. In particular, this holds for all $\mathbb{Z}_k$ orbifold theories with prime $k$. Up to nine gauge nodes, this class of theories encompasses the unique $\mathbb{Z}_3[1,1,1]$ and $\mathbb{Z}_5[1,1,3]$ orbifold theories, the two distinct $\mathbb{Z}_7$ orbifolds with configurations $\mathbb{Z}_7[1,1,5]$ and $\mathbb{Z}_7[1,2,4]$ as well as one $\mathbb{Z}_9[1,1,7]$ orbifold.

\item The $\bm{\mathbb{Z}_4[1,1,2]}$ orbifold features two independent gauge couplings $g_1$ and $g_2$ with
\begin{center}
 \begin{tabular}{c||c|c|c|c}
  $i$ & $1$ & $2$ & $3$ & $4$\\
    \hline
    $\kappa_i$ & $\kappa_1$ & $\kappa_2$ & $\kappa_1$ & $\kappa_2$\\
    $f^+_i$ & $\kappa_1$ & $\kappa_2$ & $\kappa_1$ & $\kappa_2$\\
    $f^-_i$ & $\kappa_2$ & $\kappa_1$ & $\kappa_2$ & $\kappa_1$\\
\end{tabular}
\end{center}

\item The $\bm{\mathbb{Z}_6[1,1,4]}$ orbifold features two independent gauge couplings $g_1$ and $g_2$ with
\begin{center}
 \begin{tabular}{c||c|c|c|c|c|c}
  $i$ & $1$ & $2$ & $3$ & $4$& $5$& $6$\\
    \hline
    $\kappa_i$ & $\kappa_1$ & $\kappa_2$ & $\kappa_1$ & $\kappa_2$ & $\kappa_1$ & $\kappa_2$\\
    $f^+_i$ & $\kappa_1$ & $\kappa_2$ & $\kappa_1$ & $\kappa_2$ & $\kappa_1$ & $\kappa_2$\\
    $f^-_i$ & $\kappa_2$ & $\kappa_1$ & $\kappa_2$ & $\kappa_1$& $\kappa_2$ & $\kappa_1$\\
\end{tabular}
\end{center}

\item The $\bm{\mathbb{Z}_6[1,2,3]}$ orbifold features four independent gauge couplings $g_1,g_2,g_3, g_4$ with
\begin{equation}
    \kappa_5=\sqrt{\kappa_1^2+\kappa_2^2-\kappa_4^2}\,,\qquad  \kappa_6=\sqrt{-\kappa_1^2+\kappa_3^2+\kappa_4^2}\,.
\end{equation}
Requiring reality of $\kappa_5$ and $\kappa_6$ imposes constraints similar to \eqref{eq:obstruction}. In terms of these identifications, the full table of couplings is  
\begin{center}
 \begin{tabular}{c||c|c|c|c|c|c}
  $i$ & $1$ & $2$ & $3$ & $4$& $5$& $6$\\
    \hline
    $\kappa_i$ & $\kappa_1$ & $\kappa_2$ & $\kappa_3$ & $\kappa_4$ & $\kappa_5$ & $\kappa_6$\\
    $f^+_i$ & $\kappa_4$ & $\kappa_5$ & $\kappa_6$ & $\kappa_1$ & $\kappa_2$ & $\kappa_3$\\
    $f^-_i$ & $\kappa_5$ & $\kappa_6$  & $\kappa_1$ & $\kappa_2$& $\kappa_3$ & $\kappa_4$\\
\end{tabular}
\end{center}
We will observe below that this theory has a second description as $\bm{\mathbb{Z}_2\times \mathbb{Z}_3}$ orbifold theory.

\item The $\bm{\mathbb{Z}_8[1,1,6]}$ orbifold features two independent gauge couplings $g_1$ and $g_2$ with
\begin{center}
 \begin{tabular}{c||c|c|c|c|c|c|c|c}
  $i$ & $1$ & $2$ & $3$ & $4$& $5$& $6$& $7$& $8$\\
    \hline
    $\kappa_i$ & $\kappa_1$ & $\kappa_2$ & $\kappa_1$ & $\kappa_2$ & $\kappa_1$ & $\kappa_2$& $\kappa_1$ & $\kappa_2$\\
    $f^+_i$ & $\kappa_1$ & $\kappa_2$ & $\kappa_1$ & $\kappa_2$ & $\kappa_1$ & $\kappa_2$& $\kappa_1$ & $\kappa_2$\\
    $f^-_i$ & $\kappa_2$ & $\kappa_1$ & $\kappa_2$ & $\kappa_1$& $\kappa_2$ & $\kappa_1$& $\kappa_2$ & $\kappa_1$\\
\end{tabular}
\end{center}

\item The $\bm{\mathbb{Z}_8[1,2,5]}$ orbifold features two independent gauge couplings $g_1$ and $g_2$ with
\begin{center}
 \begin{tabular}{c||c|c|c|c|c|c|c|c}
  $i$ & $1$ & $2$ & $3$ & $4$& $5$& $6$& $7$& $8$\\
    \hline
    $\kappa_i$ & $\kappa_1$ & $\kappa_2$ & $\kappa_1$ & $\kappa_2$ & $\kappa_1$ & $\kappa_2$& $\kappa_1$ & $\kappa_2$\\
    $f^+_i$ & $\kappa_2$ & $\kappa_1$ & $\kappa_2$ & $\kappa_1$ & $\kappa_2$ & $\kappa_1$& $\kappa_2$ & $\kappa_1$\\
    $f^-_i$ & $\kappa_1$ & $\kappa_2$ & $\kappa_1$ & $\kappa_2$& $\kappa_1$ & $\kappa_2$& $\kappa_1$ & $\kappa_2$\\
\end{tabular}
\end{center}

\item The $\bm{\mathbb{Z}_8[1,3,4]}$ orbifold features four independent gauge couplings $g_1,g_2,g_3, g_4$ with
\begin{center}
 \begin{tabular}{c||c|c|c|c|c|c|c|c}
  $i$ & $1$ & $2$ & $3$ & $4$& $5$& $6$& $7$& $8$\\
    \hline
    $\kappa_i$ & $\kappa_1$ & $\kappa_2$ & $\kappa_3$ & $\kappa_4$ & $\kappa_1$ & $\kappa_2$& $\kappa_3$ & $\kappa_4$\\
    $f^+_i$ & $\kappa_1$ & $\kappa_2$ & $\kappa_3$ & $\kappa_4$ & $\kappa_1$ & $\kappa_2$& $\kappa_3$ & $\kappa_4$\\
    $f^-_i$ & $\kappa_2$ & $\kappa_3$ & $\kappa_4$ & $\kappa_1$& $\kappa_2$ & $\kappa_3$& $\kappa_4$ & $\kappa_1$\\
\end{tabular}
\end{center}

\item The $\bm{\mathbb{Z}_9[1,2,6]}$ orbifold features three independent gauge couplings $g_1$, $g_2$ and $g_3$ with
\begin{center}
 \begin{tabular}{c||c|c|c|c|c|c|c|c|c}
  $i$ & $1$ & $2$ & $3$ & $4$& $5$& $6$& $7$& $8$ & $9$\\
    \hline
    $\kappa_i$ & $\kappa_1$ & $\kappa_2$ & $\kappa_3$ & $\kappa_1$ & $\kappa_2$ & $\kappa_3$& $\kappa_1$ & $\kappa_2$ & $\kappa_3$\\
    $f^+_i$ & $\kappa_1$ & $\kappa_2$ & $\kappa_3$ & $\kappa_1$ & $\kappa_2$ & $\kappa_3$& $\kappa_1$ & $\kappa_2$ & $\kappa_3$\\
    $f^-_i$ & $\kappa_2$ & $\kappa_3$ & $\kappa_1$ & $\kappa_2$& $\kappa_3$ & $\kappa_1$& $\kappa_2$ & $\kappa_3$ & $\kappa_1$\\
\end{tabular}
\end{center}

\end{itemize}

\subsection{The \texorpdfstring{$\mathbb{Z}_k\times \mathbb{Z}_\ell$}{Zk x Zl} orbifold family}\label{sec: ZkZlexamples}

Here we consider the four $\mathbb{Z}_k\times \mathbb{Z}_\ell$ orbifold theories with up to nine gauge nodes

\begin{itemize}
\item The $\bm{\mathbb{Z}_2\times \mathbb{Z}_2}$ orbifold features four independent gauge couplings $g_1,g_2,g_3, g_4$. We define auxiliary couplings
\begin{equation}
h_i=\sqrt{2-\kappa_i^2}\,.
\end{equation}
By definition \eqref{eq: definecouplings} the parameters $\kappa_i$ take values in the range $\left[ 0\, , 2 \right]$. However, the obstruction observed in \eqref{eq:obstruction} further constrains $\kappa_i\leq\sqrt{2}$, such that the auxiliary couplings $h_j$ remain real. With these definitions in place  the full table of couplings is  
\begin{center}
 \begin{tabular}{c||c|c|c|c}
  $i$ & $(1,1)$ & $(1,2)$ & $(2,1)$ & $(2,2)$\\
    \hline
    $\kappa_i$ & $\kappa_1$ & $\kappa_2$ & $\kappa_3$ & $\kappa_4$ \\
    $f^+_i$ & $h_2$ & $h_1$ & $h_4$ & $h_3$\\
    $f^-_i$ & $h_4$ & $h_3$ & $h_2$ & $h_1$\\
\end{tabular}
\end{center}

\item The $\bm{\mathbb{Z}_2\times \mathbb{Z}_3}$ orbifold features four independent gauge couplings $g_1,g_2,g_3, g_4$ with
\begin{equation}
    \kappa_5=\sqrt{-\kappa_1^2+\kappa_2^2+\kappa_4^2}\,,\qquad  \kappa_6=\sqrt{-\kappa_1^2+\kappa_3^2+\kappa_4^2}\,.
\end{equation}
The obstruction \eqref{eq:obstruction} constrains the gauge-coupling deformations such that $\kappa_5$ and $\kappa_6$ remain real. In terms of these identifications, the full table of couplings is  
\begin{center}
 \begin{tabular}{c||c|c|c|c|c|c}
  $i$ & $(1,1)$ & $(1,2)$ & $(1,3)$ & $(2,1)$ & $(2,2)$ & $(2,3)$\\
    \hline
    $\kappa_i$ & $\kappa_1$ & $\kappa_2$  & $\kappa_3$ & $\kappa_4$& $\kappa_5$ & $\kappa_6$  \\
    $f^+_i$ & $\kappa_4$ & $\kappa_5$ & $\kappa_6$ & $\kappa_1$ & $\kappa_2$ & $\kappa_3$\\
    $f^-_i$  &$\kappa_1$ & $\kappa_2$ & $\kappa_3$ & $\kappa_4$ & $\kappa_5$ & $\kappa_6$\\
\end{tabular}
\end{center}

We note that by relabelling gauge nodes and chiral multiplets, we can identify this theory with the $\bm{\mathbb{Z}_6[1,2,3]}$ orbifold discussed above. This is also confirmed by the index and full BPS spectrum. 

\item The $\bm{\mathbb{Z}_2\times \mathbb{Z}_4}$ orbifold features six independent gauge couplings $g_1,g_2,g_3, g_4,g_5,g_6$ with
\begin{equation}
    \kappa_7=\sqrt{-\kappa_1^2+\kappa_3^2+\kappa_5^2}\,,\qquad  \kappa_8=\sqrt{-\kappa_2^2+\kappa_4^2+\kappa_6^2}\,.
\end{equation}
We further define auxiliary couplings 
\begin{equation}
    h_i=\sqrt{\frac{\kappa_1^2+\kappa_2^2+\kappa_5^2+\kappa_6^2}{2}-\kappa_i^2}\,,\quad h_{ij,k}=\sqrt{\frac{\kappa_1^2+\kappa_2^2+\kappa_5^2+\kappa_6^2}{2}-\kappa_i^2-\kappa_j^2+\kappa_k^2}\,.
\end{equation}
In terms of these identifications, the full table of couplings is  
\begin{center}
 \begin{tabular}{c||c|c|c|c|c|c|c|c}
  $i$ & $(1,1)$ & $(1,2)$ & $(1,3)$ & $(1,4)$ & $(2,1)$ & $(2,2)$ & $(2,3)$ & $(2,4)$\\
    \hline
    $\kappa_i$ & $\kappa_1$ & $\kappa_2$  & $\kappa_3$ & $\kappa_4$& $\kappa_5$ & $\kappa_6$  & $\kappa_7$ & $\kappa_8$\\
    $f^+_i$ & $h_2$ & $h_1$ & $h_{25,7}$ & $h_{16,8}$ & $h_6$ & $h_5$ & $h_{16,3}$ & $h_{25,4}$ \\
    $f^-_i$  &$h_6$ & $h_5$ & $h_{16,3}$ & $h_{25,4}$& $h_2$ & $h_1$ & $h_{25,7}$ & $h_{16,8}$\\
\end{tabular}
\end{center}
Again, the obstruction \eqref{eq:obstruction} constrains the gauge-coupling deformations such that the auxiliary couplings remain real.
\item The $\bm{\mathbb{Z}_3\times \mathbb{Z}_3}$ orbifold features seven independent gauge couplings $g_1,g_2,g_3, g_4,g_5,g_6,g_7$ with
\begin{equation}
    \kappa_8=\sqrt{-\kappa_1^2+\kappa_3^2+\kappa_5^2-\kappa_6^2+\kappa_7^2}\,,\qquad  \kappa_9=\sqrt{-\kappa_2^2+\kappa_3^2-\kappa_4^2+\kappa_5^2+\kappa_7^2}\,.
\end{equation}
We further define auxiliary couplings 
\begin{equation}\begin{split}
    h_i^\pm&=\sqrt{\pm \frac{\kappa_4^2+\kappa_5^2+\kappa_6^2-\kappa_1^2-\kappa_2^2-\kappa_3^2}{3}+\kappa_i^2}\,,\\\quad h_{ij,k}^\pm&=\sqrt{\pm \frac{\kappa_4^2+\kappa_5^2+\kappa_6^2-\kappa_1^2-\kappa_2^2-\kappa_3^2}{3}+\kappa_i^2+\kappa_j^2-\kappa_k^2}\,.
\end{split}\end{equation}
In terms of these identifications, the full table of couplings is  
\begin{center}
 \begin{tabular}{c||c|c|c|c|c|c|c|c|c}
  $i$ & $(1,1)$ & $(1,2)$ & $(1,3)$ & $(2,1)$ & $(2,2)$ & $(2,3)$ & $(3,1)$ & $(3,2)$& $(3,3)$\\
    \hline
    $\kappa_i$ & $\kappa_1$ & $\kappa_2$  & $\kappa_3$ & $\kappa_4$& $\kappa_5$ & $\kappa_6$  & $\kappa_7$ & $\kappa_8$  & $\kappa_9$\\
    $f^+_i$ & $h^+_{1}$  & $h^+_{2}$ & $h^+_{3}$ & $h^+_{37,6}$ & $h^+_{18,4}$ & $h^+_{29,5}$ & $h^+_{12,5}$ & $h^+_{23,6}$ & $h^+_{13,4}$ \\
    $f^-_i$   & $h^-_{4}$  & $h^-_{5}$ & $h^-_{6}$ & $h^-_{46,3}$ & $h^-_{45,1}$ & $h^-_{56,2}$ & $h^-_{57,2}$ & $h^-_{68,3}$ & $h^-_{49,1}$\\
\end{tabular}
\end{center}
Again, the obstruction \eqref{eq:obstruction} constrains the gauge-coupling deformations such that the auxiliary couplings remain real.
\end{itemize}

To conclude this survey, we list in Table \ref{tab:Dim Marginal} the one-loop marginal operators and their subset of exactly marginal ones for all theories we considered in this appendix.

\begin{table}[ht]
 \centering
 \renewcommand{\arraystretch}{1.3}
 \begin{tabular}{c||c|c|c|c}
 & ND$[0,0]$ & 1-loop BPS multiplets & dim  & exactly marginal operators \tabularnewline
\hline
\hline 
$\mathbb{Z}_3[1,1,1]$ & -3 & $10 \times  \Ch_{0(0,0)}-13 \times \bar{\Bm}_{2(0,0)} $ & 1+2 & $\mathcal{G}_0$, $\Fm_{XYZ}$, $\Fm_{\Zb_3}$ \\ \hline 
$\mathbb{Z}_4[1,1,2]$ & 0 & $7 \times  \Ch_{0(0,0)}-7 \times \bar{\Bm}_{2(0,0)} $ & 2+1 & $\mathcal{G}_0$, $\mathcal{G}_2$, $\Fm_{XYZ}$\\ \hline 
$\mathbb{Z}_5[1,1,3]$ & 0 & $8 \times  \Ch_{0(0,0)}-8 \times \bar{\Bm}_{2(0,0)} $ & 1+1 &  $\mathcal{G}_0$, $\Fm_{XYZ}$\\ \hline 
$\mathbb{Z}_6[1,1,4]$ & -2 & $9 \times  \Ch_{0(0,0)}-11 \times \bar{\Bm}_{2(0,0)} $ & 2+1 &$\mathcal{G}_0$, $\mathcal{G}_3$, $\Fm_{XYZ}$ \\ \hline 
$\mathbb{Z}_6[1,2,3]$ & -2 & $7 \times  \Ch_{0(0,0)}-9 \times \bar{\Bm}_{2(0,0)} $ & 4+1  &$\mathcal{G}_0$, $\mathcal{G}_2$, $\mathcal{G}_3$, $\mathcal{G}_4$, $\Fm_{XYZ}$ \\ \hline 
$\mathbb{Z}_7[1,1,5]$ & 0 & $10 \times  \Ch_{0(0,0)}-10 \times \bar{\Bm}_{2(0,0)}$  & 1+1 & $\mathcal{G}_0$, $\Fm_{XYZ}$\\ \hline 
$\mathbb{Z}_7[1,2,4]$ & 0 & $ 8 \times  \Ch_{0(0,0)}-8 \times \bar{\Bm}_{2(0,0)} $ & 1+1 & $\mathcal{G}_0$, $\Fm_{XYZ}$\\ \hline 
$\mathbb{Z}_8[1,1,6]$ & 0 & $11 \times  \Ch_{0(0,0)}-11 \times \bar{\Bm}_{2(0,0)} $ & 2+1 & $\mathcal{G}_0$, $\mathcal{G}_4$, $\Fm_{XYZ}$\\ \hline 
$\mathbb{Z}_8[1,2,5]$ & 0 & $9 \times  \Ch_{0(0,0)}-9 \times \bar{\Bm}_{2(0,0)} $ & 2+1 & $\mathcal{G}_0$, $\mathcal{G}_4$, $\Fm_{XYZ}$\\ \hline 
$\mathbb{Z}_8[1,3,4]$ & 0 & $ 9 \times  \Ch_{0(0,0)}-9 \times \bar{\Bm}_{2(0,0)}$ & 4+1 & $\mathcal{G}_0$, $\mathcal{G}_2$, $\mathcal{G}_4$, $\mathcal{G}_6$, $\Fm_{XYZ}$ \\ \hline
$\mathbb{Z}_9[1,1,7]$ & 0 & $12\times  \Ch_{0(0,0)}-12 \times \bar{\Bm}_{2(0,0)}$  & 1+1 & $\mathcal{G}_0$, $\Fm_{XYZ}$\\ \hline 
$\mathbb{Z}_9[1,2,6]$ & -3 & $10 \times  \Ch_{0(0,0)}-13 \times \bar{\Bm}_{2(0,0)}$  & 3+1 & $\mathcal{G}_0$, $\mathcal{G}_3$, $\mathcal{G}_6$, $\Fm_{XYZ}$\\ \hline

$\mathbb{Z}_2 \times \mathbb{Z}_2$ & 0 & $ 5 \times  \Ch_{0(0,0)}-5 \times \bar{\Bm}_{2(0,0)}$ & 4+1 & \makecell{$\mathcal{G}_{(0,0)}$, $\mathcal{G}_{(0,1)}$, $\mathcal{G}_{(1,0)}$, $\mathcal{G}_{(1,1)}$, \\$\Fm_{XYZ}$}\\  \hline 
$\mathbb{Z}_2 \times \mathbb{Z}_3$ & -2 & $7 \times  \Ch_{0(0,0)}-9 \times \bar{\Bm}_{2(0,0)}$ & 4+1 & \makecell{$\mathcal{G}_{(0,0)}$, $\mathcal{G}_{(0,1)}$, $\mathcal{G}_{(0,2)}$, $\mathcal{G}_{(1,0)}$,\\ $\Fm_{XYZ}$}\\  \hline 
$\mathbb{Z}_2 \times \mathbb{Z}_4$ & 0 & $9 \times  \Ch_{0(0,0)}-9 \times \bar{\Bm}_{2(0,0)}$ & 6+1 & \makecell{$\mathcal{G}_{(0,0)}$, $\mathcal{G}_{(0,1)}$, $\mathcal{G}_{(0,2)}$, $\mathcal{G}_{(0,3)}$,\\ $\mathcal{G}_{(1,0)}$, $\mathcal{G}_{(1,2)}$, $\Fm_{XYZ}$}\\ \hline
$\mathbb{Z}_3 \times \mathbb{Z}_3$ & -9 & $10 \times  \Ch_{0(0,0)}-19 \times \bar{\Bm}_{2(0,0)} $ & 7+8 & \makecell{$\mathcal{G}_{(0,0)}$, $\mathcal{G}_{(0,1)}$, $\mathcal{G}_{(0,2)}$,\\ $\mathcal{G}_{(1,0)}$, $\mathcal{G}_{(2,0)}$, $\mathcal{G}_{(1,2)}$, \\
$\mathcal{G}_{(2,1)}$, $\Fm_{XYZ}$, $7\times\Fm_{\Zb_3}$}\\ 
\end{tabular}
\caption{ Table containing the net degeneracy $\text{ND}[0,0]$ \eqref{eq:Net degeneracy} of the equivalence classes $[0,0]_\pm$ \eqref{Eq:N1_Equivalence_Classes} for the theories discussed in Appendix \ref{App:survey}. We give the number of one-loop protected multiplets contributing to this net degeneracy, the actual complex dimension of the conformal manifold (split in gauge-coupling and superpotential directions) and the exactly marginal operators spanning it.  
\label{tab:Dim Marginal}}
\end{table}

\section{Deformation coefficients in the one-loop Hamiltonian}\label{app: coeffs}

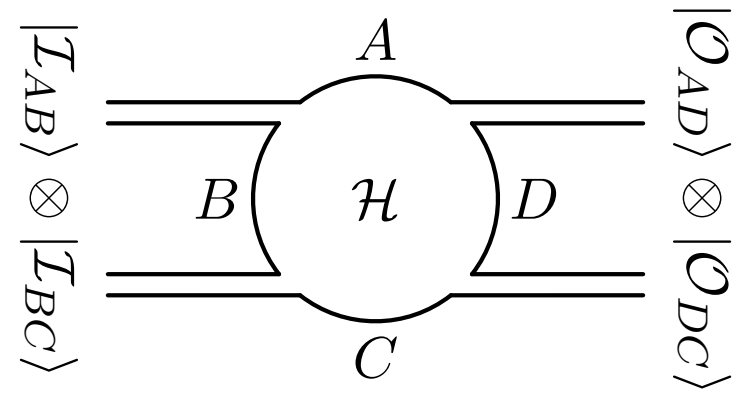
\begin{figure}[t]
    \centering
    \begin{tikzpicture}[
        x=1cm,y=1cm,
        scale=0.6,
        every node/.style={transform shape=false},
        strand/.style={line width=0.8pt,line cap=round},
        statelabel/.style={font=\normalsize,inner sep=0pt,outer sep=0pt}
    ]

        \def\R{1.35}

        \def\xL{-2.95}
        \def\xR{ 2.95}

        \node[statelabel, rotate=270] at (-3.6,0)
            {$\ket{\mathcal{ I}_{AB}}\otimes\ket{\mathcal I_{BC}}$};

        \node[statelabel, rotate=270] at (3.6,0)
            {$\ket{\mathcal O_{AD}}\otimes\ket{\mathcal O_{DC}}$};


        \coordinate (I1A) at (128:\R);
        \coordinate (I1B) at (142:\R);

        \coordinate (I2B) at (218:\R);
        \coordinate (I2C) at (232:\R);

        \coordinate (O1A) at (52:\R);
        \coordinate (O1D) at (38:\R);

        \coordinate (O2D) at (-38:\R);
        \coordinate (O2C) at (-52:\R);


        \draw[strand] (\xL,{ \R*sin(128) }) -- (I1A);
        \draw[strand] (\xL,{ \R*sin(142) }) -- (I1B);

        \draw[strand] (\xL,{ \R*sin(218) }) -- (I2B);
        \draw[strand] (\xL,{ \R*sin(232) }) -- (I2C);

        \draw[strand] (O1A) -- (\xR,{ \R*sin(52) });
        \draw[strand] (O1D) -- (\xR,{ \R*sin(38) });

        \draw[strand] (O2D) -- (\xR,{ \R*sin(-38) });
        \draw[strand] (O2C) -- (\xR,{ \R*sin(-52) });


        \draw[strand] (128:\R) arc[start angle=128,end angle=52,radius=\R];

        \draw[strand] (142:\R) arc[start angle=142,end angle=218,radius=\R];

        \draw[strand] (232:\R) arc[start angle=232,end angle=308,radius=\R];

        \draw[strand] (38:\R) arc[start angle=38,end angle=-38,radius=\R];

        \node at (90:{\R+0.4})  {$A$};
        \node at (180:{\R+0.4}) {$B$};
        \node at (270:{\R+0.4}) {$C$};
        \node at (0:{\R+0.4})   {$D$};

        \node at (0,0) {$\mathcal{H}$};

    \end{tikzpicture}
    \caption{Double-line representation of the one-loop nearest-neighbour Hamiltonian. The $A,B,C,D$ labels are placeholders for the theory-specific gauge indices partaking in the interaction. They determine which couplings $\kappa$ and $f^\pm$ appear in the deformation coefficient $d_\sigma$.}
    \label{fig: planar diagram}
\end{figure}

In this appendix we list the deformation factors $d_\sigma$ to be introduced to the one-loop Hamiltonian \eqref{eq: hamdef} once marginal deformations are turned on. As we explained in Section \ref{sec: diladef}, these depend on the reshufflings of the auxiliary $\db_i^\dagger$ oscillator alone. We allow for a general parametrisation of the couplings $\kappa_i$ \eqref{eq:kappa} and $f^\pm_i$\eqref{Eq: superpot} where the appropriate gauge indices depend on the fields participating in the two-to-two process.\footnote{In contrast to the main text we allow here for more general complex values of $f^\pm_i$ and therefore take care of appropriate conjugations.} We label the participating gauge indices as in Figure \ref{fig: planar diagram}. 
As discussed in Section \ref{sec: diladef}, some $d_\sigma$ coefficients depend on the ratio
\begin{equation}
\rho=\frac{c_\sigma(n-2,n_{12},n_{21})}{c_\sigma(n,n_{12},n_{21})}=1+\frac{c_\sigma(n,n_{12},n_{21}+2)}{c_\sigma(n,n_{12},n_{21})}
\end{equation}
in terms of the undeformed shuffling coefficient $c_\sigma$ \eqref{Eq: coeff}. Furthermore we introduced a factor $\delta_2$ which is $\delta_2=0$ when the dilatation operator is applied to single-trace operators of length $L>2$ and $\delta_2=1$ for single-trace operators of length 2. This distinction is necessary to account for the $\text{U}(1)$-channel subtraction discussed around \eqref{eq: U(1)channel}. With these definitions in place we can list the appropriate deformation factor $d_\sigma$ for every possible $\db_i^\dagger$ reshuffling process in Table 11.

\clearpage
\begin{landscape}
\begingroup
\linespread{1}\fontsize{8}{9.6}\selectfont
\setlength{\tabcolsep}{2pt}
\renewcommand{\arraystretch}{1.12}
\noindent\textbf{Table 11: Coefficients $\bm{d_\sigma}$}\par\medskip
\noindent\begin{minipage}[t]{0.49\linewidth}
\vspace{0pt}
\begin{tabular}{@{}>{\raggedright\arraybackslash}m{0.28\linewidth}>{\raggedright\arraybackslash}m{0.28\linewidth}>{\raggedright\arraybackslash}m{\dimexpr0.44\linewidth-8pt\relax}@{}}
\toprule Incoming state & Outgoing state & Coefficient $d_\sigma$ \\
\midrule
$\ket{\vphantom{\db^\dagger_1}\,0}\otimes\ket{\vphantom{\db^\dagger_1}\,0}$ & $\ket{\vphantom{\db^\dagger_1}\,0}\otimes\ket{\vphantom{\db^\dagger_1}\,0}$ & $\kappa_A^{2}$ \\ \addlinespace[2pt]
$\ket{\vphantom{\db^\dagger_1}\,0}\otimes\ket{\db^\dagger_1}$ & $\ket{\vphantom{\db^\dagger_1}\,0}\otimes\ket{\db^\dagger_1}$ & $\kappa_A^{2}$ \\ \addlinespace[2pt]
$\ket{\vphantom{\db^\dagger_1}\,0}\otimes\ket{\db^\dagger_1}$ & $\ket{\db^\dagger_1}\otimes\ket{\vphantom{\db^\dagger_1}\,0}$ & $\kappa_A\,\kappa_C$ \\ \addlinespace[2pt]
$\ket{\vphantom{\db^\dagger_1}\,0}\otimes\ket{\db^\dagger_2}$ & $\ket{\vphantom{\db^\dagger_1}\,0}\otimes\ket{\db^\dagger_2}$ & $\kappa_A^{2}$ \\ \addlinespace[2pt]
$\ket{\vphantom{\db^\dagger_1}\,0}\otimes\ket{\db^\dagger_2}$ & $\ket{\db^\dagger_2}\otimes\ket{\vphantom{\db^\dagger_1}\,0}$ & $\kappa_A\,\kappa_C$ \\ \addlinespace[2pt]
$\ket{\vphantom{\db^\dagger_1}\,0}\otimes\ket{\db^\dagger_3}$ & $\ket{\vphantom{\db^\dagger_1}\,0}\otimes\ket{\db^\dagger_3}$ & $\kappa_A^{2}$ \\ \addlinespace[2pt]
$\ket{\vphantom{\db^\dagger_1}\,0}\otimes\ket{\db^\dagger_3}$ & $\ket{\db^\dagger_3}\otimes\ket{\vphantom{\db^\dagger_1}\,0}$ & $\kappa_A\,\kappa_C$ \\ \addlinespace[2pt]
$\ket{\db^\dagger_1}\otimes\ket{\vphantom{\db^\dagger_1}\,0}$ & $\ket{\vphantom{\db^\dagger_1}\,0}\otimes\ket{\db^\dagger_1}$ & $\kappa_A\,\kappa_C$ \\ \addlinespace[2pt]
$\ket{\db^\dagger_1}\otimes\ket{\vphantom{\db^\dagger_1}\,0}$ & $\ket{\db^\dagger_1}\otimes\ket{\vphantom{\db^\dagger_1}\,0}$ & $\kappa_C^{2}$ \\ \addlinespace[2pt]
$\ket{\db^\dagger_2}\otimes\ket{\vphantom{\db^\dagger_1}\,0}$ & $\ket{\vphantom{\db^\dagger_1}\,0}\otimes\ket{\db^\dagger_2}$ & $\kappa_A\,\kappa_C$ \\ \addlinespace[2pt]
$\ket{\db^\dagger_2}\otimes\ket{\vphantom{\db^\dagger_1}\,0}$ & $\ket{\db^\dagger_2}\otimes\ket{\vphantom{\db^\dagger_1}\,0}$ & $\kappa_C^{2}$ \\ \addlinespace[2pt]
$\ket{\db^\dagger_3}\otimes\ket{\vphantom{\db^\dagger_1}\,0}$ & $\ket{\vphantom{\db^\dagger_1}\,0}\otimes\ket{\db^\dagger_3}$ & $\kappa_A\,\kappa_C$ \\ \addlinespace[2pt]
$\ket{\db^\dagger_3}\otimes\ket{\vphantom{\db^\dagger_1}\,0}$ & $\ket{\db^\dagger_3}\otimes\ket{\vphantom{\db^\dagger_1}\,0}$ & $\kappa_C^{2}$ \\ \addlinespace[2pt]
$\ket{\vphantom{\db^\dagger_1}\,0}\otimes\ket{\db^\dagger_1\db^\dagger_2}$ & $\ket{\vphantom{\db^\dagger_1}\,0}\otimes\ket{\db^\dagger_1\db^\dagger_2}$ & $\kappa_A^{2}$ \\ \addlinespace[2pt]
$\ket{\vphantom{\db^\dagger_1}\,0}\otimes\ket{\db^\dagger_1\db^\dagger_2}$ & $\ket{\db^\dagger_1}\otimes\ket{\db^\dagger_2}$ & $\kappa_A\,f^{+}_{A}$ \\ \addlinespace[2pt]
$\ket{\vphantom{\db^\dagger_1}\,0}\otimes\ket{\db^\dagger_1\db^\dagger_2}$ & $\ket{\db^\dagger_2}\otimes\ket{\db^\dagger_1}$ & $\kappa_A\,f^{-}_{D}$ \\ \addlinespace[2pt]
$\ket{\vphantom{\db^\dagger_1}\,0}\otimes\ket{\db^\dagger_1\db^\dagger_2}$ & $\ket{\db^\dagger_1\db^\dagger_2}\otimes\ket{\vphantom{\db^\dagger_1}\,0}$ & $\kappa_A\,\kappa_C$ \\ \addlinespace[2pt]
$\ket{\vphantom{\db^\dagger_1}\,0}\otimes\ket{\db^\dagger_1\db^\dagger_3}$ & $\ket{\vphantom{\db^\dagger_1}\,0}\otimes\ket{\db^\dagger_1\db^\dagger_3}$ & $\kappa_A^{2}$ \\ \addlinespace[2pt]
$\ket{\vphantom{\db^\dagger_1}\,0}\otimes\ket{\db^\dagger_1\db^\dagger_3}$ & $\ket{\db^\dagger_1}\otimes\ket{\db^\dagger_3}$ & $\kappa_A\,f^{-}_{A}$ \\ \addlinespace[2pt]
$\ket{\vphantom{\db^\dagger_1}\,0}\otimes\ket{\db^\dagger_1\db^\dagger_3}$ & $\ket{\db^\dagger_3}\otimes\ket{\db^\dagger_1}$ & $\kappa_A\,f^{+}_{D}$ \\ \addlinespace[2pt]
$\ket{\vphantom{\db^\dagger_1}\,0}\otimes\ket{\db^\dagger_1\db^\dagger_3}$ & $\ket{\db^\dagger_1\db^\dagger_3}\otimes\ket{\vphantom{\db^\dagger_1}\,0}$ & $\kappa_A\,\kappa_C$ \\ \addlinespace[2pt]
$\ket{\vphantom{\db^\dagger_1}\,0}\otimes\ket{\db^\dagger_2\db^\dagger_3}$ & $\ket{\vphantom{\db^\dagger_1}\,0}\otimes\ket{\db^\dagger_2\db^\dagger_3}$ & $\kappa_A^{2}$ \\ \addlinespace[2pt]
\bottomrule
\end{tabular}
\end{minipage}\hfill\noindent\begin{minipage}[t]{0.49\linewidth}
\vspace{0pt}
\begin{tabular}{@{}>{\raggedright\arraybackslash}m{0.28\linewidth}>{\raggedright\arraybackslash}m{0.28\linewidth}>{\raggedright\arraybackslash}m{\dimexpr0.44\linewidth-8pt\relax}@{}}
\toprule Incoming state & Outgoing state & Coefficient $d_\sigma$ \\
\midrule
$\ket{\vphantom{\db^\dagger_1}\,0}\otimes\ket{\db^\dagger_2\db^\dagger_3}$ & $\ket{\db^\dagger_2}\otimes\ket{\db^\dagger_3}$ & $\kappa_A\,f^{+}_{C}$ \\ \addlinespace[2pt]
$\ket{\vphantom{\db^\dagger_1}\,0}\otimes\ket{\db^\dagger_2\db^\dagger_3}$ & $\ket{\db^\dagger_3}\otimes\ket{\db^\dagger_2}$ & $\kappa_A\,f^{-}_{C}$ \\ \addlinespace[2pt]
$\ket{\vphantom{\db^\dagger_1}\,0}\otimes\ket{\db^\dagger_2\db^\dagger_3}$ & $\ket{\db^\dagger_2\db^\dagger_3}\otimes\ket{\vphantom{\db^\dagger_1}\,0}$ & $\kappa_A\,\kappa_C$ \\ \addlinespace[2pt]
$\ket{\db^\dagger_1}\otimes\ket{\db^\dagger_1}$ & $\ket{\db^\dagger_1}\otimes\ket{\db^\dagger_1}$ & $\kappa_B^{2}$ \\ \addlinespace[2pt]
$\ket{\db^\dagger_1}\otimes\ket{\db^\dagger_2}$ & $\ket{\vphantom{\db^\dagger_1}\,0}\otimes\ket{\db^\dagger_1\db^\dagger_2}$ & $\kappa_A\,\bar{f}^{+}_{A}$ \\ \addlinespace[2pt]
$\ket{\db^\dagger_1}\otimes\ket{\db^\dagger_2}$ & $\ket{\db^\dagger_1}\otimes\ket{\db^\dagger_2}$ & $\rho\,\kappa_B^2+(1-\rho)f^{+}_{A}\,\bar{f}^{+}_{A}$ \\ \addlinespace[2pt]
$\ket{\db^\dagger_1}\otimes\ket{\db^\dagger_2}$ & $\ket{\db^\dagger_2}\otimes\ket{\db^\dagger_1}$ & $\bar{f}^{+}_{A}\,f^{-}_{D}$ \\ \addlinespace[2pt]
$\ket{\db^\dagger_1}\otimes\ket{\db^\dagger_2}$ & $\ket{\db^\dagger_1\db^\dagger_2}\otimes\ket{\vphantom{\db^\dagger_1}\,0}$ & $\kappa_C\,\bar{f}^{+}_{A}$ \\ \addlinespace[2pt]
$\ket{\db^\dagger_1}\otimes\ket{\db^\dagger_3}$ & $\ket{\vphantom{\db^\dagger_1}\,0}\otimes\ket{\db^\dagger_1\db^\dagger_3}$ & $\kappa_A\,\bar{f}^{-}_{A}$ \\ \addlinespace[2pt]
$\ket{\db^\dagger_1}\otimes\ket{\db^\dagger_3}$ & $\ket{\db^\dagger_1}\otimes\ket{\db^\dagger_3}$ & $\rho\,\kappa_B^2+(1-\rho)f^{-}_{A}\,\bar{f}^{-}_{A}$ \\ \addlinespace[2pt]
$\ket{\db^\dagger_1}\otimes\ket{\db^\dagger_3}$ & $\ket{\db^\dagger_3}\otimes\ket{\db^\dagger_1}$ & $\bar{f}^{-}_{A}\,f^{+}_{D}$ \\ \addlinespace[2pt]
$\ket{\db^\dagger_1}\otimes\ket{\db^\dagger_3}$ & $\ket{\db^\dagger_1\db^\dagger_3}\otimes\ket{\vphantom{\db^\dagger_1}\,0}$ & $\kappa_C\,\bar{f}^{-}_{A}$ \\ \addlinespace[2pt]
$\ket{\db^\dagger_2}\otimes\ket{\db^\dagger_1}$ & $\ket{\vphantom{\db^\dagger_1}\,0}\otimes\ket{\db^\dagger_1\db^\dagger_2}$ & $\kappa_A\,\bar{f}^{-}_{B}$ \\ \addlinespace[2pt]
$\ket{\db^\dagger_2}\otimes\ket{\db^\dagger_1}$ & $\ket{\db^\dagger_1}\otimes\ket{\db^\dagger_2}$ & $f^{+}_{A}\,\bar{f}^{-}_{B}$ \\ \addlinespace[2pt]
$\ket{\db^\dagger_2}\otimes\ket{\db^\dagger_1}$ & $\ket{\db^\dagger_2}\otimes\ket{\db^\dagger_1}$ & $\rho\,\kappa_B^2+(1-\rho)\bar{f}^{-}_{B}\,f^{-}_{D}$ \\ \addlinespace[2pt]
$\ket{\db^\dagger_2}\otimes\ket{\db^\dagger_1}$ & $\ket{\db^\dagger_1\db^\dagger_2}\otimes\ket{\vphantom{\db^\dagger_1}\,0}$ & $\kappa_C\,\bar{f}^{-}_{B}$ \\ \addlinespace[2pt]
$\ket{\db^\dagger_2}\otimes\ket{\db^\dagger_2}$ & $\ket{\db^\dagger_2}\otimes\ket{\db^\dagger_2}$ & $\kappa_B^{2}$ \\ \addlinespace[2pt]
$\ket{\db^\dagger_2}\otimes\ket{\db^\dagger_3}$ & $\ket{\vphantom{\db^\dagger_1}\,0}\otimes\ket{\db^\dagger_2\db^\dagger_3}$ & $\kappa_A\,\bar{f}^{+}_{C}$ \\ \addlinespace[2pt]
$\ket{\db^\dagger_2}\otimes\ket{\db^\dagger_3}$ & $\ket{\db^\dagger_2}\otimes\ket{\db^\dagger_3}$ & $\rho\,\kappa_B^2+(1-\rho)f^{+}_{C}\,\bar{f}^{+}_{C}$ \\ \addlinespace[2pt]
$\ket{\db^\dagger_2}\otimes\ket{\db^\dagger_3}$ & $\ket{\db^\dagger_3}\otimes\ket{\db^\dagger_2}$ & $\bar{f}^{+}_{C}\,f^{-}_{C}$ \\ \addlinespace[2pt]
$\ket{\db^\dagger_2}\otimes\ket{\db^\dagger_3}$ & $\ket{\db^\dagger_2\db^\dagger_3}\otimes\ket{\vphantom{\db^\dagger_1}\,0}$ & $\kappa_C\,\bar{f}^{+}_{C}$ \\ \addlinespace[2pt]
$\ket{\db^\dagger_3}\otimes\ket{\db^\dagger_1}$ & $\ket{\vphantom{\db^\dagger_1}\,0}\otimes\ket{\db^\dagger_1\db^\dagger_3}$ & $\kappa_A\,\bar{f}^{+}_{B}$ \\ \addlinespace[2pt]
\bottomrule
\end{tabular}
\end{minipage}
\clearpage
\noindent\textbf{Table 11: Coefficients $\bm{d_\sigma}$ (cont.)}\par\medskip
\noindent\begin{minipage}[t]{0.49\linewidth}
\vspace{0pt}
\begin{tabular}{@{}>{\raggedright\arraybackslash}m{0.28\linewidth}>{\raggedright\arraybackslash}m{0.28\linewidth}>{\raggedright\arraybackslash}m{\dimexpr0.44\linewidth-8pt\relax}@{}}
\toprule Incoming state & Outgoing state & Coefficient $d_\sigma$ \\
\midrule
$\ket{\db^\dagger_3}\otimes\ket{\db^\dagger_1}$ & $\ket{\db^\dagger_1}\otimes\ket{\db^\dagger_3}$ & $\bar{f}^{+}_{B}\,f^{-}_{A}$ \\ \addlinespace[2pt]
$\ket{\db^\dagger_3}\otimes\ket{\db^\dagger_1}$ & $\ket{\db^\dagger_3}\otimes\ket{\db^\dagger_1}$ & $\rho\,\kappa_B^2+(1-\rho)\bar{f}^{+}_{B}\,f^{+}_{D}$ \\ \addlinespace[2pt]
$\ket{\db^\dagger_3}\otimes\ket{\db^\dagger_1}$ & $\ket{\db^\dagger_1\db^\dagger_3}\otimes\ket{\vphantom{\db^\dagger_1}\,0}$ & $\kappa_C\,\bar{f}^{+}_{B}$ \\ \addlinespace[2pt]
$\ket{\db^\dagger_3}\otimes\ket{\db^\dagger_2}$ & $\ket{\vphantom{\db^\dagger_1}\,0}\otimes\ket{\db^\dagger_2\db^\dagger_3}$ & $\kappa_A\,\bar{f}^{-}_{C}$ \\ \addlinespace[2pt]
$\ket{\db^\dagger_3}\otimes\ket{\db^\dagger_2}$ & $\ket{\db^\dagger_2}\otimes\ket{\db^\dagger_3}$ & $f^{+}_{C}\,\bar{f}^{-}_{C}$ \\ \addlinespace[2pt]
$\ket{\db^\dagger_3}\otimes\ket{\db^\dagger_2}$ & $\ket{\db^\dagger_3}\otimes\ket{\db^\dagger_2}$ & $\rho\,\kappa_B^2+(1-\rho)f^{-}_{C}\,\bar{f}^{-}_{C}$ \\ \addlinespace[2pt]
$\ket{\db^\dagger_3}\otimes\ket{\db^\dagger_2}$ & $\ket{\db^\dagger_2\db^\dagger_3}\otimes\ket{\vphantom{\db^\dagger_1}\,0}$ & $\kappa_C\,\bar{f}^{-}_{C}$ \\ \addlinespace[2pt]
$\ket{\db^\dagger_3}\otimes\ket{\db^\dagger_3}$ & $\ket{\db^\dagger_3}\otimes\ket{\db^\dagger_3}$ & $\kappa_B^{2}$ \\ \addlinespace[2pt]
$\ket{\db^\dagger_1\db^\dagger_2}\otimes\ket{\vphantom{\db^\dagger_1}\,0}$ & $\ket{\vphantom{\db^\dagger_1}\,0}\otimes\ket{\db^\dagger_1\db^\dagger_2}$ & $\kappa_A\,\kappa_C$ \\ \addlinespace[2pt]
$\ket{\db^\dagger_1\db^\dagger_2}\otimes\ket{\vphantom{\db^\dagger_1}\,0}$ & $\ket{\db^\dagger_1}\otimes\ket{\db^\dagger_2}$ & $\kappa_C\,f^{+}_{A}$ \\ \addlinespace[2pt]
$\ket{\db^\dagger_1\db^\dagger_2}\otimes\ket{\vphantom{\db^\dagger_1}\,0}$ & $\ket{\db^\dagger_2}\otimes\ket{\db^\dagger_1}$ & $\kappa_C\,f^{-}_{D}$ \\ \addlinespace[2pt]
$\ket{\db^\dagger_1\db^\dagger_2}\otimes\ket{\vphantom{\db^\dagger_1}\,0}$ & $\ket{\db^\dagger_1\db^\dagger_2}\otimes\ket{\vphantom{\db^\dagger_1}\,0}$ & $\kappa_C^{2}$ \\ \addlinespace[2pt]
$\ket{\db^\dagger_1\db^\dagger_3}\otimes\ket{\vphantom{\db^\dagger_1}\,0}$ & $\ket{\vphantom{\db^\dagger_1}\,0}\otimes\ket{\db^\dagger_1\db^\dagger_3}$ & $\kappa_A\,\kappa_C$ \\ \addlinespace[2pt]
$\ket{\db^\dagger_1\db^\dagger_3}\otimes\ket{\vphantom{\db^\dagger_1}\,0}$ & $\ket{\db^\dagger_1}\otimes\ket{\db^\dagger_3}$ & $\kappa_C\,f^{-}_{A}$ \\ \addlinespace[2pt]
$\ket{\db^\dagger_1\db^\dagger_3}\otimes\ket{\vphantom{\db^\dagger_1}\,0}$ & $\ket{\db^\dagger_3}\otimes\ket{\db^\dagger_1}$ & $\kappa_C\,f^{+}_{D}$ \\ \addlinespace[2pt]
$\ket{\db^\dagger_1\db^\dagger_3}\otimes\ket{\vphantom{\db^\dagger_1}\,0}$ & $\ket{\db^\dagger_1\db^\dagger_3}\otimes\ket{\vphantom{\db^\dagger_1}\,0}$ & $\kappa_C^{2}$ \\ \addlinespace[2pt]
$\ket{\db^\dagger_2\db^\dagger_3}\otimes\ket{\vphantom{\db^\dagger_1}\,0}$ & $\ket{\vphantom{\db^\dagger_1}\,0}\otimes\ket{\db^\dagger_2\db^\dagger_3}$ & $\kappa_A\,\kappa_C$ \\ \addlinespace[2pt]
$\ket{\db^\dagger_2\db^\dagger_3}\otimes\ket{\vphantom{\db^\dagger_1}\,0}$ & $\ket{\db^\dagger_2}\otimes\ket{\db^\dagger_3}$ & $\kappa_C\,f^{+}_{C}$ \\ \addlinespace[2pt]
$\ket{\db^\dagger_2\db^\dagger_3}\otimes\ket{\vphantom{\db^\dagger_1}\,0}$ & $\ket{\db^\dagger_3}\otimes\ket{\db^\dagger_2}$ & $\kappa_C\,f^{-}_{C}$ \\ \addlinespace[2pt]
$\ket{\db^\dagger_2\db^\dagger_3}\otimes\ket{\vphantom{\db^\dagger_1}\,0}$ & $\ket{\db^\dagger_2\db^\dagger_3}\otimes\ket{\vphantom{\db^\dagger_1}\,0}$ & $\kappa_C^{2}$ \\ \addlinespace[2pt]
$\ket{\vphantom{\db^\dagger_1}\,0}\otimes\ket{\db^\dagger_1\db^\dagger_2\db^\dagger_3}$ & $\ket{\vphantom{\db^\dagger_1}\,0}\otimes\ket{\db^\dagger_1\db^\dagger_2\db^\dagger_3}$ & $\kappa_A^{2}$ \\ \addlinespace[2pt]
$\ket{\vphantom{\db^\dagger_1}\,0}\otimes\ket{\db^\dagger_1\db^\dagger_2\db^\dagger_3}$ & $\ket{\db^\dagger_1}\otimes\ket{\db^\dagger_2\db^\dagger_3}$ & $(1-\rho\delta_2)\kappa_A^2$ \\ \addlinespace[2pt]
\bottomrule
\end{tabular}
\end{minipage}\hfill\noindent\begin{minipage}[t]{0.49\linewidth}
\vspace{0pt}
\begin{tabular}{@{}>{\raggedright\arraybackslash}m{0.28\linewidth}>{\raggedright\arraybackslash}m{0.28\linewidth}>{\raggedright\arraybackslash}m{\dimexpr0.44\linewidth-8pt\relax}@{}}
\toprule Incoming state & Outgoing state & Coefficient $d_\sigma$ \\
\midrule
$\ket{\vphantom{\db^\dagger_1}\,0}\otimes\ket{\db^\dagger_1\db^\dagger_2\db^\dagger_3}$ & $\ket{\db^\dagger_2}\otimes\ket{\db^\dagger_1\db^\dagger_3}$ & $(1-\rho\delta_2)\kappa_A^2$ \\ \addlinespace[2pt]
$\ket{\vphantom{\db^\dagger_1}\,0}\otimes\ket{\db^\dagger_1\db^\dagger_2\db^\dagger_3}$ & $\ket{\db^\dagger_3}\otimes\ket{\db^\dagger_1\db^\dagger_2}$ & $(1-\rho\delta_2)\kappa_A^2$ \\ \addlinespace[2pt]
$\ket{\vphantom{\db^\dagger_1}\,0}\otimes\ket{\db^\dagger_1\db^\dagger_2\db^\dagger_3}$ & $\ket{\db^\dagger_1\db^\dagger_2}\otimes\ket{\db^\dagger_3}$ & $(1-\rho\delta_2)\kappa_A^2$ \\ \addlinespace[2pt]
$\ket{\vphantom{\db^\dagger_1}\,0}\otimes\ket{\db^\dagger_1\db^\dagger_2\db^\dagger_3}$ & $\ket{\db^\dagger_1\db^\dagger_3}\otimes\ket{\db^\dagger_2}$ & $(1-\rho\delta_2)\kappa_A^2$ \\ \addlinespace[2pt]
$\ket{\vphantom{\db^\dagger_1}\,0}\otimes\ket{\db^\dagger_1\db^\dagger_2\db^\dagger_3}$ & $\ket{\db^\dagger_2\db^\dagger_3}\otimes\ket{\db^\dagger_1}$ & $(1-\rho\delta_2)\kappa_A^2$ \\ \addlinespace[2pt]
$\ket{\vphantom{\db^\dagger_1}\,0}\otimes\ket{\db^\dagger_1\db^\dagger_2\db^\dagger_3}$ & $\ket{\db^\dagger_1\db^\dagger_2\db^\dagger_3}\otimes\ket{\vphantom{\db^\dagger_1}\,0}$ & $\kappa_A^{2}$ \\ \addlinespace[2pt]
$\ket{\db^\dagger_1}\otimes\ket{\db^\dagger_1\db^\dagger_2}$ & $\ket{\db^\dagger_1}\otimes\ket{\db^\dagger_1\db^\dagger_2}$ & $\kappa_B^{2}$ \\ \addlinespace[2pt]
$\ket{\db^\dagger_1}\otimes\ket{\db^\dagger_1\db^\dagger_2}$ & $\ket{\db^\dagger_1\db^\dagger_2}\otimes\ket{\db^\dagger_1}$ & $\bar{f}^{+}_{A}\,f^{-}_{D}$ \\ \addlinespace[2pt]
$\ket{\db^\dagger_1}\otimes\ket{\db^\dagger_1\db^\dagger_3}$ & $\ket{\db^\dagger_1}\otimes\ket{\db^\dagger_1\db^\dagger_3}$ & $\kappa_B^{2}$ \\ \addlinespace[2pt]
$\ket{\db^\dagger_1}\otimes\ket{\db^\dagger_1\db^\dagger_3}$ & $\ket{\db^\dagger_1\db^\dagger_3}\otimes\ket{\db^\dagger_1}$ & $\bar{f}^{-}_{A}\,f^{+}_{D}$ \\ \addlinespace[2pt]
$\ket{\db^\dagger_1}\otimes\ket{\db^\dagger_2\db^\dagger_3}$ & $\ket{\vphantom{\db^\dagger_1}\,0}\otimes\ket{\db^\dagger_1\db^\dagger_2\db^\dagger_3}$ & $(1-\rho\delta_2)\kappa_A^2$ \\ \addlinespace[2pt]
$\ket{\db^\dagger_1}\otimes\ket{\db^\dagger_2\db^\dagger_3}$ & $\ket{\db^\dagger_1}\otimes\ket{\db^\dagger_2\db^\dagger_3}$ & $(1-\rho)(1-\delta_2)\kappa_A^2+\rho\,\kappa_B^{2}$ \\ \addlinespace[2pt]
$\ket{\db^\dagger_1}\otimes\ket{\db^\dagger_2\db^\dagger_3}$ & $\ket{\db^\dagger_2}\otimes\ket{\db^\dagger_1\db^\dagger_3}$ & $(1-\delta_2)\kappa_A^2$ \\ \addlinespace[2pt]
$\ket{\db^\dagger_1}\otimes\ket{\db^\dagger_2\db^\dagger_3}$ & $\ket{\db^\dagger_3}\otimes\ket{\db^\dagger_1\db^\dagger_2}$ & $(1-\delta_2)\kappa_A^2$ \\ \addlinespace[2pt]
$\ket{\db^\dagger_1}\otimes\ket{\db^\dagger_2\db^\dagger_3}$ & $\ket{\db^\dagger_1\db^\dagger_2}\otimes\ket{\db^\dagger_3}$ & $(1-\rho)(1-\delta_2)\kappa_A^2+\rho\,\bar{f}^{+}_{A}\,f^{+}_{C}$ \\ \addlinespace[2pt]
$\ket{\db^\dagger_1}\otimes\ket{\db^\dagger_2\db^\dagger_3}$ & $\ket{\db^\dagger_1\db^\dagger_3}\otimes\ket{\db^\dagger_2}$ & $(1-\rho)(1-\delta_2)\kappa_A^2+\rho\,\bar{f}^{-}_{A}\,f^{-}_{C}$ \\ \addlinespace[2pt]
$\ket{\db^\dagger_1}\otimes\ket{\db^\dagger_2\db^\dagger_3}$ & $\ket{\db^\dagger_2\db^\dagger_3}\otimes\ket{\db^\dagger_1}$ & $(1-\delta_2)\kappa_A^2$ \\ \addlinespace[2pt]
$\ket{\db^\dagger_1}\otimes\ket{\db^\dagger_2\db^\dagger_3}$ & $\ket{\db^\dagger_1\db^\dagger_2\db^\dagger_3}\otimes\ket{\vphantom{\db^\dagger_1}\,0}$ & $(1-\rho\delta_2)\kappa_A^2$ \\ \addlinespace[2pt]
$\ket{\db^\dagger_2}\otimes\ket{\db^\dagger_1\db^\dagger_2}$ & $\ket{\db^\dagger_2}\otimes\ket{\db^\dagger_1\db^\dagger_2}$ & $\kappa_B^{2}$ \\ \addlinespace[2pt]
$\ket{\db^\dagger_2}\otimes\ket{\db^\dagger_1\db^\dagger_2}$ & $\ket{\db^\dagger_1\db^\dagger_2}\otimes\ket{\db^\dagger_2}$ & $\bar{f}^{-}_{B}\,f^{+}_{B}$ \\ \addlinespace[2pt]
$\ket{\db^\dagger_2}\otimes\ket{\db^\dagger_1\db^\dagger_3}$ & $\ket{\vphantom{\db^\dagger_1}\,0}\otimes\ket{\db^\dagger_1\db^\dagger_2\db^\dagger_3}$ & $(1-\rho\delta_2)\kappa_A^2$ \\ \addlinespace[2pt]
$\ket{\db^\dagger_2}\otimes\ket{\db^\dagger_1\db^\dagger_3}$ & $\ket{\db^\dagger_1}\otimes\ket{\db^\dagger_2\db^\dagger_3}$ & $(1-\delta_2)\kappa_A^2$ \\ \addlinespace[2pt]
\bottomrule
\end{tabular}
\end{minipage}
\clearpage
\noindent\textbf{Table 11: Coefficients $\bm{d_\sigma}$ (cont.)}\par\medskip
\noindent\begin{minipage}[t]{0.49\linewidth}
\vspace{0pt}
\begin{tabular}{@{}>{\raggedright\arraybackslash}m{0.28\linewidth}>{\raggedright\arraybackslash}m{0.28\linewidth}>{\raggedright\arraybackslash}m{\dimexpr0.44\linewidth-8pt\relax}@{}}
\toprule Incoming state & Outgoing state & Coefficient $d_\sigma$ \\
\midrule
$\ket{\db^\dagger_2}\otimes\ket{\db^\dagger_1\db^\dagger_3}$ & $\ket{\db^\dagger_2}\otimes\ket{\db^\dagger_1\db^\dagger_3}$ & $(1-\rho)(1-\delta_2)\kappa_A^2+\rho\,\kappa_B^{2}$ \\ \addlinespace[2pt]
$\ket{\db^\dagger_2}\otimes\ket{\db^\dagger_1\db^\dagger_3}$ & $\ket{\db^\dagger_3}\otimes\ket{\db^\dagger_1\db^\dagger_2}$ & $(1-\delta_2)\kappa_A^2$ \\ \addlinespace[2pt]
$\ket{\db^\dagger_2}\otimes\ket{\db^\dagger_1\db^\dagger_3}$ & $\ket{\db^\dagger_1\db^\dagger_2}\otimes\ket{\db^\dagger_3}$ & $(1-\rho)(1-\delta_2)\kappa_A^2+\rho\,f^{-}_{B}\,\bar{f}^{-}_{B}$ \\ \addlinespace[2pt]
$\ket{\db^\dagger_2}\otimes\ket{\db^\dagger_1\db^\dagger_3}$ & $\ket{\db^\dagger_1\db^\dagger_3}\otimes\ket{\db^\dagger_2}$ & $(1-\delta_2)\kappa_A^2$ \\ \addlinespace[2pt]
$\ket{\db^\dagger_2}\otimes\ket{\db^\dagger_1\db^\dagger_3}$ & $\ket{\db^\dagger_2\db^\dagger_3}\otimes\ket{\db^\dagger_1}$ & $(1-\rho)(1-\delta_2)\kappa_A^2+\rho\,f^{+}_{D}\,\bar{f}^{+}_{D}$ \\ \addlinespace[2pt]
$\ket{\db^\dagger_2}\otimes\ket{\db^\dagger_1\db^\dagger_3}$ & $\ket{\db^\dagger_1\db^\dagger_2\db^\dagger_3}\otimes\ket{\vphantom{\db^\dagger_1}\,0}$ & $(1-\rho\delta_2)\kappa_A^2$ \\ \addlinespace[2pt]
$\ket{\db^\dagger_2}\otimes\ket{\db^\dagger_2\db^\dagger_3}$ & $\ket{\db^\dagger_2}\otimes\ket{\db^\dagger_2\db^\dagger_3}$ & $\kappa_B^{2}$ \\ \addlinespace[2pt]
$\ket{\db^\dagger_2}\otimes\ket{\db^\dagger_2\db^\dagger_3}$ & $\ket{\db^\dagger_2\db^\dagger_3}\otimes\ket{\db^\dagger_2}$ & $\bar{f}^{+}_{D}\,f^{-}_{C}$ \\ \addlinespace[2pt]
$\ket{\db^\dagger_3}\otimes\ket{\db^\dagger_1\db^\dagger_2}$ & $\ket{\vphantom{\db^\dagger_1}\,0}\otimes\ket{\db^\dagger_1\db^\dagger_2\db^\dagger_3}$ & $(1-\rho\delta_2)\kappa_A^2$ \\ \addlinespace[2pt]
$\ket{\db^\dagger_3}\otimes\ket{\db^\dagger_1\db^\dagger_2}$ & $\ket{\db^\dagger_1}\otimes\ket{\db^\dagger_2\db^\dagger_3}$ & $(1-\delta_2)\kappa_A^2$ \\ \addlinespace[2pt]
$\ket{\db^\dagger_3}\otimes\ket{\db^\dagger_1\db^\dagger_2}$ & $\ket{\db^\dagger_2}\otimes\ket{\db^\dagger_1\db^\dagger_3}$ & $(1-\delta_2)\kappa_A^2$ \\ \addlinespace[2pt]
$\ket{\db^\dagger_3}\otimes\ket{\db^\dagger_1\db^\dagger_2}$ & $\ket{\db^\dagger_3}\otimes\ket{\db^\dagger_1\db^\dagger_2}$ & $(1-\rho)(1-\delta_2)\kappa_A^2+\rho\,\kappa_B^{2}$ \\ \addlinespace[2pt]
$\ket{\db^\dagger_3}\otimes\ket{\db^\dagger_1\db^\dagger_2}$ & $\ket{\db^\dagger_1\db^\dagger_2}\otimes\ket{\db^\dagger_3}$ & $(1-\delta_2)\kappa_A^2$ \\ \addlinespace[2pt]
$\ket{\db^\dagger_3}\otimes\ket{\db^\dagger_1\db^\dagger_2}$ & $\ket{\db^\dagger_1\db^\dagger_3}\otimes\ket{\db^\dagger_2}$ & $(1-\rho)(1-\delta_2)\kappa_A^2+\rho\,f^{+}_{B}\,\bar{f}^{+}_{B}$ \\ \addlinespace[2pt]
$\ket{\db^\dagger_3}\otimes\ket{\db^\dagger_1\db^\dagger_2}$ & $\ket{\db^\dagger_2\db^\dagger_3}\otimes\ket{\db^\dagger_1}$ & $(1-\rho)(1-\delta_2)\kappa_A^2+\rho\,f^{-}_{D}\,\bar{f}^{-}_{D}$ \\ \addlinespace[2pt]
$\ket{\db^\dagger_3}\otimes\ket{\db^\dagger_1\db^\dagger_2}$ & $\ket{\db^\dagger_1\db^\dagger_2\db^\dagger_3}\otimes\ket{\vphantom{\db^\dagger_1}\,0}$ & $(1-\rho\delta_2)\kappa_A^2$ \\ \addlinespace[2pt]
$\ket{\db^\dagger_3}\otimes\ket{\db^\dagger_1\db^\dagger_3}$ & $\ket{\db^\dagger_3}\otimes\ket{\db^\dagger_1\db^\dagger_3}$ & $\kappa_B^{2}$ \\ \addlinespace[2pt]
$\ket{\db^\dagger_3}\otimes\ket{\db^\dagger_1\db^\dagger_3}$ & $\ket{\db^\dagger_1\db^\dagger_3}\otimes\ket{\db^\dagger_3}$ & $\bar{f}^{+}_{B}\,f^{-}_{B}$ \\ \addlinespace[2pt]
$\ket{\db^\dagger_3}\otimes\ket{\db^\dagger_2\db^\dagger_3}$ & $\ket{\db^\dagger_3}\otimes\ket{\db^\dagger_2\db^\dagger_3}$ & $\kappa_B^{2}$ \\ \addlinespace[2pt]
$\ket{\db^\dagger_3}\otimes\ket{\db^\dagger_2\db^\dagger_3}$ & $\ket{\db^\dagger_2\db^\dagger_3}\otimes\ket{\db^\dagger_3}$ & $\bar{f}^{-}_{D}\,f^{+}_{C}$ \\ \addlinespace[2pt]
$\ket{\db^\dagger_1\db^\dagger_2}\otimes\ket{\db^\dagger_1}$ & $\ket{\db^\dagger_1}\otimes\ket{\db^\dagger_1\db^\dagger_2}$ & $f^{+}_{A}\,\bar{f}^{-}_{B}$ \\ \addlinespace[2pt]
$\ket{\db^\dagger_1\db^\dagger_2}\otimes\ket{\db^\dagger_1}$ & $\ket{\db^\dagger_1\db^\dagger_2}\otimes\ket{\db^\dagger_1}$ & $\kappa_B^{2}$ \\ \addlinespace[2pt]
\bottomrule
\end{tabular}
\end{minipage}\hfill\noindent\begin{minipage}[t]{0.49\linewidth}
\vspace{0pt}
\begin{tabular}{@{}>{\raggedright\arraybackslash}m{0.28\linewidth}>{\raggedright\arraybackslash}m{0.28\linewidth}>{\raggedright\arraybackslash}m{\dimexpr0.44\linewidth-8pt\relax}@{}}
\toprule Incoming state & Outgoing state & Coefficient $d_\sigma$ \\
\midrule
$\ket{\db^\dagger_1\db^\dagger_2}\otimes\ket{\db^\dagger_2}$ & $\ket{\db^\dagger_2}\otimes\ket{\db^\dagger_1\db^\dagger_2}$ & $f^{-}_{D}\,\bar{f}^{+}_{D}$ \\ \addlinespace[2pt]
$\ket{\db^\dagger_1\db^\dagger_2}\otimes\ket{\db^\dagger_2}$ & $\ket{\db^\dagger_1\db^\dagger_2}\otimes\ket{\db^\dagger_2}$ & $\kappa_B^{2}$ \\ \addlinespace[2pt]
$\ket{\db^\dagger_1\db^\dagger_2}\otimes\ket{\db^\dagger_3}$ & $\ket{\vphantom{\db^\dagger_1}\,0}\otimes\ket{\db^\dagger_1\db^\dagger_2\db^\dagger_3}$ & $(1-\rho\delta_2)\kappa_A^2$ \\ \addlinespace[2pt]
$\ket{\db^\dagger_1\db^\dagger_2}\otimes\ket{\db^\dagger_3}$ & $\ket{\db^\dagger_1}\otimes\ket{\db^\dagger_2\db^\dagger_3}$ & $(1-\rho)(1-\delta_2)\kappa_A^2+\rho\,f^{+}_{A}\,\bar{f}^{+}_{C}$ \\ \addlinespace[2pt]
$\ket{\db^\dagger_1\db^\dagger_2}\otimes\ket{\db^\dagger_3}$ & $\ket{\db^\dagger_2}\otimes\ket{\db^\dagger_1\db^\dagger_3}$ & $(1-\rho)(1-\delta_2)\kappa_A^2+\rho\,f^{-}_{D}\,\bar{f}^{-}_{D}$ \\ \addlinespace[2pt]
$\ket{\db^\dagger_1\db^\dagger_2}\otimes\ket{\db^\dagger_3}$ & $\ket{\db^\dagger_3}\otimes\ket{\db^\dagger_1\db^\dagger_2}$ & $(1-\delta_2)\kappa_A^2$ \\ \addlinespace[2pt]
$\ket{\db^\dagger_1\db^\dagger_2}\otimes\ket{\db^\dagger_3}$ & $\ket{\db^\dagger_1\db^\dagger_2}\otimes\ket{\db^\dagger_3}$ & $(1-\rho)(1-\delta_2)\kappa_A^2+\rho\,\kappa_B^{2}$ \\ \addlinespace[2pt]
$\ket{\db^\dagger_1\db^\dagger_2}\otimes\ket{\db^\dagger_3}$ & $\ket{\db^\dagger_1\db^\dagger_3}\otimes\ket{\db^\dagger_2}$ & $(1-\delta_2)\kappa_A^2$ \\ \addlinespace[2pt]
$\ket{\db^\dagger_1\db^\dagger_2}\otimes\ket{\db^\dagger_3}$ & $\ket{\db^\dagger_2\db^\dagger_3}\otimes\ket{\db^\dagger_1}$ & $(1-\delta_2)\kappa_A^2$ \\ \addlinespace[2pt]
$\ket{\db^\dagger_1\db^\dagger_2}\otimes\ket{\db^\dagger_3}$ & $\ket{\db^\dagger_1\db^\dagger_2\db^\dagger_3}\otimes\ket{\vphantom{\db^\dagger_1}\,0}$ & $(1-\rho\delta_2)\kappa_A^2$ \\ \addlinespace[2pt]
$\ket{\db^\dagger_1\db^\dagger_3}\otimes\ket{\db^\dagger_1}$ & $\ket{\db^\dagger_1}\otimes\ket{\db^\dagger_1\db^\dagger_3}$ & $f^{-}_{A}\,\bar{f}^{+}_{B}$ \\ \addlinespace[2pt]
$\ket{\db^\dagger_1\db^\dagger_3}\otimes\ket{\db^\dagger_1}$ & $\ket{\db^\dagger_1\db^\dagger_3}\otimes\ket{\db^\dagger_1}$ & $\kappa_B^{2}$ \\ \addlinespace[2pt]
$\ket{\db^\dagger_1\db^\dagger_3}\otimes\ket{\db^\dagger_2}$ & $\ket{\vphantom{\db^\dagger_1}\,0}\otimes\ket{\db^\dagger_1\db^\dagger_2\db^\dagger_3}$ & $(1-\rho\delta_2)\kappa_A^2$ \\ \addlinespace[2pt]
$\ket{\db^\dagger_1\db^\dagger_3}\otimes\ket{\db^\dagger_2}$ & $\ket{\db^\dagger_1}\otimes\ket{\db^\dagger_2\db^\dagger_3}$ & $(1-\rho)(1-\delta_2)\kappa_A^2+\rho\,\bar{f}^{-}_{C}\,f^{-}_{A}$ \\ \addlinespace[2pt]
$\ket{\db^\dagger_1\db^\dagger_3}\otimes\ket{\db^\dagger_2}$ & $\ket{\db^\dagger_2}\otimes\ket{\db^\dagger_1\db^\dagger_3}$ & $(1-\delta_2)\kappa_A^2$ \\ \addlinespace[2pt]
$\ket{\db^\dagger_1\db^\dagger_3}\otimes\ket{\db^\dagger_2}$ & $\ket{\db^\dagger_3}\otimes\ket{\db^\dagger_1\db^\dagger_2}$ & $(1-\rho)(1-\delta_2)\kappa_A^2+\rho\,f^{+}_{D}\,\bar{f}^{+}_{D}$ \\ \addlinespace[2pt]
$\ket{\db^\dagger_1\db^\dagger_3}\otimes\ket{\db^\dagger_2}$ & $\ket{\db^\dagger_1\db^\dagger_2}\otimes\ket{\db^\dagger_3}$ & $(1-\delta_2)\kappa_A^2$ \\ \addlinespace[2pt]
$\ket{\db^\dagger_1\db^\dagger_3}\otimes\ket{\db^\dagger_2}$ & $\ket{\db^\dagger_1\db^\dagger_3}\otimes\ket{\db^\dagger_2}$ & $(1-\rho)(1-\delta_2)\kappa_A^2+\rho\,\kappa_B^{2}$ \\ \addlinespace[2pt]
$\ket{\db^\dagger_1\db^\dagger_3}\otimes\ket{\db^\dagger_2}$ & $\ket{\db^\dagger_2\db^\dagger_3}\otimes\ket{\db^\dagger_1}$ & $(1-\delta_2)\kappa_A^2$ \\ \addlinespace[2pt]
$\ket{\db^\dagger_1\db^\dagger_3}\otimes\ket{\db^\dagger_2}$ & $\ket{\db^\dagger_1\db^\dagger_2\db^\dagger_3}\otimes\ket{\vphantom{\db^\dagger_1}\,0}$ & $(1-\rho\delta_2)\kappa_A^2$ \\ \addlinespace[2pt]
$\ket{\db^\dagger_1\db^\dagger_3}\otimes\ket{\db^\dagger_3}$ & $\ket{\db^\dagger_3}\otimes\ket{\db^\dagger_1\db^\dagger_3}$ & $f^{+}_{D}\,\bar{f}^{-}_{D}$ \\ \addlinespace[2pt]
$\ket{\db^\dagger_1\db^\dagger_3}\otimes\ket{\db^\dagger_3}$ & $\ket{\db^\dagger_1\db^\dagger_3}\otimes\ket{\db^\dagger_3}$ & $\kappa_B^{2}$ \\ \addlinespace[2pt]
\bottomrule
\end{tabular}
\end{minipage}
\clearpage
\noindent\textbf{Table 11: Coefficients $\bm{d_\sigma}$ (cont.)}\par\medskip
\noindent\begin{minipage}[t]{0.49\linewidth}
\vspace{0pt}
\begin{tabular}{@{}>{\raggedright\arraybackslash}m{0.28\linewidth}>{\raggedright\arraybackslash}m{0.28\linewidth}>{\raggedright\arraybackslash}m{\dimexpr0.44\linewidth-8pt\relax}@{}}
\toprule Incoming state & Outgoing state & Coefficient $d_\sigma$ \\
\midrule
$\ket{\db^\dagger_2\db^\dagger_3}\otimes\ket{\db^\dagger_1}$ & $\ket{\vphantom{\db^\dagger_1}\,0}\otimes\ket{\db^\dagger_1\db^\dagger_2\db^\dagger_3}$ & $(1-\rho\delta_2)\kappa_A^2$ \\ \addlinespace[2pt]
$\ket{\db^\dagger_2\db^\dagger_3}\otimes\ket{\db^\dagger_1}$ & $\ket{\db^\dagger_1}\otimes\ket{\db^\dagger_2\db^\dagger_3}$ & $(1-\delta_2)\kappa_A^2$ \\ \addlinespace[2pt]
$\ket{\db^\dagger_2\db^\dagger_3}\otimes\ket{\db^\dagger_1}$ & $\ket{\db^\dagger_2}\otimes\ket{\db^\dagger_1\db^\dagger_3}$ & $(1-\rho)(1-\delta_2)\kappa_A^2+\rho\,f^{+}_{B}\,\bar{f}^{+}_{B}$ \\ \addlinespace[2pt]
$\ket{\db^\dagger_2\db^\dagger_3}\otimes\ket{\db^\dagger_1}$ & $\ket{\db^\dagger_3}\otimes\ket{\db^\dagger_1\db^\dagger_2}$ & $(1-\rho)(1-\delta_2)\kappa_A^2+\rho\,f^{-}_{B}\,\bar{f}^{-}_{B}$ \\ \addlinespace[2pt]
$\ket{\db^\dagger_2\db^\dagger_3}\otimes\ket{\db^\dagger_1}$ & $\ket{\db^\dagger_1\db^\dagger_2}\otimes\ket{\db^\dagger_3}$ & $(1-\delta_2)\kappa_A^2$ \\ \addlinespace[2pt]
$\ket{\db^\dagger_2\db^\dagger_3}\otimes\ket{\db^\dagger_1}$ & $\ket{\db^\dagger_1\db^\dagger_3}\otimes\ket{\db^\dagger_2}$ & $(1-\delta_2)\kappa_A^2$ \\ \addlinespace[2pt]
$\ket{\db^\dagger_2\db^\dagger_3}\otimes\ket{\db^\dagger_1}$ & $\ket{\db^\dagger_2\db^\dagger_3}\otimes\ket{\db^\dagger_1}$ & $(1-\rho)(1-\delta_2)\kappa_A^2+\rho\,\kappa_B^{2}$ \\ \addlinespace[2pt]
$\ket{\db^\dagger_2\db^\dagger_3}\otimes\ket{\db^\dagger_1}$ & $\ket{\db^\dagger_1\db^\dagger_2\db^\dagger_3}\otimes\ket{\vphantom{\db^\dagger_1}\,0}$ & $(1-\rho\delta_2)\kappa_A^2$ \\ \addlinespace[2pt]
$\ket{\db^\dagger_2\db^\dagger_3}\otimes\ket{\db^\dagger_2}$ & $\ket{\db^\dagger_2}\otimes\ket{\db^\dagger_2\db^\dagger_3}$ & $f^{+}_{B}\,\bar{f}^{-}_{C}$ \\ \addlinespace[2pt]
$\ket{\db^\dagger_2\db^\dagger_3}\otimes\ket{\db^\dagger_2}$ & $\ket{\db^\dagger_2\db^\dagger_3}\otimes\ket{\db^\dagger_2}$ & $\kappa_B^{2}$ \\ \addlinespace[2pt]
$\ket{\db^\dagger_2\db^\dagger_3}\otimes\ket{\db^\dagger_3}$ & $\ket{\db^\dagger_3}\otimes\ket{\db^\dagger_2\db^\dagger_3}$ & $f^{-}_{B}\,\bar{f}^{+}_{C}$ \\ \addlinespace[2pt]
$\ket{\db^\dagger_2\db^\dagger_3}\otimes\ket{\db^\dagger_3}$ & $\ket{\db^\dagger_2\db^\dagger_3}\otimes\ket{\db^\dagger_3}$ & $\kappa_B^{2}$ \\ \addlinespace[2pt]
$\ket{\db^\dagger_1\db^\dagger_2\db^\dagger_3}\otimes\ket{\vphantom{\db^\dagger_1}\,0}$ & $\ket{\vphantom{\db^\dagger_1}\,0}\otimes\ket{\db^\dagger_1\db^\dagger_2\db^\dagger_3}$ & $\kappa_A^{2}$ \\ \addlinespace[2pt]
$\ket{\db^\dagger_1\db^\dagger_2\db^\dagger_3}\otimes\ket{\vphantom{\db^\dagger_1}\,0}$ & $\ket{\db^\dagger_1}\otimes\ket{\db^\dagger_2\db^\dagger_3}$ & $(1-\rho\delta_2)\kappa_A^2$ \\ \addlinespace[2pt]
$\ket{\db^\dagger_1\db^\dagger_2\db^\dagger_3}\otimes\ket{\vphantom{\db^\dagger_1}\,0}$ & $\ket{\db^\dagger_2}\otimes\ket{\db^\dagger_1\db^\dagger_3}$ & $(1-\rho\delta_2)\kappa_A^2$ \\ \addlinespace[2pt]
$\ket{\db^\dagger_1\db^\dagger_2\db^\dagger_3}\otimes\ket{\vphantom{\db^\dagger_1}\,0}$ & $\ket{\db^\dagger_3}\otimes\ket{\db^\dagger_1\db^\dagger_2}$ & $(1-\rho\delta_2)\kappa_A^2$ \\ \addlinespace[2pt]
$\ket{\db^\dagger_1\db^\dagger_2\db^\dagger_3}\otimes\ket{\vphantom{\db^\dagger_1}\,0}$ & $\ket{\db^\dagger_1\db^\dagger_2}\otimes\ket{\db^\dagger_3}$ & $(1-\rho\delta_2)\kappa_A^2$ \\ \addlinespace[2pt]
$\ket{\db^\dagger_1\db^\dagger_2\db^\dagger_3}\otimes\ket{\vphantom{\db^\dagger_1}\,0}$ & $\ket{\db^\dagger_1\db^\dagger_3}\otimes\ket{\db^\dagger_2}$ & $(1-\rho\delta_2)\kappa_A^2$ \\ \addlinespace[2pt]
$\ket{\db^\dagger_1\db^\dagger_2\db^\dagger_3}\otimes\ket{\vphantom{\db^\dagger_1}\,0}$ & $\ket{\db^\dagger_2\db^\dagger_3}\otimes\ket{\db^\dagger_1}$ & $(1-\rho\delta_2)\kappa_A^2$ \\ \addlinespace[2pt]
$\ket{\db^\dagger_1\db^\dagger_2\db^\dagger_3}\otimes\ket{\vphantom{\db^\dagger_1}\,0}$ & $\ket{\db^\dagger_1\db^\dagger_2\db^\dagger_3}\otimes\ket{\vphantom{\db^\dagger_1}\,0}$ & $\kappa_A^{2}$ \\ \addlinespace[2pt]
$\ket{\db^\dagger_1}\otimes\ket{\db^\dagger_1\db^\dagger_2\db^\dagger_3}$ & $\ket{\db^\dagger_1}\otimes\ket{\db^\dagger_1\db^\dagger_2\db^\dagger_3}$ & $\kappa_C^{2}$ \\ \addlinespace[2pt]
\bottomrule
\end{tabular}
\end{minipage}\hfill\noindent\begin{minipage}[t]{0.49\linewidth}
\vspace{0pt}
\begin{tabular}{@{}>{\raggedright\arraybackslash}m{0.28\linewidth}>{\raggedright\arraybackslash}m{0.28\linewidth}>{\raggedright\arraybackslash}m{\dimexpr0.44\linewidth-8pt\relax}@{}}
\toprule Incoming state & Outgoing state & Coefficient $d_\sigma$ \\
\midrule
$\ket{\db^\dagger_1}\otimes\ket{\db^\dagger_1\db^\dagger_2\db^\dagger_3}$ & $\ket{\db^\dagger_1\db^\dagger_2}\otimes\ket{\db^\dagger_1\db^\dagger_3}$ & $\kappa_C\,\bar{f}^{+}_{A}$ \\ \addlinespace[2pt]
$\ket{\db^\dagger_1}\otimes\ket{\db^\dagger_1\db^\dagger_2\db^\dagger_3}$ & $\ket{\db^\dagger_1\db^\dagger_3}\otimes\ket{\db^\dagger_1\db^\dagger_2}$ & $\kappa_C\,\bar{f}^{-}_{A}$ \\ \addlinespace[2pt]
$\ket{\db^\dagger_1}\otimes\ket{\db^\dagger_1\db^\dagger_2\db^\dagger_3}$ & $\ket{\db^\dagger_1\db^\dagger_2\db^\dagger_3}\otimes\ket{\db^\dagger_1}$ & $\kappa_A\,\kappa_C$ \\ \addlinespace[2pt]
$\ket{\db^\dagger_2}\otimes\ket{\db^\dagger_1\db^\dagger_2\db^\dagger_3}$ & $\ket{\db^\dagger_2}\otimes\ket{\db^\dagger_1\db^\dagger_2\db^\dagger_3}$ & $\kappa_C^{2}$ \\ \addlinespace[2pt]
$\ket{\db^\dagger_2}\otimes\ket{\db^\dagger_1\db^\dagger_2\db^\dagger_3}$ & $\ket{\db^\dagger_1\db^\dagger_2}\otimes\ket{\db^\dagger_2\db^\dagger_3}$ & $\kappa_C\,\bar{f}^{-}_{C}$ \\ \addlinespace[2pt]
$\ket{\db^\dagger_2}\otimes\ket{\db^\dagger_1\db^\dagger_2\db^\dagger_3}$ & $\ket{\db^\dagger_2\db^\dagger_3}\otimes\ket{\db^\dagger_1\db^\dagger_2}$ & $\kappa_C\,\bar{f}^{+}_{D}$ \\ \addlinespace[2pt]
$\ket{\db^\dagger_2}\otimes\ket{\db^\dagger_1\db^\dagger_2\db^\dagger_3}$ & $\ket{\db^\dagger_1\db^\dagger_2\db^\dagger_3}\otimes\ket{\db^\dagger_2}$ & $\kappa_A\,\kappa_C$ \\ \addlinespace[2pt]
$\ket{\db^\dagger_3}\otimes\ket{\db^\dagger_1\db^\dagger_2\db^\dagger_3}$ & $\ket{\db^\dagger_3}\otimes\ket{\db^\dagger_1\db^\dagger_2\db^\dagger_3}$ & $\kappa_C^{2}$ \\ \addlinespace[2pt]
$\ket{\db^\dagger_3}\otimes\ket{\db^\dagger_1\db^\dagger_2\db^\dagger_3}$ & $\ket{\db^\dagger_1\db^\dagger_3}\otimes\ket{\db^\dagger_2\db^\dagger_3}$ & $\kappa_C\,\bar{f}^{+}_{C}$ \\ \addlinespace[2pt]
$\ket{\db^\dagger_3}\otimes\ket{\db^\dagger_1\db^\dagger_2\db^\dagger_3}$ & $\ket{\db^\dagger_2\db^\dagger_3}\otimes\ket{\db^\dagger_1\db^\dagger_3}$ & $\kappa_C\,\bar{f}^{-}_{D}$ \\ \addlinespace[2pt]
$\ket{\db^\dagger_3}\otimes\ket{\db^\dagger_1\db^\dagger_2\db^\dagger_3}$ & $\ket{\db^\dagger_1\db^\dagger_2\db^\dagger_3}\otimes\ket{\db^\dagger_3}$ & $\kappa_A\,\kappa_C$ \\ \addlinespace[2pt]
$\ket{\db^\dagger_1\db^\dagger_2}\otimes\ket{\db^\dagger_1\db^\dagger_2}$ & $\ket{\db^\dagger_1\db^\dagger_2}\otimes\ket{\db^\dagger_1\db^\dagger_2}$ & $\kappa_B^{2}$ \\ \addlinespace[2pt]
$\ket{\db^\dagger_1\db^\dagger_2}\otimes\ket{\db^\dagger_1\db^\dagger_3}$ & $\ket{\db^\dagger_1}\otimes\ket{\db^\dagger_1\db^\dagger_2\db^\dagger_3}$ & $\kappa_C\,f^{+}_{A}$ \\ \addlinespace[2pt]
$\ket{\db^\dagger_1\db^\dagger_2}\otimes\ket{\db^\dagger_1\db^\dagger_3}$ & $\ket{\db^\dagger_1\db^\dagger_2}\otimes\ket{\db^\dagger_1\db^\dagger_3}$ & $\rho\,\kappa_B^2+(1-\rho)f^{+}_{A}\,\bar{f}^{+}_{A}$ \\ \addlinespace[2pt]
$\ket{\db^\dagger_1\db^\dagger_2}\otimes\ket{\db^\dagger_1\db^\dagger_3}$ & $\ket{\db^\dagger_1\db^\dagger_3}\otimes\ket{\db^\dagger_1\db^\dagger_2}$ & $f^{+}_{A}\,\bar{f}^{-}_{A}$ \\ \addlinespace[2pt]
$\ket{\db^\dagger_1\db^\dagger_2}\otimes\ket{\db^\dagger_1\db^\dagger_3}$ & $\ket{\db^\dagger_1\db^\dagger_2\db^\dagger_3}\otimes\ket{\db^\dagger_1}$ & $\kappa_A\,f^{+}_{A}$ \\ \addlinespace[2pt]
$\ket{\db^\dagger_1\db^\dagger_2}\otimes\ket{\db^\dagger_2\db^\dagger_3}$ & $\ket{\db^\dagger_2}\otimes\ket{\db^\dagger_1\db^\dagger_2\db^\dagger_3}$ & $\kappa_C\,f^{-}_{C}$ \\ \addlinespace[2pt]
$\ket{\db^\dagger_1\db^\dagger_2}\otimes\ket{\db^\dagger_2\db^\dagger_3}$ & $\ket{\db^\dagger_1\db^\dagger_2}\otimes\ket{\db^\dagger_2\db^\dagger_3}$ & $\rho\,\kappa_B^2+(1-\rho)f^{-}_{C}\,\bar{f}^{-}_{C}$ \\ \addlinespace[2pt]
$\ket{\db^\dagger_1\db^\dagger_2}\otimes\ket{\db^\dagger_2\db^\dagger_3}$ & $\ket{\db^\dagger_2\db^\dagger_3}\otimes\ket{\db^\dagger_1\db^\dagger_2}$ & $\bar{f}^{+}_{D}\,f^{-}_{C}$ \\ \addlinespace[2pt]
$\ket{\db^\dagger_1\db^\dagger_2}\otimes\ket{\db^\dagger_2\db^\dagger_3}$ & $\ket{\db^\dagger_1\db^\dagger_2\db^\dagger_3}\otimes\ket{\db^\dagger_2}$ & $\kappa_A\,f^{-}_{C}$ \\ \addlinespace[2pt]
$\ket{\db^\dagger_1\db^\dagger_3}\otimes\ket{\db^\dagger_1\db^\dagger_2}$ & $\ket{\db^\dagger_1}\otimes\ket{\db^\dagger_1\db^\dagger_2\db^\dagger_3}$ & $\kappa_C\,f^{-}_{A}$ \\ \addlinespace[2pt]
\bottomrule
\end{tabular}
\end{minipage}
\clearpage
\noindent\textbf{Table 11: Coefficients $\bm{d_\sigma}$ (cont.)}\par\medskip
\noindent\begin{minipage}[t]{0.49\linewidth}
\vspace{0pt}
\begin{tabular}{@{}>{\raggedright\arraybackslash}m{0.28\linewidth}>{\raggedright\arraybackslash}m{0.28\linewidth}>{\raggedright\arraybackslash}m{\dimexpr0.44\linewidth-8pt\relax}@{}}
\toprule Incoming state & Outgoing state & Coefficient $d_\sigma$ \\
\midrule
$\ket{\db^\dagger_1\db^\dagger_3}\otimes\ket{\db^\dagger_1\db^\dagger_2}$ & $\ket{\db^\dagger_1\db^\dagger_2}\otimes\ket{\db^\dagger_1\db^\dagger_3}$ & $\bar{f}^{+}_{A}\,f^{-}_{A}$ \\ \addlinespace[2pt]
$\ket{\db^\dagger_1\db^\dagger_3}\otimes\ket{\db^\dagger_1\db^\dagger_2}$ & $\ket{\db^\dagger_1\db^\dagger_3}\otimes\ket{\db^\dagger_1\db^\dagger_2}$ & $\rho\,\kappa_B^2+(1-\rho)f^{-}_{A}\,\bar{f}^{-}_{A}$ \\ \addlinespace[2pt]
$\ket{\db^\dagger_1\db^\dagger_3}\otimes\ket{\db^\dagger_1\db^\dagger_2}$ & $\ket{\db^\dagger_1\db^\dagger_2\db^\dagger_3}\otimes\ket{\db^\dagger_1}$ & $\kappa_A\,f^{-}_{A}$ \\ \addlinespace[2pt]
$\ket{\db^\dagger_1\db^\dagger_3}\otimes\ket{\db^\dagger_1\db^\dagger_3}$ & $\ket{\db^\dagger_1\db^\dagger_3}\otimes\ket{\db^\dagger_1\db^\dagger_3}$ & $\kappa_B^{2}$ \\ \addlinespace[2pt]
$\ket{\db^\dagger_1\db^\dagger_3}\otimes\ket{\db^\dagger_2\db^\dagger_3}$ & $\ket{\db^\dagger_3}\otimes\ket{\db^\dagger_1\db^\dagger_2\db^\dagger_3}$ & $\kappa_C\,f^{+}_{C}$ \\ \addlinespace[2pt]
$\ket{\db^\dagger_1\db^\dagger_3}\otimes\ket{\db^\dagger_2\db^\dagger_3}$ & $\ket{\db^\dagger_1\db^\dagger_3}\otimes\ket{\db^\dagger_2\db^\dagger_3}$ & $\rho\,\kappa_B^2+(1-\rho)f^{+}_{C}\,\bar{f}^{+}_{C}$ \\ \addlinespace[2pt]
$\ket{\db^\dagger_1\db^\dagger_3}\otimes\ket{\db^\dagger_2\db^\dagger_3}$ & $\ket{\db^\dagger_2\db^\dagger_3}\otimes\ket{\db^\dagger_1\db^\dagger_3}$ & $f^{+}_{C}\,\bar{f}^{-}_{D}$ \\ \addlinespace[2pt]
$\ket{\db^\dagger_1\db^\dagger_3}\otimes\ket{\db^\dagger_2\db^\dagger_3}$ & $\ket{\db^\dagger_1\db^\dagger_2\db^\dagger_3}\otimes\ket{\db^\dagger_3}$ & $\kappa_A\,f^{+}_{C}$ \\ \addlinespace[2pt]
$\ket{\db^\dagger_2\db^\dagger_3}\otimes\ket{\db^\dagger_1\db^\dagger_2}$ & $\ket{\db^\dagger_2}\otimes\ket{\db^\dagger_1\db^\dagger_2\db^\dagger_3}$ & $\kappa_C\,f^{+}_{B}$ \\ \addlinespace[2pt]
$\ket{\db^\dagger_2\db^\dagger_3}\otimes\ket{\db^\dagger_1\db^\dagger_2}$ & $\ket{\db^\dagger_1\db^\dagger_2}\otimes\ket{\db^\dagger_2\db^\dagger_3}$ & $f^{+}_{B}\,\bar{f}^{-}_{C}$ \\ \addlinespace[2pt]
$\ket{\db^\dagger_2\db^\dagger_3}\otimes\ket{\db^\dagger_1\db^\dagger_2}$ & $\ket{\db^\dagger_2\db^\dagger_3}\otimes\ket{\db^\dagger_1\db^\dagger_2}$ & $\rho\,\kappa_B^2+(1-\rho)f^{+}_{B}\,\bar{f}^{+}_{D}$ \\ \addlinespace[2pt]
$\ket{\db^\dagger_2\db^\dagger_3}\otimes\ket{\db^\dagger_1\db^\dagger_2}$ & $\ket{\db^\dagger_1\db^\dagger_2\db^\dagger_3}\otimes\ket{\db^\dagger_2}$ & $\kappa_A\,f^{+}_{B}$ \\ \addlinespace[2pt]
$\ket{\db^\dagger_2\db^\dagger_3}\otimes\ket{\db^\dagger_1\db^\dagger_3}$ & $\ket{\db^\dagger_3}\otimes\ket{\db^\dagger_1\db^\dagger_2\db^\dagger_3}$ & $\kappa_C\,f^{-}_{B}$ \\ \addlinespace[2pt]
$\ket{\db^\dagger_2\db^\dagger_3}\otimes\ket{\db^\dagger_1\db^\dagger_3}$ & $\ket{\db^\dagger_1\db^\dagger_3}\otimes\ket{\db^\dagger_2\db^\dagger_3}$ & $\bar{f}^{+}_{C}\,f^{-}_{B}$ \\ \addlinespace[2pt]
$\ket{\db^\dagger_2\db^\dagger_3}\otimes\ket{\db^\dagger_1\db^\dagger_3}$ & $\ket{\db^\dagger_2\db^\dagger_3}\otimes\ket{\db^\dagger_1\db^\dagger_3}$ & $\rho\,\kappa_B^2+(1-\rho)f^{-}_{B}\,\bar{f}^{-}_{D}$ \\ \addlinespace[2pt]
$\ket{\db^\dagger_2\db^\dagger_3}\otimes\ket{\db^\dagger_1\db^\dagger_3}$ & $\ket{\db^\dagger_1\db^\dagger_2\db^\dagger_3}\otimes\ket{\db^\dagger_3}$ & $\kappa_A\,f^{-}_{B}$ \\ \addlinespace[2pt]
$\ket{\db^\dagger_2\db^\dagger_3}\otimes\ket{\db^\dagger_2\db^\dagger_3}$ & $\ket{\db^\dagger_2\db^\dagger_3}\otimes\ket{\db^\dagger_2\db^\dagger_3}$ & $\kappa_B^{2}$ \\ \addlinespace[2pt]
$\ket{\db^\dagger_1\db^\dagger_2\db^\dagger_3}\otimes\ket{\db^\dagger_1}$ & $\ket{\db^\dagger_1}\otimes\ket{\db^\dagger_1\db^\dagger_2\db^\dagger_3}$ & $\kappa_A\,\kappa_C$ \\ \addlinespace[2pt]
$\ket{\db^\dagger_1\db^\dagger_2\db^\dagger_3}\otimes\ket{\db^\dagger_1}$ & $\ket{\db^\dagger_1\db^\dagger_2}\otimes\ket{\db^\dagger_1\db^\dagger_3}$ & $\kappa_A\,\bar{f}^{+}_{A}$ \\ \addlinespace[2pt]
$\ket{\db^\dagger_1\db^\dagger_2\db^\dagger_3}\otimes\ket{\db^\dagger_1}$ & $\ket{\db^\dagger_1\db^\dagger_3}\otimes\ket{\db^\dagger_1\db^\dagger_2}$ & $\kappa_A\,\bar{f}^{-}_{A}$ \\ \addlinespace[2pt]
$\ket{\db^\dagger_1\db^\dagger_2\db^\dagger_3}\otimes\ket{\db^\dagger_1}$ & $\ket{\db^\dagger_1\db^\dagger_2\db^\dagger_3}\otimes\ket{\db^\dagger_1}$ & $\kappa_A^{2}$ \\ \addlinespace[2pt]
\bottomrule
\end{tabular}
\end{minipage}\hfill\noindent\begin{minipage}[t]{0.49\linewidth}
\vspace{0pt}
\begin{tabular}{@{}>{\raggedright\arraybackslash}m{0.28\linewidth}>{\raggedright\arraybackslash}m{0.28\linewidth}>{\raggedright\arraybackslash}m{\dimexpr0.44\linewidth-8pt\relax}@{}}
\toprule Incoming state & Outgoing state & Coefficient $d_\sigma$ \\
\midrule
$\ket{\db^\dagger_1\db^\dagger_2\db^\dagger_3}\otimes\ket{\db^\dagger_2}$ & $\ket{\db^\dagger_2}\otimes\ket{\db^\dagger_1\db^\dagger_2\db^\dagger_3}$ & $\kappa_A\,\kappa_C$ \\ \addlinespace[2pt]
$\ket{\db^\dagger_1\db^\dagger_2\db^\dagger_3}\otimes\ket{\db^\dagger_2}$ & $\ket{\db^\dagger_1\db^\dagger_2}\otimes\ket{\db^\dagger_2\db^\dagger_3}$ & $\kappa_A\,\bar{f}^{-}_{C}$ \\ \addlinespace[2pt]
$\ket{\db^\dagger_1\db^\dagger_2\db^\dagger_3}\otimes\ket{\db^\dagger_2}$ & $\ket{\db^\dagger_2\db^\dagger_3}\otimes\ket{\db^\dagger_1\db^\dagger_2}$ & $\kappa_A\,\bar{f}^{+}_{D}$ \\ \addlinespace[2pt]
$\ket{\db^\dagger_1\db^\dagger_2\db^\dagger_3}\otimes\ket{\db^\dagger_2}$ & $\ket{\db^\dagger_1\db^\dagger_2\db^\dagger_3}\otimes\ket{\db^\dagger_2}$ & $\kappa_A^{2}$ \\ \addlinespace[2pt]
$\ket{\db^\dagger_1\db^\dagger_2\db^\dagger_3}\otimes\ket{\db^\dagger_3}$ & $\ket{\db^\dagger_3}\otimes\ket{\db^\dagger_1\db^\dagger_2\db^\dagger_3}$ & $\kappa_A\,\kappa_C$ \\ \addlinespace[2pt]
$\ket{\db^\dagger_1\db^\dagger_2\db^\dagger_3}\otimes\ket{\db^\dagger_3}$ & $\ket{\db^\dagger_1\db^\dagger_3}\otimes\ket{\db^\dagger_2\db^\dagger_3}$ & $\kappa_A\,\bar{f}^{+}_{C}$ \\ \addlinespace[2pt]
$\ket{\db^\dagger_1\db^\dagger_2\db^\dagger_3}\otimes\ket{\db^\dagger_3}$ & $\ket{\db^\dagger_2\db^\dagger_3}\otimes\ket{\db^\dagger_1\db^\dagger_3}$ & $\kappa_A\,\bar{f}^{-}_{D}$ \\ \addlinespace[2pt]
$\ket{\db^\dagger_1\db^\dagger_2\db^\dagger_3}\otimes\ket{\db^\dagger_3}$ & $\ket{\db^\dagger_1\db^\dagger_2\db^\dagger_3}\otimes\ket{\db^\dagger_3}$ & $\kappa_A^{2}$ \\ \addlinespace[2pt]
$\ket{\db^\dagger_1\db^\dagger_2}\otimes\ket{\db^\dagger_1\db^\dagger_2\db^\dagger_3}$ & $\ket{\db^\dagger_1\db^\dagger_2}\otimes\ket{\db^\dagger_1\db^\dagger_2\db^\dagger_3}$ & $\kappa_C^{2}$ \\ \addlinespace[2pt]
$\ket{\db^\dagger_1\db^\dagger_2}\otimes\ket{\db^\dagger_1\db^\dagger_2\db^\dagger_3}$ & $\ket{\db^\dagger_1\db^\dagger_2\db^\dagger_3}\otimes\ket{\db^\dagger_1\db^\dagger_2}$ & $\kappa_A\,\kappa_C$ \\ \addlinespace[2pt]
$\ket{\db^\dagger_1\db^\dagger_3}\otimes\ket{\db^\dagger_1\db^\dagger_2\db^\dagger_3}$ & $\ket{\db^\dagger_1\db^\dagger_3}\otimes\ket{\db^\dagger_1\db^\dagger_2\db^\dagger_3}$ & $\kappa_C^{2}$ \\ \addlinespace[2pt]
$\ket{\db^\dagger_1\db^\dagger_3}\otimes\ket{\db^\dagger_1\db^\dagger_2\db^\dagger_3}$ & $\ket{\db^\dagger_1\db^\dagger_2\db^\dagger_3}\otimes\ket{\db^\dagger_1\db^\dagger_3}$ & $\kappa_A\,\kappa_C$ \\ \addlinespace[2pt]
$\ket{\db^\dagger_2\db^\dagger_3}\otimes\ket{\db^\dagger_1\db^\dagger_2\db^\dagger_3}$ & $\ket{\db^\dagger_2\db^\dagger_3}\otimes\ket{\db^\dagger_1\db^\dagger_2\db^\dagger_3}$ & $\kappa_C^{2}$ \\ \addlinespace[2pt]
$\ket{\db^\dagger_2\db^\dagger_3}\otimes\ket{\db^\dagger_1\db^\dagger_2\db^\dagger_3}$ & $\ket{\db^\dagger_1\db^\dagger_2\db^\dagger_3}\otimes\ket{\db^\dagger_2\db^\dagger_3}$ & $\kappa_A\,\kappa_C$ \\ \addlinespace[2pt]
$\ket{\db^\dagger_1\db^\dagger_2\db^\dagger_3}\otimes\ket{\db^\dagger_1\db^\dagger_2}$ & $\ket{\db^\dagger_1\db^\dagger_2}\otimes\ket{\db^\dagger_1\db^\dagger_2\db^\dagger_3}$ & $\kappa_A\,\kappa_C$ \\ \addlinespace[2pt]
$\ket{\db^\dagger_1\db^\dagger_2\db^\dagger_3}\otimes\ket{\db^\dagger_1\db^\dagger_2}$ & $\ket{\db^\dagger_1\db^\dagger_2\db^\dagger_3}\otimes\ket{\db^\dagger_1\db^\dagger_2}$ & $\kappa_A^{2}$ \\ \addlinespace[2pt]
$\ket{\db^\dagger_1\db^\dagger_2\db^\dagger_3}\otimes\ket{\db^\dagger_1\db^\dagger_3}$ & $\ket{\db^\dagger_1\db^\dagger_3}\otimes\ket{\db^\dagger_1\db^\dagger_2\db^\dagger_3}$ & $\kappa_A\,\kappa_C$ \\ \addlinespace[2pt]
$\ket{\db^\dagger_1\db^\dagger_2\db^\dagger_3}\otimes\ket{\db^\dagger_1\db^\dagger_3}$ & $\ket{\db^\dagger_1\db^\dagger_2\db^\dagger_3}\otimes\ket{\db^\dagger_1\db^\dagger_3}$ & $\kappa_A^{2}$ \\ \addlinespace[2pt]
$\ket{\db^\dagger_1\db^\dagger_2\db^\dagger_3}\otimes\ket{\db^\dagger_2\db^\dagger_3}$ & $\ket{\db^\dagger_2\db^\dagger_3}\otimes\ket{\db^\dagger_1\db^\dagger_2\db^\dagger_3}$ & $\kappa_A\,\kappa_C$ \\ \addlinespace[2pt]
$\ket{\db^\dagger_1\db^\dagger_2\db^\dagger_3}\otimes\ket{\db^\dagger_2\db^\dagger_3}$ & $\ket{\db^\dagger_1\db^\dagger_2\db^\dagger_3}\otimes\ket{\db^\dagger_2\db^\dagger_3}$ & $\kappa_A^{2}$ \\ \addlinespace[2pt]
$\ket{\db^\dagger_1\db^\dagger_2\db^\dagger_3}\otimes\ket{\db^\dagger_1\db^\dagger_2\db^\dagger_3}$ & $\ket{\db^\dagger_1\db^\dagger_2\db^\dagger_3}\otimes\ket{\db^\dagger_1\db^\dagger_2\db^\dagger_3}$ & $\kappa_A^{2}$ \\ \addlinespace[2pt]
\bottomrule
\end{tabular}
\end{minipage}
\endgroup
\end{landscape}

\clearpage

\bibliography{biblio.bib}
\bibliographystyle{JHEP}

\end{document}